# SPIN RELAXATION IN PHASE SPACE

YURI P. KALMYKOV[1], WILLIAM T. COFFEY[2], and SERGUEY V. TITOV[3]

[1]*Laboratoire de Mathématiques et Physique (LAMPS, EA 4217), Université de Perpignan Via Domitia, 54, Avenue Paul Alduy, F-66860 Perpignan, France*

[2]*Department of Electronic and Electrical Engineering, Trinity College, Dublin 2, Ireland*

[3]*Kotel'nikov Institute of Radio Engineering and Electronics of the Russian Academy of Sciences, Vvedenskii Square 1, Fryazino, Moscow Region, 141190, Russian Federation*

## CONTENTS

# I. INTRODUCTION

The interpretation of spin relaxation experiments comprises a fundamental problem of condensed phase physics and chemistry yielding a well-defined means of extracting information concerning the spin dynamics in gases, liquids and solids so providing a bridge between microscopic and macroscopic physics. For example, since the number of spins in a sample roughly corresponds to the number of atoms, on an *atomic* level, nuclear magnetic and related spin resonance experiments, etc. examine the time evolution of the individual *elementary* spins [1,2] of nuclei, electrons, muons, etc., while on *mesoscales* the time evolution of magnetic *molecular clusters* (i.e., spins 15-25 $\mu_B$) exhibiting relatively large quantum effects is relevant to the fabrication of molecular magnets [3]. On *nanoscales* single domain ferromagnetic particles (giant spins $10^4$ - $10^5$ $\mu_B$) with a given orientation of the particle moment and permanent magnetization exist. These have spawned very extensive magnetic recording industries, the particles commonly used being on or near the microsize scale. Also on a *nanoscale* level, we have magnetic fluids composed of single domain ferromagnetic particles in a colloidal suspension. Here relaxation experiments detect [4,5] both the Arrhenius or solid state like (Néel) mechanism [6] of relaxation of the magnetization which may overcome via thermal agitation anisotropy potential barriers inside the particle and the Debye orientational relaxation [7,8] due to physical Brownian rotation of the suspended particles in the presence of thermal agitation. Here quantum effects are expected to be much smaller. Finally, on the *bulk macroscopic scale* permanent magnets ($10^{20}$ $\mu_B$) exist, i.e., *multi-domain* systems, where magnetization reversal occurs via the *macroscopic* processes of nucleation, propagation and annihilation of domain walls. Thus, a *well-defined spin number scale* ranging from the bulk macroscopic down to individual atom and spins naturally occurs [9] (see Fig. 1).

On an atomic level spin relaxation experiments in nuclear magnetic or electron spin resonance are usually interpreted via the phenomenological Bloch [10] equations and their later modifications [1,2] pertaining to the relaxation of elementary spins subjected to an external magnetic field and interacting with an environment constituting a heat reservoir at constant temperature *T*. These simple linear equations of motion for the nuclear magnetization were originally proposed on phenomenological grounds. The main assumption is that the effects of the heat bath can be described by two time constants, the so-called relaxation times. They provide a substantially correct quantitative description for liquid samples [1]. Microscopic theories of the relaxation in quantum spin systems have been developed by Bloch [11], Bloembergen, Purcell and Pound [12], and other authors (see, e.g., [13-15]). These microscopic theories have provided general evolution equations for the density matrix operator of the spin system allowing one, in principle, to evaluate all desired observables such as relaxation times, etc.

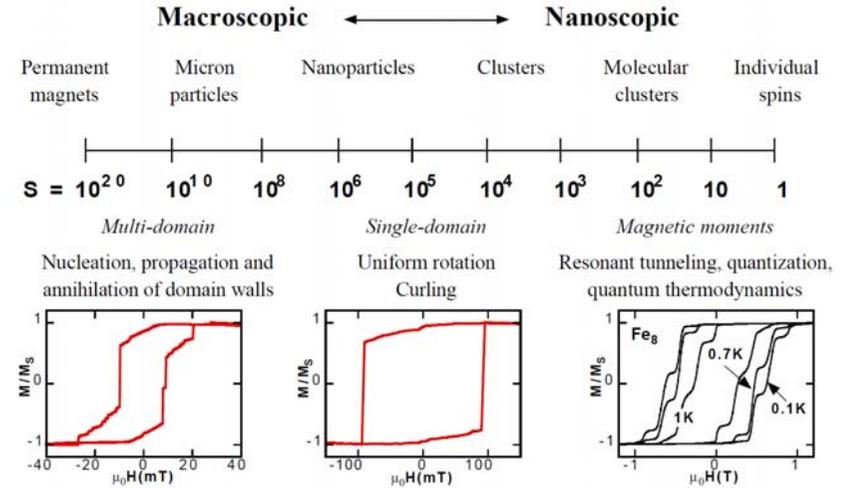

**Figure 1.** (Color on line) The transition from macroscopic to nanoscopic magnets. The hysteresis loops are typical examples of magnetization reversal via nucleation, propagation and annihilation of domain walls (left), via uniform rotation (middle), and via quantum tunneling (right). Reprinted from W. Wernsdorfer, Adv. Chem. Phys. **118**, 99 (2001) with the permission of John Wiley & Sons.

On meso- and nano-scales both the behavior of the hysteresis loop and the reversal time of the magnetization of molecular clusters and nanomagnets are essential for observation of the transition from the macroscopic to the nanoscale level as quantum effects are likely to occur as the spin number decreases (see Fig. 1). In this region, the magnetization in molecular nanomagnets may reverse both due to thermal agitation and quantum tunneling as may be observed in the corresponding hysteresis loop [9]. In contrast, in single domain ferromagnetic nanoparticles (originally encountered in paleomagnetism in the context of past reversals of the Earth's magnetic field, where depending on the volume of the particle, the relaxation time may vary from nanoseconds to millions of years), the magnetization reversal is treated classically by assuming *uniform* rotation of the magnetization vector as conjectured by Néel [16] and Stoner and Wohlfarth [17]. The relaxation time epochs represent the transition from Langevin paramagnetic behavior of nanoparticles (superparamagnetism) with no hysteresis involved via the magnetic after effect stage, where the magnetization reversal time is of the order of the time of a measurement, to stable ferromagnetism. There a given ferromagnetic state corresponds to one of many possible such metastable states in which the magnetization vector is held in a preferred orientation. In the hypothesis of uniform or coherent rotation, the exchange interactions render all atomic spins collinear and the magnitude of the magnetization vector is constant in space. This hypothesis should hold for fine (nanometric) magnetic particles, where domain walls cannot form in the sample because it is energetically unfavorable.

The *static* magnetization properties of single domain particles are usually treated via the method given by Stoner and Wohlfarth [17]. Their procedure simply consists in *minimizing the free energy of the particle*, i.e., the sum of the Zeeman and anisotropy energies with respect to the polar and azimuthal angles specifying the orientation of the magnetization for each value of the applied field. The calculation always leads to hysteresis because in certain field ranges two or more minima exist and thermally agitated transitions between them are neglected. The value of the applied field, at which the magnetization reverses, is called the *switching field* and the *angular dependence of that field with respect to the easy axis of the magnetization* yields the well-known Stoner-Wohlfarth astroids [17,18]. These were originally given for uniaxial shape anisotropy only, which is the anisotropy of the magnetostatic energy of the sample induced by its nonspherical shape. The astroids represent a parametric plot of the parallel versus the perpendicular component of the switching field, which in the uniaxial case is the field that destroys the bistable structure of the free energy. The astroid concept was later generalized to *arbitrary* effective anisotropy by Thiaville [19] including any given magnetocrystalline and surface anisotropy. He proposed a geometrical method for the calculation of the energy of a particle allowing one to determine the switching field for all values of the applied magnetic field yielding the critical switching field surface analogous to the Stoner-Wohlfarth astroids. A knowledge [9] of the switching field surface permits one to determine the effective anisotropy of the particle and all other parameters such as the frequencies of oscillations in the wells of the potential, i.e., the ferromagnetic resonance frequency, etc. We reiterate that these *static* calculations all ignore thermal effects on the switching field, i.e., transitions between the minima of the potential are neglected so that they are strictly only valid at zero temperature.

As far as the magnetization dynamics of fine particles at finite temperatures are concerned, Néel [16] determined the magnetization relaxation time, i.e., the time of reversal of the magnetization of the particle, due to thermal agitation over its internal magnetocrystalline anisotropy barrier from the inverse escape rate over the barriers using transition state theory (TST) [20] as specialized to magnetic moments, viz.,

$$\tau \sim f_0^{-1} e^{\beta \Delta V}. \tag{1}$$

Here $\Delta V$ is the barrier height, $\beta = (kT)^{-1}$, $k$ is Boltzmann's constant, $T$ is the absolute temperature, and $f_0$ is the so-called attempt frequency associated with the frequency of the *gyromagnetic precession* of the magnetization in the effective field of the magnetocrystalline anisotropy. It follows that, by varying the volume or the temperature of the particles, $\tau$ can be made to vary from $10^{-9}$ s to *millions of years* ( $f_0^{-1}$ is often taken as small as $10^{-10}$–$10^{-11}$ s in practice). The presence of the exponential factor in Eq. (1) indicates that, in order to approach the zero remanence (corresponding to thermal equilibrium), a sufficient number of particles (magnetic moments) must be reversed by thermal activation over the energy barrier. Thus, his treatment, given in detail for uniaxial anisotropy only, is confined to a *discrete* set of orientations for the magnetic moment of the particle. Moreover, the equilibrium distribution is all that is ever required since the disturbance to the Boltzmann distribution in the wells of the magnetocrystalline anisotropy potential due to the escape of the magnetization over the barrier is ignored. Besides the overbarrier relaxation process, Bean and Livingston [21] have suggested that in a single domain particle the magnetization may also reverse by *macroscopic quantum tunneling* (macroscopic since a giant spin is involved) through the magnetocrystalline-Zeeman energy potential barrier. In general, the magnetization may reverse by quantum tunneling at *very low temperatures* [22], which may be observed in the behavior of the Stoner-Wohlfarth astroids and associated hysteresis loops. Hence, in order to distinguish tunneling reversal from reversal by thermal agitation, systematic ways of introducing quantum effects into the magnetization reversal of nano-mesoscale magnets, which in general may be restated as quantum effects in parameters characterizing the decay of metastable states in spin systems, are required.

In the context of thermal effects in the magnetization reversal of classical spins, we have mentioned that the original dynamical calculations of Néel for single domain particles utilize classical TST. In the more recent treatment formulated by Brown [23,24] (now known as the Néel-Brown model [6,9]), which explicitly treats the system as a gyromagnetic one and which includes nonequilibrium effects due to the loss of magnetization at the barrier, the time evolution of the magnetization of the particle $\mathbf{M}(t)$ is described by a classical (magnetic) Langevin equation. This is the phenomenological Landau-Lifshitz [25] or Gilbert equation [26,27] augmented by torques due to random white noise magnetic fields $\mathbf{h}(t)$ characterizing the giant spin-bath interaction, viz.,

$$\dot{\mathbf{u}} = \gamma \left[ \mathbf{u} \times \left( \mathbf{H}_{\text{ef}} + \mathbf{h} \right) \right] - \alpha \left[ \mathbf{u} \times \dot{\mathbf{u}} \right]. \tag{2}$$

Here $\mathbf{u} = \mathbf{M}/M_S$ is a unit vector in the direction of $\mathbf{M}$, $M_S$ is the saturation magnetization, assumed constant, $\gamma$ is the gyromagnetic-type constant, $\alpha$ is the dimensionless damping constant, $\mathbf{H}_{\text{ef}} = -\mu_0^{-1} \partial V / \partial \mathbf{M}$ is the effective magnetic field, $\mu_0 = 4\pi \cdot 10^{-7}$ JA$^{-2}$m$^{-1}$ is the permeability of free space in SI units, and $V$ is the free energy per unit volume comprising the *non-separable* Hamiltonian of the magnetic anisotropy and Zeeman energy densities. Thus the only variable is the orientation of $\mathbf{M} = M_S \mathbf{u}$ which is specified by the polar and azimuthal angles $\vartheta$ and $\varphi$ of the spherical polar coordinate system (see Fig. 45 in Appendix D). The stochastic differential equation (2) for the isotropic Brownian motion of the classical spin containing both precessional and aligment terms is then used to derive the Fokker-Planck equation accompanying Eq. (2), governing the time evolution of the distribution function $W(\vartheta,\varphi,t)$ of magnetization orientations on the surface of a sphere of constant radius $M_S$. The relevant Fokker-Planck equation is [5,6]

$$\frac{\partial W}{\partial t} = \mathrm{L}_{\mathrm{FP}} W, \qquad (3)$$

where the Fokker-Planck operator $\mathrm{L}_{\mathrm{FP}}$ again comprising both precessional and alignment terms is defined as

$$\mathrm{L}_{\mathrm{FP}} W = \frac{1}{2\tau_N}\left\{\frac{\beta v}{\alpha}\left(\mathbf{u}\cdot\left[\frac{\partial V}{\partial \mathbf{u}}\times\frac{\partial W}{\partial \mathbf{u}}\right]\right) + \beta v\left(\frac{\partial}{\partial \mathbf{u}}\cdot W\frac{\partial V}{\partial \mathbf{u}}\right) + \Delta W\right\}. \qquad (4)$$

Here $\Delta$ is the Laplacian operator on the surface of the unit sphere, $\tau_N = (\alpha+\alpha^{-1})v\beta\mu_0 M_S/(2\gamma)$ is the characteristic free diffusion time of $\mathbf{M}(t)$, and $v$ is the volume of a typical particle. A detailed discussion of the assumptions made in deriving the Fokker-Planck and Gilbert equations is given elsewhere (e.g., [5,6]; see also Appendix D). Now Brown's model [23,24] rooted in a magnetic Langevin equation, allows one to treat the relaxation processes in classical spin systems (single domain particles, etc.) using powerful computational techniques which have been developed in the theory of the Brownian motion. These include continued fractions, mean first passage times, escape rate theory, etc. [5,6]. For example, by using the Kramers escape rate method [28,29], as ingeniously adapted to magnetic relaxation by Brown [23,24], we can evaluate the reversal time of the magnetization over wide ranges of temperature and damping [6]. Thus, we can compare theoretical predictions with experimental data on superparamagnetic relaxation, which we saw plays a fundamental role in information storage, paleomagnetism, biotechnology, etc.

Many quantum, semiclassical, and classical methods for the description of spin relaxation and resonance already exist. These include the reduced density matrix evolution equation [2,30,31], the stochastic Liouville equation [32,33], the Langevin equation [5,6], besides the phase space (generalized coherent state) [34-40] treatment. The latter comprises the extension of Wigner's phase space representation of the density operator [41-48] (originally developed to obtain quantum corrections to the classical distributions for point particles in the phase space of positions and momenta) to the description of spin systems, see, e.g., Refs. 35-40, 49-51. Furthermore, phase-space representations of quantum mechanical evolution equations provide a formal means of treating quantum effects in dynamical systems transparently linking to the classical representations, facilitating the calculation of quantum corrections to classical distribution functions. These representations generally based on the coherent state representation of the density matrix introduced by Glauber and Sudarshan and widely used in quantum optics [45,46] when applied to spin systems (e.g., [35-40,52-69]) allow one to analyze quantum spin relaxation using [70] a master equation for a *quasiprobability distribution function* $W_S(\vartheta,\varphi,t)$ of spin orientations in a phase (here configuration) space $(\vartheta,\varphi)$. The mapping of the quantum spin dynamics onto $c$-number quasiprobability density evolution equations clearly shows how these equations reduce to the Fokker-Planck equation in the classical limit [40,62]. The phase-space distribution function for spins was originally introduced by Stratonovich [49] for closed systems and further developed both for closed and open spin systems [35-40,52-69]. It is entirely analogous to the Wigner distribution $W(q,p,t)$ in the phase space of positions and momenta $(q,p)$ of a particle [41-47], which is a certain (overlap) Fourier transform corresponding to a quasiprobability representation of the density matrix operator $\hat{\rho}(t)$. However, particular differences arise because of the angular momentum commutation relations. The phase-space distribution function $W_S(\vartheta,\varphi,t)$ of spin orientations in a configuration space, just as the Wigner function $W(q,p,t)$ for the translational motion of a particle, enables one to calculate the expected value of a spin operator $\hat{A}$ in Hilbert space via the *corresponding c-number* (or Weyl) symbol $A(\vartheta,\varphi)$. Hence, quantum mechanical averages may be determined in a classical-like manner. The Wigner representation contains only such features as are common to both quantum and classical statistical mechanics and formally represents quantum mechanics as a statistical theory on classical phase space [42,47]. Therefore, it is especially suitable for the development of semiclassical methods of solution, for example, for the purpose of analysis of two interacting systems, where one is treated quantum mechanically while the other is treated by using classical theory. Here the Wigner representation allows one to easily obtain quantum corrections to the classical results, which is naturally suited to the calculation of quantum corrections to the latter. The formalism is relatively easy to implement because master equations governing the time evolution of phase-space distributions enable powerful computational techniques originally developed for the solution of classical Fokker-Planck equations for the rotational Brownian motion of classical magnetic dipoles (e.g., continued fractions, mean first passage times, etc. [5,71]) to be seamlessly carried over into the quantum domain [62,65,72,73].

Despite the undoubted merits of the phase-space description of spin relaxation [35-40], it appears that the formulation has been relatively underexploited outside the realm of quantum optics [45,46]. Thus, applications to other spin systems are few including spin relaxation problems, etc. This may in part be due to the relatively complex mathematical manipulations, which are involved in comparison with the phase-space formulation for particles with separable and additive Hamiltonians. Therefore, it is now our purpose to provide in the spirit of the *Advances in Chemical Physics* a reasonably comprehensive and didactic account of the phase-space description of spin relaxation and other allied topics. It is also our purpose to show by using Wigner's quasiprobability distribution function formalism, how the existing methods of analysis of the Brownian motion of classical spins may be extended to include quantum effects. Thus, a semiclassical theory of spin relaxation will ensue. Moreover, we shall demonstrate how the results of calculations of relaxation times, dynamic susceptibilities, etc. for various model systems via the phase-space formalism are in complete agreement with those obtained from the solution of quantum master equations for the density matrix. The results are significant particularly because they will elucidate the role played by quantum effects in the various spin relaxation processes in nanomagnets as well as providing a basis for a theory of macroscopic quantum tunneling for which the semiclassical approach is ideally suited.

The remaining contents can be summarized as follows. In Section II, we recall the principal features of both the density matrix in Hilbert space and the phase-space formalisms as applied to spin relaxation phenomena. In particular, we present a detailed derivation of the evolution equation for the density matrix and the corresponding differential-recurrence equation for the statistical moments (average polarization operators) in the Markov approximation. Furthermore, we introduce the Wigner quasiprobability distribution function for particles and illustrate its application to both TST and Brownian motion. Then we describe the basic aspects of the phase-space formalism for spins and derive the master equation for the spin phase-space quasiprobability distributions and the differential-recurrence equation for the statistical moments (average spherical harmonics). Finally, we treat in detail the stationary solutions of phase-space master equations for various model spin systems with arbitrary spin number $S$ (e.g., spins in a uniform external field, uniaxial nanomagnets in longitudinal and transverse fields, biaxial and cubic systems) and we calculate quantum corrections to both switching field curves and to the spin TST reversal time. Section III is devoted to both the derivation of and the solution of phase-space master equations for axially symmetric problems concerning both noninteracting spins in a dc magnetic field and uniaxial nanomagnets subjected to a dc magnetic field. We evaluate characteristic relaxation times and dynamic susceptibility for these spin systems besides treating nonlinear longitudinal relaxation of spins in superimposed ac and dc magnetic fields and investigating quantum effects in dynamic magnetic hysteresis and stochastic resonance. In Section IV, we derive and solve the statistical moment equations for average spherical harmonics for an arbitrary spin Hamiltonian. As a particular example, we calculate the characteristic relaxation times and dynamic susceptibility of a uniaxial nanomagnet subjected to a dc bias field of *arbitrary* orientation with respect to the easy axis. Throughout, we shall compare the results obtained via the phase-space formalism with the corresponding classical limit, $S \to \infty$. In order to facilitate a better understanding of the spin phase-space formalism and its connection to the theory of relaxation of classical spins, the appendixes contain a detailed account of the properties of spin and polarization operators (Appendix A) and spherical harmonics (Appendix B). Appendix C describes in detail the intricate calculations involved in transforming the reduced density matrix evolution equation into a master equation in phase space for a uniaxial nanomagnet. The principal results of the classical theory of the Brownian motion of magnetic moments are summarized in Appendix D, while definitions and derivations of characteristic relaxation times and correlation functions are given in Appendix E.

## II. DENSITY MATRIX AND PHASE-SPACE FORMULATIONS OF RELAXATION PHENOMENA IN SPIN SYSTEMS

### A. Density matrix formulation of spin relaxation and resonance

*1. General equations*

In classical mechanics, we consider the behavior of any given mechanical system of interest as it changes in time from one precisely defined state to another. In statistical mechanics, we have some knowledge of the system but not enough for a complete specification of the precise state. For this purpose, we consider the average behavior of a collection of systems of the same structure as the one of actual interest, but distributed over a range of different possible states. We speak of such a collection as an *ensemble* of systems. Thus, we represent the instantaneous state of a dynamical system of $N$ degrees of freedom by a phase point $(\mathbf{q},\mathbf{p})$ in a $2N$ dimensional phase space $\Gamma$ of all the coordinates $\mathbf{q} = (q_1,...,q_N)$ and momenta $\mathbf{p} = (p_1,...,p_N)$ of the system. The instantaneous state of any system in an ensemble can then be regarded as being specified by the position of a *representative point* in the phase space, and the condition of the ensemble as a whole can be described by a "cloud" of density $f(\mathbf{q},\mathbf{p},t)$ of such representative points, one for each system in the ensemble. The behavior of the ensemble over time can then be associated with the "streaming" motion of the representative points as they describe trajectories in the phase space, in accordance with the laws of classical mechanics. The probability density function $f(\mathbf{q},\mathbf{p},t)$ defined such that $f(\mathbf{q},\mathbf{p},t)d\mathbf{q}d\mathbf{p}$ is the probability at time $t$ that the phase point will be inside a volume element $d\mathbf{q}d\mathbf{p}$ in phase space obeys the classical Liouville equation, namely, [74-76]

$$\frac{\partial f}{\partial t} = -\{H,f\} = -i\mathrm{L}f , \quad (5)$$

where $H = K + U$ is the Hamiltonian (total energy of the system), $K$ and $U$ are, respectively, kinetic and potential energies, $\{H,f\}$ defines the classical Poisson bracket, viz.,

$$\{H,f\} = \sum_{i=1}^{N}\left(\frac{\partial f}{\partial q_i}\frac{\partial H}{\partial p_i} - \frac{\partial f}{\partial p_i}\frac{\partial H}{\partial q_i}\right)$$

and $L = -i\{H, \ \}$ is the classical Liouville operator. From the principle of the *conservation of density in phase space*, it follows [74] that when we consider the rate of change of density in the neighborhood of any selected moving phase point instead of in the neighborhood of a fixed point that the hydrodynamical derivative

$$\frac{df}{dt} = 0, \qquad (6)$$

i.e., the density of the phase points is a constant along a phase space trajectory at all times, viz.

$$f(\mathbf{q}(0), \mathbf{p}(0), t=0) = f(\mathbf{q}(t), \mathbf{p}(t), t). \qquad (7)$$

Now, the evolution equation for a dynamical variable $A(\mathbf{q},\mathbf{p})$ is given by

$$\frac{dA}{dt} = \sum_{i=1}^{N}\left(\frac{\partial A}{\partial q_i}\frac{\partial H}{\partial p_i} - \frac{\partial A}{\partial p_i}\frac{\partial H}{\partial q_i}\right) = \{H, A\} = i L A. \qquad (8)$$

The expectation value of $A(\mathbf{q},\mathbf{p})$ at time *t* is then defined as

$$\langle A \rangle(t) = \int A(\mathbf{q},\mathbf{p}) f(\mathbf{q},\mathbf{p},t) d\mathbf{q} d\mathbf{p} \qquad (9)$$

with the normalization

$$\int f(\mathbf{q},\mathbf{p},t) d\mathbf{q} d\mathbf{p} = 1. \qquad (10)$$

In the quantum mechanics, on the other hand, such a probability density function is replaced by a density matrix $\hat{\rho}$ in Hilbert space to play a role somewhat similar to the density *f* in the classical statistical mechanics [74,77]. For convenience, we briefly recall the basic equations of the density matrix formalism [30,76].

The *state* of a quantum many-particle system characterized by the Hamiltonian operator $\hat{H}$ is represented by a wave function $\Psi$ and the corresponding state vector $|\Psi\rangle$ in Hilbert space. The wave function $\Psi$ satisfies the Schrödinger equation

$$\frac{\partial \Psi}{\partial t} = -\frac{i}{\hbar}\hat{H}\Psi, \qquad (11)$$

where $\hbar = h/(2\pi)$ and *h* is Planck's constant. Now, in the quantum mechanics, a dynamical variable *A* always corresponds to a Hermitian operator $\hat{A}$ and the expectation value of $\hat{A}$ is formally defined as the scalar (inner) product

$$\langle \hat{A} \rangle(t) = \langle \Psi(t) | \hat{A} | \Psi(t) \rangle, \qquad (12)$$

where $\langle \Psi(t) | = |\Psi(t)\rangle^{\dagger}$ and the symbol † denotes the Hermitian conjugate. Equation (12) implies that the wave function $\Psi(t)$ must be normalized to unity, i.e., $\langle \Psi(t) | \Psi(t) \rangle = 1$. Furthermore the wave function may be expanded in the complete orthonormal basis of eigenstates $\{\phi_n\}$ of the Hamiltonian operator $\hat{H}$ as

$$\Psi(t) = \sum_n c_n(t) \phi_n, \qquad (13)$$

where $c_n(t)$ is given by the scalar product $c_n(t) = \langle \phi_n | \Psi(t) \rangle$ with the normalization $\sum_n |c_n(t)|^2 = 1$, which follows from normalization of $\Psi(t)$. In this case, one says that the system is in the *pure* state. Now, the expectation value $\langle \hat{A} \rangle(t) = \langle \Psi(t) | \hat{A} | \Psi(t) \rangle$ of a dynamical variable *A* represented by an operator $\hat{A}$ is given by [76]

$$\langle \hat{A} \rangle(t) = \sum_{n,n'} c_{n'}(t) c_n^*(t) A_{nn'} \equiv \sum_{n,n'} \rho_{n'n}(t) A_{nn'}, \qquad (14)$$

where the coefficients $A_{nn'} = \langle \phi_n | \hat{A} | \phi_{n'} \rangle$ and $\rho_{n'n}(t) = c_{n'}(t) c_n^*(t)$ are, respectively, the matrix elements of the operators $\hat{A}$ and the density operator $\hat{\rho}$ in the orthonormal basis of the eigenstates $\{\phi_n\}$ and the asterisks denote the complex conjugate. For a pure state, the density operator $\hat{\rho}$ can be written as the outer product [76]

$$\begin{aligned}\hat{\rho}(t) &= |\Psi(t)\rangle\langle\Psi(t)| \\ &= \sum_{n,n'} c_{n'}(t) c_n^*(t) |\phi_{n'}\rangle\langle\phi_n|.\end{aligned} \qquad (15)$$

The density matrix operator defined by Eq. (15) is Hermitian, i.e., $\hat{\rho}^{\dagger}(t) = \hat{\rho}(t)$, and idempotent, that is, it satisfies the condition $\hat{\rho}^2(t) = \hat{\rho}(t)$.

The density matrix operator defined above for a quantum system in a *pure* state can also be applied with some modifications to a quantum system in a *mixed* state, i.e., to a system, whose actual state $\psi(t) = \sum_n c_n(t) \phi_n$ is not known completely and only the probabilities $P_n = |c_n(t)|^2$ to be in any of the different states *n* can be evaluated [76]. An example is a system in thermal equilibrium, whose eigenfunctions constitute the orthonormal basis $\{\phi_n\}$ with probabilities $P_n$ to be found in the different states *n* given by the Boltzmann distribution and all microscopic states compatible with these probabilities are assumed to be equally probable; here $\rho_{n'n} = P_n \delta_{n'n}$. The most obvious difference between pure and mixed states is encountered in their diagonal representation, where the pure state density matrix will have only one nonzero element on its diagonal (equal to 1) while the mixed state density matrix must on the other hand have at least two nonzero elements (whose sum is 1) [76].

For our purposes, the most relevant properties of the density matrix operator $\hat{\rho}$, which are valid both for pure and mixed states are the following: (i) $\hat{\rho}$ is a *positive definite operator*, i.e., $\langle \psi(t) | \hat{\rho} | \psi(t) \rangle \geq 0$ yielding the average probability to find the system in the state $|\psi(t)\rangle$; (ii) $\hat{\rho}$ is *Hermitian* and its diagonal elements $\rho_{nn}$ are real, positive and represent the *average* probability that a system chosen at random from the ensemble would be found in the state specified by *n*. Moreover, $\hat{\rho}$ satisfies the normalization condition [74]

$$\text{Tr}(\hat{\rho}) = 1, \tag{16}$$

where $\text{Tr}(\hat{\rho})$ denotes its trace. Furthermore, the quantum mechanical analog of the classical expression Eq. (9) for the mean value of a dynamical variable *A* is

$$\langle \hat{A} \rangle = \text{Tr}(\hat{\rho}\hat{A}). \tag{17}$$

Thus, the integrals over all phase space $\Gamma$ in the classical Eqs. (9) and (10) are replaced by the traces in the quantum Eqs. (16) and (17). Now, the time evolution of the density operator $\hat{\rho}$ is described by the quantum Liouville equation [74,76]

$$\frac{\partial \hat{\rho}}{\partial t} = -\frac{i}{\hbar}\left[\hat{H}, \hat{\rho}\right] = -i\text{L}\hat{\rho}, \tag{18}$$

where $\hat{H}$ is the Hamiltonian operator, the square brackets denote the commutator, viz.,

$$\left[\hat{H}, \hat{\rho}\right] = \hat{H}\hat{\rho} - \hat{\rho}\hat{H},$$

and $\text{L} = \hbar^{-1}\left[\hat{H}, \ \right]$ is the quantum Liouville operator. The formal solution of the operator Eq. (18) is [76,78]

$$\hat{\rho}(t) = e^{-\frac{i}{\hbar}\hat{H}t}\hat{\rho}(0)e^{\frac{i}{\hbar}\hat{H}t}. \tag{19}$$

Finally, the quantum evolution equation for an operator $\hat{A}$ is given by

$$\frac{d\hat{A}}{dt} = \frac{i}{\hbar}\left[\hat{H}, \hat{A}\right] = i\text{L}\hat{A}. \tag{20}$$

Equations (18) and (20) are quantum analogs of Eqs. (5) and (8), respectively. We emphasize [74] that the quantum Liouville Eq. (18) for the density matrix $\hat{\rho}$ has the same mathematical form as the classical Liouville equation (5) for the distribution function *f*.

Now the object of our interest is the dynamics of a spin characterized by the spin number *S* in contact with a thermal bath or environment, which is only of relevance to us insofar as it influences the spin dynamics (a detailed account of properties of relevant spin operators is given in Appendix A). Moreover, it is impossible to follow every variable of the composite spin-environment taken as a whole [79]. Hence, we desire a closed evolution equation for a *reduced density matrix*, where only *relevant* variables, i.e., those belonging to the subspace comprising the spin, appear explicitly. Then by tracing Eq. (18) over the bath variables, we shall ultimately have a master or reduced density-matrix evolution equation [76]. In other words, we can average out the environment of the spin so yielding a *statistical* description of the spin alone [69]. The ideas being entirely analogous to those used to derive an evolution equation for the single-particle distribution function in the classical statistical mechanics.

In order to treat the spin dynamics in a dissipative environment via the density matrix formalism, we loosely follow arguments advanced by Nitzan [76]. Thus, it is supposed that the overall Hamiltonian $\hat{H}$ may be decomposed into

$$\hat{H} = \hat{H}_S + \hat{H}_B + \hat{H}_{SB}, \tag{21}$$

where the operators $\hat{H}_S$, $\hat{H}_{SB}$, and $\hat{H}_B$ are the Hamiltonians of the spin alone, the spin–bath interactions, and the bath, respectively. The time evolution equation for the *overall* system-bath density matrix $\hat{\rho}_{SB}(t)$ of a spin system (*S*) in contact with a heat bath (*B*) may now be written taking account of the Liouville Eq. (18) as

$$\frac{\partial \hat{\rho}_{SB}}{\partial t} = -\frac{i}{\hbar}\left[\hat{H}_S, \hat{\rho}_{SB}\right] - \frac{i}{\hbar}\left[\hat{H}_B, \hat{\rho}_{SB}\right] - \frac{i}{\hbar}\left[\hat{H}_{SB}, \hat{\rho}_{SB}\right]. \tag{22}$$

Now our ultimate objective is to obtain an evolution equation for a *reduced* density matrix, describing the spin relaxation in contact with the thermal bath. Therefore, it is natural to first simplify Eq. (22) by considering the time evolution of the overall density matrix in the *interaction* representation, which is defined for an arbitrary operator $\hat{A}(t)$ as [76]

$$\hat{A}^I(t) = e^{i\hat{H}_0 t/\hbar}\hat{A}(t)e^{-i\hat{H}_0 t/\hbar}, \tag{23}$$

where $\hat{H}_0 = \hat{H}_S + \hat{H}_B$. Next taking the time derivative of $\hat{\rho}_{SB}^I(t)$ yields

$$\frac{\partial \hat{\rho}_{SB}^I}{\partial t} = \frac{i}{\hbar}\left[\hat{H}_0, \hat{\rho}_{SB}^I\right] + e^{i H_0 t/\hbar}\frac{\partial \hat{\rho}_{SB}}{\partial t}e^{-i H_0 t/\hbar}. \tag{24}$$

Therefore, substituting Eq. (22) into Eq. (24), we have the evolution equation for $\hat{\rho}_{SB}^I(t)$ in the interaction representation [76]

$$\frac{\partial \hat{\rho}_{SB}^I}{\partial t} = -\frac{i}{\hbar}\left[\hat{H}_{SB}^I, \hat{\rho}_{SB}^I\right]. \tag{25}$$

Now recall that we desire the effect of the thermal environment on the dynamical behavior of a given spin and that the dynamics of the spin alone is given by $\hat{H}_S$. Therefore, a quite natural viewpoint is to consider the spin–bath interaction Hamiltonian $\hat{H}_{SB}$ as a perturbation of the *bare* dynamics of the spin. In order to formulate this idea, projection operators are used [75,76]. In particular, for problems involving a system interacting with its equilibrium thermal environment, a projection operator $\hat{P}$ is a particular operator projecting the overall system–bath density operator $\hat{\rho}_{SB}$ onto a product of the reduced density operator of the system $\hat{\rho}_S$ and the equilibrium density operator of the bath $\hat{\rho}_B^{eq}$, namely,

$$\hat{P}\hat{\rho}_{SB} = \hat{\rho}_B^{eq}\text{Tr}_B(\hat{\rho}_{SB}) = \hat{\rho}_B^{eq}\hat{\rho}_S. \tag{26}$$

Here the operator $\hat{\rho}_S = \text{Tr}_B(\hat{\rho}_{SB})$ resulting from tracing over the bath variables designates a *reduced density operator* in contrast to the *overall* system-bath operator $\hat{\rho}_{SB}(t)$ and our task is to then find an evolution equation for $\hat{\rho}_S$. In order to achieve this because by definition $\text{Tr}_B(\hat{\rho}_B^{eq}) = 1$, we will have $\hat{P}^2 = \hat{P}$, i.e., $\hat{P}$ is idempotent while the complementary projector is just $\hat{Q} = 1 - \hat{P}$. Moreover, we have the commutation relations

$$\hat{P}\left[\hat{H}_S, \hat{\rho}_{SB}\right] = \left[\hat{H}_S, \hat{P}\hat{\rho}_{SB}\right] = \hat{\rho}_B^{eq}\left[\hat{H}_S, \hat{\rho}_S\right] \tag{27}$$

and

$$\hat{P}\left[\hat{H}_B, \hat{\rho}_{SB}\right] = 0. \tag{28}$$

Now Eqs. (26)-(28) are also valid in the interaction representation implying that we can rewrite Eq. (25) as the two projected equations [76]

$$\frac{\partial}{\partial t}\hat{P}\hat{\rho}_{SB}^I = -\frac{i}{\hbar}\hat{P}\left[\hat{H}_{SB}^I, (\hat{P} + \hat{Q})\hat{\rho}_{SB}^I\right], \tag{29}$$

$$\frac{\partial}{\partial t}\hat{Q}\hat{\rho}_{SB}^I = -\frac{i}{\hbar}\hat{Q}\left[\hat{H}_{SB}^I, (\hat{P} + \hat{Q})\hat{\rho}_{SB}^I\right]. \tag{30}$$

Next, we use the projection Eq. (26) to further rewrite these equations as

$$\frac{\partial}{\partial t}\hat{\rho}_S^I = -\frac{i}{\hbar}\text{Tr}_B\left(\left[\hat{H}_{SB}^I, \hat{\rho}_B^{eq}\hat{\rho}_S^I\right]\right) - \frac{i}{\hbar}\text{Tr}_B\left(\left[\hat{H}_{SB}^I, \hat{Q}\hat{\rho}_{SB}^I\right]\right), \tag{31}$$

$$\frac{\partial}{\partial t}\hat{Q}\hat{\rho}_{SB}^I = -\frac{i}{\hbar}\hat{Q}\left[\hat{H}_{SB}^I, \hat{\rho}_B^{eq}\hat{\rho}_S^I\right] - \frac{i}{\hbar}\hat{Q}\left[\hat{H}_{SB}^I, \hat{Q}\hat{\rho}_{SB}^I\right]. \tag{32}$$

Formal integration of the complementary projection equation Eq. (32) then yields

$$\hat{Q}\hat{\rho}_{SB}^I(t) = \hat{Q}\hat{\rho}_{SB}^I(0) - \frac{i}{\hbar}\int_0^t \hat{Q}\left[\hat{H}_{SB}^I, \hat{\rho}_B^{eq}\hat{\rho}_S^I(t')\right]dt' - \frac{i}{\hbar}\int_0^t \hat{Q}\left[\hat{H}_{SB}^I, \hat{Q}\hat{\rho}_{SB}^I(t')\right]dt'. \tag{33}$$

This integral equation can be solved in iterative fashion by successively inserting $\hat{Q}\hat{\rho}_{SB}^I(t)$ into the integrand in the third term on the right-hand side. Thus we have a perturbation expansion for that quantity as increasing powers of the system-bath Hamiltonian in the interaction representation $\hat{H}_{SB}^I$ by continuing this procedure repeatedly in the usual manner of perturbation theory. Now if we neglect all the highest orders of $\hat{H}_{SB}^I$, we can use a simplified version of Eq. (33), viz.,

$$\hat{Q}\hat{\rho}_{SB}^I(t) = -\frac{i}{\hbar}\int_0^t \hat{Q}\left[\hat{H}_{SB}^I, \hat{\rho}_B^{eq}\hat{\rho}_S^I(t')\right]dt'. \tag{34}$$

In so doing we have disregarded the initial correlation term $\hat{Q}\hat{\rho}_I(0)$ in Eq. (33). The latter approximation is tantamount to assuming that $\hat{\rho}_{SB}^I(0)$ is in P space, that is, *initially the system and the bath are uncorrelated* and also that the bath is in thermal equilibrium, or at least that the effect of initial correlations decays rapidly relative to the timescale on which the system is observed [76]. Inserting the (truncation) *Ansatz* Eq. (34) into Eq. (31) then leads to the *closed equation in the interaction representation*

$$\frac{\partial\hat{\rho}_S^I(t)}{\partial t} = -\frac{i}{\hbar}\text{Tr}_B\left(\left[\hat{H}_{SB}^I, \hat{\rho}_B^{eq}\hat{\rho}_S^I(t)\right]\right) - \frac{1}{\hbar^2}\int_0^t \text{Tr}_B\left(\left[\hat{H}_{SB}^I, \hat{Q}\left[\hat{H}_{SB}^I, \hat{\rho}_B^{eq}\hat{\rho}_S^I(t')\right]\right]\right)dt'. \tag{35}$$

Now returning to the Schrödinger representation, Eq. (35) can be rewritten in that representation as the *reduced* equation

$$\frac{\partial\hat{\rho}_S}{\partial t} = -\frac{i}{\hbar}\left[\hat{H}_S + \hat{\bar{V}}, \hat{\rho}_S\right] + \text{St}(\hat{\rho}_S), \tag{36}$$

where

$$\hat{\bar{V}} = \text{Tr}_B\left(\hat{H}_{SB}\hat{\rho}_B^{eq}\right), \tag{37}$$

$$\text{St}(\hat{\rho}_S) = -\frac{1}{\hbar^2}e^{-i\hat{H}_0 t/\hbar}\text{Tr}_B\left(\left[\hat{H}_{SB}^I, \int_0^t (1-\hat{P})\left[\hat{H}_{SB}^I, \hat{\rho}_B^{eq}\hat{\rho}_S^I(t')\right]dt'\right]\right)e^{i\hat{H}_0 t/\hbar}. \tag{38}$$

By inspection of Eq. (36) it is apparent that the operator $\hat{\bar{V}}$ now has a simple interpretation. It is just a mean potential that corrects the spin Hamiltonian $\hat{H}_S$ for the *average* effect of the bath. Such corrections are very important in chemical physics, for example, in determining solvent shifts of spectral lines. The shifts occur because the average solvent interaction may influence in a different way the energies of the ground and excited states of a solvent molecule. However, it is also clear that such *average* interactions can only affect the system eigenstates and energy levels, and therefore cannot cause relaxation [76]. Thus relaxation phenomena are solely associated with the collision (relaxation) operator kernel $\text{St}(\hat{\rho}_S)$ in Eq. (36). Hence, in addressing spin relaxation, we shall disregard the operator $\hat{\bar{V}}$ or else if unconvinced one could consider a renormalized system Hamiltonian $\hat{H}_S$, which includes the energy shifts associated with the average effect of the bath.

Next we consider the term containing the projection operator $\hat{P}$ involved in the integrand of Eq. (38). Investigation of this term may be simplified if we assume that the interaction $\hat{H}_{SB}^I$ is merely the *product* of system and bath operators, that is, $\hat{H}_{SB}^I = \hat{V}_S^I \hat{V}_B^I$. Considering the relevant term in Eq. (38), we then have [76]

$$\text{Tr}_B\left(\left[\hat{H}_{SB}^I, \hat{P}\left[\hat{H}_{SB}^I, \hat{\rho}_B^{eq}\hat{\rho}_S^I(t')\right]\right]\right)$$
$$= \text{Tr}_B\left(\hat{V}_B^I\hat{\rho}_B^{eq}\right)\text{Tr}_B\left(\hat{V}_B^I\hat{\rho}_B^{eq}\right)\left[\hat{V}_S^I, \left[\hat{V}_S^I, \hat{\rho}_S^I(t')\right]\right] \tag{39}$$
$$= \left(\hat{\bar{V}}^B\right)^2\left[\hat{V}_S^I, \left[\hat{V}_S^I, \hat{\rho}^I(t')\right]\right],$$

where $\hat{\bar{V}}^B = \text{Tr}_B\left(\hat{V}_B^I\hat{\rho}_B^{eq}\right)$. Now the mean value of the bath operator $\hat{\bar{V}}^B$ is zero for isotropic bath operators (however, it gives rise to corrections to bath correlation functions for *anisotropic* baths). Therefore, assuming for simplicity that the bath is isotropic and neglecting this term in Eq. (38), we have the desired reduced density matrix evolution equation describing the spin relaxation in contact with a thermal bath, viz.,

$$\frac{\partial \hat{\rho}_S}{\partial t} + \frac{i}{\hbar}\left[\hat{H}_S, \hat{\rho}_S\right] = \mathrm{St}(\hat{\rho}_S), \qquad (40)$$

where $\mathrm{St}(\hat{\rho}_S)$ is given by with an obvious change of variable in Eq. (38) [76]

$$\mathrm{St}(\hat{\rho}_S) = -\frac{1}{\hbar^2}\int_0^t \mathrm{Tr}_B\left(\left[\hat{H}_{SB}, e^{-i\hat{H}_0\tau/\hbar}\left[\hat{H}_{SB}, \hat{\rho}_B^{eq}\hat{\rho}_S(t-\tau)\right]e^{i\hat{H}_0\tau/\hbar}\right]\right)d\tau. \qquad (41)$$

The reduced nature of the evolution Eq. (40) manifests itself in the *appearance of memory* meaning that the time evolution of $\hat{\rho}_S(t)$ at time $t$ is determined not only by $\hat{\rho}_S(t)$ but also by its past history, namely, $\hat{\rho}_S(t-\tau)$. The nonlocal temporal (or non-Markovian) behavior appears because the system evolves at time $t$ in response to the state of the bath at that time which in turn is determined by the history of the system–bath interaction. Now we saw that the closed Eq. (40) ultimately results from a *low order* perturbation expansion of the system–bath interaction [cf. Eq. (34)], so its validity is expected to be limited to *weak system–bath coupling*. Furthermore, the neglect of initial spin–bath correlations, as expressed by dropping the term $\hat{Q}\hat{\rho}_{SB}^I(0)$ in Eq. (33) constitutes yet another approximation, or, more precisely, a *restriction on the choice of the initial nonequilibrium state*.

*2. Collision kernel in the Markov approximation*

The collision kernel $\mathrm{St}(\hat{\rho})$ in Eq. (41) can be further considerably simplified in the limit, where *the thermal bath dynamics are much faster than those of the spin* [76]. However, in order to implement this condition in the evolution equation (40), we require an *explicit* form for the spin-bath interaction Hamiltonian $\hat{H}_{SB}$. Various models for spin-bath interactions have been discussed in detail, e.g., in Refs. [12-15,76,80-82]. Here, we suppose for simplicity that the spin-bath interaction Hamiltonian operator $\hat{H}_{SB}$ has the rudimentary Zeeman form (see, e.g., [35-37]; in our notation)

$$\hat{H}_{SB} = -\gamma\hbar\left(\hat{\mathbf{S}}\cdot\hat{\mathbf{h}}\right) = -\gamma\hbar\sum_{\mu=-1}^{1}\hat{h}^\mu \hat{S}_\mu, \qquad (42)$$

where the operator $\hat{\mathbf{h}} = (\hat{h}_X, \hat{h}_Y, \hat{h}_Z)$ represents the random noise field characterizing collisional damping due to the bath,

$$\hat{S}_0 = \hat{S}_Z, \quad \hat{S}_{\pm 1} = \mp\frac{1}{\sqrt{2}}\left(\hat{S}_X \pm i\hat{S}_Y\right) \qquad (43)$$

and

$$\hat{h}^0 = \hat{h}_Z, \quad \hat{h}^{\pm 1} = \mp\frac{1}{\sqrt{2}}\left(\hat{h}_X \mp i\hat{h}_Y\right) \qquad (44)$$

are, respectively, the covariant and contravariant spherical components of the spin operator $\hat{\mathbf{S}}$ and field operator $\hat{\mathbf{h}}$ (the properties of $\hat{\mathbf{S}}$ and $\hat{S}_\mu$ are explained in detail in Appendix A). We select the interaction operator $\hat{H}_{SB}$ in the form of Eq. (42) merely because that particular choice renders Eqs. (40) and (41) as a direct quantum analog of the Brownian rotation of a magnetic dipole [5] [cf. Eq. (2)]. Alternative forms of the Hamiltonian $\hat{H}_{SB}$ are discussed, for example, in Refs. [76,80-82]. Then using the properties of the bath correlation functions, viz.,

$$\left\langle \hat{h}^\mu(t_1)\hat{h}^{-\nu}(t_2)\right\rangle = \delta_{\mu\nu}\mathrm{Tr}_B\left(\hat{h}^\mu(t_1)\hat{h}^{-\nu}(t_2)\rho_B^{eq}\right)$$

($\delta_{\mu\nu}$ is Kronecker's delta), the collision kernel Eq. (41) further simplifies to

$$\mathrm{St}(\hat{\rho}_S) = \gamma^2 \sum_{\mu=-1}^{1}\int_0^t\left\{\left\langle \hat{h}^\mu\hat{h}^{-\mu}(\tau)\right\rangle\left[e^{-\frac{i}{\hbar}\hat{H}_S\tau}\hat{S}_{-\mu}\hat{\rho}_S(t-\tau)e^{\frac{i}{\hbar}\hat{H}_S\tau},\hat{S}_\mu\right]\right.$$
$$\left.+\left\langle \hat{h}^\mu(\tau)\hat{h}^{-\mu}\right\rangle\left[\hat{S}_{-\mu},e^{-\frac{i}{\hbar}\hat{H}_S\tau}\hat{\rho}_S(t-\tau)\hat{S}_\mu e^{\frac{i}{\hbar}\hat{H}_S\tau}\right]\right\}d\tau. \qquad (45)$$

Now, in general, the elementary collision kernel Eq. (45) although now vastly simplified still describes the *non-Markovian* behavior of the spin as coupled to the bath, which is determined by the history of the spin-bath interaction. In order to avoid these memory effects, we may use the so-called Markovian limit [76], whereby the bath dynamics are supposed *much faster* than those of the spin. A similar *Ansatz* is essentially made in the classical theory of the Brownian motion [5]. Moreover, because the correlation functions $C_{\mu,-\mu}(\pm\tau)$ characterizing the properties of the equilibrium bath, viz.,

$$C_{\mu,-\mu}(\tau) = (-1)^\mu \gamma^2 \left\langle \hat{h}^\mu(0)\hat{h}^{-\mu}(\tau)\right\rangle \qquad (46)$$

and

$$C_{\mu,-\mu}(-\tau) = (-1)^\mu \gamma^2 \left\langle \hat{h}^\mu(\tau)\hat{h}^{-\mu}(0)\right\rangle \qquad (47)$$

*decay to zero much faster than any characteristic system time scale* of the system [76], we can extend the upper limit of integration in Eq. (45) to infinity yielding the collision kernel in the more appealing form

$$\mathrm{St}(\hat{\rho}_S) = \sum_{\mu=-1}^{1}(-1)^\mu\int_0^\infty\left\{C_{\mu,-\mu}(\tau)\left[e^{-\frac{i}{\hbar}\hat{H}_S\tau}\hat{S}_{-\mu}e^{\frac{i}{\hbar}\hat{H}_S\tau}\hat{\rho}_S(t-\tau),\hat{S}_\mu\right]\right.$$
$$\left.+C_{\mu,-\mu}(-\tau)\left[\hat{S}_{-\mu},\hat{\rho}_S(t-\tau)e^{-\frac{i}{\hbar}\hat{H}_S\tau}\hat{S}_\mu e^{\frac{i}{\hbar}\hat{H}_S\tau}\right]\right\}d\tau, \qquad (48)$$

which is amenable to further simplification. However, tempting as it may seem, the rapidly decaying noise correlation functions $C_{\mu,-\mu}(\tau)$ and $C_{\mu,-\mu}(-\tau)$ in the collision kernel Eq. (48) cannot simply be replaced by Dirac-delta functions $\sim \delta(\tau)$ in order to avoid memory effects, because in the quantum case *the bath dynamics may be slower than the phase oscillations in the system* [76]. For example, in the free or bare system, i.e., without coupling to the reservoir, so that $\mathrm{St}(\hat{\rho}_S) = 0$, the solution of the evolution Eq. (40) is

$$\hat{\rho}_S(t) = e^{-\frac{i}{\hbar}\hat{H}_S t}\hat{\rho}_S(0)e^{\frac{i}{\hbar}\hat{H}_S t}. \tag{49}$$

Hence quantities involving fast phase oscillations of $\hat{\rho}_S(t)$ due to the phase factors $e^{\pm i\hat{H}_S t/\hbar}$ embodied in Eq.(49), or its remaining signature in the presence of system-bath coupling (i.e., when $\mathrm{St}(\hat{\rho}_S) \neq 0$), cannot simply be taken outside the integrals in the collision kernel Eq. (45). Therefore to circumvent this problem we again use the interaction representation of $\hat{\rho}_S(t)$, namely,

$$\hat{\rho}_S(t) = e^{-\frac{i}{\hbar}\hat{H}_S t}\hat{\rho}_S^I(t)e^{\frac{i}{\hbar}\hat{H}_S t}.$$

Thus, the collision operator $\mathrm{St}(\hat{\rho})$ becomes in that representation

$$\mathrm{St}(\hat{\rho}_S) = \sum_{\mu=-1}^{1}(-1)^\mu \int_0^\infty \left\{ C_{\mu,-\mu}(\tau)\left[ e^{-\frac{i}{\hbar}\hat{H}_S \tau}\hat{S}_{-\mu}e^{-\frac{i}{\hbar}\hat{H}_S(t-\tau)}\hat{\rho}_S^I(t-\tau)e^{\frac{i}{\hbar}\hat{H}_S(t-\tau)}e^{\frac{i}{\hbar}\hat{H}_S \tau}, \hat{S}_\mu \right] \right.$$
$$\left. + C_{\mu,-\mu}(-\tau)\left[ \hat{S}_{-\mu}, e^{-\frac{i}{\hbar}\hat{H}_S \tau}e^{-\frac{i}{\hbar}\hat{H}_S(t-\tau)}\hat{\rho}_S^I(t-\tau)e^{\frac{i}{\hbar}\hat{H}_S(t-\tau)}\hat{S}_\mu e^{\frac{i}{\hbar}\hat{H}_S \tau}\right] \right\}d\tau. \tag{50}$$

In order to further simplify Eq. (50), we next assume that the relaxation of the noise correlation functions $C_{\mu,-\mu}(\tau)$ and $C_{\mu,-\mu}(-\tau)$ to zero as $\tau \to \infty$ is *fast* relative to the timescale on which the time shifted density matrix $\hat{\rho}_S^I(t-\tau)$ in the interaction representation changes. We then make the approximation $\hat{\rho}_S^I(t-\tau) \approx \hat{\rho}_S^I(t)$ and then finally return to $\hat{\rho}_S(t)$. Thus, we have our final simplified expression for the collision kernel

$$\mathrm{St}(\hat{\rho}_S) = \sum_{\mu=-1}^{1}(-1)^\mu \int_0^\infty \left\{ C_{\mu,-\mu}(\tau)\left[ e^{-\frac{i}{\hbar}\hat{H}_S \tau}\hat{S}_{-\mu}e^{\frac{i}{\hbar}\hat{H}_S \tau}\hat{\rho}_S(t), \hat{S}_\mu \right] \right.$$
$$\left. + C_{\mu,-\mu}(-\tau)\left[ \hat{S}_{-\mu}, \hat{\rho}_S(t)e^{-\frac{i}{\hbar}\hat{H}_S \tau}\hat{S}_\mu e^{\frac{i}{\hbar}\hat{H}_S \tau}\right] \right\}d\tau, \tag{51}$$

which we note now involves only $\hat{\rho}_S(t)$ and not the shifted expression $\hat{\rho}_S(t-\tau)$.

The approximate Eq. (51) for the collision kernel corresponds to the traditional Redfield equation derived in [14]. This can be proved as follows. We first express $\mathrm{St}(\hat{\rho}_S)$ in the representation defined by the eigenstates $|\alpha\rangle$ of the system Hamiltonian $\hat{H}_S$ and then use the various commutation relations in Eq. (51) as written in terms of their matrix elements, viz.,

$$\left[ e^{-\frac{i}{\hbar}\hat{H}_S \tau}\hat{S}_{-\mu}e^{\frac{i}{\hbar}\hat{H}_S \tau}\hat{\rho}_S(t), \hat{S}_\mu \right]_{\alpha\alpha'} = \sum_{\varepsilon,\varepsilon'}\left[\hat{S}_{-\mu}\right]_{\alpha\varepsilon}\left[\hat{S}_\mu\right]_{\varepsilon'\alpha'}\rho_{\varepsilon\varepsilon'}(t)e^{-\frac{i}{\hbar}(E_\alpha-E_\varepsilon)\tau}$$
$$-\sum_{\varepsilon',\varepsilon}\left[\hat{S}_\mu\right]_{\alpha\varepsilon'}\left[\hat{S}_{-\mu}\right]_{\varepsilon'\varepsilon}\rho_{\varepsilon\alpha'}(t)e^{-\frac{i}{\hbar}(E_{\varepsilon'}-E_\varepsilon)\tau}, \tag{52}$$

$$\left[ \hat{S}_{-\mu}, \hat{\rho}_S(t)e^{-\frac{i}{\hbar}\hat{H}_S \tau}\hat{S}_\mu e^{\frac{i}{\hbar}\hat{H}_S \tau}\right]_{\alpha\alpha'} = \sum_{\varepsilon,\varepsilon'}\left[\hat{S}_{-\mu}\right]_{\alpha\varepsilon}\left[\hat{S}_\mu\right]_{\varepsilon'\alpha'}\rho_{\varepsilon\varepsilon'}(t)e^{\frac{i}{\hbar}(E_{\alpha'}-E_{\varepsilon'})\tau}$$
$$-\sum_{\varepsilon',\varepsilon}\left[\hat{S}_\mu\right]_{\varepsilon'\varepsilon}\left[\hat{S}_{-\mu}\right]_{\varepsilon\alpha'}\rho_{\alpha\varepsilon'}(t)e^{\frac{i}{\hbar}(E_\varepsilon-E_{\varepsilon'})\tau}. \tag{53}$$

Next by introducing the matrix elements $R_{\alpha\alpha'\varepsilon\varepsilon'}$ of the relaxation operator $\mathrm{St}(\hat{\rho}_S)$ via

$$\left[\mathrm{St}(\hat{\rho}_S)\right]_{\alpha\alpha'} = \sum_{\varepsilon\varepsilon'}R_{\alpha\alpha'\varepsilon\varepsilon'}\rho_{\varepsilon\varepsilon'},$$

then from Eq. (51) we have $R_{\alpha\alpha'\varepsilon\varepsilon'}$ in the eigenvector representation of the Hamiltonian $\hat{H}_S$, rendered as the Redfield form

$$R_{\alpha\alpha'\varepsilon\varepsilon'} = \sum_{\mu=-1}^{1}(-1)^\mu \left\{ \left( \tilde{C}_{\mu,-\mu}(\omega_{\alpha\varepsilon}) + \tilde{C}_{-\mu,\mu}^*(\omega_{\alpha'\varepsilon'})\right)\left[\hat{S}_{-\mu}\right]_{\alpha\varepsilon}\left[\hat{S}_\mu\right]_{\varepsilon'\alpha'} \right.$$
$$-\delta_{\varepsilon'\alpha'}\sum_\lambda e^{\beta\hbar\omega_{\varepsilon\lambda}}\tilde{C}_{\mu,-\mu}^*(\omega_{\varepsilon\lambda})\left[\hat{S}_\mu\right]_{\alpha\lambda}\left[\hat{S}_{-\mu}\right]_{\lambda\varepsilon} \tag{54}$$
$$\left. -\delta_{\alpha\varepsilon}\sum_\lambda e^{\beta\hbar\omega_{\varepsilon'\lambda}}\tilde{C}_{-\mu,\mu}(\omega_{\varepsilon'\lambda})\left[\hat{S}_\mu\right]_{\varepsilon'\lambda}\left[\hat{S}_{-\mu}\right]_{\lambda\alpha'} \right\}.$$

Here

$$\omega_{\alpha\beta} = \frac{E_\alpha - E_\beta}{\hbar}$$

are the transition frequencies, $E_\alpha, E_\varepsilon$ are the energy eigenvalues, while the Fourier transforms $\tilde{C}_{\nu,\mu}(\omega)$ of the noise correlation functions (spectral density) are

$$\tilde{C}_{\nu,\mu}(\omega) = \int_0^\infty C_{\nu,\mu}(\tau)e^{-i\omega\tau}d\tau$$

with

$$C_{-\mu,\mu}(-\tau) = C_{\mu,-\mu}^*(\tau) \text{ and } \tilde{C}_{\mu,-\mu}^*(\omega) = e^{-\beta\hbar\omega}\tilde{C}_{\mu,-\mu}(-\omega).$$

The latter relation between the spectral densities can be proved as follows (with $\hbar\omega + E_l = E_k$). We have

$$e^{-\beta\hbar\omega}\tilde{C}_{\mu,-\mu}(-\omega) = Z_B^{-1} e^{-\beta\hbar\omega} \int_0^\infty \sum_{k,l} e^{iE_l\tau/\hbar} h_{lk}^\mu e^{-iE_k\tau/\hbar} h_{kl}^{-\mu} e^{-\beta E_l} e^{i\omega\tau} d\tau$$
$$= Z_B^{-1} \int_0^\infty \sum_{k,l} e^{iE_l\tau/\hbar} \left(h_{lk}^{-\mu}\right)^* e^{-iE_k\tau/\hbar} \left(h_{kl}^\mu\right)^* e^{-\beta E_k} e^{i\omega\tau} d\tau = \tilde{C}_{\mu,-\mu}^*(\omega). \quad (55)$$

The Redfield equation (54) describes the time evolution of the reduced density matrix of a system coupled to an equilibrium bath. The effect of the bath enters via the matrix elements of the "relaxation operator" $R_{\alpha\alpha'\varepsilon\varepsilon'}$. Equation (54) for this relaxation operator is written in the basis of eigenstates of the Hamiltonian $\hat{H}_S$. That equation has been obtained using three approximations: (i) the neglect of initial correlations; (ii) the assumption of weak coupling, and (iii) the assumption of a distinct timescale separation between the (fast) bath variables and the (slow) system (spin) variables, which have been used to get the final Markovian form. The dynamics of the bath or environment merely enters through the bath correlation functions Eqs. (46) and (47). These functions are properties of the *equilibrium* bath only, regardless of whatever system it may be coupled to. We reiterate that although we have assumed that the bath dynamics are *fast on the timescale of the system dynamics*, the details of its dynamics do indeed matter, that is, we could not have simply assumed that the correlation functions are $C_{\mu,-\mu}(\tau) = 2D_\mu \delta(\tau)$ as in the classical case in the white noise approximation for random fields. The reason again is the fact that the *bath dynamics are usually slow relative to the phase oscillations* (which are obviously related to the inverse spacing between energy levels) in the system. We emphasize that solutions of Redfield-like equations for the density matrix even for simple systems (such as harmonic oscillators) do not necessarily satisfy the positivity property [76,83] (for some extreme, nevertheless physically acceptable initial conditions, this *breakdown of positivity* arises from the *omission of memory effects* in the early time evolution [76]). However, in regard to relaxation of spin systems *close to thermal equilibrium*, the Redfield form is always positively defined; moreover, it also has a well-defined classical limit (likewise invariably positively defined). Furthermore, for a wide range of initial distributions (in particular, for those considered here), the solutions of Redfield-like equations *invariably preserve positivity* hence under these conditions their use appears justifiable on physical grounds alone [76,84].

Now the collision kernel also satisfies the equilibrium condition [14]

$$\sum_\alpha \left[\text{St}(\hat{\rho}_{eq})\right]_{\alpha\alpha} = \sum_{\alpha,\varepsilon} R_{\alpha\alpha\varepsilon\varepsilon} \rho_{\varepsilon\varepsilon}^{eq} = 0, \quad (56)$$

where $\rho_{\varepsilon\varepsilon'}^{eq} = \rho_{\varepsilon\varepsilon}^{eq} \delta_{\varepsilon\varepsilon'}$. In order to demonstrate this, we use Eqs. (52) and (53). We have

$$\sum_\alpha \left[\text{St}(\hat{\rho}_{eq})\right]_{\alpha\alpha} = \sum_{\alpha,\varepsilon} \sum_{\mu=-1}^1 (-1)^\mu \int_0^\infty \Big\{ C_{\mu,-\mu}(\tau) \Big(\left[\hat{S}_{-\mu}\right]_{\alpha\varepsilon}\left[\hat{S}_\mu\right]_{\varepsilon\alpha} \rho_{\varepsilon\varepsilon}^{eq}(t) e^{-i(E_\alpha - E_\varepsilon)\tau/\hbar}$$
$$-\left[\hat{S}_\mu\right]_{\alpha\varepsilon}\left[\hat{S}_{-\mu}\right]_{\varepsilon\alpha} \rho_{\alpha\alpha}^{eq}(t) e^{-i(E_\varepsilon - E_\alpha)\tau/\hbar}\Big) + C_{\mu,-\mu}(-\tau)$$
$$\times \Big(\left[\hat{S}_{-\mu}\right]_{\alpha\varepsilon}\left[\hat{S}_\mu\right]_{\varepsilon\alpha} \rho_{\varepsilon\varepsilon}^{eq}(t) e^{i(E_\alpha - E_\varepsilon)\tau/\hbar} - \left[\hat{S}_\mu\right]_{\alpha\varepsilon}\left[\hat{S}_{-\mu}\right]_{\varepsilon\alpha} \rho_{\alpha\alpha}^{eq}(t) e^{i(E_\varepsilon - E_\alpha)\tau/\hbar}\Big)\Big\} d\tau.$$

Then by interchanging indices $\alpha \leftrightarrow \varepsilon$ in the second and fourth sums we have Eq. (56). Next, we can formally define diffusion matrix coefficients $D_\mu$ as

$$D_\mu = \tilde{C}_\mu^{sym}(\omega) \text{sech}(\beta\hbar\omega/2), \quad (57)$$

where the symmetrized spectral density

$$\tilde{C}_\mu^{sym}(\omega) = \frac{1}{2}\left[\tilde{C}_{\mu,-\mu}(-\omega) + \tilde{C}_{\mu,-\mu}^*(\omega)\right]$$

determines the spectrum of the *symmetrized* bath correlation functions [85] which are

$$C_\mu^{sym}(\tau) = \frac{(-1)^\mu}{2} \gamma^2 \left(\langle \hat{h}^\mu \hat{h}^{-\mu}(\tau)\rangle + \langle \hat{h}^{-\mu}(\tau)\hat{h}^\mu\rangle\right). \quad (58)$$

Thus, with this definition, we have for the matrix elements $R_{\alpha\alpha'\varepsilon\varepsilon'}$ in the eigenvector representation

$$R_{\alpha\alpha'\varepsilon\varepsilon'} \approx \sum_{\mu=-1}^1 (-1)^\mu D_\mu \Big\{ e^{\beta\hbar\omega_{\varepsilon\alpha}/2}\left[\hat{S}_{-\mu}\right]_{\alpha\varepsilon}\left[\hat{S}_\mu\right]_{\varepsilon'\alpha'} + e^{\beta\hbar\omega_{\varepsilon'\alpha'}/2}\left[\hat{S}_{-\mu}\right]_{\alpha\varepsilon}\left[\hat{S}_\mu\right]_{\varepsilon'\alpha'}$$
$$-\sum_\lambda \Big(\delta_{\varepsilon'\alpha'} e^{\beta\hbar\omega_{\varepsilon\lambda}/2}\left[\hat{S}_{-\mu}\right]_{\alpha\lambda}\left[\hat{S}_\mu\right]_{\lambda\varepsilon} - \delta_{\alpha\varepsilon} e^{\beta\hbar\omega_{\varepsilon'\lambda}/2}\left[\hat{S}_{-\mu}\right]_{\varepsilon'\lambda}\left[\hat{S}_\mu\right]_{\lambda\alpha'}\Big)\Big\}. \quad (59)$$

In the operator notation, Eq. (59) then yields the collision kernel operator in the *symmetrized* form [86]

$$\text{St}(\hat{\rho}_S) = \sum_{\mu=-1}^1 (-1)^\mu D_\mu \left(\left[\hat{S}_\mu, \hat{\rho}_S e^{\beta\hat{H}_S/2}\hat{S}_{-\mu}e^{-\beta\hat{H}_S/2}\right] + \left[e^{-\beta\hat{H}_S/2}\hat{S}_{-\mu}e^{\beta\hat{H}_S/2}\hat{\rho}_S, \hat{S}_\mu\right]\right). \quad (60)$$

The vital difference between Eqs. (59) and Eq. (60) is that the latter is valid for an *arbitrary state* representation whereas Eq. (59) originates in contrast in the *energy state* representation [76].

Now the collision kernel $\text{St}(\hat{\rho}_S)$ in the operator form Eq. (60) satisfies certain basic requirements constituting important consistency checks. These are as follows:

(i) $\text{St}(\hat{\rho}_S)$ is also Hermitian if $\hat{\rho}_S(t)$ and $\hat{H}_S$ are Hermitian;

(ii) the equilibrium density matrix $\hat{\rho}_{eq} = e^{-\beta\hat{H}_S}/\text{Tr}(e^{-\beta\hat{H}_S})$ renders $\text{St}(\hat{\rho}_{eq}) = 0$ ensuring that the spin system reaches a Boltzmann distribution at equilibrium while

(iii) the matrix exponents $e^{\pm\beta\hat{H}_S/2}$ in the collision kernel operator Eq. (60) ensure detailed balance;

(iv) $\text{St}(\hat{\rho}_S)$ in Eq. (60) is invariant under rotations of the coordinate system;

(v) the Hubbard model [86] embodied in Eq. (60) can be generalized to time-dependent Hamiltonians (see the next subsection); finally,

(vi) spin density matrix evolution Eq. (40) with $\text{St}(\hat{\rho}_S)$ in the form of Eq. (60) effectively constitutes the direct analog of the classical Fokker-Planck equation (3) for rotational diffusion of a classical magnetic dipole.

The relaxation operator in the form of Eq. (60) is a reasonable approximation in the high temperature limit [35,36]. Essentially, it follows from the equation of motion of the reduced density matrix in the rotating-wave approximation (familiar in quantum optics, where counter rotating, rapidly oscillating terms, are averaged out [27]) provided the spin – bath interactions are taken in the weak coupling limit and for Ohmic damping. In which case, the correlation time $\tau_c$ characterizing the bath may be regarded as very short so that the stochastic process originating in the bath is Markovian ($\gamma H \tau_c \ll 1$, where $H$ is the averaged amplitude of the random magnetic field). If the approximation rendered by Eq. (60) is invalid (e.g., throughout the very low temperature region), alternative models for the spin-bath interactions should be used [76,80-82,85-89]. Nevertheless, we shall persevere with the model as just described because despite its drawbacks our objective is merely to understand in semiclassical fashion how quantum effects alter the rotational Brownian motion and longitudinal relaxation of a classical giant spin at finite $S$. Moreover, that model as well as providing a qualitative description of the spin relaxation of a variety of systems can also be regarded as the direct analog of the Fokker-Planck equation formalism used by Brown [23,24] and others (e.g., [5,6,27,90-94]) to treat relaxation of classical macrospins (see Appendix D).

We remark that the collision kernel Eq. (60) can be further simplified in the high temperature limit, $\beta \hbar \omega_{\alpha\alpha} \ll 1$, by supposing that the diffusion matrix coefficients $D_\mu$ are *frequency-independent*, i.e.,

$$D_\mu = D_{-\mu} \approx \tilde{C}_\mu^{sym}(0). \tag{61}$$

In the time domain, this approximation corresponds to the representation of the bath correlation functions $C_\mu^{sym}(\tau)$ Eq. (58) as the sum of two delta functions, viz.,

$$C_\mu^{sym}(\tau) \approx D_\mu \left[ \delta(\tau + i\beta\hbar/2) + \delta(\tau - i\beta\hbar/2) \right]. \tag{62}$$

In the classical limit $\hbar \to 0$, Eq. (62) reduces to the familiar result for the classical bath correlation functions in the white noise approximation for random fields, namely, $C_\mu^{cl}(\tau) = 2D_\mu \delta(\tau)$.

### 3. Time-dependent Hamiltonian

The derivation given above assumes that the spin Hamiltonian does not depend explicitly on the time [14]. However, Hubbard [86] also considered the more general case, viz., the time-dependent operator $\hat{H}_S = \hat{H}_S(t)$ and in so doing proposed the corresponding form for the collision kernel (in our notation)

$$\text{St}(\hat{\rho}_S) = \sum_{\mu=-1}^{1} \sum_{r} (-1)^\mu e^{i\omega_r^{-\mu}t} D_\mu(\omega_r^{-\mu}) \left\{ e^{\beta\hbar\omega_r^{-\mu}/2} \left[ \hat{S}_\mu, \hat{\rho}_S \hat{U}^{-1}(t) \hat{S}_{-\mu}^r \hat{U}(t) \right] \right. \\ \left. + e^{-\beta\hbar\omega_r^{-\mu}/2} \left[ \hat{U}^{-1}(t) \hat{S}_{-\mu}^r \hat{U}(t) \hat{\rho}_S, \hat{S}_\mu \right] \right\}, \tag{63}$$

where the $\hat{S}_{\mu'}^r$ are the coefficients in the series expansion of the time-dependent spin operators $\hat{S}_{\mu'}(t) = \hat{U}(t) \hat{S}_{\mu'} \hat{U}^{-1}(t)$, namely,

$$\hat{S}_{\mu'}(t) = \sum_r \hat{S}_{\mu'}^r e^{i\omega_r^{\mu'}t}, \tag{64}$$

where $\omega_r^{\mu'}$ represents a parameter, the operator $\hat{U}(t)$ is defined as

$$\hat{U}(t) = e^{\frac{i}{\hbar} \int_0^t \hat{H}_S(t')dt'}, \tag{65}$$

and $D_\mu(\omega)$ is the correlation function of the bath given by Eq. (57). Then reconverting the result to operator form [see Eq. (64)], we have for the collision kernel

$$\text{St}(\hat{\rho}_S) = \sum_{\mu=-1}^{1} \sum_r (-1)^\mu D_\mu e^{i\omega_r^{-\mu}t}$$
$$\times \left( e^{\beta\hbar\omega_r^{-\mu}/2} \left[ \hat{S}_\mu, \hat{\rho}_S \hat{U}^{-1}(t) \hat{S}_{-\mu}^r U(t) \right] + e^{-\beta\hbar\omega_r^{-\mu}/2} \left[ \hat{U}^{-1}(t) \hat{S}_{-\mu}^r \hat{U}(t) \hat{\rho}_S, \hat{S}_\mu \right] \right)$$
$$= \sum_{\mu=-1}^{1} (-1)^\mu D_\mu \left( \left[ \hat{S}_\mu, \hat{\rho}_S \hat{U}^{-1}(t) \hat{U}(t - i\beta\hbar/2) \hat{S}_{-\mu} \hat{U}^{-1}(t - i\beta\hbar/2) \hat{U}(t) \right] \right. \\ \left. + \left[ \hat{U}^{-1}(t) \hat{U}(t + i\beta\hbar/2) \hat{S}_{-\mu} \hat{U}^{-1}(t + i\beta\hbar/2) \hat{U}(t) \hat{\rho}_S, \hat{S}_\mu \right] \right). \tag{66}$$

Next we consider typical products like $\hat{U}^{-1}(t) \hat{U}(t \pm i\beta\hbar/2)$ given by

$$\hat{U}^{-1}(t) \hat{U}(t \pm i\beta\hbar/2) = e^{\frac{i}{\hbar} \int_t^{t \pm i\beta\hbar/2} \hat{H}_S(t')dt'}. \tag{67}$$

In the high temperature limit, we have for the integral

$$\frac{i}{\hbar} \int_t^{t \pm i\beta\hbar/2} \hat{H}_S(t')dt' \approx \mp \frac{\beta}{2} \hat{H}_S(t). \tag{68}$$

Here we have supposed that the operator $\hat{H}_S(t)$ does not alter significantly during the small time increments $\Delta t \sim \beta\hbar/2 \ll 1$ in Eq. (68). Thus, we can simply take the value of that operator at time $t$ and, consequently, may place it outside the integral. By treating in like manner all other such time-dependent functions in Eq. (66), we have an equation similar to the previous Eq. (60).

Thus, the Hubbard form of the collision kernel Eq. (66) with time dependent Hamiltonian $\hat{H}_S(t)$ simplifies to

$$\text{St}(\hat{\rho}_S) = \sum_{\mu=-1}^{1} (-1)^\mu D_\mu \left( \left[ \hat{S}_\mu, \hat{\rho}_S e^{\beta \hat{H}_S(t)/2} \hat{S}_{-\mu} e^{-\beta \hat{H}_S(t)/2} \right] + \left[ e^{-\beta \hat{H}_S(t)/2} \hat{S}_{-\mu} e^{\beta \hat{H}_S(t)/2} \hat{\rho}_S, \hat{S}_\mu \right] \right). \quad (69)$$

As discussed above the form of the collision kernel given by Eq. (69) corresponds to the high temperature limit and short correlation time of the Markovian approximation.

*4. Method of statistical moments*

One of the most powerful techniques for the solution of the evolution equations governing the relaxation dynamics of classical and quantum spins ultimately is the method of statistical moments [5,30,71,95]. In order to summarize the principal equations of that method for quantum spins (the corresponding classical results are presented in Appendix D), we first recall that the density matrix $\hat{\rho}_S$ of the particles of spin $S$ is represented by a $(2S+1) \times (2S+1)$ square matrix [95]. We also recall that in order to describe spin states of a particle the so-called polarization operators $\hat{T}_{LM}^{(S)}$ are widely used [30,95] (the main properties of $\hat{T}_{LM}^{(S)}$ are explained in Appendix A). Because the polarization operators constitute an orthonormal basis in the space of $(2S+1) \times (2S+1)$ square matrixes, the density matrix $\hat{\rho}_S$ may be expanded into a finite sum of the polarization operator $\hat{T}_{LM}^{(S)}$ as [30,95]

$$\hat{\rho}_S(t) = \sum_{L=0}^{2S} \sum_{M=-L}^{L} a_{L,M}(t) \hat{T}_{LM}^{(S)}, \quad (70)$$

where the expansion coefficients $a_{L,M}(t)$ defined as

$$a_{L,M}(t) = \left\langle \hat{T}_{LM}^{\dagger(S)} \right\rangle(t) = \text{Tr}\left( \hat{\rho}_S(t) \hat{T}_{LM}^{\dagger(S)} \right). \quad (71)$$

are simply the *statistical moments* representing expectation values of the operators $\hat{T}_{LM}^{\dagger(S)}$ in a state described by the density matrix $\hat{\rho}_S$. Equation (71) follows immediately from the orthogonality properties of the operators $\{\hat{T}_{LM}^{(S)}\}$ given by Eq. (A34) in Appendix A.

Now, the *formal* solution of the spin density matrix evolution Eq. (40) with an arbitrary collision kernel $\text{St}(\hat{\rho}_S)$ can be written as the operator series Eq. (70). Therefore to find the statistical moments $\left\langle \hat{T}_{LM}^{\dagger(S)} \right\rangle(t)$, we simply substitute Eq. (70) into Eq. (40). We have

$$\sum_{L=0}^{2S} \sum_{M=-L}^{L} \hat{T}_{LM}^{(S)} \frac{d}{dt} a_{L,M}(t) = \sum_{L=0}^{2S} \sum_{M=-L}^{L} \left( -\frac{i}{\hbar} \left[ \hat{H}_S, \hat{T}_{LM}^{(S)} \right] + \text{St}(\hat{T}_{LM}^{(S)}) \right) a_{L,M}(t), \quad (72)$$

where the collision operator is

$$\text{St}(\hat{T}_{LM}^{(S)}) = \sum_{\mu=-1}^{1} (-1)^\mu D_\mu \left( \left[ \hat{S}_\mu, \hat{T}_{LM}^{(S)} e^{\frac{\beta}{2} \hat{H}_S} \hat{S}_{-\mu} e^{-\frac{\beta}{2} \hat{H}_S} \right] + \left[ e^{-\frac{\beta}{2} \hat{H}_S} \hat{S}_{-\mu} e^{\frac{\beta}{2} \hat{H}_S} \hat{T}_{LM}^{(S)}, \hat{S}_\mu \right] \right).$$

Now the term

$$-\frac{i}{\hbar} \left[ \hat{H}_S, \hat{T}_{LM}^{(S)} \right] + \text{St}(\hat{T}_{LM}^{(S)})$$

in Eq. (72) is itself just a matrix operator and can also be formally expanded in terms of the polarization operator series (see Appendix A), yielding (the expansion coefficients which are as usual defined by the trace of that operator)

$$-\frac{i}{\hbar} \left[ \hat{H}_S, \hat{T}_{LM}^{(S)} \right] + \text{St}(\hat{T}_{LM}^{(S)}) = \sum_{L'=0}^{2S} \sum_{M'=-L}^{L} \text{Tr}\left\{ \left( -\frac{i}{\hbar} \left[ \hat{H}_S, \hat{T}_{LM}^{(S)} \right] + \text{St}(\hat{T}_{LM}^{(S)}) \right) \hat{T}_{L'M'}^{\dagger(S)} \right\} \hat{T}_{L'M'}^{(S)}. \quad (73)$$

Consequently, by substituting the polarization operator expansion Eq. (73) into Eq. (72) and equating (by orthogonality) terms with the same $\hat{T}_{LM}^{(S)}$, we have the formal evolution equations for the statistical moments $a_{L,M}(t)$, viz.,

$$\frac{d}{dt} a_{L,M}(t) = \sum_{L'=0}^{2S} \sum_{M'=-L}^{L} g_{L'M';LM} a_{L',M'}(t), \quad (74)$$

where the $g_{LM;L'M'}$ are formally defined as

$$g_{LM;L'M'} = \text{Tr}\left\{ -\frac{i}{\hbar} \left[ \hat{H}_S, \hat{T}_{LM}^{(S)} \right] \hat{T}_{L'M'}^{\dagger(S)} + \text{St}(\hat{T}_{LM}^{(S)}) \hat{T}_{L'M'}^{\dagger(S)} \right\}. \quad (75)$$

The coefficients $g_{L'M';LM}$ used in Eq. (74) can be obtained from the coefficients $g_{LM;L'M'}$ defined in Eq. (75) by the replacements $L \leftrightarrow L'$ and $M \leftrightarrow M'$. Now Eqs. (74) and (75) are valid for an *arbitrary* collision kernel operator $\text{St}(\hat{\rho}_S)$. In particular cases, e.g., for $\text{St}(\hat{\rho}_S)$ given by the symmetrized Hubbard form Eq. (60), Eq. (75) becomes

$$g_{L,M;L',M'} = \text{Tr}\left\{ \left( -\frac{i}{\hbar} \left[ \hat{H}_S, \hat{T}_{LM}^{(S)} \right] \right. \right.$$
$$\left. \left. + \sum_{\mu=-1}^{1} (-1)^\mu D_\mu \left( \left[ \hat{S}_\mu, \hat{T}_{LM}^{(S)} e^{\frac{\beta}{2} \hat{H}_S} \hat{S}_{-\mu} e^{-\frac{\beta}{2} \hat{H}_S} \right] + \left[ e^{-\frac{\beta}{2} \hat{H}_S} \hat{S}_{-\mu} e^{\frac{\beta}{2} \hat{H}_S} \hat{T}_{LM}^{(S)}, \hat{S}_\mu \right] \right) \right) \hat{T}_{L'M'}^{\dagger(S)} \right\}. \quad (76)$$

Nowadays, the explicit calculation of the coefficients $g_{LM;L'M'}$ for a particular collision operator is best accomplished using MATHEMATICA. In general we will always have a *finite* set of differential-recurrence equations for the averages $\left\langle \hat{T}_{LM}^{\dagger(S)} \right\rangle(t)$ defined by Eq. (71), viz.,

$$\frac{d}{dt} \left\langle \hat{T}_{LM}^{\dagger(S)} \right\rangle(t) = \sum_{L'=0}^{2S} \sum_{M'=-L}^{L} g_{LM;L'M'} \left\langle \hat{T}_{L'M'}^{\dagger(S)} \right\rangle(t), \quad (77)$$

or because $\hat{T}_{LM}^{\dagger(S)} = (-1)^M \hat{T}_{L-M}^{(S)}$ we also have

$$\frac{d}{dt} \left\langle \hat{T}_{LM}^{(S)} \right\rangle(t) = \sum_{L',M'} (-1)^{M'-M} g_{L'-M';L-M} \left\langle \hat{T}_{L'M'}^{(S)} \right\rangle(t). \quad (78)$$

The differential-recurrence Eqs. (77) and (78) constitute the *general* solution of the problem in the operator representation for *arbitrary spin* Hamiltonians. Explicit forms of the coefficients $g_{LM;L'M'}$ resulting from the polarization operator expansion of the Liouville and collision terms in the density matrix evolution equation, e.g. Eq.(76), for specific cases of the Hamiltonian $\hat{H}_S$ and particular examples of solutions of the differential-recurrence equations Eqs. (77) and (78) will be given later. Moreover, Eqs. (77) and (78) are very useful for the purpose of obtaining differential recurrence relations for phase-space observables as we demonstrate in Sections II.C and III.A.

Equation (77) may be formally solved using standard matrix inversion techniques. This is accomplished by first noting that according to Eq. (77), the behavior of any selected average of the polarization operators $a_{L',M'}(t) = \langle \hat{T}_{L'M'}^{\dagger(S)} \rangle(t)$ say is coupled to that of all the others so forming a *finite* hierarchy of differential-recurrence equations because the index $L$ ranges only between 0 and $2S$. Now the solution of such a multi-term recurrence relation may always be obtained [5] by rewriting it as a first-order linear matrix differential equation with constant coefficients. This is accomplished by first defining a supercolumn vector $\mathbf{C}(t)$ such that

$$\mathbf{C}(t) = \begin{pmatrix} \mathbf{c}_1(t) \\ \mathbf{c}_2(t) \\ \vdots \\ \mathbf{c}_{2S}(t) \end{pmatrix}, \quad \mathbf{c}_L(t) = \begin{pmatrix} a_{L,-L}(t) \\ a_{L,-L+1}(t) \\ \vdots \\ a_{L,L}(t) \end{pmatrix}, \tag{79}$$

whereupon the evolution Eq. (77) of the average polarization operators becomes the linear homogeneous matrix differential equation

$$\frac{d}{dt}\mathbf{C}(t) + \mathbf{X}\mathbf{C}(t) = 0, \tag{80}$$

where $\mathbf{X}$ is the $4S(S+1) \times 4S(S+1)$ transition supermatrix with matrix elements

$$(\mathbf{X})_{L,L'} = \mathbf{G}_{L,L'}^S, \tag{81}$$

and

$$\left[\mathbf{G}_{L,L'}^S\right]_{M,M'} = -g_{L'M';LM}. \tag{82}$$

Here we have utilized the evolution equation for $a_{0,0}(t)$, namely, $\partial_t a_{0,0}(t) = 0$ with the trivial solution $a_{0,0}(t) = const$. The formal solution of Eq. (80) for the desired column vector is [96]

$$\mathbf{C}(t) = e^{-\mathbf{X}t}\mathbf{C}(0) = \mathbf{U}e^{-\mathbf{\Lambda}t}\mathbf{U}^{-1}\mathbf{C}(0), \tag{83}$$

where $\mathbf{\Lambda} = \mathbf{U}^{-1}\mathbf{X}\mathbf{U}$ is a diagonal matrix with elements composed of all the eigenvalues $\lambda_k$ of the transition matrix $\mathbf{X}$ and $\mathbf{U}$ is a right eigenvector matrix composed of all the eigenvectors of $\mathbf{X}$. Now having calculated $\mathbf{C}(t)$ from the matrix Eq. (83), we have all the statistical moments

$$a_{L,M}(t) = \langle \hat{T}_{LM}^{\dagger(S)} \rangle(t) = (-1)^M \langle \hat{T}_{L-M}^{(S)} \rangle(t).$$

Next, recalling that the spin operators $\hat{S}_X$, $\hat{S}_Y$, and $\hat{S}_Z$ may always be expressed in terms of the $\hat{T}_{10}^{(S)}$ and $\hat{T}_{1\pm1}^{(S)}$ via Eq. (A21) of Appendix A, we can evaluate the average components of the spin operators in terms of the average polarization operators, viz.,

$$\langle \hat{S}_X \rangle(t) = a\left[\langle \hat{T}_{1-1}^{(S)} \rangle(t) - \langle \hat{T}_{11}^{(S)} \rangle(t)\right], \tag{84}$$

$$\langle \hat{S}_Y \rangle(t) = ia\left[\langle \hat{T}_{1-1}^{(S)} \rangle(t) + \langle \hat{T}_{11}^{(S)} \rangle(t)\right], \tag{85}$$

$$\langle \hat{S}_Z \rangle(t) = \sqrt{2}a\langle \hat{T}_{10}^{(S)} \rangle(t), \tag{86}$$

where

$$a = \sqrt{\frac{S(S+1)(2S+1)}{6}}.$$

We remark that the differential-recurrence Eq. (77) can also be solved by the matrix continued fraction method [5,71] (see also Appendix D).

The density-matrix method which we have illustrated via the evolution equation for the polarization operators has hitherto constituted the usual approach to the treatment of spin relaxation and resonance phenomena. However, relaxation and resonance of spins interacting with a bath can also be treated in classical-like fashion via the relevant quasiprobability distribution function $W_S(\vartheta,\varphi,t)$ of spin orientations in a phase space (here configuration space) $(\vartheta,\varphi)$; $\vartheta$ and $\varphi$ are the polar and azimuthal angles (see, e.g., [35-40,52-70]). The quasiprobability distribution function $W_S(\vartheta,\varphi,t)$ is considered in the following sections.

### B. Quasiprobability distribution functions for particles

In order to introduce the concept of a quasiprobability distribution function we recall that, in general, a classical dynamical system of one degree of freedom may be described by a phase space probability distribution function $W(q,p,t)$ yielding the probability $W(q,p,t)dpdq$ that the system is in a volume element $dqdp$ centered around the phase space point $(q,p)$ of coordinate $q$ and momentum $p$. However, in the quantum mechanical description of a dynamical system (because of the uncertainty principle) the phase space coordinates $(q,p)$ cannot take definite values simultaneously. Therefore, the concept of such a function does not exist for a quantum system because the idea of a *sharp* phase point and a collection or ensemble of such sharp points has of itself no meaning. Nevertheless, it is still possible to utilize certain mathematical constructs called *quasiprobability distributions*, closely resembling the classical phase-space distribution functions. Such quasiprobability distributions have proven [41-47] to be very useful in a variety of physical applications as they provide fruitful insights into the connection between classical and quantum mechanics allowing one to express quantum mechanical averages in a form which is very similar to that of classical averages. Thus, they are ideally suited to the study of the quantum-classical correspondence. Furthermore, they provide a useful tool for introducing quantum corrections to classical models of dissipation such as many body collisions, Brownian motion, escape rate theory, etc. (see, e.g., Refs. [47,87,97-100]) as well as the connection between decoherence and the quantum to classical transition [101]. The first of these quasiprobability distributions was introduced by Wigner [41] in 1932 in order to study in semiclassical fashion quantum corrections to the Maxwell-Boltzmann distribution of classical statistical mechanics which *inter alia* elucidated the role played by tunneling effects at high temperatures in reaction rate theory [97-100]. In principle, the Wigner distribution function was meant to be a reformulation, using the concept of a quasiprobability distribution in *phase space*, of Schrödinger's wave mechanics, which describes quantum states in *configuration space*.

Before proceeding, we recall the properties of the coordinate and momentum operators $\hat{q}$ and $\hat{p} = -i\hbar \partial / \partial q$, which are used to describe a quantum-mechanical system in Hilbert space with one degree of freedom [78]:

(i) the operators $\hat{q}$ and $\hat{p}$ are *non-commutative*, i.e.,

$$[\hat{q}, \hat{p}] = \hat{q}\hat{p} - \hat{p}\hat{q} = i\hbar \ ; \quad (87)$$

(ii) the *eigenvectors* $|q\rangle$ and $|p\rangle$ of the operators $\hat{q}$ and $\hat{p}$ obey the conditions:

$$\hat{q}|q\rangle = q|q\rangle, \quad (88)$$

$$\hat{p}|p\rangle = p|p\rangle, \quad (89)$$

$$\langle q'|q''\rangle = \delta(q' - q''), \quad (90)$$

$$\langle p'|p''\rangle = \delta(p' - p''), \quad (91)$$

$$\langle q|p\rangle = \frac{1}{\sqrt{2\pi\hbar}} e^{iqp/\hbar}, \quad (92)$$

$$\langle p|q\rangle = \frac{1}{\sqrt{2\pi\hbar}} e^{-iqp/\hbar}, \quad (93)$$

$$\int |q\rangle\langle q| dq = \hat{I}, \quad (94)$$

$$\int |p\rangle\langle p| dp = \hat{I}, \quad (95)$$

where $\hat{I}$ is the unity operator and $\delta$ is the Dirac-delta function;

(iii) the wave functions in the *coordinate* and *momentum* representations are defined via a state vector $|\psi(t)\rangle$ in Hilbert space as

$$\psi(q,t) = \langle q|\psi(t)\rangle, \quad (96)$$

$$\psi(p,t) = \langle p|\psi(t)\rangle; \quad (97)$$

and, finally,

(iv) the probability densities of the coordinate and momentum are given by

$$W(q,t) = |\langle q|\psi(t)\rangle|^2, \quad (98)$$

$$W(p,t) = |\langle p|\psi(t)\rangle|^2. \quad (99)$$

Now for purposes of exposition we consider following Puri [51] a one dimensional dynamical system described classically by a phase-space distribution function $W(q,p,t)$. Then the classical statistical average of any function $A(q,p)$ is by definition

$$\langle A(q,p)\rangle_{cl} = \int W(q,p,t) A(q,p) dqdp \ . \quad (100)$$

By analogy the quantum mechanical description of a system is contained in its Hilbert space density operator $\hat{\rho}(t)$, which determines the quantum statistical average of any function $A(\hat{q},\hat{p})$ of the coordinate and momentum operators $\hat{q}$ and $\hat{p}$ by the relation

$$\langle A(\hat{q},\hat{p})\rangle = \text{Tr}[\hat{\rho}(t) A(\hat{q},\hat{p})]. \quad (101)$$

Now the classical distribution function $W(q,p,t)$ may always be expressed in terms of the averages of a *complete* set of functions of $q$ and $p$. This fact suggests that we may be able to construct a quantum analog of the classical distribution function by expressing the latter distribution in terms of the average of a suitably chosen complete set of functions and then relabeling those classical averages as quantum mechanical ones. In order to explore this possibility, we rewrite a typical classical distribution in integral form as [51]

$$W(q,p,t) = \int_{-\infty}^{\infty}\int_{-\infty}^{\infty} \delta(p-p')\delta(q-q')W(q',p',t)dq'dp'$$
$$= \frac{1}{4\pi^2}\int_{-\infty}^{\infty}\int_{-\infty}^{\infty}\int_{-\infty}^{\infty}\int_{-\infty}^{\infty} e^{ik(q-q')}e^{il(p-p')}W(q',p',t)dq'dp'dkdl, \quad (102)$$

which by definition is

$$W(q,p,t) = \frac{1}{4\pi^2}\int_{-\infty}^{\infty}\int_{-\infty}^{\infty} e^{ikq}e^{ilp}\langle e^{-ikq}e^{-ilp}\rangle_{cl} dkdl \ . \quad (103)$$

Thus, we have expressed the classical distribution function $W(q,p,t)$ in terms of the *average* of a complete set of functions of $e^{-ikq}$ and $e^{-ilp}$ or in statistical terminology as the inverse Fourier transform of the characteristic function $\langle e^{-ikq}e^{-ilp}\rangle_{cl}$. To construct the quantum analog of $W(q,p,t)$, we must first replace the *classical dynamical variables* $(q,p)$ by the Hilbert space quantum operators $(\hat{q}, \hat{p})$ and then replace the classical average $\langle e^{-ikq}e^{-ilp}\rangle_{cl}$ or characteristic function in the integrand of Eq. (103) by the quantum average as defined by Eq. (101). Thus, we have a quantum analog of the classical phase-space distribution function

$$W^{\mathrm{T}}(q,p,t) = \frac{1}{4\pi^2}\int_{-\infty}^{\infty}\int_{-\infty}^{\infty} e^{ikq}e^{ilp}\left\langle \mathrm{T}(e^{-ik\hat{q}}, e^{-il\hat{p}})\right\rangle_{qm} dk dl \quad (104)$$

called a *quasiprobability distribution*. In writing Eq. (104) we formally wrote the product $e^{-ikq}e^{-ilp}$ as an ordering operator $\mathrm{T}(e^{-ik\hat{q}}, e^{-il\hat{p}})$ because owing to the *non-commutativity* of the Hilbert space operators $\hat{q}$ and $\hat{p}$ several *different* quantum operator forms exist which all may be considered as *quantum analogs* of the *unique* classical product $e^{-ikq}e^{-ilp}$. The different operator forms must then be formally described by an ordering operator $\mathrm{T}(e^{-ik\hat{q}}, e^{-il\hat{p}})$. Hence, *different choices* of $\mathrm{T}(e^{-ik\hat{q}}, e^{-il\hat{p}})$ ultimately lead to *different quantum phase-space distribution functions* $W^{\mathrm{T}}(q,p,t)$ [47].

The name *quasiprobability* is used to emphasize that such a distribution represents a merely mathematical construct and is not in itself a true phase-space distribution function as no such joint distribution function can exist for a quantum system [47]. Now in order to systematically investigate various operator orderings $\mathrm{T}(e^{-ik\hat{q}}, e^{-il\hat{p}})$, it is convenient to express the quantum operators $\hat{q}, \hat{p}$ in terms of the creation and annihilation operators $\hat{a}$ and $\hat{a}^\dagger$ defined as [40]

$$\hat{a} = \frac{1}{\sqrt{2\hbar}}(\hat{q}+i\hat{p}), \quad \hat{a}^\dagger = \frac{1}{\sqrt{2\hbar}}(\hat{q}-i\hat{p}) \quad (105)$$

so that

$$\hat{q} = \sqrt{\frac{\hbar}{2}}(\hat{a}+\hat{a}^\dagger), \quad \hat{p} = i\sqrt{\frac{\hbar}{2}}(\hat{a}^\dagger - \hat{a}) \quad (106)$$

and then consider the operator $\mathrm{T}(e^{i\xi\hat{a}}, e^{i\xi^*\hat{a}^\dagger})$ which may be rewritten as [51]

$$\mathrm{T}(e^{i\xi\hat{a}}, e^{i\xi^*\hat{a}^\dagger}) = e^{s|\xi|^2/2} e^{i\xi\hat{a}+i\xi^*\hat{a}^\dagger}. \quad (107)$$

Here $s$ is a complex number. The ordering for $s=0$ usually called the *Weyl ordering* or the *symmetric ordering*, corresponds to the *Wigner function*. The quasiprobability distribution function for $s=1$ is known as the *P-function* $W^P(q,p,t)$ while that for $s=-1$ is known as the *Q-function* $W^Q(q,p,t)$ [40,51,54]. The *Q*- and *P*-representations, and the Wigner function representation arise from three different aims. In the *Q*-representation, it is desired to create a quasiprobability density for the quantum system by using the *diagonal matrix elements of the density operator* for this purpose. On the other hand, the *P*-representation arises from the desire to represent the density operator as an *ensemble of coherent states* [78]. Finally, the Wigner function yields a *joint* quasi-distribution for *canonically conjugate* variables $p$ and $q$, which in many respects resembles a classical probability distribution. All of these phase space representations have their particular advantages and disadvantages [102]. At first, we restrict ourselves to the Wigner function for particles. For spins, the *Q*- and *P*-representations as well as the Wigner function representation will be considered in detail in the next section.

*1. The Wigner distribution function for particles*

By directly proceeding from a statistical view point, Moyal [103] has shown how the Wigner function follows naturally from inversion of a characteristic function $M(\mu,v)$ for canonically conjugate variables such as position and momentum in a state described by a state vector $|\psi(t)\rangle$ in the Hilbert space. Moyal [103], in view of his strong background in statistics, starts by defining a *characteristic function operator* [cf. Eq. (103) et seq.] (coordinates and momenta are once more used instead of the creation and annihilation operators), namely,

$$\mathbf{M}_{\hat{q},\hat{p}}(\mu,v) = e^{i\mu\hat{q}+iv\hat{p}}. \quad (108)$$

The characteristic function $M(\mu,v)$ in a state $|\psi(t)\rangle$ in Hilbert space is then given by definition as the scalar product

$$M(\mu,v) = \left\langle e^{i\mu\hat{q}+iv\hat{p}}\right\rangle = \mathrm{Tr}\left(\hat{\rho} e^{i\mu\hat{q}+iv\hat{p}}\right)$$
$$= \left\langle\psi(t)|e^{i\mu\hat{q}+iv\hat{p}}|\psi(t)\right\rangle = \int_{-\infty}^{\infty} \psi^*(q,t) e^{i\mu\hat{q}+iv\hat{p}} \psi(q,t) dq, \quad (109)$$

where $\psi(q,t) = \langle q|\psi(t)\rangle$ is the wave function in the coordinate representation and $\hat{q}$ and $\hat{p} = -i\hbar\partial/\partial q$ are the canonically conjugate coordinate and momentum operators. Because the noncommuting operators $\hat{p}$ and $\hat{q}$ satisfy the Baker-Campbell-Hausdorff identity [43]

$$e^{\hat{A}+\hat{B}} = e^{\hat{A}} e^{\hat{B}} e^{-[\hat{A},\hat{B}]/2}, \quad (110)$$

the characteristic function operator $\mathbf{M}_{\hat{q},\hat{p}}(\mu,v)$ then becomes

$$\mathbf{M}_{\hat{q},\hat{p}}(\mu,v) = e^{i\hbar\mu v/2} e^{i\mu\hat{q}} e^{iv\hat{p}} = e^{-i\hbar\mu v/2} e^{iv\hat{p}} e^{i\mu\hat{q}} \quad (111)$$

since we have Eq. (87) for the commutator of the operators $\hat{q}$ and $\hat{p}$. Next, since we may eliminate $\hat{p}$ via [78]

$$e^{iv\hat{p}}\psi(q) = \psi(q) + \hbar v\frac{\partial}{\partial q}\psi(q) + \ldots = \psi(q+v\hbar), \quad (112)$$

we have the simplified expression

$$M(\mu,v) = \int_{-\infty}^{\infty} \psi^*(q,t) e^{i\mu(q+v\hbar/2)} \psi(q+v\hbar,t) dq, \quad (113)$$

which with the replacement $q \to q - v\hbar/2$ yields the characteristic function as the overlap integral

$$M(\mu,v) = \int_{-\infty}^{\infty} \psi^*(q-u,t) e^{i\mu q} \psi(q+u,t) dq . \tag{114}$$

where $u = v\hbar/2$. The phase space quasi-distribution $W(q,p,t)$ is thus by Fourier inversion

$$\begin{aligned} W(q,p,t) &= \frac{1}{4\pi^2} \int_{-\infty}^{\infty}\int_{-\infty}^{\infty} M(\mu,v) e^{-i\mu q - ivp} d\mu dv \\ &= \frac{1}{2\hbar\pi^2} \int_{-\infty}^{\infty}\int_{-\infty}^{\infty}\int_{-\infty}^{\infty} e^{-i\mu(q-x)} \psi^*(x-u,t) e^{-ivp} \psi(x+u,t) du dx d\mu \\ &= \frac{1}{\pi\hbar} \int_{-\infty}^{\infty} \psi^*(q-u,t) e^{-ivp} \psi(q+u,t) du, \end{aligned} \tag{115}$$

which is real but not *everywhere positive* so that it is indeed a quasiprobability distribution. Here we have used the definition

$$\frac{1}{2\pi} \int_{-\infty}^{\infty} e^{-i\mu(q-x)} d\mu = \delta(q-x) . \tag{116}$$

Equation (115) alias the Wigner function $W(q,p,t)$ (which represents the Fourier transform of the overlap function, namely, a type of spatial autocorrelation function of the wave function $\psi$ in coordinate space [77]) holds if $\psi$ evolves according to the Schrödinger equation

$$i\hbar \frac{\partial \psi}{\partial t} = -\frac{\hbar^2}{2m} \frac{\partial^2 \psi}{\partial q^2} + V(q)\psi \tag{117}$$

for a particle of mass $m$ moving in a potential $V(q)$. Now integration of $W(q,p,t)$ with respect to the momentum $p$ yields

$$\int_{-\infty}^{\infty} W(q,p,t) dp = |\psi(q,t)|^2 , \tag{118}$$

i.e., the correct quantum mechanical probability for the coordinate $q$. Conversely, integration of $W$ with respect to $q$, yields the correct quantum mechanical probability for the momentum $p$, viz.,

$$\int_{-\infty}^{\infty} W(q,p,t) dq = |\Phi(p)|^2 , \tag{119}$$

where $\Phi(p)$ is the wave function of the momentum given by

$$\Phi(p) = \int_{-\infty}^{\infty} \psi(q,t) e^{-ipq} dq .$$

It follows from Eq. (118) that if the wave function $\psi(q,t)$ is normalized to unity then

$$\int_{-\infty}^{\infty}\int_{-\infty}^{\infty} W(q,p,t) dq dp = 1 . \tag{120}$$

Now by recalling [77] the definition of the density matrix operator $\hat{\rho}$ for a pure state, viz.,

$$\hat{\rho}(q_1,q_2,t) = \langle q_1 | \psi(t) \rangle \langle \psi(t) | q_2 \rangle = \psi(q_1,t) \psi^*(q_2,t),$$

and introducing the replacements $q = (q_1 + q_2)/2$ and $y/2 = u = (q_1 - q_2)/2$, one can finally write Eq. (115) as [77]

$$\begin{aligned} W(q,p,t) &= \frac{1}{2\pi\hbar} \int_{-\infty}^{\infty} \hat{\rho}(q+y/2, q-y/2, t) e^{-ipy/\hbar} dy \\ &= \frac{1}{2\pi\hbar} \int_{-\infty}^{\infty} \hat{\rho}(q-y/2, q+y/2, t) e^{ipy/\hbar} dy. \end{aligned} \tag{121}$$

Moreover, the inverse transformation is given by the Weyl transform yielding the mapping from the phase space back to operators in Hilbert space [47], viz.,

$$\hat{\rho}(t) = \frac{\hbar}{2\pi} \int_{-\infty}^{\infty}\int_{-\infty}^{\infty}\int_{-\infty}^{\infty}\int_{-\infty}^{\infty} W(q,p,t) e^{ix(\hat{q}-q)+iy(\hat{p}-p)} dq dp dx dy . \tag{122}$$

We remark in passing that in contrast to Eq. (115), the definition of the Wigner function $W(q,p,t)$ via the density operator $\hat{\rho}(t)$, Eq. (121), is valid both for *pure* and mixed states. The only difference is in the definition of $\hat{\rho}(t)$. For a pure state $|\psi(t)\rangle$, the density operator is given by $\hat{\rho}(t) = |\psi(t)\rangle\langle\psi(t)|$, while for a mixed state, the density operator is defined as [76]

$$\hat{\rho}(t) = \sum_n P_n |\psi_n(t)\rangle\langle\psi_n(t)|, \tag{123}$$

where $P_n$ is the probability of each state $|\psi_n(t)\rangle$. The Wigner function of a *mixed* state $W(q,p,t)$ is then given by the following relation [76]

$$W(q,p,t) = \sum_n P_n W_n(q,p,t), \tag{124}$$

where $W_n(q,p,t)$ is the Wigner function for the state $|\psi_n(t)\rangle$. In particular, the definition, Eq. (124), applies to a canonical ensemble of particles at temperature $T$, where $P_n = e^{-\beta E_n}/Z$ is the probability to find the particle in an energy state $E_n$, and $Z$ is the partition function.

The Wigner function $W(q,p,t)$ exhibits most of the properties of a classical phase-space distribution, including the fact that the expectation value $\langle \hat{A} \rangle(t) = \text{Tr}(\hat{\rho}\hat{A})$ of a quantum operator $\hat{A}$ may be calculated in *classical fashion via the corresponding Weyl symbol* $A(q,p)$ as [47]

$$\langle \hat{A} \rangle(t) = \int_{-\infty}^{\infty}\int_{-\infty}^{\infty} W(q,p,t) A(q,p) dq dp , \tag{125}$$

where

$$A(q,p) = \int_{-\infty}^{\infty} e^{-ipy/\hbar} \langle q+y/2|\hat{A}|q-y/2\rangle dy \tag{126}$$

is the Weyl transform of the operator $\hat{A}$ expressed in the position $q$ basis. Equivalently, the Weyl transform can be expressed in terms of matrix elements of the operator $\hat{A}$ in the momentum $p$ basis as

$$A(q,p) = \int_{-\infty}^{\infty} e^{-iqy/\hbar} \langle p+y/2|\hat{A}|p-y/2\rangle dy. \tag{127}$$

The Weyl transform implies that the trace of the product of two operators $\hat{A}$ and $\hat{B}$ is given by the integral over the phase space $(q,p)$ of the product of their Weyl transforms $A(q,p)$ and $B(q,p)$, viz., [47]

$$\mathrm{Tr}\langle \hat{B}\hat{A}\rangle = \frac{1}{2\pi\hbar}\int_{-\infty}^{\infty}\int_{-\infty}^{\infty} B(q,p)A(q,p)dq\,dp. \tag{128}$$

Equations (121) and (128) (for $\hat{B} \equiv \hat{\rho}$) yield immediately Eq. (125).

We may now determine from the overlap Fourier transform Eq. (115) the partial differential equation governing the time evolution of the Wigner function. Taking the derivative of Eq. (115), we have

$$\frac{\partial W}{\partial t} = \frac{1}{\pi\hbar}\int_{-\infty}^{\infty}\left[\psi^*(q+u,t)\frac{\partial \psi(q-u,t)}{\partial t} + \frac{\partial \psi^*(q+u,t)}{\partial t}\psi(q-u,t)\right]e^{2ipu/\hbar}du. \tag{129}$$

Now since the wave function $\psi$ in coordinate space evolves according to the Schrödinger equation, Eq. (117), substitution of that equation into Eq. (129) gives

$$\begin{aligned}\frac{\partial W}{\partial t} &= -\frac{i}{2m\pi}\int_{-\infty}^{\infty}\left\{\psi(q-u)\frac{\partial^2\psi^*(q+u)}{\partial u^2} - \psi^*(q+u)\frac{\partial^2\psi(q-u)}{\partial u^2}\right.\\ &\quad\left.-\frac{2m}{\hbar^2}\psi(q-u)\psi^*(q+u)[V(q+u)-V(q-u)]\right\}e^{2ipu/\hbar}du \\ &= -\frac{p}{m}\frac{\partial W}{\partial q} + \frac{i}{\pi\hbar^2}\int_{-\infty}^{\infty}\psi(q-u)\psi^*(q+u)[V(q+u)-V(q-u)]e^{2ipu/\hbar}du.\end{aligned} \tag{130}$$

Here we have utilized the derivative property

$$\frac{\partial^2\psi(q\pm u)}{\partial q^2} = \frac{\partial^2\psi(q\pm u)}{\partial u^2}$$

and have performed one partial integration with respect to $u$ in the terms which do not involve the potential functions $V(q\pm u)$. Equation (130) is identical to the classical Liouville equation in the force-free case $V(q)=0$ and concurs with that given by Hillery et al. [44]. We now expand $V(q+u)-V(q-u)$ in Eq. (130) in a Taylor series about the point $q$ yielding

$$V(q+u)-V(q-u) = 2\sum_{n=1}^{\infty}\frac{u^{2n-1}}{(2n-1)!}\frac{\partial^{2n-1}V}{\partial q^{2n-1}}. \tag{131}$$

Substituting Eq. (131) into Eq. (130) we have

$$\frac{\partial W}{\partial t} + \frac{p}{m}\frac{\partial W}{\partial q} = \frac{2i}{\pi\hbar}\int_{-\infty}^{\infty}\sum_{n=1}^{\infty}\frac{u^{2n-1}}{(2n-1)!}\frac{\partial^{2n-1}V}{\partial q^{2n-1}}\psi^*(q+u)\psi(q-u)e^{2ipu/\hbar}du. \tag{132}$$

Next because of the relation

$$u^{2n-1}e^{2ipu/\hbar} = \left(\frac{\hbar}{2i}\right)^{2n-1}\frac{\partial^{2n-1}}{\partial p^{2n-1}}e^{2ipu/\hbar}, \tag{133}$$

we have

$$\begin{aligned}&\frac{2i}{\pi\hbar}\sum_{n=1}^{\infty}\frac{1}{(2n-1)!}\left(\frac{\hbar}{2i}\right)^{2n-1}\frac{\partial^{2n-1}V}{\partial q^{2n-1}}\frac{\partial^{2n-1}}{\partial p^{2n-1}}\int_{-\infty}^{\infty}\psi^*(q+u)\psi(q-u)e^{2ipu/\hbar}du \\ &= \sum_{n=1}^{\infty}\frac{1}{(2n-1)!}\left(\frac{\hbar}{2i}\right)^{2n-2}\frac{\partial^{2n-1}V}{\partial q^{2n-1}}\frac{\partial^{2n-1}W}{\partial p^{2n-1}} \\ &= -\frac{\partial V}{\partial q}\frac{\partial W}{\partial p} - \sum_{r=1}^{\infty}\frac{(i\hbar/2)^{2r}}{(2r+1)!}\frac{\partial^{2r+1}V}{\partial q^{2r+1}}\frac{\partial^{2r+1}W}{\partial p^{2r+1}}.\end{aligned} \tag{134}$$

Thus, we have the time evolution equation for the Wigner distribution function $W(q,p,t)$, viz.,

$$\frac{\partial W}{\partial t} + \hat{M}_W W = 0, \tag{135}$$

where the operator $\hat{M}_W$ is defined as

$$\hat{M}_W W = \frac{p}{m}\frac{\partial W}{\partial q} - \frac{\partial V}{\partial q}\frac{\partial W}{\partial p} - \sum_{r=1}^{\infty}\frac{(i\hbar/2)^{2r}}{(2r+1)!}\frac{\partial^{2r+1}V}{\partial q^{2r+1}}\frac{\partial^{2r+1}W}{\partial p^{2r+1}}. \tag{136}$$

Equation (135) often known as the Wigner-Moyal equation [77] is a quantum analog of the classical Liouville equation. Equations (135) and (136) also hold if the system is in a mixed state represented by a density matrix $\hat{\rho}$.

Now Wigner [41] originally calculated quantum correction terms to the classical stationary distribution functions for a system with $n$ degrees of freedom. However, for illustrative purposes, we only consider a system with $n = 1$. As an example, we determine the stationary (equilibrium) solution of Eq. (135) for an assembly of noninteracting particles each of mass $m$ moving in a potential $V(q)$ at temperature $T$. Each particle is characterized by the energy

$$\varepsilon(q,p) = \frac{p^2}{2m} + V(q). \tag{137}$$

Following Wigner [41], we develop the stationary distribution function $W_{eq}(q,p)$ in a power series

$$W_{eq}(q,p) = W_0(q,p) + \hbar^2 W_2(q,p) + \hbar^4 W_4(q,p) + ..., \tag{138}$$

where $W_0(q,p) = e^{-\beta\varepsilon(q,p)}$ is the (unnormalized) Maxwell-Boltzmann distribution. By substituting Eq. (138) into Eq. (135), the function $W_2(q,p)$ and $W_4(q,p)$ are easily evaluated. Thus, we have [48]

$$W_{eq}(q,p) = e^{-\beta\varepsilon(q,p)}\left\{1 + \Lambda\left(\frac{\beta p^2}{m}V^{(2)} - 3V^{(2)} + \beta\left(V^{(1)}\right)^2\right)\right.$$
$$+ 3\Lambda^2\left[\frac{\beta^2}{6}\left(V^{(1)}\right)^4 + \frac{5}{2}\left(V^{(2)}\right)^2 - \frac{9\beta}{5}V^{(2)}\left(V^{(1)}\right)^2\right.$$
$$+ 2V^{(3)}V^{(1)} - \frac{3}{2\beta}V^{(4)} + \frac{p^2}{m}\left(V^{(4)} - \frac{2\beta}{5}V^{(3)}V^{(1)}\right)$$
$$\left.\left.+ \frac{1}{3}V^{(2)}\left(\beta V^{(1)}\right)^2 - \frac{9\beta}{5}\left(V^{(2)}\right)^2 + \frac{p^4}{m^2}\left(\frac{1}{6}\left(\beta V^{(2)}\right)^2 - \frac{\beta V^{(4)}}{10}\right)\right]\right\} + \ldots,$$
(139)

where $\Lambda = \hbar^2\beta^2/(24m)$ is a characteristic quantum parameter and $V^{(m)} \equiv d^mV/dq^m$. The above equations are written explicitly to $o(\hbar^4)$. In like manner, higher order quantum correction terms to the Wigner stationary distribution $W_{eq}(q,p)$ may be calculated. Thus, $W_{eq}(q,p)$ can be given, in principle to any desired degree $r$ of $\hbar^{2r}$. In general, quantum effects give rise to *non-Gaussian behavior* of the equilibrium phase-space distribution function $W_{eq}(q,p)$ and it is no longer separable in the position and momentum variables.

Obviously, the calculation of the equilibrium Wigner distribution $W_{eq}(q,p)$ is a tedious task for an arbitrary potential $V(q)$. However, in some cases, $W_{eq}(q,p)$ can be found in closed form (various methods for the calculation of Wigner functions are described, e.g., in Refs. 43, 44, 46, and 47). A famous example is the quantum harmonic oscillator, where the potential is

$$V(q) = \frac{1}{2}m\omega_0^2 q^2 \tag{140}$$

($\omega_0$ is the angular frequency of the oscillator). Here perturbation theory may be avoided because the evolution equation for the Wigner function Eq. (135) for the potential (140) now coincides with the corresponding classical Liouville equation and the unnormalized equilibrium Wigner function $W_{eq}(q,p)$ can be written in the exact Gaussian form [43,44,48,104]

$$W_{eq}(q,p) = \frac{\text{sech}(\beta\hbar\omega_0/2)}{2\pi\hbar}e^{-\frac{q^2}{2\langle q^2\rangle_{eq}} - \frac{p^2}{2\langle p^2\rangle_{eq}}}, \tag{141}$$

where

$$\langle q^2\rangle_{eq} = \frac{\hbar}{2m\omega_0}\coth\frac{\beta\hbar\omega_0}{2} \tag{142}$$

and

$$\langle p^2\rangle_{eq} = \frac{m\hbar\omega_0}{2}\coth\frac{\beta\hbar\omega_0}{2}. \tag{143}$$

The Wigner function $W_{eq}(q,p)$ from Eq. (141) is in fact a *superposition* of the Wigner functions $W_n(q,p)$ for the pure states of a harmonic oscillator, viz., [43,44,48]

$$W_{eq}(q,p) = \sum_{n=0}^{\infty}W_n(q,p)e^{-\beta\hbar\omega_0(n+1/2)}. \tag{144}$$

Here $W_n(q,p)$ and the corresponding eigenfunctions of the harmonic oscillator $\psi_n(q)$ are given by the well-known analytic equations [46]

$$W_n(q,p) = \frac{1}{\pi\hbar}\int_{-\infty}^{\infty}\psi_n(q+u)\psi_n^*(q-u)e^{-2ipu/\hbar}du$$
$$= \frac{(-1)^n}{\pi\hbar}e^{-\frac{p^2+m^2\omega_0^2 q^2}{m\hbar\omega_0}}L_n\left[\frac{2(p^2+m^2\omega_0^2 q^2)}{m\hbar\omega_0}\right], \tag{145}$$

$$\psi_n(q) = \frac{1}{\sqrt{2^n n!}}\left(\frac{m\omega_0}{\pi\hbar}\right)^{1/4}e^{-\frac{m\omega_0 q^2}{2\hbar}}H_n\left(q\sqrt{\frac{m\omega_0}{\hbar}}\right), \tag{146}$$

where $H_n(z)$ and $L_n(z)$ are the Hermite and Laguerre polynomials, respectively [105]. For illustration, the Wigner functions $W_n$ of a harmonic oscillator for the pure states $n = 1, 2, 3$ and 4, Eq. (145), are shown in Fig. 2. In contrast to the behavior of the equilibrium Wigner function $W_{eq}(q,p)$, Eq. (141), the Wigner functions $W_n(q,p)$ of the pure states can take on *negative* values so that it is impossible to interpret the Wigner function $W_n$ as a true probability distribution.

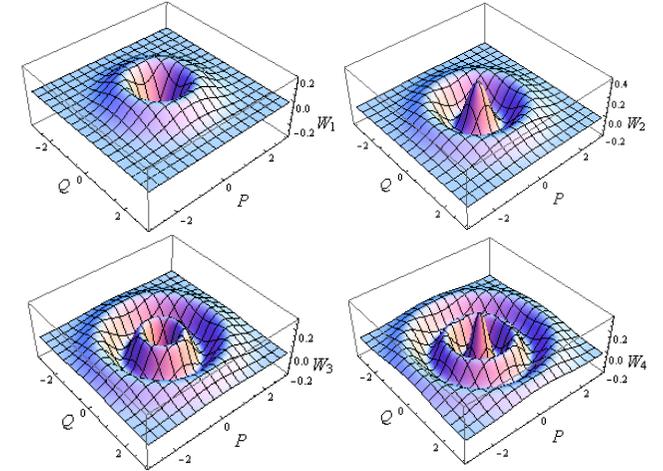

**Figure 2.** (Color on line) 3D plots of the Wigner functions $W_n$ of a harmonic oscillator vs. the normalized coordinate $Q = q\sqrt{m\omega_0/\hbar}$ and momentum $P = p/\sqrt{m\hbar\omega_0}$ for the pure states $n = 1$, 2, 3, and 4.

Equilibrium Wigner functions have also been calculated for various simple quantum systems such as a particle in an infinite square well [104], the Morse oscillator [106,107], the anharmonic quartic oscillator [108,109], the inverted harmonic oscillator [110], etc.

## 2. Application to Transition State Theory

The simplest description of thermally activated escape of a particle with the total energy $\varepsilon(q,p)$ given by Eq. (137) over a potential barrier may be given in terms of transition state theory (TST). In the simplest form of TST, two assumptions are made [5,20]. First thermal equilibrium prevails in the well (for example, through the action of Maxwell's demon who keeps replenishing the particles at the source) so that the metastable state is represented by a canonical equilibrium distribution (unlike in the Kramers [28] treatment of the escape rate, where nonequilibrium effects due to the loss of particles from the well are accounted for using the theory of Brownian motion, automatically leading to friction dependence of the transmission coefficient). Secondly, a particle is supposed never to return to the well once it has crossed the potential barrier. The first assumption means that friction, i.e., dissipation to the bath does not affect the escape rate. Thus, the system in effect is a closed classical one. Nevertheless, according to Mel'nikov [111] the results of classical TST should also be applicable in a *wide range of dissipation* for which thermal noise is sufficiently strong to thermalize the escaping particles yet not so strong as to affect particle motion across the top of the potential barrier, i.e., a Maxwell-Boltzmann distribution still holds there. In the context of the Kramers model, this is the so called *intermediate damping* case (cf. Fig. 1.13.2 of [5]). In the treatment of Kramers, however, which explicitly involves an *open* classical system with fluctuation-dissipation due to the bath described by the Brownian motion *Stosszahlansatz*, he shows that for sufficiently weak friction the escape rate is suppressed because of the *depletion of the well population* while for strong friction the escape rate is also suppressed due to the *slowing down of the particle motion at the barrier top*.

The suggestion that quantum mechanical tunneling might play a significant role in some chemical reactions was first made in 1927 by Hund [99] almost at the inception of quantum mechanics. The first guess at a quantum transition state theory appears to have been made by Wigner who proposed a quantum generalization of the classical TST [97-99], where the reaction rate $\Gamma$ is given via the flux-over-population method [20]

$$\Gamma \sim \frac{I_C}{Z_A}, \qquad (147)$$

where

$$Z_A = \iint_{well} W_{eq}(q,p)\,dpdq \qquad (148)$$

and

$$I_C = \iint_{barrier} J_C(q,p)\,dpdq \qquad (149)$$

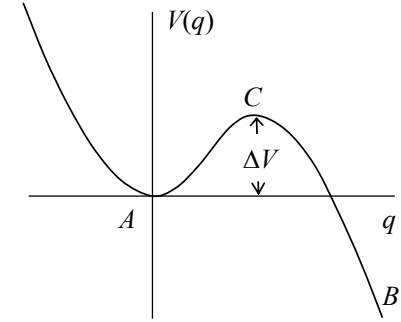

**Figure 3.** Single well potential function as the simplest example of escape over a barrier. Particles are initially trapped in the well near the point $A$ by a high potential barrier at the point $C$. They very rapidly thermalize in the well. Due to thermal agitation, however, very few may attain enough energy to escape over the barrier into region $B$ whence they never return. The height of the barrier $\Delta V = V(q_C) - V(q_A)$ is supposed to be large compared with the thermal energy $kT$.

are, respectively, the well partition function and the total current of the particles at the top of the barrier point $C$ (see Fig. 3). Here the current density at the barrier point $J_C$ is given by

$$J_C(q,p) = \theta(p)\delta(q-q_C)\dot{q}W_{eq}(q,p), \qquad (150)$$

where $\theta(x)$ and $\delta(x)$ are the unit step and Dirac delta functions, respectively, and $\dot{q} = \partial \varepsilon / \partial p = p/m$ is the particle velocity. Now we assume that near the summit point $C$ and near the bottom point $A$ of the well the potential $V(q)$ may be described by inverted harmonic oscillator and harmonic oscillator potentials, respectively, namely,

$$V(q) \approx \begin{cases} V(q_C) - \dfrac{m\omega_C^2}{2}(q-q_C)^2 \\ V(q_A) + \dfrac{m\omega_A^2}{2}(q-q_A)^2 \end{cases} \qquad (151)$$

where

$$\omega_C = \sqrt{|V''(q_C)|/m} \text{ and } \omega_A = \sqrt{V''(q_A)/m}.$$

Thus, near the bottom of the well, $W_{eq}(q,p)$ is approximated by that of a harmonic oscillator with $\omega_0 = \omega_A$ [see Eq. (141) above] meaning that the well partition function $Z_A$ can be evaluated as [48]

$$Z_A \approx \frac{e^{-\beta V(q_A)}}{2\pi\hbar \cosh(\beta\hbar\omega_A/2)} \int_{-\infty}^{\infty}\int_{-\infty}^{\infty} e^{-\frac{2}{\hbar\omega_A}\left(\frac{m}{2}\omega_A^2 q^2 + \frac{1}{2m}p^2\right)\tanh\left(\frac{\beta\hbar\omega_A}{2}\right)} dq\,dp \qquad (152)$$
$$= \frac{e^{-\beta V(q_A)}}{2\sinh(\hbar\omega_A\beta/2)}.$$

Here the limits of integration with respect to $q$ may be formally extended to $\pm$ infinity without significant error since the particles are almost all at $A$. Near the top of the barrier, the Wigner function $W_{eq}(q,p)$ is approximated by that of an *inverted* harmonic oscillator with $\omega_0 = i\omega_C$ in Eq. (141) so that at $C$ it is given by [48]

$$W_{eq}^C(q_C,p) \approx \frac{e^{-\beta V(q_C) - \frac{p^2 \tan(\beta\hbar\omega_C/2)}{m\hbar\omega_C}}}{2\pi\hbar \cos(\beta\hbar\omega_C/2)}. \qquad (153)$$

Thus, we have from Eqs. (149) and (153)

$$I_C \approx \frac{e^{-\beta V(q_C)}}{2\pi\hbar m \cos(\beta\hbar\omega_C/2)} \int_0^{\infty} p e^{-\frac{p^2 \tan(\beta\hbar\omega_C/2)}{m\hbar\omega_C}} dp \qquad (154)$$
$$= \frac{\omega_C e^{-\beta V(q_C)}}{4\pi \sin(\beta\hbar\omega_C/2)}.$$

Finally, substituting Eqs. (152) and (154) into Eq. (147), we obtain [48]

$$\Gamma \approx \frac{\omega_A}{2\pi} \Xi e^{-\beta \Delta V}, \qquad (155)$$

where $\Delta V = V(q_C) - V(q_A)$ is the barrier height and

$$\Xi = \frac{\omega_C}{\omega_A}\frac{\sinh(\hbar\beta\omega_A/2)}{\sin(\hbar\beta\omega_C/2)} = 1 + \frac{\beta^2\hbar^2}{24}\left(\omega_C^2 + \omega_A^2\right) + \dots \qquad (156)$$

is the quantum correction to the classical TST result. The lowest order quantum correction to this pre-exponential factor was first obtained by Wigner [97] (see also [98]). He emphasized that the quantum factor $\Xi$ represents an *effective lowering* of the potential barrier so enhancing the escape rate. According to Wigner [97], Eq. (155) constitutes the quantum correction to classical TST at high temperatures. In the context of quantum dissipation, one may infer that Eq. (155) also represents the extension of the intermediate damping Kramers escape rate (for which classical TST provides a reasonably accurate approximation) to include quantum effects. An important feature of Eq. (155) not appearing in the first order in $\hbar^2$ approximation is that the prefactor $\Xi$ *diverges* at a crossover temperature $T_c$ given by

$$T_c = \frac{\hbar\omega_C}{2\pi k}.$$

The divergence occurs because the parabolic (or inverted oscillator) approximation for the potential is only valid near the top of the barrier. However, at very low temperatures $T \ll T_c$, where the particle is near the bottom of the well, the parabolic approximation to the barrier shape is no longer sufficient [111]. In contrast for $T > T_c$, transitions near the barrier top dominate so that the parabolic approximation is accurate [111]. Moreover, the simple approximation appearing on the right-hand side of Eq. (156) should hold with a reasonable degree of accuracy. This approximation also appears to be in substantial agreement with the experimental results of Bouchaud *et al.* [112].

We have demonstrated how the quantum escape rate in the absence of dissipation, namely Eq. (155), may be obtained by Wigner's perturbation method. However, that result may be obtained in a more succinct fashion without using perturbation theory by recalling that the rate constant may be written as [99,113]

$$\Gamma = \frac{1}{Z_A}\int_{-\infty}^{\infty} w(\varepsilon) e^{-\beta\varepsilon} d\varepsilon, \qquad (157)$$

where the quantity

$$w(\varepsilon) = \frac{1}{1 + e^{-2\pi(\varepsilon - V_C)/(\hbar\omega_C)}} \qquad (158)$$

is the quantum transmission coefficient (ignoring dissipation) of a parabolic barrier [78]. We again approximate the potential near the top of the barrier by that of an inverted harmonic oscillator, which holds good at the barrier as well as at a small distance below it. Hence, one can regard the integral in Eq. (157) as having infinite limits. Thus, one finds on evaluating that integral with these limits that the escape rate is given by

$$\Gamma = \frac{\hbar\omega_C}{2Z_A \sin(\beta\hbar\omega_C/2)} e^{-\beta V_C}. \qquad (159)$$

Then using Eq. (152), we have the quantum TST escape rate Eq. (155) from Eq. (159).

Quantum TST as formulated for particles with separable and additive Hamiltonians in the manner just described has also been applied [114] in the context of magnetization reversal to the escape rate of the giant spin model of single domain ferromagnetic particles. This model describes a ferromagnetic particle with uniaxial anisotropy with external fields applied parallel and perpendicular to the anisotropy axis. Tunneling in such a model will be caused by the transverse field [114]. Now, the Hamiltonian of the model (which is not separable and additive as the canonical variables are now the polar angles $\vartheta$ and $\varphi$ specifying the orientation of the magnetization vector) may be mapped [114] onto that of a mechanical particle moving in a double well potential. Hence, the quantum TST rate described above which is a close approximation to the exact escape rate in the intermediate damping region may also be used to study thermally assisted tunneling of the magnetization of a single domain ferromagnetic particle. Nevertheless, the complete solution of the foregoing problem [64] will involve the extension of Wigner's phase-space formalism to spin systems and will be given in Sec. II.D.6.

*3. Application to quantum Brownian motion*

The classical theory of the Brownian motion in a potential is ubiquitous in many areas of physics and chemistry, particularly those dealing with the nature of metastable states and the rates at which these states decay. Typical examples are current-voltage characteristics of Josephson junctions, the rate of condensation of a supersaturated vapor, dielectric and Kerr effect relaxation in liquids and nematic liquid crystals, dynamic light scattering, chemical reaction rate theory in condensed phases, superparamagnetic relaxation, polymer dynamics, nuclear fission and fusion and so on [5,20,115,116]. All these phenomena in one way or another depend on the nucleation and growth of some characteristic disturbance within a metastable system, e.g., condensation of a saturated vapor is initiated by the formation of a sufficiently large droplet of the liquid. If this droplet is big enough it will be more likely to grow than to dissipate and will bring about condensation of the entire sample [115]. In many cases, particularly at low temperatures, a theory of dissipation based on the classical Brownian motion may be inadequate because it ignores quantum effects. Quantum noise arising from quantum fluctuations is also important in nanoscale and biological systems. We mention [116] the noise assisted tunneling and transfer of electrons and quasi-particles. The characteristics of such quantum noise vary strongly with temperature and at high temperatures a crossover to Johnson-Nyquist noise which is essentially governed by the classical Brownian motion takes place. Yet another aspect of the subject, which has come to the fore in recent years, is the quantum mechanics of macroscopic quantum variables such as the decay of a zero voltage state in a biased Josephson junction, flux quantum transitions in a SQUID [116] and the possible reversal by quantum tunneling of the magnetization of a single domain ferromagnetic particle. It has been conjectured by Bean and Livingston [21] that the magnetization may reverse by quantum tunneling through the internal magnetocrystalline anisotropy barrier of the particle instead of by the conventional mechanism of thermally agitated jumping over the barrier, namely, Néel relaxation [16].

All these considerations necessitate the development of a theory of quantum Brownian motion particularly a theory, which directly addresses the issue of the quantum-classical correspondence [117] in terms of a quantum analog of the classical Fokker-Planck equation. Such an evolution equation will allow dynamical parameters such as escape rates, correlation times, susceptibilities, etc. to be calculated in terms of the eigensolutions of that equation in a manner analogous to those of the Fokker-Planck equation. Moreover, it would be possible to compare asymptotic solutions for parameters such as escape rates yielded by reaction rate theory with the corresponding quantities calculated from the quantum master equation. The description of quantum mechanics in terms of phase space distributions as advanced by Wigner is also an ideal starting point for the formulation of semiclassical quantum master equations of *open* quantum systems [102]. In particular, for quantum Brownian motion, the evolution equation for $W(q, p, t)$, which we shall call the *Wigner-Fokker-Planck* equation for the translational Brownian motion of a particle in a potential $V(q)$, can be derived by proceeding to the high temperature limit. This procedure is equivalent to treating the system as a quantum mechanical particle embedded in a classical bath [118] and is effected by regarding the Brownian particle as bilinearly coupled to a bath of harmonic oscillators in thermal equilibrium at temperature $T$ [119]. The most convenient way of characterizing the influence of the bath is by means of the *spectral density* characterizing the coupling to the bath of oscillators. The oscillators constituting the string or transmission line then represent the normal modes of the bath. We remark that the effect of friction is regarded as entirely equivalent to the bilinear coupling to these normal modes. Quantization of the bath of oscillators [119] then yields the following semiclassical master equation for the Wigner distribution function $W(q, p, t)$, which has intuitive appeal [121-129]

$$\frac{\partial W}{\partial t} + \hat{M}_W W = \text{St}(W). \quad (160)$$

In this equation the operator $\hat{M}_W$ is given by Eq. (136) and the collision kernel operator St($W$) is

$$\text{St}(W) = \frac{\partial}{\partial p}\left(D_p pW + D_{pp}\frac{\partial W}{\partial p} + D_{qp}\frac{\partial W}{\partial q}\right) + \frac{\partial}{\partial q}\left(D_{qq}\frac{\partial W}{\partial q}\right). \quad (161)$$

Here $D_p, D_{pp}, D_{qp}, D_{qq}$ are coordinate, momentum, and time dependent parameters (diffusion coefficients). The left-hand side of Eq. (160) again comprising the Wigner-Moyal operator is of course the quantum analog of the Liouville equation, while St($W$) characterizes the interaction of the Brownian particle with the thermal bath at temperature $T$; the collision kernel St($W$) being the analog of the collision kernel (*Stosszahlansatz*) in the classical kinetic theory. Conditions for the validity of the master Eq. (160) are discussed elsewhere (e.g., Refs. 104 and 130). In the classical limit, $\hbar \to 0$, Eq. (160) reduces to the Klein-Kramers (Fokker-Planck) equation for the translational Brownian motion of a particle in a potential $V(q)$, namely,

$$\frac{\partial W}{\partial t}+\frac{p}{m}\frac{\partial W}{\partial q}-\frac{\partial V}{\partial q}\frac{\partial W}{\partial p}=\frac{\zeta}{m}\frac{\partial}{\partial p}\left(pW+\frac{m}{\beta}\frac{\partial W}{\partial p}\right) \qquad (162)$$

and the coefficients $D_p, D_{pp}, D_{qp}, D_{qq}$ become

$$D_p=\frac{\zeta}{m},\ D_{pp}=\frac{\zeta}{\beta},\ D_{qp}=D_{qq}=0, \qquad (163)$$

where $\zeta$ is the drag coefficient of a Brownian particle.

Now being in possession of the functional form of the master Eq. (160) for a Brownian particle, the next crucial step is to determine $D_p, D_{pp}, D_{qp}, D_{qq}$. With this end in mind we shall select the extension to the semiclassical case of a simple heuristic idea originally used by Einstein, Smoluchowski, Langevin, and Kramers to determine diffusion coefficients in the classical theory of the Brownian motion [5,71]. Thus in order to obtain the explicit form of $D_p, D_{pp}, D_{qp}, D_{qq}$ in Eq. (161), we first recall Wigner's results for the unnormalized equilibrium distribution $W_{eq}(q,p)$ developed in a power series in $\hbar^2$ Eq. (139). This equilibrium distribution, being a stationary solution of the Wigner-Moyal Eq. (135), must be the equilibrium solution of the generic master Eq. (160), i.e., it must also satisfy $\text{St}(W_{eq})=0$. Hence if $W_{eq}(q,p)$ from Eq. (139) is to satisfy $\text{St}(W_{eq})=0$, the coefficients $D_p$, $D_{qp}$, and $D_{qq}$ should remain as in Eq. (163) and only $D_{pp}$ must be altered to read [48,131] (for simplicity, we retain only the leading quantum correction terms)

$$D_{pp}=\frac{\zeta}{\beta}+\frac{\zeta\beta\hbar^2}{12m}\frac{\partial^2 V}{\partial q^2}+.... \qquad (164)$$

Thus, the explicit form of the Wigner Fokker-Planck equation up to $o(\hbar^2)$ is [48,131]

$$\frac{\partial W}{\partial t}+\frac{p}{m}\frac{\partial W}{\partial q}-\frac{\partial V}{\partial q}\frac{\partial W}{\partial p}+\frac{\hbar^2}{24}\frac{\partial^3 V}{\partial q^3}\frac{\partial^3 W}{\partial p^3}+...=\frac{\zeta}{m}\frac{\partial}{\partial p}\left[pW+\left(\frac{m}{\beta}+\frac{\beta\hbar^2}{12}\frac{\partial^2 V}{\partial q^2}+...\right)\frac{\partial W}{\partial p}\right]. \qquad (165)$$

The phase-space master equation (165) has been derived in [48,131] assuming that $D_p, D_{pp}, D_{qp}, D_{qq}$ in Eq. (161) are time-independent. We observe that in the high temperature limit, $\beta\to 0$, Eq. (165) obviously reduces to the Caldeira-Leggett master equation [119], where the collision kernel $\text{St}(W)$ has the same form as in the classical Fokker-Planck equation (162). The evolution equation for the density matrix $\hat{\rho}$ in Hilbert space corresponding to Eq. (165) is again to order $\hbar^2$ [132]

$$\frac{\partial\hat{\rho}}{\partial t}+\frac{i}{\hbar}\left[\hat{H},\hat{\rho}\right]=-\frac{\zeta}{m}\left(\frac{i}{2\hbar}[\hat{q},\hat{p}\hat{\rho}+\hat{\rho}\hat{p}]+\frac{m}{\beta\hbar^2}[\hat{q},[\hat{q},\hat{\rho}]]+\frac{\beta}{6}\left[\hat{q},\left[\partial_q\hat{V},\hat{\rho}\right]\right]\right). \qquad (166)$$

We also observe that the imposition of the Wigner phase-space distribution $W_{eq}(q,p)$ as the equilibrium solution of Eq. (160) so yielding a diffusion coefficient $D_{pp}$, which depends on the derivatives of the potential, appears to be the quantum analog of the *Ansatz* of a Maxwell-Boltzmann stationary distribution for the classical Klein-Kramers equation, Eq. (162). Furthermore, the condition $\text{St}(W_{eq})=0$ is equivalent to the property of the collision kernel $\text{St}(W)$ in the classical kinetic theory, whereby the reduced or single particle phase-space distribution function $W(q,p,t)$ obeys the kinetic equation

$$\frac{\partial W}{\partial t}+\frac{p}{m}\frac{\partial W}{\partial q}-\frac{\partial V}{\partial q}\frac{\partial W}{\partial p}=\text{St}(W). \qquad (167)$$

Here the equilibrium Maxwell-Boltzmann distribution function $W_{eq}(q,p)\sim e^{-\beta\varepsilon(q,p)}$ always satisfies the condition $\text{St}(W_{eq})=0$. In particular, this is so for the Fokker-Planck equation Eq. (162). In the quantum case, this idea has been used before, e.g., by Gross and Lebowitz [133] in formulating quantum kinetic models of impulsive collisions. According to [133], for a system with a Hamiltonian $\hat{H}$, the equation governing the time behavior of the density matrix $\hat{\rho}$ is given by the (reduced) Eq. (40), where the collision kernel operator $\text{St}(\hat{\rho})$ satisfies the condition $\text{St}(\hat{\rho}_{eq})=0$, where $\hat{\rho}_{eq}$ is the equilibrium density matrix. The condition $\text{St}(\hat{\rho}_{eq})=0$ has also been used by Redfield [14] [see Eq. (56)] in calculating the matrix elements of the relaxation operator $\text{St}(\hat{\rho})$. Now we have evaluated $D_p, D_{pp}, D_{qp}, D_{qq}$ for frequency independent damping, meaning that in Eq. (160) they are *independent* of the time [124,127]. In the high temperature limit, this approximation may be used in a wide range of the model parameters both in the limits of weak and strong damping (a detailed discussion of the validity of this approximation is given by Grabert [134]). However, for the parameter range, where it is invalid, e.g., throughout the very low temperature region, other methods should be used.

The evolution Eq. (165) obtained for an arbitrary potential $V(q)$ simplifies substantially for the harmonic potential, Eq. (140), i.e., the quantum Brownian oscillator model, viz., [48]

$$\frac{\partial W}{\partial t}+\frac{p}{m}\frac{\partial W}{\partial q}-m\omega_0^2 q\frac{\partial W}{\partial p}=\frac{\zeta}{m}\frac{\partial}{\partial p}\left(pW+D_{pp}\frac{\partial W}{\partial p}\right), \qquad (168)$$

where our heuristic generalization of the Einstein procedure yields $D_{pp}$ in closed form

$$D_{pp}=\left\langle p^2\right\rangle_{eq}=\frac{m\hbar\omega_0}{2}\coth\frac{\beta\hbar\omega_0}{2}.$$

The master Eq. (168) coincides in all respects with that of Agarwal [121], who first developed a detailed theory of the Brownian motion of a quantum oscillator for the *weak-coupling* case [see his equation (2.11) with $\lambda=0$]. Furthermore, Eq. (168) is the same as the Fokker-Planck equation (here the Klein-Kramers equation) for a classical Brownian oscillator [48] except the diffusion coefficient $D_{pp}$ is altered to include the quantum effects. It is known, however, that according to the theory of quantum dissipation for a quantum Brownian oscillator, both the stationary distribution as well as the corresponding averages $\langle q^2 \rangle$ and $\langle p^2 \rangle$ *depend on damping* (appropriate equations are given in Chapter 6 of [120]). In the approximation of Ohmic damping with Drude's regularization, these equations read [120]

$$\langle q^2 \rangle = \frac{1}{m\beta} \sum_{n=-\infty}^{\infty} \frac{1}{\omega_0^2 + v_n^2 + \frac{\omega_D \zeta / m}{1 + \omega_D / v_n}} \tag{169}$$

and

$$\langle p^2 \rangle = \frac{m}{\beta} \sum_{n=-\infty}^{\infty} \frac{\omega_0^2 + \frac{\omega_D \zeta / m}{1 + \omega_D / v_n}}{\omega_0^2 + v_n^2 + \frac{\omega_D \zeta / m}{1 + \omega_D / v_n}}, \tag{170}$$

where $v_n = 2\pi |n| / (\hbar \beta)$ and $\omega_D$ is a cutoff frequency (a Drude regularization is necessary as in pure Ohmic damping $\langle p^2 \rangle$ diverges [120]). However, both of these equations reduce to Eqs. (143) either for vanishing damping ($\zeta / m \to 0$) or in the high temperature limit ($\beta \hbar \omega_0 \to 0$). Moreover, the difference between the damping dependent and damping independent equations is negligible for $\zeta / (m\omega_0) < 0.1$ (which is simply the condition for the existence of damped oscillations and/or narrow spectral lines). Furthermore, for $\beta \hbar \zeta / m \leq 1$, Eq. (168) (i.e., the Agarwal model) may be used as an approximate description of the kinetics of a quantum oscillator.

To summarize the merit of the phase-space formalism for the quantum Brownian motion in a potential is that it originates in the master Eq. (160). This equation is a partial differential equation in phase space akin to the Fokker-Planck equation and so operators are not involved. Moreover, the main advantage of the phase space approach now becomes apparent, namely it provides a master equation that may be solved using the methods [6,71] associated with the classical theory of the Brownian motion in a potential, allowing one to study the quantum–classical correspondence for dissipative systems (see, e.g.,[131,132,135-141]). Many other examples of the use of the Wigner function representation of the density matrix in various applications in physics and chemistry may be found in the books [42,46,47,77] and references cited therein.

We now turn our attention to phase-space representations for spins demonstrating how the Wigner-Moyal formulation of quantum mechanics as a statistical theory on classical phase space can be applied successfully to spinning particles.

### C. Quasiprobability distribution functions for spins

By way of background to the discussion which follows, we recall that in providing a phase space description of spin systems, Stratonovich in 1956 [49] introduced the quasiprobability (Wigner) distribution function $W_S(\vartheta, \varphi, t)$ for the spin orientations in the configuration space of the polar and the azimuthal angles $(\vartheta, \varphi)$. This idea formed part of a general discussion of *c*-number quasiprobability distributions for quantum systems in a representation space based on the *symmetry properties* of the underlying group. Examples are the Heisenberg-Weyl group for particles and the SU(2) group for rotations. The *c*-number representation for spins is especially important in treating spin relaxation phenomena. There *the spin orientation distribution is defined as the linear invertible bijective map onto the representation space comprised of the trace of the product of the system density matrix and the irreducible tensor operators with matrix elements in the spherical basis representation given via the Clebsch-Gordan coefficients*. Alternative quasiprobability distribution functions for spins have also been proposed using the spin coherent-state representation of the density matrix [40,50-54,57-60] introduced by Glauber and Sudarshan and commonly used in quantum optics (see, e.g., [45,46,51]). Moreover, Várilly and Gracia-Bondía [50] have shown that the spin coherent-state representation approach is equivalent to the Stratonovich formalism.

In view of the importance of and the generality of Stratonovich's study of representation distributions for quantum systems, we shall summarize the general principles underlying such representations as given by him. Then we shall apply them to derive the Wigner and other representation distributions for systems with symmetries described by the SU(2) rotation group.

Stratonovich [49] defines the "representation distribution" in the representation space M by the following requirements:

1. The space, in which the "representation distribution" is defined, has a *classical* meaning, for example, phase space $(q, p)$ or the space of orientations $(\vartheta, \varphi)$.

2. The representation distribution can be expressed *linearly* in terms of the density matrix $\hat{\rho}$. This requirement is directly related to the linearity of the whole apparatus of quantum theory, i.e., it is connected with the statistical interpretation of the theory. The density matrix $\hat{\rho}$ like any other quantum operator $\hat{A}$, has associated with it a (*c*-number) function in the representation space, viz.,

$$\rho(M, t) = \text{Tr}\{\hat{\rho}(t) \hat{w}(M)\}, \tag{171}$$

and, in general,

$$A(M) = \text{Tr}\{\hat{A} \hat{w}(M)\}, \tag{172}$$

comprising the *direct* mapping of the quantum operator $\hat{A}$ onto the representation space via the kernel $\hat{w}$, which is an operator depending on the point $M$ as a parameter. For example, the point of the representation space could be a point on the unit sphere $(\vartheta, \varphi)$.

3. The representation distribution must be real, i.e., in general, to a Hermitian operator $\hat{A}$ there must correspond a real *c*-number (Weyl symbol) $A(\vartheta, \varphi)$. This requirement amounts to the condition that the bijective operator $\hat{w}(M)$ must be Hermitian for all points $M$.

4. Statistical averaging of the *c*-number $A(M)$ over the representation distribution *must* give the same results as the rule for averaging of Hilbert space operators, namely,

$$\langle \hat{A} \rangle(t) = \mathrm{Tr}\left(\hat{\rho}(t)\hat{A}\right) = \int A(M)\rho(M,t)dM. \quad (173)$$

Moreover, we have as a representation of a Hilbert space operator $\hat{A}$ the *inverse* map

$$\hat{A} = \int A(M)\hat{w}(M)dM \quad (174)$$

as can be easily verified by forming $\mathrm{Tr}\left(\hat{\rho}\hat{A}\right)$ via Eq. (174) and then using Eq. (171) thereby yielding Eq. (173).

These transformation rules tell one how to get the operator relation from the *c*-number representation and vice-versa. Thus condition 4 is equivalent to the requirement that the *direct* Eq. (172) and *inverse* maps Eq. (174) are accomplished via the *same kernel* $\hat{w}(M)$, i.e., the mapping given by Eq. (172) is *bijective* (one to one onto).

By regarding given operators say $\hat{A}, \hat{B}, \ldots$ as elements of a complex Euclidean space with a scalar product given by the trace

$$\left(\hat{A}, \hat{B}\right) = \mathrm{Tr}\left(\hat{A}\hat{B}^\dagger\right),$$

we can then introduce an orthonormal basis of operators $\{\hat{A}_1, \hat{A}_2, \ldots\}$ with orthogonality relation

$$\mathrm{Tr}\left(\hat{A}_i \hat{A}_j^\dagger\right) = \delta_{ij}. \quad (175)$$

Thus, the kernel operator $\hat{w}(M)$ may be represented by its expansion in the orthonormal basis of operators as

$$\hat{w}(M) = \sum_i \hat{A}_i^\dagger A_i(M), \quad (176)$$

where by definition the *c*-number expansion coefficients $A_i(M)$ are given by the direct map [cf. Eq. (172)]

$$A_i(M) = \mathrm{Tr}\left\{\hat{A}_i \hat{w}(M)\right\}. \quad (177)$$

The requirement given by Eq. (175) is entirely equivalent to the *c*-number orthogonality relation

$$\int A_i(M) A_j^*(M) dM = \delta_{ij}. \quad (178)$$

According to Eq. (173) the normalization condition for the density matrix $\hat{\rho}$, viz. $\mathrm{Tr}(\hat{\rho}) = 1$, must become

$$\int \rho(M,t)\mathrm{Tr}\{\hat{w}(M)\}dM = 1. \quad (179)$$

Thus, as the normalized distribution we must take a *c*-number function $W(M,t)$ given by

$$W(M,t) = \mathrm{Tr}\{\hat{w}(M)\}\rho(M,t) \quad (180)$$

and the averaging rule Eq. (173) becomes

$$\langle \hat{A} \rangle(t) = \int W(M,t) A(M) \left[\mathrm{Tr}\{\hat{w}(M)\}\right]^{-1} dM. \quad (181)$$

Stratonovich's abstract treatment presented above may be used to derive specific distributions. The Wigner distribution function $W(q,p,t)$ defined by Eq. (121) is an important explicit example. In this instance, Eq. (121) can also be rewritten in terms of a bijective map $\hat{w}$ using the density matrix $\hat{\rho}(q_1, q_2, t)$ in the $(q_1, q_2)$ notation [77] as

$$W(q,p,t) = \int_{-\infty}^{\infty}\int_{-\infty}^{\infty} \hat{\rho}(q_1,q_2,t)\hat{w}(q,p,q_1,q_2)dq_1 dq_2, \quad (182)$$

where the kernel $\hat{w}(q,p,q_1,q_2)$ is now given by [49]

$$\hat{w}(q,p,q_1,q_2) = \frac{1}{\pi\hbar} e^{i(q_1-q_2)p/\hbar} \delta\left(q - \frac{q_1+q_2}{2}\right). \quad (183)$$

Thus, the Wigner distribution function $W(q,p,t)$ is derived merely by applying the principles of homogeneity and equivalence of directions embodied in the symmetries of the Heisenberg-Weyl group combined with the notion of a classical phase space. This group-theoretic argument should be compared with the intuitive method of Wigner who appears to have arrived at his distribution by *ad hoc* reasoning insofar as that distribution yields the correct marginal probabilities for either the positions or the momentum.

*1. Spin phase-space distribution functions*

Now Stratonovich [49] originally introduced the spin phase-space distribution function for zero dissipation, i.e., for closed systems. This function was further developed both for closed and open spin systems (e.g., [35-39, 50-60]) and is entirely analogous to the translational Wigner distribution $W(q,p,t)$ in phase space $(q,p)$, which is the quasiprobability representation of the density operator except that certain differences arise because of the angular momentum commutation relations. The basic ideas may be summarized as follows [51]. First, we recall that the classical distribution function $W(\vartheta, \varphi, t)$ of magnetic moment orientations on the surface of a unit sphere can be expanded in an *infinite* series of the spherical harmonics $Y_{LM}(\vartheta,\varphi)$ as [5] [cf. Eq. (B12) from Appendix B]

$$W(\vartheta,\varphi,t) = \sum_{L=0}^{\infty} \sum_{M=-L}^{L} Y_{LM}(\vartheta,\varphi) \langle Y_{LM}^* \rangle(t), \tag{184}$$

where the expansion coefficients (statistical moments) $\langle Y_{LM}^* \rangle(t)$ are defined by

$$\langle Y_{LM}^* \rangle(t) = \int_0^\pi \int_0^{2\pi} Y_{LM}^*(\vartheta,\varphi) W(\vartheta,\varphi,t) \sin\vartheta \, d\vartheta \, d\varphi \tag{185}$$

(the asterisk denotes the complex conjugate) because $Y_{LM}(\vartheta,\varphi)$ are orthogonal and constitute a complete set in configuration space $(\vartheta,\varphi)$. Various definitions and properties of the spherical harmonics $Y_{LM}(\vartheta,\varphi)$ are discussed in detail in Appendix B. In particular, the spherical harmonics $Y_{LM}(\vartheta,\varphi)$, which are components of some irreducible tensor of rank $L$, may formally be defined by the commutation relations [95]

$$\left[ \hat{L}_\mu, Y_{LM}(\vartheta,\varphi) \right] = \sqrt{l(l+1)} C_{LM1\mu}^{LM+\mu} Y_{LM+\mu}(\vartheta,\varphi) \tag{186}$$

yielding three relations $(\mu = 0, \pm 1)$, viz., [95]

$$\hat{L}_0 Y_{LM}(\vartheta,\varphi) = M Y_{LM}(\vartheta,\varphi), \tag{187}$$

$$\hat{L}_{\pm 1} Y_{LM}(\vartheta,\varphi) = \mp \sqrt{\frac{L(L+1) - M(M \pm 1)}{2}} Y_{LM \pm 1}(\vartheta,\varphi), \tag{188}$$

where $\hat{L}_\mu$ are the components of the orbital angular momentum operator $\hat{\mathbf{L}}$ in the spherical basis given by the differential operators [95]

$$\hat{L}_0 = -i \frac{\partial}{\partial \varphi}, \tag{189}$$

$$\hat{L}_{\pm 1} = -\frac{e^{\pm i\varphi}}{\sqrt{2}} \left( \frac{\partial}{\partial \vartheta} \pm i \cot\vartheta \frac{\partial}{\partial \varphi} \right). \tag{190}$$

The angular momentum operators $\hat{L}_\mu$ satisfy the same commutation relations as the spherical components $\hat{S}_\mu$ of the spin operator $\hat{\mathbf{S}}$ given by Eq. (A8) from Appendix A. Now the quantum quasiprobability distribution function for spins $W_S(\vartheta,\varphi,t)$ may be obtained in the manner of Eqs. (100) *et seq.* merely by replacing the *classical* average $\langle Y_{LM}^* \rangle(t)$ in Eq. (184) by the *quantum mechanical expectation values* of appropriate operators, which must transform under rotation of the coordinate system in exactly the same way as the spherical harmonics $Y_{LM}(\vartheta,\varphi)$. Therefore, we must seek *specific* operators in Hilbert space corresponding to the spherical harmonics $Y_{LM}(\vartheta,\varphi)$ such that they have commutation relations with the spin spherical component operators $\hat{S}_\mu$ which are the same as those, Eq. (186), between the spherical harmonics $Y_{LM}(\vartheta,\varphi)$ and the spherical component operators $\hat{L}_\mu$ of the angular momentum operator $\hat{\mathbf{L}}$ [51]. The particular operators with these commutation properties are the polarization operators $\hat{T}_{LM}^{(S)}$ [51,95]; see Eq. (A30) in Appendix A. This correspondence become obvious by comparing Eqs. (187) and (188) for the spherical harmonics with the commutation Eq. (A30) from Appendix A for the polarization and spin operators yielding [95]

$$[\hat{S}_0, \hat{T}_{LM}^{(S)}] = M \hat{T}_{LM}^{(S)} \tag{191}$$

and

$$[\hat{S}_{\pm 1}, \hat{T}_{LM}^{(S)}] = \mp \sqrt{\frac{L(L+1) - M(M \pm 1)}{2}} \hat{T}_{LM \pm 1}^{(S)}. \tag{192}$$

We may go into more details in the following manner. We have seen above that the density matrix $\hat{\rho}_S$ of the spin characterized by the spin number $S$ may be expanded as a finite series of the polarization operators $\hat{T}_{LM}^{(S)}$, Eq. (70), namely,

$$\hat{\rho}_S(t) = \sum_{L=0}^{2S} \sum_{M=-L}^{L} \hat{T}_{LM}^{(S)} \langle \hat{T}_{LM}^{\dagger(S)} \rangle(t). \tag{193}$$

Therefore, in order to construct the quasiprobability distribution function $W_S(\vartheta,\varphi,t)$ of the polar and the azimuthal angles $(\vartheta,\varphi)$ corresponding to the density matrix $\hat{\rho}_S$ from Eq. (193), we first, in the light of the previous paragraph, formally express $W_S(\vartheta,\varphi,t)$ as a *finite* series of spherical harmonics $Y_{LM}(\vartheta,\varphi)$ in the representation space, viz.,

$$\frac{2S+1}{4\pi} W_S(\vartheta,\varphi,t) = \sum_{L=0}^{2S} \sum_{M=-L}^{L} Y_{LM}(\vartheta,\varphi) \langle Y_{LM}^* \rangle(t), \tag{194}$$

where the expansion coefficients $\langle Y_{LM}^* \rangle(t)$ are given by

$$\langle Y_{LM}^* \rangle(t) = \frac{2S+1}{4\pi} \int_0^\pi \int_0^{2\pi} Y_{LM}^*(\vartheta,\varphi) W_S(\vartheta,\varphi,t) \sin\vartheta \, d\vartheta \, d\varphi \tag{195}$$

due to the orthogonality property of the $Y_{LM}(\vartheta,\varphi)$ given by Eq. (B11) (see Appendix B). Equation (194) represents a quantum analog of the expansion Eq. (184) of the *classical* distribution function $W(\vartheta,\varphi,t)$ of spin orientations in configuration space in terms of spherical harmonics. Thus, following the above argument, it appears that both the quasiprobability (Wigner) distribution function on the sphere $W_S(\vartheta,\varphi,t)$ and the corresponding kernel operator $\hat{w}(\vartheta,\varphi)$ of the Wigner-Stratonovich bijective map onto phase space defined by Eq. (171), viz.,

$$W_S(\vartheta,\varphi,t) = \text{Tr}\{\hat{\rho}_S(t) \hat{w}(\vartheta,\varphi)\}, \tag{196}$$

can be obtained by merely replacing the averages of spherical harmonics $\langle Y_{LM}^* \rangle(t)$ over the representation space $(\vartheta,\varphi)$ in Eq. (194) by the quantum mechanical expectation values of the polarization operators $\left\langle \hat{T}_{LM}^{\dagger(S)} \right\rangle(t)$. However, the replacement must be achieved in order to preserve commutation relations according to the prescription indicated by Eq. (195), i.e., we must have

$$\left\langle Y_{LM}^*(\vartheta,\varphi) \right\rangle(t) = \frac{2S+1}{4\pi} \Omega_{L,M}^S \left\langle \hat{T}_{LM}^{\dagger(S)} \right\rangle(t), \quad (197)$$

where $\Omega_{L,M}^S$ is a constant to be determined. Substitution of Eq. (197) into Eq. (194) immediately yields the desired Eq. (194) in representation space in the finite series form

$$W_S(\vartheta,\varphi,t) = \mathrm{Tr}\{\hat{\rho}_S \hat{w}(\vartheta,\varphi)\} = \sum_{L=0}^{2S} \sum_{M=-L}^{L} \Omega_{L,M}^S Y_{LM}(\vartheta,\varphi) \left\langle \hat{T}_{LM}^{\dagger(S)} \right\rangle(t). \quad (198)$$

Equation (198) implies that *different* forms of the quasiprobability distribution function $W_S(\vartheta,\varphi,t)$ exist corresponding to different choices of the constant $\Omega_{L,M}^S$ in Eq. (197) analogous to Eq. (107) for the translational case. Thus, we have obtained quasiprobability distribution functions of spins *by identifying the average of the spherical harmonic $Y_{LM}(\vartheta,\varphi)$ over $(\vartheta,\varphi)$ as an average of the polarization operator $\hat{T}_{LM}^{(S)}$ via the spin density matrix $\hat{\rho}_S$* [51].

Now in order to find explicit equations for the prefactor $\Omega_{L,M}^S$ with the ultimate aim of introducing three different quasiprobability distribution functions analogous to those already described for particles [cf. Eq. (107)], we first consider the SU(2) coherent (or minimum uncertainty) states. Many spin relaxation problems can be dealt with in terms of the interaction of an assembly of spins with electromagnetic fields or with those arising from internal anisotropy potentials. In these problems, a particular set of spin states has to be selected. The choice of a particular representation is motivated rather by convenience than by necessity. In this Section, we define the coherent spin states and discuss their properties following Ref. [52]. The coherent state representation of the density matrix when applied to spin systems allows one to analyze spin relaxation phenomena using a quasiprobability distribution function $W_S(\vartheta,\varphi,t)$ of spin orientations in a phase (here configuration) space of the polar and the azimuthal angles $(\vartheta,\varphi)$.

In quantum mechanics, the spin operator $\hat{\mathbf{S}}$ is usually represented by a set of three square $(2S+1)\times(2S+1)$ matrixes $\hat{S}_X$, $\hat{S}_Y$, and $\hat{S}_Z$ with $S$ being the spin number, while the basis spin functions $\chi_{Sm} = |S,m\rangle$, which describe the states with definite spin $S$ and spin projection $m$ onto the Z-axis, are eigenfunctions of the spin operators $\hat{\mathbf{S}}^2$ and $\hat{S}_Z$ [95] (the properties of the spin operators $\hat{\mathbf{S}}$, $\hat{S}_Z$, etc. and the spin functions $\chi_{Sm}$ are described in Appendix A). Now, the spin coherent states $|\psi_\kappa\rangle$ for a single particle of spin $S$ can be defined as [52] (in our notation)

$$|\psi_\kappa\rangle = \frac{1}{\sqrt{N}} e^{\kappa \hat{S}_{-1}} \chi_{SS}. \quad (199)$$

Here $\kappa$ is complex valued, $N$ is a normalization factor, $\hat{S}_{-1} = (\hat{S}_X - i\hat{S}_Y)/\sqrt{2}$ is the spin spherical component operator, and $\chi_{SS}$ is the spin ground state such that

$$\hat{S}_Z \chi_{SS} = S \chi_{SS}.$$

Expanding the exponential operator $e^{\kappa \hat{S}_{-1}}$ in Eq. (199) yields in turn the expansion of the state $|\psi_\kappa\rangle$ in the orthonormal basis of the spin eigenstates $\chi_{Sm}$, viz., [52]

$$|\psi_\kappa\rangle = \frac{1}{\sqrt{N}} \sum_{p=0}^{\infty} \left(\kappa \hat{S}_{-1}\right)^p \chi_{SS} = \frac{1}{\sqrt{N}} \sum_{p=0}^{2S} \kappa^p \sqrt{\frac{p!(2S)!}{2^p (2S-p)!}} \chi_{SS-p}. \quad (200)$$

Here we have utilized the fact that the operator $\hat{S}_{-1}$ acting on the ground state $\chi_{SS}$ creates spin deviations of the form [52]

$$\left(\hat{S}_{-1}\right)^p \chi_{SS} = \sqrt{\frac{p!(2S)!}{2^p (2S-p)!}} \chi_{SS-p}, \quad 0 \le p \le 2S. \quad (201)$$

Now the normalization factor $N$ is determined by the equality

$$\langle \psi_\kappa | \psi_\kappa \rangle = N^{-1} \sum_{p=0}^{2S} \frac{(2S)! |\kappa|^{2p}}{2^p p!(2S-p)!}$$
$$= N^{-1} \left(1 + |\kappa|^2/2\right)^{2S} = 1, \quad (202)$$

so that the normalized state $|\psi_\kappa\rangle$ is then given by

$$|\psi_\kappa\rangle = \left(1 + |\kappa|^2/2\right)^{-S} \sum_{p=0}^{2S} \kappa^p \sqrt{\frac{p!(2S)!}{2^p (2S-p)!}} \chi_{SS-p}. \quad (203)$$

However, the coherent spin states are *not* orthogonal so that the scalar product of two states $|\psi_\kappa\rangle$ and $|\psi_\lambda\rangle$ is

$$\langle \psi_\lambda | \psi_\kappa \rangle = \left(1 + |\lambda|^2/2\right)^{-S} \left(1 + |\kappa|^2/2\right)^{-S} \sum_{p=0}^{2S} \frac{(2S)! \lambda^{*p} \kappa^p}{2^p p!(2S-p)!}$$
$$= \frac{\left(2 + \lambda^* \kappa\right)^{2S}}{\left(2 + |\lambda|^2\right)^S \left(2 + |\kappa|^2\right)^S} \quad (204)$$

and so

$$\left|\langle \psi_\lambda | \psi_\kappa \rangle\right|^2 = \left(1 - \frac{2|\lambda - \kappa|^2}{\left(2 + |\lambda|^2\right)\left(2 + |\kappa|^2\right)}\right)^{2S}. \quad (205)$$

Now, if we write in Eq. (203) that

$$\kappa = \sqrt{2} e^{i\varphi} \tan(\vartheta/2), \quad 0 \le \vartheta < \pi, 0 \le \varphi < 2\pi, \quad (206)$$

then the normalized states $|\psi_\kappa\rangle = |S,\vartheta,\varphi\rangle$ can be written as (with $m = S - p$)

$$|S,\vartheta,\varphi\rangle = \cos^{2S}\frac{\vartheta}{2} e^{\sqrt{2}e^{i\varphi}\tan(\vartheta/2)\hat{S}_{-1}}\chi_{SS}$$

$$= \cos^{2S}\frac{\vartheta}{2}\sum_{m=-S}^{S}\sqrt{\frac{(2S)!}{(S+m)!(S-m)!}}\left(e^{i\varphi}\tan\frac{\vartheta}{2}\right)^{S-m}\chi_{Sm}. \qquad (207)$$

Now the spin coherent state vector $|S,\vartheta,\varphi\rangle$ defined by Eq. (207) can also be written in the *equivalent* form

$$|S,\vartheta,\varphi\rangle = \tilde{\chi}_{SS}(\vartheta,\varphi)e^{iS\varphi},$$

where the $\tilde{\chi}_{S\lambda}(\vartheta,\varphi)$ constitute the helicity basis functions defined as [95]

$$\tilde{\chi}_{S\lambda}(\vartheta,\varphi) = \sum_{m=-S}^{S} D_{m\lambda}^{S}(\varphi,\vartheta,0)\chi_{Sm},$$

and $D_{m\lambda}^{S}(\varphi,\vartheta,\psi)$ are the Wigner $D$ functions [95]. Both forms of the state vector $|S,\vartheta,\varphi\rangle$ are equivalent due to the trigonometric identity [95]

$$D_{m\pm S}^{S}(\varphi,\vartheta,0)e^{\pm iS\varphi} = \sqrt{\frac{(2S)!}{(S+m)!(S-m)!}}\cos^{2S}\frac{\vartheta}{2}\left[\pm e^{\pm i\varphi}\tan\frac{\vartheta}{2}\right]^{S\mp m}.$$

The states $|S,\vartheta,\varphi\rangle$ so defined form *a complete set* with the completeness relation given by

$$\frac{2S+1}{4\pi}\int_{0}^{\pi}\int_{0}^{2\pi}|S,\vartheta,\varphi\rangle\langle S,\vartheta,\varphi|\sin\vartheta d\vartheta d\varphi$$

$$= \frac{2S+1}{2}\sum_{m=-S}^{S}\frac{(2S)!\chi_{Sm}\chi_{Sm}^{\dagger}}{(S+m)!(S-m)!}\int_{0}^{\pi}\cos^{4S}\frac{\vartheta}{2}\tan^{2(S-m)}\frac{\vartheta}{2}\sin\vartheta d\vartheta \qquad (208)$$

$$= \sum_{m=-S}^{S}\chi_{Sm}\chi_{Sm}^{\dagger} = \hat{I},$$

where $\hat{I}$ is the $(2S+1)\times(2S+1)$ identity matrix. Furthermore, the overlap of two states $|S,\vartheta,\varphi\rangle$ and $|S,\vartheta',\varphi'\rangle$ is [52]

$$\langle S,\vartheta',\varphi'|S,\vartheta,\varphi\rangle = \left[\cos\frac{\vartheta}{2}\cos\frac{\vartheta'}{2}+e^{i(\varphi-\varphi')}\sin\frac{\vartheta}{2}\sin\frac{\vartheta'}{2}\right]^{2S}, \qquad (209)$$

and so

$$|\langle S,\vartheta',\varphi'|S,\vartheta,\varphi\rangle| = 2^{-S}\left[1+(\mathbf{u}\cdot\mathbf{u}')\right]^{S}, \qquad (210)$$

where $\mathbf{u}$ and $\mathbf{u}'$ are unit vectors in the directions specified by spherical polar coordinates $(\vartheta,\varphi)$ and $(\vartheta',\varphi')$, respectively.

Now, the matrix elements $\langle S,\vartheta,\varphi|\hat{T}_{LM}^{(S)}|S,\vartheta,\varphi\rangle$ of the polarization operator $\hat{T}_{LM}^{(S)}$ in the coherent state representation may be expressed via the spherical harmonics $Y_{LM}(\vartheta,\varphi)$ by using the trigonometric expansion of the latter Eq. (B18) from Appendix B and Eq. (A17) from Appendix A yielding the matrix elements in the following compact closed form

$$\langle S,\vartheta,\varphi|\hat{T}_{LM}^{(S)}|S,\vartheta,\varphi\rangle$$

$$= (2S)!\cos^{4S}\frac{\vartheta}{2}\sum_{m,m'=-S}^{S}\frac{e^{-i(m-m')\varphi}\left(\tan\frac{\vartheta}{2}\right)^{2S-m-m'}\left(\chi_{Sm'}^{\dagger},\hat{T}_{LM}^{(S)}\chi_{Sm}\right)}{\sqrt{(S+m')!(S-m')!(S+m)!(S-m)!}} \qquad (211)$$

$$= \sqrt{\frac{4\pi}{2S+1}}C_{SSL0}^{SS}Y_{LM}(\vartheta,\varphi),$$

where $C_{SSL0}^{SS}$ is the Clebsch-Gordan coefficient [95] given by

$$C_{SSL0}^{SS} = \frac{(2S)!\sqrt{2S+1}}{\sqrt{(2S-L)!(2S+L+1)!}}. \qquad (212)$$

Thus, taking into account Eqs. (211) and (212), the definitions of the operators of the Cartesian components of the spin $\hat{S}_X$, $\hat{S}_Y$, and $\hat{S}_Z$ via the polarization operators $\hat{T}_{10}^{(S)}$ and $\hat{T}_{1\pm1}^{(S)}$, Eq. (A21) from Appendix A, and the definitions of the spherical harmonics $Y_{10}(\vartheta,\varphi)$ and $Y_{1\pm1}(\vartheta,\varphi)$, Eq. (B5) from Appendix B, we have the matrix elements of $\hat{S}_X$, $\hat{S}_Y$, and $\hat{S}_Z$ in the coherent state representation as

$$\langle S,\vartheta,\varphi|\hat{S}_Z|S,\vartheta,\varphi\rangle = S\cos\vartheta, \qquad (213)$$

$$\langle S,\vartheta,\varphi|\hat{S}_X|S,\vartheta,\varphi\rangle = S\sin\vartheta\cos\varphi, \qquad (214)$$

$$\langle S,\vartheta,\varphi|\hat{S}_Y|S,\vartheta,\varphi\rangle = S\sin\vartheta\sin\varphi, \qquad (215)$$

in turn yielding the matrix elements of the spin operator $\hat{\mathbf{S}}$ as

$$\langle S,\vartheta,\varphi|\hat{\mathbf{S}}|S,\vartheta,\varphi\rangle = S\mathbf{u}. \qquad (216)$$

Hence the spin operator $\hat{\mathbf{S}}$ in the coherent state representation is the direct quantum analogue of the classical magnetic dipole vector $\boldsymbol{\mu} = \mu\mathbf{u}$.

Now, we shall use the coherent spin state representation of the spin density matrix to derive transformation kernels for the spin $Q$-, $P$-, and Wigner quasiprobability distributions. Proceeding we first consider the so-called *Q-function representation*. The $Q$-function representation distribution corresponding to a spin density matrix $\hat{\rho}_S$ in Hilbert space may be defined in a manner analogous to that for particles [cf. Eq. (107) with $s=-1$] via the *diagonal* matrix elements of the density matrix in the spin coherent state representation [51]. Thus, we may then form the matrix elements $\langle S,\vartheta,\varphi|\hat{\rho}_S|S,\vartheta,\varphi\rangle$ of the density matrix $\hat{\rho}_S$ as defined by the polarization operator expansion Eq. (70) in the spin coherent state representation $|S,\vartheta,\varphi\rangle$ as the linear combination [mindful that in Eq. (70) $a_{L,M}(t) = \langle\hat{T}_{LM}^{\dagger(S)}\rangle(t)$]

$$\langle S,\vartheta,\varphi|\hat{\rho}_S|S,\vartheta,\varphi\rangle = \sum_{L=0}^{2S}\sum_{M=-L}^{L}\langle S,\vartheta,\varphi|\hat{T}_{LM}^{(S)}|S,\vartheta,\varphi\rangle\langle\hat{T}_{LM}^{\dagger(S)}\rangle(t), \qquad (217)$$

where the $\langle S,\vartheta,\varphi|\hat{T}_{LM}^{(S)}|S,\vartheta,\varphi\rangle$ are given by Eq. (211). Then taking account of the replacement prescription for averages of polarization operators of the density matrix $\hat{\rho}_S$ embodied in Eq. (197) and using (211), we have from Eq. (217) the matrix elements $\langle S,\vartheta,\varphi|\hat{\rho}_S|S,\vartheta,\varphi\rangle$ of $\hat{\rho}_S$ rendered as the finite sum of spherical harmonics

$$\langle S,\vartheta,\varphi|\hat{\rho}_S|S,\vartheta,\varphi\rangle = \sqrt{\frac{4\pi}{2S+1}}\sum_{L=0}^{2S}\sum_{M=-L}^{L}\left(\Omega_{L,M}^S\right)^{-1}C_{SSL0}^{SS}Y_{LM}(\vartheta,\varphi)\langle Y_{LM}^*\rangle(t). \quad (218)$$

Now, if we let the prefactor

$$\Omega_{L,M}^S = \sqrt{\frac{4\pi}{2S+1}}C_{SSL0}^{SS}, \quad (219)$$

then in accordance with the representation distribution Eq. (194) we have the final form of the matrix elements of the density operator in the spin coherent state representation

$$\langle S,\vartheta,\varphi|\hat{\rho}_S|S,\vartheta,\varphi\rangle = \sum_{L=0}^{2S}\sum_{M=-L}^{L}Y_{LM}(\vartheta,\varphi)\langle Y_{LM}^*\rangle(t), \quad (220)$$

that is, we have with the formal Eq. (194) the phase space representation distribution $W_S^Q(\vartheta,\varphi,t)$ defined as

$$W_S^Q(\vartheta,\varphi,t) = \frac{4\pi}{2S+1}\langle S,\vartheta,\varphi|\hat{\rho}_S|S,\vartheta,\varphi\rangle.$$

This quasiprobability distribution function $W_S^Q(\vartheta,\varphi,t)$ (called the Q-function) thus comprises the *diagonal* matrix elements of the density operator $\hat{\rho}_S$ in the *spin coherent state representation* and is finally given explicitly (returning to averages of the polarization operators) by

$$W_S^Q(\vartheta,\varphi,t) = \sqrt{\frac{4\pi}{2S+1}}\sum_{L=0}^{2S}\sum_{M=-L}^{L}C_{SSL0}^{SS}Y_{LM}(\vartheta,\varphi)\langle \hat{T}_{LM}^{\dagger(S)}\rangle(t). \quad (221)$$

The normalization of $W_S^Q(\vartheta,\varphi,t)$ in Eq. (221) can be found because by orthogonality we have

$$\int_0^\pi\int_0^{2\pi}W_S^Q(\vartheta,\varphi,t)\sin\vartheta d\vartheta d\varphi = \frac{4\pi}{2S+1}. \quad (222)$$

Next, we consider the *P-function* $W_S^P(\vartheta,\varphi,t)$, which is defined in terms of the density matrix $\hat{\rho}_S$ in the coherent state representation via the inverse Wigner-Stratonovich map [51]

$$\hat{\rho}_S(t) = \frac{2S+1}{4\pi}\int_0^\pi\int_0^{2\pi}W_S^P(\vartheta,\varphi,t)|S,\vartheta,\varphi\rangle\langle S,\vartheta,\varphi|\sin\vartheta d\vartheta d\varphi. \quad (223)$$

However, the definition Eq. (223) implies that the average polarization operator

$$\langle \hat{T}_{LM}^{\dagger(S)}\rangle(t) = \mathrm{Tr}\left(\hat{\rho}_S(t)\hat{T}_{LM}^{\dagger(S)}\right)$$

is given by

$$\langle \hat{T}_{LM}^{\dagger(S)}\rangle(t) = \frac{2S+1}{4\pi}\int_0^\pi\int_0^{2\pi}W_S^P(\vartheta,\varphi,t)\mathrm{Tr}\left(|S,\vartheta,\varphi\rangle\langle S,\vartheta,\varphi|\hat{T}_{LM}^{\dagger(S)}\right)\sin\vartheta d\vartheta d\varphi$$

$$= \frac{2S+1}{4\pi}\int_0^\pi\int_0^{2\pi}W_S^P(\vartheta,\varphi,t)\langle S,\vartheta,\varphi|\hat{T}_{LM}^{\dagger(S)}|S,\vartheta,\varphi\rangle\sin\vartheta d\vartheta d\varphi \quad (224)$$

$$= \sqrt{\frac{2S+1}{4\pi}}C_{SSL0}^{SS}\int_0^\pi\int_0^{2\pi}W_S^P(\vartheta,\varphi,t)Y_{LM}^*(\vartheta,\varphi)\sin\vartheta d\vartheta d\varphi$$

$$= \sqrt{\frac{4\pi}{2S+1}}C_{SSL0}^{SS}\langle Y_{LM}^*\rangle(t).$$

Here we have used the fact that

$$\mathrm{Tr}\left(|S,\vartheta,\varphi\rangle\langle S,\vartheta,\varphi|\hat{T}_{LM}^{\dagger(S)}\right) = \langle S,\vartheta,\varphi|\hat{T}_{LM}^{\dagger(S)}|S,\vartheta,\varphi\rangle,$$

which in turn based on Eqs.(A17), (A27), and (A34) from Appendix A and the identity Eq.(211). Comparing Eq. (224) and the general prescription given by Eq. (197) then yields in this case

$$\Omega_{L,M}^S = \left(\sqrt{\frac{4\pi}{2S+1}}C_{SSL0}^{SS}\right)^{-1}. \quad (225)$$

For purposes of symmetry and in order to satisfy the normalization condition Eq. (222), we define the *P*-function $W_S^P(\vartheta,\varphi,t)$ in a form like Eq. (221)

$$W_S^P(\vartheta,\varphi,t) = \sqrt{\frac{4\pi}{2S+1}}\sum_{L=0}^{2S}\sum_{M=-L}^{L}\left(C_{SSL0}^{SS}\right)^{-1}Y_{LM}(\vartheta,\varphi)\langle \hat{T}_{LM}^{\dagger(S)}\rangle(t) \quad (226)$$

which differs from Eq. (221) only because of the reciprocal of the Clebsch-Gordan coefficients involved in the summand.

Finally, the *Wigner function for spins* is defined as the function, which is its *own conjugate*, i.e., the function obtained by setting [51]

$$\Omega_{L,M}^S = 1. \quad (227)$$

Again for purposes of symmetry and in order to satisfy the normalization relation Eq. (222), we define the Wigner function $W_S^W(\vartheta,\varphi,t)$ as

$$W_S^W(\vartheta,\varphi,t) = \sqrt{\frac{4\pi}{2S+1}}\sum_{L=0}^{2S}\sum_{M=-L}^{L}Y_{LM}(\vartheta,\varphi)\langle \hat{T}_{LM}^{\dagger(S)}\rangle(t). \quad (228)$$

Now, by introducing the parameter $s$, which takes the values $s = -1, 0, +1$ for $Q$-, Wigner-, and P-functions, respectively, we can finally rewrite Eqs. (221), (226), and (228) as the single mapping [59]

$$W_S^{(s)}(\vartheta,\varphi,t) = \mathrm{Tr}\{\hat{\rho}_S(t)\hat{w}_s(\vartheta,\varphi)\}$$

$$= \sqrt{\frac{4\pi}{2S+1}}\sum_{L=0}^{2S}\sum_{M=-L}^{L}\left(C_{SSL0}^{SS}\right)^{-s}Y_{LM}(\vartheta,\varphi)\langle \hat{T}_{LM}^{\dagger(S)}\rangle(t), \quad (229)$$

where $\hat{w}_s(\vartheta,\varphi)$ is the Wigner-Stratonovich kernel of the bijective transformation given by

$$\hat{w}_s(\vartheta,\varphi)=\sqrt{\frac{4\pi}{2S+1}}\sum_{L=0}^{2S}\sum_{M=-L}^{L}\left(C_{SSL0}^{SS}\right)^{-s}Y_{LM}^{*}(\vartheta,\varphi)\hat{T}_{LM}^{(S)} \qquad (230)$$

such that [cf. Eqs. (175)-(180)] $\hat{w}_s(\vartheta,\varphi)=\hat{w}_s^{\dagger}(\vartheta,\varphi)$, $\text{Tr}(\hat{w}_s)=1$, and

$$\frac{2S+1}{4\pi}\int_0^{\pi}\int_0^{2\pi}\hat{w}_s(\vartheta,\varphi)\sin\vartheta\, d\vartheta\, d\varphi=\hat{I}.$$

Now, due to the orthogonality properties of the spherical harmonics Eq. (B11) from Appendix B, all the phase-space distributions embodied in Eq. (229) can also be written in compact form as the finite series

$$\frac{2S+1}{4\pi}W_S^{(s)}(\vartheta,\varphi,t)=\sum_{L=0}^{2S}\sum_{M=-L}^{L}Y_{LM}(\vartheta,\varphi)\left\langle Y_{LM}^{*}\right\rangle^{(s)}(t), \qquad (231)$$

where

$$\left\langle Y_{LM}^{*}\right\rangle^{(s)}(t)=\frac{2S+1}{4\pi}\int_0^{\pi}\int_0^{2\pi}Y_{LM}^{*}(\vartheta,\varphi)W_S^{(s)}(\vartheta,\varphi,t)\sin\vartheta\, d\vartheta\, d\varphi$$
$$=\sqrt{\frac{4\pi}{2S+1}}\left(C_{SSL0}^{SS}\right)^{-s}\left\langle \hat{T}_{LM}^{\dagger(S)}\right\rangle(t). \qquad (232)$$

The quantum distribution function $(2S+1)W_S^{(s)}(\vartheta,\varphi,t)/4\pi$ given by the general finite series of Eq. (231) clearly has a similar form to the Fourier expansion of the classical orientational distribution $W(\vartheta,\varphi,t)$, Eq. (184), and reduces to it in the classical limit, $S\to\infty$. Furthermore, the quantum distribution Eq. (231) in representation space is a general result valid for an *arbitrary* spin system described by a spin density matrix $\hat{\rho}_S(t)$. Therefore, the phase space representation allows one (as already discussed for particles) to describe spin systems via a *quasiprobability density function* $W_S^{(s)}(\vartheta,\varphi,t)$ of spin orientations in the phase (here configuration) space $(\vartheta,\varphi)$. The advantage of such a mapping of the density matrix onto a $c$-number quasiprobability density function $W_S^{(s)}(\vartheta,\varphi,t)$ as extensively used in quantum optics (see, e.g., Refs. 45 and 46) is that it is possible to learn *how $W_S^{(s)}(\vartheta,\varphi,t)$ evolves as a function of $S$*. In addition for large spins, $(2S+1)W_S^{(s)}(\vartheta,\varphi,t)/4\pi$ as defined by Eq. (231) reduces to the classical distribution $W(\vartheta,\varphi,t)$, Eq. (184), thereby naturally linking the quantum and classical regimes.

Now, knowing the Wigner-Stratonovich kernel $\hat{w}_s(\vartheta,\varphi)$ of the bijective map, we can calculate the Weyl symbol $A^{(s)}(\vartheta,\varphi)$ of any spin operator $\hat{A}$ as

$$A^{(s)}(\vartheta,\varphi)=\text{Tr}\left(\hat{A}\hat{w}_s(\vartheta,\varphi)\right). \qquad (233)$$

Conversely, as we shall immediately justify via Eqs. (236) - (238) below, the operator $\hat{A}$ can be reconstructed from its Weyl symbol $A^{(s)}(\vartheta,\varphi)$ via the inverse Wigner-Stratonovich map

$$\hat{A}=\frac{2S+1}{4\pi}\int_0^{\pi}\int_0^{2\pi}A^{(-s)}(\vartheta,\varphi)\hat{w}_s(\vartheta,\varphi)\sin\vartheta\, d\vartheta\, d\varphi. \qquad (234)$$

In particular, via the inverse Eq. (234) the density matrix $\hat{\rho}_S(t)$ (like any other quantum operator) may then be neatly expressed as

$$\hat{\rho}_S(t)=\frac{2S+1}{4\pi}\int_0^{\pi}\int_0^{2\pi}W_S^{(-s)}(\vartheta,\varphi,t)\hat{w}_s(\vartheta,\varphi)\sin\vartheta\, d\vartheta\, d\varphi. \qquad (235)$$

Thus knowing the phase-space distribution $W_S^{(s)}(\vartheta,\varphi,t)$, the density matrix $\hat{\rho}_S(t)$ in Hilbert space can then be directly reconstructed via the particular inverse Wigner-Stratonovich map Eq. (235). This procedure may be justified, as we shall now prove, because the Weyl symbols of any two spin operators $\hat{A}$ and $\hat{B}$ will provide the overlap relation [cf. Eq. (128)]

$$\frac{2S+1}{4\pi}\int_0^{\pi}\int_0^{2\pi}A^{(s)}(\vartheta,\varphi)B^{(-s)}(\vartheta,\varphi)\sin\vartheta\, d\vartheta\, d\varphi=\text{Tr}\left(\hat{A}\hat{B}\right), \qquad (236)$$

leading to Eqs. (234) and (235). To justify Eq. (236), we substitute the Weyl symbols of the operators $\hat{A}$ and $\hat{B}$ defined by Eq. (233) into Eq. (236), then we use the series expression for the kernel Eq. (230) of the bijective map and the orthogonality property of spherical harmonics Eq. (B11) from Appendix B. Thus, we have from Eq. (233) and (236)

$$\frac{2S+1}{4\pi}\int_0^{\pi}\int_0^{2\pi}A^{(s)}(\vartheta,\varphi)B^{(-s)}(\vartheta,\varphi)\sin\vartheta\, d\vartheta\, d\varphi$$
$$=\sum_{L,L'=0}^{2S}\sum_{M,M'=-L}^{L}\left(\frac{C_{SSL'0}^{SS}}{C_{SSL0}^{SS}}\right)^s\text{Tr}\left(\hat{A}\hat{T}_{LM}^{(S)}\right)\text{Tr}\left(\hat{B}\hat{T}_{L'M'}^{\dagger(S)}\right)$$
$$\times\int_0^{\pi}\int_0^{2\pi}Y_{LM}^{*}(\vartheta,\varphi)Y_{L'M'}(\vartheta,\varphi)\sin\vartheta\, d\vartheta\, d\varphi$$
$$=\sum_{L=0}^{2S}\sum_{M=-L}^{L}\text{Tr}\left(\hat{A}\hat{T}_{LM}^{(S)}\right)\text{Tr}\left(\hat{B}\hat{T}_{LM}^{\dagger(S)}\right) \qquad (237)$$
$$=\sum_{L=0}^{2S}\sum_{M=-L}^{L}\sum_{m,n,i,j=-S}^{S}A_{mi}B_{nj}\left[\hat{T}_{LM}^{(S)}\right]_{im}\left[\hat{T}_{LM}^{\dagger(S)}\right]_{jn}$$
$$=\sum_{m,n,i,j=-S}^{S}A_{mi}B_{nj}\sum_{L=0}^{2S}\sum_{M=-L}^{L}(-1)^M\frac{2L+1}{2S+1}C_{SnL-M}^{Sj}C_{SmLM}^{Si}$$
$$=\sum_{m,n=-S}^{S}A_{mn}B_{nm}=\text{Tr}\left(\hat{A}\hat{B}\right).$$

Here we have used Eq. (A17) from Appendix A for the matrix elements of the polarization operators $\left[\hat{T}_{LM}^{(S)}\right]_{im}$ and $\left[\hat{T}_{LM}^{\dagger(S)}\right]_{jn}$ as well as a property of Clebsch-Gordan coefficients [95], viz.,

$$\sum_{L=0}^{2S}\sum_{M=-L}^{L}(-1)^M\frac{2L+1}{2S+1}C_{SnL-M}^{Sj}C_{SmLM}^{Si}$$
$$=\sum_{L=0}^{2S}\sum_{M=-L}^{L}C_{S-jSn}^{LM}C_{S-mSi}^{LM}=\delta_{in}\delta_{mj}.$$

Thus, according to Eq. (236), with obvious replacements the *average* value $\langle \hat{A} \rangle = \text{Tr}(\hat{\rho}_S \hat{A})$ of a spin operator $\hat{A}$ is given by the integral

$$\langle \hat{A} \rangle = \frac{2S+1}{4\pi} \int_0^\pi \int_0^{2\pi} A^{(-s)}(\vartheta,\varphi) W_S^{(s)}(\vartheta,\varphi,t) \sin\vartheta \, d\vartheta \, d\varphi, \qquad (238)$$

thereby transparently leading to Eqs. (234) and (235). By definition, the observable $\langle \hat{A} \rangle$ calculated from Eq. (238) is *independent* of the value of the parameter $s$ chosen, i.e., the Q-, Wigner-, and P-functions will all yield identical results for $\langle \hat{A} \rangle$ as they must do.

In summary, the one to one correspondence between the quantum state in the Hilbert space and a real representation space function first envisaged for the closed system in the spin context by Stratonovich [49], formally represents the quantum mechanics of a spin as a statistical theory in the representation space of polar angles $(\vartheta,\varphi)$. This is accomplished essentially in the manner of Wigner [41] who we recall formally represented the quantum mechanics of a particle with Hamiltonian

$$\hat{H} = \frac{1}{2m}\hat{p}^2 + V(\hat{q})$$

as a statistical theory in phase space with the canonical variables $(p,q)$. Clearly, the average value of a quantum spin operator may be calculated from Eq. (238) like in classical mechanics. Thus the Stratonovich representation for spins [49] just as the Wigner representation for particles is well suited to the development of semiclassical methods of solution allowing one to obtain quantum corrections for finite $S$ in a manner closely analogous to the classical case, $S \to \infty$ (see, e.g., [5]). We emphasize that besides spin relaxation of assemblies of noninteracting spins in contact with the thermal bath, the phase-space formalism can also be applied to related problems such as spin waves, interacting spins with Heisenberg coupling, etc. (see e.g., [51,52,142]).

### 2. Weyl symbols of some spin operators

Any spin operator $\hat{A}$ is associated via Eq. (238) with its Weyl symbol (c-number function) $A^{(s)}(\vartheta,\varphi)$ in the representation space. In this section, we evaluate the Weyl symbols of the spin operators $\hat{S}_X$, $\hat{S}_Y$, $\hat{S}_Z$, and $\hat{\mathbf{S}}$ and also those of some other model spin Hamiltonians $\hat{H}_S$. Using the definition of the spin operators in terms of the polarization operators given by Eq. (A21) from Appendix A, the Weyl symbols $S_X^{(s)}(\vartheta,\varphi)$, $S_Y^{(s)}(\vartheta,\varphi)$, $S_Z^{(s)}(\vartheta,\varphi)$, and $\mathbf{S}^{(s)}(\vartheta,\varphi)$ of the corresponding spin operators $\hat{S}_X$, $\hat{S}_Y$, $\hat{S}_Z$, and $\hat{\mathbf{S}}$ can be calculated from the mapping Eq. (233) for $s=0,\pm 1$ via the orthogonality property of the polarization operators defined by Eq. (A34). Thus, we have the simple maps from Hilbert space onto phase space [59] [cf. Eqs. (213)-(215)]

$$\hat{S}_X \to S_X^{(s)}(\vartheta,\varphi) = \sqrt{\frac{2\pi S(S+1)}{3}} \sum_{L=0}^{2S} \sum_{M=-L}^{L} \left(C_{SSL0}^{SS}\right)^{-s} Y_{LM}^*(\vartheta,\varphi) \text{Tr}\left\{(\hat{T}_{1-1}^{(S)} - \hat{T}_{11}^{(S)})\hat{T}_{LM}^{(S)}\right\}$$

$$= \sqrt{\frac{2\pi}{3}} S\left(\frac{S+1}{S}\right)^{(s+1)/2} \left[Y_{1-1}(\vartheta,\varphi) - Y_{11}(\vartheta,\varphi)\right] \qquad (239)$$

$$= S\left(\frac{S+1}{S}\right)^{(s+1)/2} \sin\vartheta\cos\varphi,$$

$$\hat{S}_Y \to S_Y^{(s)}(\vartheta,\varphi) = i\sqrt{\frac{2\pi S(S+1)}{3}} \sum_{L=0}^{2S} \sum_{M=-L}^{L} \left(C_{SSL0}^{SS}\right)^{-s} Y_{LM}^*(\vartheta,\varphi) \text{Tr}\left\{(\hat{T}_{1-1}^{(S)} + \hat{T}_{11}^{(S)})\hat{T}_{LM}^{(S)}\right\}$$

$$= i\sqrt{\frac{2\pi}{3}} S\left(\frac{S+1}{S}\right)^{(s+1)/2} \left[Y_{1-1}(\vartheta,\varphi) + Y_{11}(\vartheta,\varphi)\right] \qquad (240)$$

$$= S\left(\frac{S+1}{S}\right)^{(s+1)/2} \sin\vartheta\sin\varphi,$$

$$\hat{S}_Z \to S_Z^{(s)}(\vartheta,\varphi) = \sqrt{\frac{4\pi S(S+1)}{3}} \sum_{L=0}^{2S} \sum_{M=-L}^{L} \left(C_{SSL0}^{SS}\right)^{-s} Y_{LM}^*(\vartheta,\varphi) \text{Tr}\left\{\hat{T}_{10}^{(S)}\hat{T}_{LM}^{(S)}\right\}$$

$$= \sqrt{\frac{4\pi}{3}} S\left(\frac{S+1}{S}\right)^{(s+1)/2} Y_{10}(\vartheta,\varphi) \qquad (241)$$

$$= S\left(\frac{S+1}{S}\right)^{(s+1)/2} \cos\vartheta.$$

Clearly, these phase space mappings bear a close resemblance to the corresponding classical quantities. Moreover, we have the overall compact form of the phase-space mapping of the spin operator $\hat{\mathbf{S}}$, viz.,

$$\hat{\mathbf{S}} \to \mathbf{S}^{(s)}(\vartheta,\varphi) = S\left(\frac{S+1}{S}\right)^{(s+1)/2} \begin{pmatrix} \sin\vartheta\cos\varphi \\ \sin\vartheta\sin\varphi \\ \cos\vartheta \end{pmatrix}$$

$$= S\left(\frac{S+1}{S}\right)^{(s+1)/2} \mathbf{u}. \qquad (242)$$

Next since the magnetic moment operator $\hat{\boldsymbol{\mu}}$ is defined via the spin operator $\hat{\mathbf{S}}$ as $\hat{\boldsymbol{\mu}} = \gamma\hbar\hat{\mathbf{S}}$, it then follows that the *Weyl symbol* $\boldsymbol{\mu}^{(s)} = \mu^{(s)}\mathbf{u}$ *of* $\hat{\boldsymbol{\mu}}$ *has essentially the form of the magnetic moment vector* $\boldsymbol{\mu}$ *for a classical spin* $[\mu^{(s)} = \gamma\hbar S(1+S^{-1})^{(s+1)/2}]$.

As further examples of the mapping procedure onto phase space, we evaluate the Weyl symbols for the uniaxial, biaxial, cubic, and mixed anisotropy Hamiltonians defined, respectively, as

$$\beta\hat{H}_S^{un} = -\frac{\sigma}{S^2}\hat{S}_Z^2, \qquad (243)$$

$$\beta\hat{H}_S^{bi} = -\frac{\sigma}{S^2}\hat{S}_Z^2 + \frac{\delta}{S^2}\left(\hat{S}_X^2 - \hat{S}_Y^2\right), \qquad (244)$$

$$\beta \hat{H}_S^{cub} = -\frac{\sigma_c}{2S^4}\left(\hat{S}_X^4 + \hat{S}_Y^4 + \hat{S}_Z^4\right), \tag{245}$$

$$\beta \hat{H}_S^{mix} = -\frac{\sigma_1}{S^2}\hat{S}_Z^2 - \frac{\sigma_2}{S^4}\hat{S}_Z^4 + \frac{\chi}{S^4}\left(\hat{S}_{+1}^4 + \hat{S}_{-1}^4\right), \tag{246}$$

where $\sigma$, $\delta$, $\sigma_c$, $\sigma_1$, $\sigma_2$, and $\chi$ are dimensionless anisotropy parameters. The uniaxial Hamiltonian Eq. (243) is commonly used, e.g., to describe the magnetic properties of the dodecanuclear manganese molecular cluster Mn12 with $S = 10$, $\sigma_1 T / S^2 = 0.6 \div 0.7\,\text{K}$ [143]. The biaxial anisotropy Hamiltonian $\hat{H}_S^{bi}$, Eq. (244), is commonly used to describe the magnetic properties of an octanuclear iron(III) molecular cluster Fe8 [9,144] with $S = 10$, $\sigma T / S^2 = 0.275\,\text{K}$ and $\delta T /(S^2) = 0.046\,\text{K}$. The cubic anisotropy Hamiltonian Eq. (245) contributes to the mixed anisotropy Hamiltonian Eq. (246), which is commonly used, e.g., to describe more accurately the magnetic properties of the dodecanuclear manganese molecular cluster Mn12 with $S = 10$, $\sigma_1 /(\beta S^2) = 0.56\,K$, $\sigma_2 /(\beta S^4) = 1.1 \cdot 10^{-3}\,K$, and $\chi /(\beta S^4) = \pm 3 \cdot 10^{-5}\,K$ [144].

Here for simplicity, we only evaluate the Weyl symbols for the Hamiltonians Eqs. (243)-(246) for the Q-function corresponding to $s = -1$ (for $s = 0$ and $s = +1$ the calculations can be accomplished in like manner). Thus, the Weyl symbols $H_S^{un}(\vartheta,\varphi)$, $H_S^{bi}(\vartheta,\varphi)$, $H_S^{cub}(\vartheta,\varphi)$, and $H_S^{mix}(\vartheta,\varphi)$ corresponding to the Hamiltonians Eqs. (243)-(246) can now be calculated from the general finite series representation of the kernel $\hat{w}_s$ [cf. Eq. (230)]

$$H_S^{(-1)}(\vartheta,\varphi) = \text{Tr}\left\{\hat{H}_S \hat{w}_{-1}(\vartheta,\varphi)\right\}$$
$$= \sqrt{\frac{4\pi}{2S+1}} \text{Tr}\left\{\hat{H}_S \sum_{L=0}^{2S}\sum_{M=-L}^{L} C_{SSL0}^{SS} Y_{LM}^*(\vartheta,\varphi)\hat{T}_{LM}^{(S)}\right\} \tag{247}$$

[cf. the mapping Eq. (233) for the phase-space representation of an arbitrary spin operator $\hat{A}$]. Hence, we obtain from the Hamiltonians given by Eqs. (243)-(246) and the general mapping Eq. (247) after some algebra involving both products of the polarization operators and their orthogonality relations as described in Appendix A the following explicit maps onto phase space

$$\beta \hat{H}_S^{un} \underset{s=-1}{\to} \beta H_S^{un}(\vartheta,\varphi) = -\sigma \frac{S-1/2}{S}\cos^2\vartheta - \frac{\sigma}{2S}, \tag{248}$$

$$\beta \hat{H}_S^{bi} \underset{s=-1}{\to} \beta H_S^{bi}(\vartheta,\varphi)$$
$$= -\frac{S-1/2}{S}\left(\sigma\cos^2\vartheta - \delta\cos 2\varphi \sin^2\vartheta\right) - \frac{\sigma}{2S}, \tag{249}$$

$$\beta \hat{H}_S^{cub} \underset{s=-1}{\to} \beta H_S^{cub}(\vartheta,\varphi)$$
$$= -\sigma_c \frac{2S^3 + 3S - 1}{4S^3} \tag{250}$$
$$+\sigma_c \frac{(S-1/2)(S-1)(S-3/2)}{4S^3}\left(\sin^2 2\vartheta + \sin^4\vartheta \sin^2 2\varphi\right),$$

$$\beta \hat{H}_S^{mix} \underset{s=-1}{\to} \beta H_S^{mix}(\vartheta,\varphi)$$
$$= -\frac{1}{4S}\left(2\sigma_1 + \sigma_2 \frac{3S-1}{S^2}\right)$$
$$-\frac{S-1/2}{S}\left(\sigma_1 + \sigma_2 \frac{3S-2}{S^2}\right)\cos^2\vartheta \tag{251}$$
$$-\frac{(S-1/2)(S-1)(S-3/2)}{2S^3}\left[2\sigma_2 \cos^4\vartheta - \chi \sin^4\vartheta \cos 4\varphi\right].$$

Again, the Weyl symbols of these quantum Hamiltonians bear a close resemblance to the classical free energies of the corresponding magnetocrystalline anisotropies (see, e.g., [6]). All these Weyl symbols will be used below.

*3. Master equation and statistical moment equations for spin relaxation in phase space*

By transforming the reduced density operator evolution Eq. (40) into phase space via the Wigner-Stratonovich map Eqs. (229) and (235), the phase space evolution (master) equation for $W_S^{(s)}(\vartheta,\varphi,t)$ may be formally written as

$$\frac{\partial W_S^{(s)}}{\partial t} = \text{L}_S W_S^{(s)}, \tag{252}$$

where $\text{L}_S$ is the phase-space differential operator *corresponding* to the operator in Hilbert space

$$-\frac{i}{\hbar}[\hat{H}_S,\hat{\rho}_S] + \text{St}(\hat{\rho}_S)$$

in the density matrix evolution Eq. (40). Although the operator $\text{L}_S$ will have, in general, a very complicated form even for axial symmetry, nevertheless the phase space master Eq. (252) still has some obvious advantages over the density matrix evolution Eq. (40) because it is now possible to treat the spin relaxation for arbitrary $S$ like that of classical spins [e.g., Eqs. (239)-(242)] (see Appendix C for a specific nontrivial example). Indeed, in the classical limit, the phase-space evolution equation (252) reduces to the Fokker-Planck equation (3) for the distribution function $W_{cl}(\vartheta,\varphi,t)$ of the orientations of classical spins thereby naturally linking the quantum and classical regimes. We reiterate that the analogy between the quantum and classical formulations for spins again enables powerful methods of solution of classical Fokker-Planck equations for the rotational Brownian motion of classical magnetic dipoles (e.g., continued fractions, mean first passage times, etc. [5]) to be used in the quantum domain [62-70].

As shown in Sec. II.A.4, the formal solution of the evolution Eq. (40) for the reduced density matrix can be written using the polarization operators $\hat{T}_{LM}^{(S)}$ and the statistical moments $\left\langle \hat{T}_{LM}^{\dagger(S)} \right\rangle(t)$ as the finite linear combination given by Eq. (70) [30,95]. The statistical moments $\left\langle \hat{T}_{LM}^{\dagger(S)} \right\rangle(t)$ can then be evaluated (usually after lengthy operator algebra) from the differential-recurrence Eq. (77). Now the statistical moment method can also be applied in analogous fashion to the phase space master equation (252) because the phase-space distribution $W_S^{(s)}(\vartheta,\varphi,t)$ may be written for arbitrary $S$ in terms of a finite linear combination of the spherical harmonics, Eq. (231). Then by using Eq. (232) relating the average spherical harmonics $\left\langle Y_{LM}^* \right\rangle^{(s)}(t)$ and the average polarization operators $\left\langle \hat{T}_{LM}^{\dagger(S)} \right\rangle(t)$, the differential-recurrence equations for $\left\langle Y_{LM}^* \right\rangle^{(s)}(t)$ can be obtained by simple algebraic transformation from the differential-recurrence Eq. (77) so that the latter becomes

$$\frac{d}{dt}\left\langle Y_{LM}^* \right\rangle^{(s)}(t) = \sum_{L',M'} p_{L'M';LM} \left\langle Y_{L'M'}^* \right\rangle^{(s)}(t). \tag{253}$$

Here the coefficients

$$p_{L'M';LM} = \int_0^\pi \int_0^{2\pi} Y_{LM}(\vartheta,\varphi) \mathrm{L}_S Y_{L'M'}^*(\vartheta,\varphi) \sin\vartheta \, d\vartheta \, d\varphi \tag{254}$$

by definition constitute the matrix elements of the phase-space operator $\mathrm{L}_S$ given by

$$p_{L'M';LM} = \left( \frac{C_{SSL0}^{SS}}{C_{SSL'0}^{SS}} \right)^{-s} g_{L'M';LM}, \tag{255}$$

where the coefficients $g_{L'M';LM}$ which depend upon the precise form of the Hamiltonian $\hat{H}_S$ are defined by the averages indicated by Eq. (75). For classical spins, $S\to\infty$, the explicit equation for $p_{L'M';LM}$ for an arbitrary free energy has been derived in Ref. 145 (see Appendix D). Equation (253) is just a phase space correspondent of Eq. (77) which governs the evolution of the average polarization operators $\left\langle \hat{T}_{LM}^{\dagger(S)} \right\rangle(t)$. Now Eq. (253) written as a matrix differential equation can be solved either by direct matrix diagonalization, involving the calculation of the eigenvalues and eigenvectors of the system matrix, or by the computationally efficient (matrix) continued fraction method [5,71]. We remark that due to the identity $Y_{lm}^* = (-1)^m Y_{l-m}$ [95], the conjugate Eq. (253) can also be rewritten as an evolution equation for the statistical moments $\left\langle Y_{LM} \right\rangle^{(s)}(t)$, viz.,

$$\frac{d}{dt}\left\langle Y_{LM} \right\rangle^{(s)}(t) = \sum_{L',M'} (-1)^{M'-M} p_{L'-M';L-M} \left\langle Y_{L'M'} \right\rangle^{(s)}(t). \tag{256}$$

Now the phase-space evolution Eq. (253) for the average spherical harmonics $\left\langle Y_{LM}^* \right\rangle^{(s)}(t)$ and its correspondent Eq. (77) for the average polarization operators $\left\langle \hat{T}_{LM}^{\dagger(S)} \right\rangle(t)$ are equivalent. Thus having determined $\left\langle Y_{LM}^* \right\rangle^{(s)}(t)$ from the phase-space Eqs. (231) and (252), we can also evaluate the density matrix $\hat{\rho}$ from the polarization operator expansion Eqs. (70) and (71) without formally solving its evolution Eq. (40). Vice versa, having calculated $\left\langle \hat{T}_{LM}^{\dagger(S)} \right\rangle(t)$ from the density matrix Eqs. (40) and (70), we also have the phase-space distribution $W_S^{(s)}(\vartheta,\varphi,t)$ from Eqs. (231) and (232) without solving the phase space evolution Eq. (252).

According to the finite series phase space representation Eq. (231), all the statistical moments $\left\langle Y_{LM} \right\rangle^{(s)}(t)$ are required (in general) to evaluate the phase-space distribution $W_S^{(s)}(\vartheta,\varphi,t)$ for given $S$. However, for the calculation of *particular* observables only a *few* moments may in practice be necessary. For example, in evaluating the *average* spin operators $\left\langle \hat{S}_X \right\rangle(t)$, $\left\langle \hat{S}_Y \right\rangle(t)$, and $\left\langle \hat{S}_Z \right\rangle(t)$ only the spherical harmonic averages $\left\langle Y_{10} \right\rangle^{(s)}(t)$ and $\left\langle Y_{1\pm1} \right\rangle^{(s)}(t)$ are required according to the Weyl symbols Eqs. (239)-(241), namely, [cf. Eqs. (84)-(86) for these averages in terms of polarization operators]

$$\left\langle \hat{S}_X \right\rangle(t) = \sqrt{\frac{2\pi}{3}} S \left( \frac{S+1}{S} \right)^{(s+1)/2} \left[ \left\langle Y_{1-1} \right\rangle^{(s)}(t) - \left\langle Y_{11} \right\rangle^{(s)}(t) \right], \tag{257}$$

$$\left\langle \hat{S}_Y \right\rangle(t) = i\sqrt{\frac{2\pi}{3}} S \left( \frac{S+1}{S} \right)^{(s+1)/2} \left[ \left\langle Y_{1-1} \right\rangle^{(s)}(t) + \left\langle Y_{11} \right\rangle^{(s)}(t) \right], \tag{258}$$

$$\left\langle \hat{S}_Z \right\rangle(t) = \sqrt{\frac{4\pi}{3}} S \left( \frac{S+1}{S} \right)^{(s+1)/2} \left\langle Y_{10} \right\rangle^{(s)}(t). \tag{259}$$

The above results are formal and general. Specific applications of Eqs. (252)-(256) are given in Sections III.A.2 and IV. Next, we shall demonstrate how the phase space distributions for *particular* spin systems can be determined. Notice that the representation space analysis for spins is intrinsically more complicated than that for particles because spin and polarization operators are involved. Hence, the Wigner correspondents must be evaluated from first principles for a given Hamiltonian as we shall now illustrate.

### D. Equilibrium phase-space distribution functions for spins

Here we shall demonstrate how to obtain both analytically and numerically *equilibrium* time-independent quasiprobability distribution functions for spin systems with various time-independent Hamiltonians $\hat{H}_S$, where the equilibrium density matrix $\hat{\rho}_S$ is given by

$$\hat{\rho}_S = \frac{1}{Z_S} e^{-\beta \hat{H}_S} \qquad (260)$$

with the partition function $Z_S = \text{Tr}\left(e^{-\beta \hat{H}_S}\right)$. We recall that according to Eq. (70) the density matrix $\hat{\rho}_S$ for an arbitrary Hamiltonian $\hat{H}_S$ can be written as a finite series of polarization operators, viz.,

$$\hat{\rho}_S = \sum_{L=0}^{2S} \sum_{M=-L}^{L} \hat{T}_{LM}^{(S)} \left\langle \hat{T}_{LM}^{\dagger(S)} \right\rangle_{eq}, \qquad (261)$$

while the general Eq. (231) yields the corresponding equilibrium phase space distributions $W_S^{(s)}(\vartheta,\varphi)$ as a finite series of spherical harmonics, viz.,

$$\frac{2S+1}{4\pi} W_S^{(s)}(\vartheta,\varphi) = \sum_{L=0}^{2S} \sum_{M=-L}^{L} Y_{LM}(\vartheta,\varphi) \left\langle Y_{LM}^* \right\rangle_{eq}^{(s)}, \qquad (262)$$

where averaged spherical harmonics are related to averaged polarization operators via

$$\left\langle Y_{LM}^* \right\rangle_{eq}^{(s)} = \sqrt{\frac{4\pi}{2S+1}} \left(C_{SSL0}^{SS}\right)^{-s} \left\langle \hat{T}_{LM}^{\dagger(S)} \right\rangle_{eq}. \qquad (263)$$

Hence, $W_S^{(s)}(\vartheta,\varphi)$ can also be written as a finite series of the averaged polarization operators thus it can always be determined (analytically or numerically) for a given Hamiltonian $\hat{H}_S$.

Following Ref. [65,66], we shall evaluate analytically (for small $S$) or numerically (for large $S$) from Eqs. (262) and (263) the equilibrium phase-space distributions $W_S^{(s)}(\vartheta,\varphi)$. In order to implement this procedure for a particular Hamiltonian:

(i) First, we write the density matrix operator $\hat{\rho}_S$ from Eqs. (260) and (261) for the given *effective anisotropy-Zeeman energy* Hamiltonian expressed either in terms of polarization operators $\hat{T}_{L,M}^{(S)}$ (as will be needed to implement step (ii) below) or, using Eqs. (A21) and (A28) from Appendix A, in terms of the spin operators $\hat{S}_i$ ($i = X, Y, Z$).

(ii) Next, we calculate for the given Hamiltonian the averaged polarization operator $\left\langle \hat{T}_{LM}^{\dagger(S)} \right\rangle_{eq} = \text{Tr}\left(\hat{\rho}_S \hat{T}_{LM}^{\dagger(S)}\right)$ using the operator expansion method described in Appendix A [Eq. (A20) *et seq.*].

(iii) Then, we can write from Eq. (263) the Fourier coefficients $\left\langle Y_{LM}^* \right\rangle_{eq}^{(s)}$ connecting the average of a spherical harmonic to that of a polarization operator.

(iv) Thus, we obtain the phase-space distribution $W_S^{(s)}(\vartheta,\varphi)$ for the chosen Hamiltonian from the formal finite series Eq. (262) for any particular $S$.

All the calculations for a given Hamiltonian $\hat{H}_S$, which are tedious, can be accomplished using MATHEMATICA.

Initially we evaluate the equilibrium Wigner ($s = 0$), Q- ($s = -1$) and P- ($s = 1$) phase-space distributions $W_S^{(s)}(\vartheta,\varphi)$ for the very simple case of a spin with spin number $S$ and magnetic moment $\mu = \gamma\hbar S/\mu_0$ in an external *constant* field **H** applied along the Z-axis (essentially these distributions correspond to the usual treatment of quantum paramagnetism). Consequently the spin Hamiltonian $\hat{H}_S$ is just

$$\beta\hat{H}_S = -\frac{\xi}{S} \hat{S}_Z, \qquad (264)$$

where $\xi = \beta\mu_0\mu H$ is the dimensionless external field parameter. Hence we shall see that the Q-function $W_S^{(-1)}(\vartheta,\varphi)$ alone satisfies the *nonnegativity* condition, viz., $W_S^{(-1)} \geq 0$, required of a true probability density function. The quasiprobability densities $W_S^{(1)}$ and $W_S^{(0)}$ do not satisfy this condition (because they may take on negative values). Thus in future determinations of a phase-space representation, we shall usually restrict ourselves to the (Q-) function $W_S^{(-1)}(\vartheta,\varphi)$ as all other functions can be treated in like manner. First we shall determine the equilibrium Q- ($s = -1$) phase-space distributions for an assembly of noninteracting spins in an external *constant* field **H** applied in an *arbitrary* direction in space. Here the Hamiltonian of a spin is

$$\beta\hat{H}_S = -\frac{\xi}{S}\left(\gamma_X \hat{S}_X + \gamma_Y \hat{S}_Y + \gamma_Z \hat{S}_Z\right), \qquad (265)$$

where $\gamma_X, \gamma_Y, \gamma_Z$ are the direction cosines of the field **H**. Next, we shall treat various magnetic anisotropies so establishing one or more preferred orientations of the magnetization of an assembly of spins. In particular, we shall consider a uniaxial paramagnet in an external magnetic field with *arbitrary* orientation so that

$$\beta\hat{H}_S = -\frac{\xi}{S}\left(\gamma_X \hat{S}_X + \gamma_Y \hat{S}_Y + \gamma_Z \hat{S}_Z\right) - \frac{\sigma}{S^2} \hat{S}_Z^2 \qquad (266)$$

as well as two particular cases of Eq. (266), namely, a uniaxial nanomagnet in both a longitudinal and a transverse external field with

$$\beta\hat{H}_S = -\frac{\xi}{S} \hat{S}_Z - \frac{\sigma}{S^2} \hat{S}_Z^2 \qquad (267)$$

and

$$\beta\hat{H}_S = -\frac{\xi}{S} \hat{S}_X - \frac{\sigma}{S^2} \hat{S}_Z^2, \qquad (268)$$

respectively, where $\sigma$ is the dimensionless anisotropy parameter. In the classical limit, $S \to \infty$, the latter Hamiltonian corresponds to the nonaxially symmetric problem of a uniaxial nanomagnet with two equivalent ground states of magnetization separated by a magnetocrystalline anisotropy energy barrier (the $\sigma$ term) in the presence of an applied transverse field. The transverse field in the quantum case will *enhance* the tunneling probability. Finally, we shall consider biaxial and cubic-like systems with Hamiltonians

$$\beta \hat{H}_S = -\frac{\sigma}{S^2}\hat{S}_Z^2 - \frac{\delta}{S^2}(\hat{S}_X^2 - \hat{S}_Y^2), \quad (269)$$

($\delta$ is a dimensionless biaxiality parameter) and

$$\beta \hat{H}_S = -\frac{\xi}{S}\hat{S}_Z - \frac{\sigma_c}{2S^4}(\hat{S}_X^4 + \hat{S}_Y^4 + \hat{S}_Z^4), \quad (270)$$

where $\sigma_c$ is the dimensionless cubic anisotropy parameter, which may be either positive or negative.

Having determined the equilibrium quasiprobability distributions $W_S^{(s)}(\vartheta,\varphi)$ corresponding to these Hamiltonians $\hat{H}_S$, our second purpose is to calculate the magnetization reversal time via the quantum generalization of TST (previously treated for classical spins by Néel [16]) permitting one to estimate temperature effects in the astroids and hysteresis loops within the limitations imposed by quantum TST (moderate damping, etc.) [5,6]. Finally, we shall calculate the Stoner-Wohlfarth magnetization curves (represented in switching field astroid form) as a function of spin number $S$ for nonaxially symmetric potentials. This calculation will generalize Thiaville's geometrical method [19] (for the construction of switching field curves for such potentials) to include quantum effects due to *finite* spin number. Thus, one may study the behavior of the astroids in the interesting magnetic cluster – single domain nanoparticle transition region. In the magnetic context, explicit equations for the equilibrium phase space distributions have already been obtained for an assembly of noninteracting spins in a uniform magnetic field [35,36] and for spins in the simplest uniaxial potential of the magnetocrystalline anisotropy and Zeeman energy [65].

### 1. Spins in a uniform external field

First, we evaluate the equilibrium phase-space distributions for the axially symmetric situation pertaining to a spin with spin number $S$ (integer or half-integer) in an external dc field **H** applied along the Z-axis. The spin Hamiltonian $\hat{H}_S$ is then given by Eq. (264), namely, [2]

$$\hat{H}_S = -\gamma \hbar H \hat{S}_Z \quad (271)$$

with the eigenenergies

$$E_m = -\gamma \hbar H m \text{ with } m = -S, -S+1, ..., S,$$

the distance between adjacent energy levels being $\gamma \hbar H$. Now in equilibrium, the phase-space distributions $W_S^{(s)}(\vartheta)$ are independent of the azimuthal angle $\varphi$ and, according to Eq. (262), can be expressed by the series [65]

$$W_S^{(s)}(\vartheta) = \sqrt{\frac{4\pi}{2S+1}} \sum_{L=0}^{2S} \left(C_{SSL0}^{SS}\right)^{-s} Y_{L0}(\vartheta,\varphi)\left\langle \hat{T}_{L0}^{\dagger(S)}\right\rangle_{eq}$$

$$= \frac{1}{Z_S(2S+1)} \sum_{L=0}^{2S} (2L+1)\left(C_{SSL0}^{SS}\right)^{-s} P_L(\cos\vartheta) \sum_{m=-S}^{S} C_{SmL0}^{Sm} e^{\xi m/S}, \quad (272)$$

where $P_L(z)$ are the Legendre polynomials [105], $\xi = \beta\gamma\hbar HS$ is the dimensionless field parameter, and the partition function $Z_S$ is given by

$$Z_S = \sum_{m=-S}^{S} e^{\xi m/S} = \frac{\sinh\left[(S+\tfrac{1}{2})\xi/S\right]}{\sinh\left(\tfrac{1}{2}\xi/S\right)}. \quad (273)$$

In writing Eq. (272), we have noticed Eq. (B4) from Appendix B and have utilized the explicit expression for the matrix elements $\rho_{m'm}$ of the density matrix $\hat{\rho}_S$, viz.,

$$\rho_{m'm} = \frac{\delta_{mm'}}{Z_S} e^{\xi m/S} \quad (274)$$

and Eq. (A17) from Appendix A for $\hat{T}_{LM}^{\dagger(S)}$ so yielding the closed form expression for the average polarization operators in Eq. (272), viz.,

$$\left\langle \hat{T}_{L0}^{\dagger(S)}\right\rangle_{eq} = \frac{1}{Z_S}\sqrt{\frac{2L+1}{2S+1}} \sum_{m=-S}^{S} C_{SmLM}^{Sm} e^{\xi m/S}. \quad (275)$$

Furthermore, the finite series in Eq. (272) for $s = -1$ can be summed (after some algebra, which is best accomplished via MATHEMATICA). Thus the $Q$-function $W_S^{(-1)}(\vartheta)$ can finally be written for arbitrary $S$ in the (known) concise closed form [36], viz.,

$$W_S^{(-1)}(\vartheta) = \frac{1}{Z_S}\left[\cosh\frac{\xi}{2S} + \cos\vartheta\sinh\frac{\xi}{2S}\right]^{2S}. \quad (276)$$

Moreover, using the general rule for the calculation of the expected value of the spin operator $\hat{S}_Z$ via the corresponding (*c*-number) function Eq. (241), the average longitudinal component of the spin at equilibrium is (because the integral over $\varphi$ is $2\pi$)

$$\left\langle \hat{S}_Z\right\rangle_{eq} = (S+\tfrac{1}{2})S\left(\frac{S+1}{S}\right)^{(s+1)/2} \int_0^\pi \cos\vartheta W_S^{(-s)}(\vartheta)\sin\vartheta d\vartheta$$

$$= \frac{S}{Z_S}\left(\frac{S+1}{S}\right)^{(s+1)/2}\left(\frac{(2S)!\sqrt{2S+1}}{\sqrt{(2S-1)!(2S+2)!}}\right)^s \sum_{m=-S}^{S} C_{Sm10}^{Sm} e^{\xi m/S} \quad (277)$$

$$= \frac{1}{Z_S}\sum_{m=-S}^{S} m e^{\xi m/S} = SB_S(\xi),$$

where $B_S(x)$ is the Brillouin function defined as [2]

$$B_S(x) = \frac{2S+1}{2S}\coth\left(\frac{2S+1}{2S}x\right) - \frac{1}{2S}\coth\left(\frac{x}{2S}\right) \quad (278)$$

and we have utilized the Clebsch-Gordan identity [95]

$$C_{Sm10}^{Sm} = \frac{m}{\sqrt{S(S+1)}}.$$

Obviously, $\langle \hat{S}_Z \rangle_{eq}$ from Eq. (277) has the all-important feature that it is independent of the parameter $s$. Notice for future reference that setting $s = 1$ unlike in the Weyl symbol Eq. (241) means that we are utilizing the $Q$-distribution, where $s = -1$. Moreover, Eq. (277) is in complete agreement with the established result for the equilibrium magnetization of an assembly of noninteracting spins in a uniform magnetic field [1,2]. Furthermore, in the classical limit,

$$S \to \infty \text{ and } \mu = \gamma \hbar S / \mu_0 = const, \tag{279}$$

the distribution $W_S^{(s)}(\vartheta)$ tends to the Boltzmann distribution for classical magnetic dipoles

$$(S + 1/2) W_S^{(s)}(\vartheta) \to Z_{cl}^{-1} e^{\xi \cos \vartheta}, \tag{280}$$

while the Brillouin function $B_S(\xi)$ tends to the Langevin function $L(\xi) = \coth \xi - 1/\xi$, viz.,

$$B_S(\xi) \to \frac{1}{Z_{cl}} \int_0^\pi \cos \vartheta e^{\xi \cos \vartheta} \sin \vartheta d\vartheta = L(\xi). \tag{281}$$

Here $Z_{cl}$ is the classical partition function given by

$$Z_{cl} = \int_0^\pi e^{\xi \cos \vartheta} \sin \vartheta d\vartheta = 2 \frac{\sinh \xi}{\xi}. \tag{282}$$

Experimental studies of the magnetization of various paramagnetic atoms and molecules indicate that they agree closely with the Brillouin function Eq. (278) (see, e.g., Refs. 146 and 147). In Fig. 4, the magnetization $M = \mu B_S(\xi)$ per molecule in units of $\mu_B$ as a function of applied field $H$ at $T = 2$ K is shown for the three isotropic high-spin molecules [Cr{(CN)Cu(tren)}$_6$](ClO$_4$)$_{21}$ ($S = 9/2$), [Cr{(CN)Ni(tetren)}$_6$](ClO$_4$)$_9$ ($S = 15/2$), [Cr{(CN)Mn(tren)}$_6$](ClO$_4$)$_{21}$ ($S = 27/2$), which are labelled as CrCu$_6$, CrNi$_6$, and CrMn$_6$, respectively. These consist of clusters of metal ions ordered in a crystal lattice and coupled only via Heisenberg ferromagnetic or antiferromagnetic interactions between spins inside the molecule. In all cases, $M$ increases with the applied field reaching a saturation value of $(9/2)g\mu_B$, $(15/2)g\mu_B$, and $(27/2)g\mu_B$ in CrCu$_6$, CrNi$_6$, and CrMn$_6$, respectively ($g \simeq 2$ is Landé's factor).

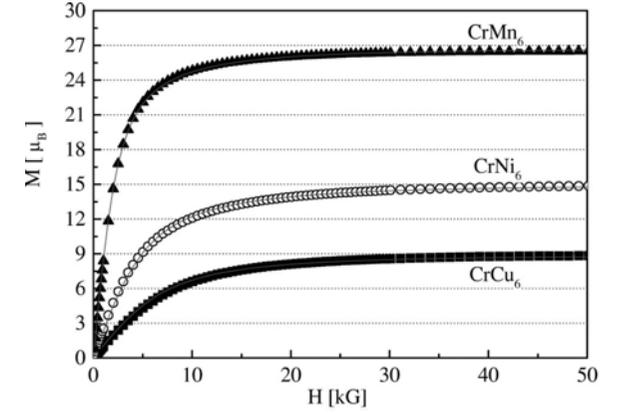

**Figure 4.** The magnetization $M = g\mu_B S B_S(\beta \mu_0 g \mu_B SH)$ per molecule in units of Bohr magnetons $\mu_B$ of CrCu$_6$, CrNi$_6$, and CrMn$_6$ vs. the external applied field $H$ at $T = 2$ K. The experimental data (symbols) are accurately described by the Brillouin functions (solid lines). Reproduced by permission of the American Physical Society from Z. Salman, A. Keren, P. Mendels, V. Marvaud, A. Scuiller, M. Verdaguer, J. S. Lord, and C. Baines, *Phys. Rev. B* 132403, **65** (2002).

The three phase-space equilibrium distributions $W_S^{(s)}(\vartheta)$ embodied in the finite series representation Eq. (272) are shown for comparison purposes in Fig. 5 for $s = 0, \pm 1$. The foregoing example then amply demonstrates that the $Q$-function $W_S^{(-1)}(\vartheta, \varphi)$ alone satisfies the *nonnegativity* condition, viz., $W_S^{(-1)} \geq 0$, required of a true probability density function. The other quasiprobability distribution functions $W_S^{(1)}$ and $W_S^{(0)}$ violate this condition (because they may take on negative values). From now on for purposes of convenience, we shall consider $W_S^{(-1)}(\vartheta, \varphi, t)$ only omitting everywhere the superscript $(-1)$ in $W_S^{(-1)}(\vartheta, \varphi, t)$ [nevertheless all results may be easily generalized for $W_S^{(1)}(\vartheta, \varphi, t)$ and $W_S^{(0)}(\vartheta, \varphi, t)$, which can be treated in like manner].

Next we calculate the phase-space $Q$-function distribution (henceforth this will be given the generic title "Wigner function") for spins in an external uniform field **H** of an *arbitrary* orientation rather than just applied along the Z axis so that the Hamiltonian is given by Eq. (265). This operator which now pertains to a *nonaxially symmetric* problem can be rewritten in terms of the spherical spin operators $\hat{S}_\mu$ (see Appendix A) as

$$\beta \hat{H}_S = -\frac{\xi}{S} \sum_{\mu=-1}^{1} \gamma^\mu \hat{S}_\mu,$$

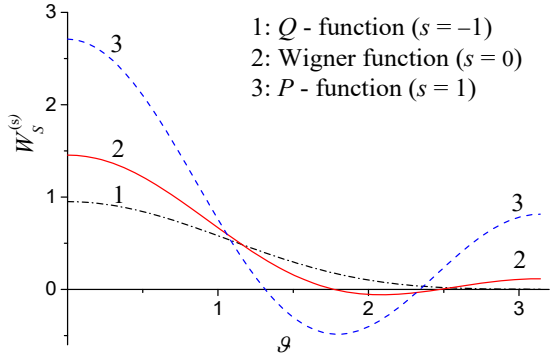

**Figure 5.** (Color on line) $W_S^{(s)}(\vartheta)$ vs. the polar angle $\vartheta$ for $s = 0, \pm 1$, $\xi = 3$ and $S = 1$.

where in terms of direction cosines $\gamma^{\pm 1} = \mp 2^{-1/2}(\gamma_X \mp i\gamma_Y)$, $\gamma^0 = \gamma_Z$, and the matrix elements of the Hamiltonian operator $\hat{H}_S$ can again be given in closed form, viz.,

$$\left[\beta\hat{H}_S\right]_{m'm} = -\frac{\xi}{S}\left(\gamma^{-1}\left[\hat{S}_{-1}\right]_{m-1m'}\delta_{m-1m} + \gamma^0 m\delta_{mm'} + \gamma^1\left[\hat{S}_1\right]_{m+1m'}\delta_{m+1m'}\right), \qquad (283)$$

where the matrix elements of the spherical spin operator $\hat{S}_{\pm 1}$ are given by Eq. (A11) from Appendix A, namely,

$$\left[\hat{S}_{\pm 1}\right]_{m\pm 1m} = \mp\sqrt{(S\mp m)(S\pm m+1)/2}. \qquad (284)$$

Furthermore, for small $S$, the equilibrium density matrix $\hat{\rho}_S = e^{-\beta\hat{H}_S}/Z_S$ can be written in closed form as a finite series of the spin operators. For example, for $S = \frac{1}{2}$, $S = 1$, etc., one has

$$\hat{\rho}_{1/2} = \frac{1}{Z_{1/2}}\left(\hat{I}\cosh\xi + 2\sum_{\mu=-1}^{1}\gamma^\mu\hat{S}_\mu\sinh\xi\right), \qquad (285)$$

$$\hat{\rho}_1 = \frac{1}{Z_1}\left[\hat{I} + \sum_{\mu=-1}^{1}\gamma^\mu\hat{S}_\mu\sinh\xi + 2\left(\sum_{\mu=-1}^{1}\gamma^\mu\hat{S}_\mu\right)^2\sinh^2\frac{\xi}{2}\right], \qquad (286)$$

etc., where $\hat{I}$ is the identity matrix.

The corresponding $Q$-distribution $W_S(\vartheta,\varphi)$ can then be calculated in the finite series form Eq. (262). In turn, this finite series can then be summed (after tedious algebra, which is again best accomplished via MATHEMATICA) so that the distribution $W_S(\vartheta,\varphi)$ can ultimately be written for arbitrary spin $S$ in concise closed form [66], viz., [cf. Eq. (276)]

$$W_S(\vartheta,\varphi) = \frac{1}{Z_S}\left[\cosh\frac{\xi}{2S} + F(\vartheta,\varphi)\sinh\frac{\xi}{2S}\right]^{2S}, \qquad (287)$$

where

$$F(\vartheta,\varphi) = \gamma_X\sin\vartheta\cos\varphi + \gamma_Y\sin\vartheta\sin\varphi + \gamma_Z\cos\vartheta \qquad (288)$$

and

$$Z_S = \frac{2S+1}{4\pi}\int_0^\pi\int_0^{2\pi}\left[\cosh\frac{\xi}{2S} + F(\vartheta,\varphi)\sinh\frac{\xi}{2S}\right]^{2S}\sin\vartheta d\vartheta d\varphi$$
$$= \frac{\sinh\left[(S+\tfrac{1}{2})\xi/S\right]}{\sinh(\tfrac{1}{2}\xi/S)} \qquad (289)$$

is the partition function [Eq. (289) concurs with Eq. (273)]. Moreover, for the three specific cases represented by the following direction cosines

$$\gamma_X = 1, \ \gamma_Y = 0, \ \gamma_Z = 0,$$

$$\gamma_X = 0, \ \gamma_Y = 1, \ \gamma_Z = 0,$$

and

$$\gamma_X = 0, \ \gamma_Y = 0, \ \gamma_Z = 1,$$

the phase-space distribution Eq. (287) reduces to the equations already given by Takahashi and Shibata [35,36] and reproduces Eq. (276). Now, the equilibrium average $\langle(\hat{\boldsymbol{\mu}}\cdot\mathbf{H})\rangle_{eq}$ is then from Eqs. (287) and (288) [cf. Eq. (277)]

$$\langle(\hat{\boldsymbol{\mu}}\cdot\mathbf{H})\rangle_{eq} = \mu H\frac{(2S+1)(S+1)}{4\pi}\int_0^\pi\int_0^{2\pi}F(\vartheta,\varphi)W_S(\vartheta,\varphi)\sin\vartheta d\vartheta d\varphi = \mu H B_S(\xi), \qquad (290)$$

where $B_S(x)$ is the Brillouin function defined by Eq. (278). In the classical limit $S\to\infty$, the equilibrium distribution $W_S(\vartheta,\varphi)$ given by Eq. Eq. (287) tends to the Boltzmann distribution

$$\frac{2S+1}{4\pi}W_S(\vartheta,\varphi) \to \frac{1}{Z_{cl}}e^{\xi F(\vartheta,\varphi)}, \qquad (291)$$

while the the equilibrium average $\langle(\hat{\boldsymbol{\mu}}\cdot\mathbf{H})\rangle_{eq}$ tends to the Langevin function

$$\langle(\hat{\boldsymbol{\mu}}\cdot\mathbf{H})\rangle_{eq} \to \frac{\mu H}{Z_{cl}}\int_0^\pi\int_0^{2\pi}\cos\vartheta e^{\xi F(\vartheta,\varphi)}\sin\vartheta d\vartheta d\varphi = \mu HL(\xi) = \mu H\left(\coth\xi - \frac{1}{\xi}\right), \qquad (292)$$

where $Z_{cl}$ is the classical partition function given by

$$Z_{cl} = \int_0^\pi\int_0^{2\pi}e^{\xi F(\vartheta,\varphi)}\sin\vartheta d\vartheta d\varphi = 4\pi\frac{\sinh\xi}{\xi}. \qquad (293)$$

Clearly, the above calculations represent quantum and classical treatments of paramagnetism [2]. Here, quantum effects as identified via the Brillouin function Eq. (278) become important at small $S$ when that function must be used instead of the Langevin function given by Eq. (291) which is valid in the classical limit.

## 2. Uniaxial nanomagnet in an external field

Now we calculate the equilibrium Wigner function $W_S(\vartheta,\varphi)$ for a *uniaxial* nanomagnet of arbitrary spin number $S$ in an external magnetic field of an *arbitrary* orientation. First, we briefly consider the more general nonaxially symmetric case of a spin in an external *constant* field $\mathbf{H}$ with the Hamiltonian operator $\hat{H}_S$ given by Eq. (266); then we specialize it to a longitudinal field. In the general case, the matrix elements of $\hat{H}_S$ can again be given via the matrix elements of the spherical spin operators $\hat{S}_\mu$ $(\mu=0,\pm1)$ and $\hat{S}_0^2$, viz. [cf. Eq. (283) with superimposed anisotropy term],

$$\left[\beta\hat{H}_S\right]_{m'm} = \left[-\frac{\xi}{S}\sum_{\mu=-1}^{1}\gamma^\mu \hat{S}_\mu - \frac{\sigma}{S^2}\hat{S}_0^2\right]_{m'm}$$
$$= -\frac{\xi}{S}\left(\gamma^{-1}\left[\hat{S}_{-1}\right]_{m-1m}\delta_{m-1m'} + \gamma^1\left[\hat{S}_1\right]_{m+1m}\delta_{m+1m'}\right) - \left(m^2\frac{\sigma}{S^2} + \gamma^0 m\frac{\xi}{S}\right)\delta_{mm'}, \quad (294)$$

where the matrix elements $\left[\hat{S}_{\pm1}\right]_{m\pm1m}$ are defined by Eq. (284). The corresponding phase-space distribution $W_S(\vartheta,\varphi)$ can then be calculated in the finite series form Eqs. (262) and (263). Now motivated by the form of the classical potential, the results of the calculation of an "effective" free energy potential defined by $\beta V(\vartheta,\varphi) = -\ln W_S(\vartheta,\varphi)$ are shown in Fig. 6 for various values of $S$. The effective potential $V(\vartheta,\varphi)$ has two nonequivalent minima (the minimum at $\vartheta = \pi$ is masked in these plots) and one saddle point in the plane $\varphi = 0$; the potential shape and barrier heights strongly depend on the spin number $S$. Moreover, in the classical limit, $S \to \infty$, $V(\vartheta,\varphi)$ tends to the normalized classical free energy $V_{cl}(\vartheta,\varphi)$ given by

$$\beta V_{cl}(\vartheta,\varphi) = -\sigma\left\{\cos^2\vartheta + 2h\left[(\gamma_X\cos\varphi + \gamma_Y\sin\varphi)\sin\vartheta + \gamma_Z\cos\vartheta\right]\right\}, \quad (295)$$

which is also shown in Fig. 6 for the purpose of comparison.

However, the general treatment above considerably simplifies for a longitudinal field (ubiquitous in magnetic applications) so that $\gamma_X = 0$, $\gamma_Y = 0$ and $\gamma_Z = 1$ [see Eq. (267)] and so the problem becomes axially symmetric. Here the density matrix $\hat{\rho}_S$ is diagonal with matrix elements $\rho_{mm'}$ given explicitly by [143,148]

$$\rho_{mm'} = \frac{\delta_{mm'}}{Z_S}e^{\frac{\sigma m^2}{S^2} + \frac{\xi m}{S}}, \quad (296)$$

where the partition function $Z_S$ is

$$Z_S = \sum_{m=-S}^{S} e^{\frac{\sigma m^2}{S^2} + \frac{\xi m}{S}}. \quad (297)$$

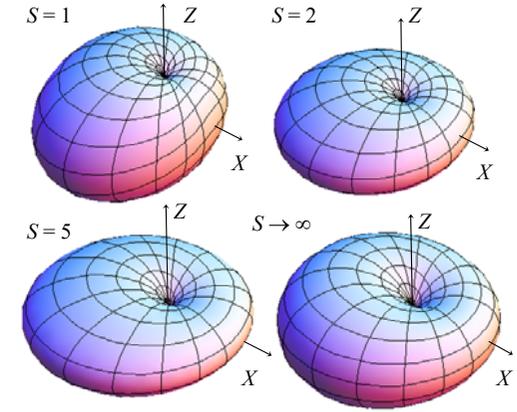

**Figure 6.** (Color on line) 3D plot of the "effective" free energy potential $\beta V(\vartheta,\varphi)$ for a uniaxial nanomagnet in an external field for various values of $S = 1, 2, 5,$ and $S \to \infty$ [classical limit; Eq. (295)] and the parameters $\sigma = 5$, $h = \xi/2\sigma = 0.2$, $\gamma_Z = 1/2$, $\gamma_Y = 0$, and $\gamma_X = \sqrt{3}/2$ (i.e., the field $\mathbf{H}$ is in the *XZ*-plane and directed at an angle $\pi/3$ to the *Z*-axis)

The explicit matrix elements $\rho_{mm'}$ from Eq. (296) can then be used to evaluate the averages $\left\langle \hat{T}_{LM}^{\dagger(S)}\right\rangle_{eq}$ in Eq. (263) as [cf. Eq. (275)]

$$\left\langle \hat{T}_{LM}^{\dagger(S)}\right\rangle_{eq} = \frac{\delta_{M0}}{Z_S}\sqrt{\frac{2L+1}{2S+1}}\sum_{m=-S}^{S} C_{SmLM}^{Sm} e^{\frac{\sigma m^2}{S^2} + \frac{\xi m}{S}}. \quad (298)$$

Furthermore, due to the symmetry about the *Z*-axis, the phase-space distribution function $W_S(\vartheta,\varphi) \equiv W_S(\vartheta)$ is independent of the azimuthal angle $\varphi$ so that the Wigner function Eq. (262) for this important problem simplifies to the series of Legendre polynomials [65]

$$W_S(\vartheta) = \sum_{L=0}^{2S}\frac{L+1/2}{S+1/2}\left\langle P_L\right\rangle_{eq} P_L(\cos\vartheta). \quad (299)$$

Here $\left\langle P_L\right\rangle_{eq}$ are the equilibrium averages of the Legendre polynomials $P_L(\cos\vartheta)$ given by

$$\left\langle P_L\right\rangle_{eq} = (S+1/2)\int_0^\pi P_L(\cos\vartheta)W_S(\vartheta)\sin\vartheta\, d\vartheta$$
$$= \frac{(2S)!\sqrt{2S+1}}{Z_S\sqrt{(2S-L)!(2S+L+1)!}}\sum_{m=-S}^{S} C_{SmL0}^{Sm} e^{\frac{\sigma m^2}{S^2} + \frac{\xi m}{S}} \quad (300)$$

and we have used Eq. (298) and Eq. (B4) from Appendix B. The equilibrium statistical moment $\left\langle P_1\right\rangle_{eq}$ then yields the average longitudinal component of the spin as

$$\langle \hat{S}_Z \rangle_{eq} = (S+1)\langle P_1 \rangle_{eq} = (S+1)\langle \cos\vartheta \rangle_{eq}$$
$$= \frac{1}{Z_S} \sum_{m=-S}^{S} m\, e^{\frac{\sigma m^2}{S^2} + \frac{\xi m}{S}}, \quad (301)$$

which concurs with the well-known result for the equilibrium magnetization for arbitrary $S$ [143,148]. Here we have used the Weyl symbol of the operator $\hat{S}_Z$ given by Eq. (241) with $s=1$. From the explicit expressions for the $P_L(\cos\vartheta)$ [105] in Eq. (299), we then have explicit trigonometric forms for the distribution functions, e.g., for $S = 1/2, 1, 3/2, 2$, etc. [65]

$$W_{1/2}(\vartheta) = \frac{e^\sigma}{Z_{1/2}} f_\xi(\vartheta),$$

$$W_1(\vartheta) = \frac{e^\sigma}{Z_1}\left[\left(f_\xi^{(1)}(\vartheta)\right)^2 + \frac{1}{2}\left(e^{-\sigma}-1\right)\sin^2\vartheta\right],$$

$$W_{3/2}(\vartheta) = \frac{e^{9\sigma/4}}{Z_{3/2}}\left[\left(f_\xi^{(3/2)}(\vartheta)\right)^3 + \frac{3}{4}\left(e^{-8\sigma/9}-1\right)f_\xi^{(3/2)}(\vartheta)\sin^2\vartheta\right],$$

$$W_2(\vartheta) = \frac{e^\sigma}{Z_2}\left[\left(f_\xi^{(2)}(\vartheta)\right)^4 + \left(e^{-3\sigma/4}-1\right)\left(f_\xi^{(2)}(\vartheta)\right)^2\sin^2\vartheta + \frac{1}{8}\left(3e^{-\sigma}-4e^{-3\sigma/4}+1\right)\sin^4\vartheta\right],$$

etc., where the function

$$f_\xi^{(S)}(\vartheta) = \cosh\frac{\xi}{2S} + \cos\vartheta\sinh\frac{\xi}{2S}. \quad (302)$$

For *arbitrary* $S$, the trigonometric series in $\left(f_\xi^{(S)}(\vartheta)\right)^{2(S-m)}\sin^{2m}\vartheta$ for the distribution $W_S(\vartheta)$ can be rewritten in general form as

$$W_S(\vartheta) = \frac{e^\sigma}{Z_S}\sum_{n=0}^{[S]} b_n \left(f_\xi^{(S)}(\vartheta)\right)^{2(S-n)}\sin^{2n}\vartheta, \quad (303)$$

where $[S]$ means the whole part of $S$ and the leading coefficients $b_n$ are

$$b_0 = 1,$$

$$b_1 = \frac{S}{2}\left[e^{-(2S-1)\sigma/S^2}-1\right],$$

$$b_2 = \frac{S}{16}\left[(2S-1)e^{-4(S-1)\sigma/S^2} - 4(S-1)e^{-(2S-1)\sigma/S^2} + 2S-3\right],$$

etc.

The distribution $W_S(\vartheta)$ given by Eq. (299) is shown in Fig. 7 as a function of the polar angle $\vartheta$. The maxima of $W_S(\vartheta)$ occur at $\vartheta=0$ and $\vartheta=\pi$, where

$$W_S(0) = \frac{1}{Z_S}e^{\xi+\sigma} \text{ and } W_S(\pi) = \frac{1}{Z_S}e^{-\xi+\sigma}, \quad (304)$$

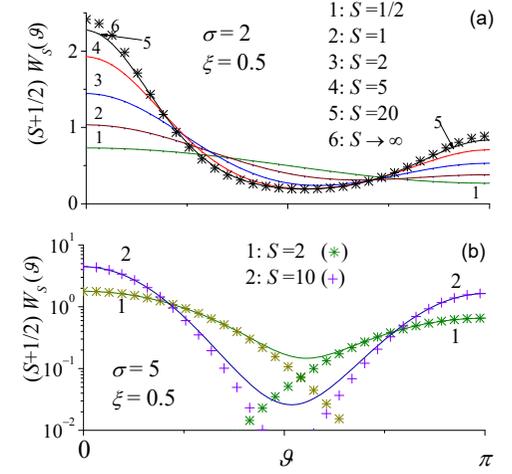

**Figure 7.** (Color on line) (a) $(S+1/2)W_S(\vartheta)$ vs. $\vartheta$ for $\sigma=2$, $\xi=0.5$, and various values of $S$ including the classical limit, $S\to\infty$ (asterisks), Eq. (305). (b) The distribution $(S+1/2)W_S(\vartheta)$ (solid lines) for $\sigma=5$, $\xi=0.5$, and $S=2$ and 10. Crosses (×) and stars (∗): Eq. (306).

respectively, meaning classically speaking that the spins are concentrated at the bottom of the wells, where the minima of the potential energy occur. In the classical limit, $S\to\infty$, $W_S(\vartheta)$ from Eq. (299) tends to the usual Boltzmann distribution for a uniaxial nanomagnet in a longitudinal field, i.e.,

$$(S+1/2)W_S(\vartheta) \to \frac{1}{Z_{cl}}e^{\xi\cos\vartheta + \sigma\cos^2\vartheta}, \quad (305)$$

where

$$Z_{cl} = \int_0^\pi e^{\xi\cos\vartheta + \sigma\cos^2\vartheta}\sin\vartheta\, d\vartheta$$

is the classical partition function. Clearly from Fig. 7a, the deviations of the quantum distribution $(S+1/2)W_S(\vartheta)$ from the classical Boltzmann distribution Eq. (305) become pronounced only for *small* spin numbers $S<10$ while as $S$ increases, the distribution $(S+1/2)W_S(\vartheta)$ tends to the classical expression Eq. (305) (e.g., for $S=20$, the differences between the two distributions Eqs. (303) and (305) do not exceed 10 percent; see the curve 5 in Fig. 7a). Due to the biasing effect of the external field, the maxima are unequal in height. Moreover, in the low temperature limit, the dynamics of the spin in the vicinity of the maxima $\vartheta=0$ and $\vartheta=\pi$ represent precession in the effective magnetic field with characteristic angular frequencies $\omega^+$ and $\omega^-$, respectively,

$$\omega^{\pm} = \frac{1}{\beta\hbar}\left[\pm\frac{\xi}{S} + (2S-1)\frac{\sigma}{S^2}\right],$$

so that the distribution $W_S(\vartheta)$ can be approximated by

$$W_S(\vartheta) \approx \begin{cases} Z_S^{-1} e^{-\sigma(1-S^{-1})}\left(f_{\beta\hbar\omega^+}^{(S)}(\vartheta)\right)^{2S}, & (\vartheta \leq 1) \\ Z_S^{-1} e^{-\sigma(1-S^{-1})}\left(f_{\beta\hbar\omega^-}^{(S)}(\vartheta)\right)^{2S}, & (\pi-\vartheta \leq 1) \end{cases}$$
$$= \begin{cases} W_S(0) e^{-\xi/S - \sigma(1-S^{-1})}\left(f_{\beta\hbar\omega^+}^{(S)}(\vartheta)\right)^{2S}, & (\vartheta \leq 1) \\ W_S(\pi) e^{\xi/S - \sigma(1-S^{-1})}\left(f_{\beta\hbar\omega^-}^{(S)}(\vartheta)\right)^{2S}, & (\pi-\vartheta \leq 1) \end{cases} \quad (306)$$

For $\sigma = 0$, i.e., for *a spin in a uniform external magnetic field*, when the Hamiltonian becomes simply $\beta\hat{H}_S = -\xi\hat{S}_Z/S$, Eq. (299) reduces to our previous result Eq. (276) [35,36], viz.,

$$W_S(\vartheta) = \frac{1}{Z_S}\left(f_\xi^{(S)}(\vartheta)\right)^{2S}, \quad (307)$$

where $Z_S$ is defined by Eq. (273). As may be seen in Fig. 7b, the "oscillator" function $f$ from Eq. (307) describes with a very high degree of accuracy the behavior of $W_S(\vartheta)$ near $\vartheta = 0$ and $\vartheta = \pi$ as expected from intuitive reasoning. The above distribution Eq. (307) represents a quantum analog of the Boltzmann distribution Eq. (280) for classical magnetic dipoles $\mu$ precessing in the uniform magnetic field $\mathbf{H}$ with the precession angular frequency $\omega_0 = \gamma H$.

For another particular case, viz., $\xi = 0$, i.e., for a uniaxial spin system alone with Hamiltonian $\beta\hat{H}_S = -\sigma\hat{S}_Z^2/S^2$, the equilibrium phase-space distribution $W_S(\vartheta)$ from Eq. (299) simplifies to

$$W_S(\vartheta) = \frac{1}{Z_S}\sum_{L=0}^{2S}\frac{(2S)!(2L+1)P_L(\cos\vartheta)}{\sqrt{(2S+1)(2S-L)!(2S+L+1)!}}\sum_{m=-S}^{S} C_{SmL0}^{Sm} e^{\sigma m^2/S^2}, \quad (308)$$

where the partition function

$$Z_S = \sum_{m=-S}^{S} e^{\sigma m^2/S^2}. \quad (309)$$

Near $\vartheta = 0$ and $\vartheta = \pi/2$, the leading terms of the series expansion of the equilibrium distribution $W(\vartheta)$ from Eq. (308) (i.e., in $\sin^2\vartheta$ and $\cos^2\vartheta$, respectively) are [65]

$$W_S(\vartheta) = W_S(0)\left\{1 + \frac{S}{2}\left[e^{-(2S-1)\sigma/S^2} - 1\right]\sin^2\vartheta + ...\right\}, \quad (310)$$

$$W_S(\vartheta) = W_S(\pi/2)\left[1 + A\cos^2\vartheta + ...\right], \quad (311)$$

where

$$W_S(0) = \frac{e^{\sigma}}{Z_S}, \quad (312)$$

$$W_S(\pi/2) = \frac{(2S)!}{2^{2S} Z_S}\sum_{m=-S}^{S}\frac{e^{\sigma m^2/S^2}}{(S+m)!(S-m)!}, \quad (313)$$

$$A = \frac{\sum_{m=-S+1}^{S-1} e^{m^2\sigma/S^2}\frac{e^{(1-2m)\sigma/S^2} + e^{(1+2m)\sigma/S^2} - 2}{(S-1+m)!(S-1-m)!}}{2\sum_{m=-S}^{S}\frac{e^{m^2\sigma/S^2}}{(S+m)!(S-m)!}}. \quad (314)$$

*3. Uniaxial nanomagnet in a transverse field*

As a further example, we calculate the Wigner function of a uniaxial nanomagnet in a *transverse* external field with the *nonaxially symmetric* Hamiltonian $\hat{H}_S$ Eq. (268) otherwise known as the Lipkin-Meshkov Hamiltonian [149]. For small $S$, the density matrix $\hat{\rho}_S = e^{-\beta\hat{H}_S}/Z_S$ can again be calculated in closed form using MATHEMATICA. For example, for $S = 1/2$, $S = 1$, etc., we have in terms of the spin operators [66]

$$\hat{\rho}_{1/2} = \frac{e^{\sigma}}{Z_{1/2}}\left[\hat{I}\cosh\xi + 2\hat{S}_X\sinh\xi\right], \quad (315)$$

$$\hat{\rho}_1 = \frac{1}{Z_1}\left[\left(e^{\sigma} - \sigma A\right)\hat{I} + A\left(\xi\hat{S}_X + \sigma\hat{S}_Z^2\right) + \left(e^{\sigma/2}\cosh\sqrt{\xi^2 + \frac{\sigma^2}{4}} - e^{\sigma} + \frac{\sigma A}{2}\right)\hat{S}_X^2\right], \quad (316)$$

etc., where

$$Z_{1/2} = 2e^{\sigma}\cosh(\xi),$$

$$Z_1 = e^{\sigma} + 2e^{\sigma/2}\cosh\sqrt{\xi^2 + \sigma^2/4},$$

$$A = \frac{e^{\sigma/2}}{\sqrt{\xi^2 + \sigma^2/4}}\sinh\sqrt{\xi^2 + \sigma^2/4}.$$

The corresponding equations for the phase-space distribution $W_S(\vartheta,\varphi)$ which can be obtained from the general expressions Eqs. (262) and (263) are given by [66]

$$W_{1/2}(\vartheta,\varphi) = \frac{e^{\sigma}}{Z_{1/2}}\left[\cosh\xi + \sinh\xi\sin\vartheta\cos\varphi\right], \quad (317)$$

$$W_1(\vartheta,\varphi) = \frac{1}{2Z_1}\left\{e^{\sigma}\left(1 - \sin^2\vartheta\cos^2\varphi\right) + e^{\sigma/2}\cosh\sqrt{\xi^2 + \frac{\sigma^2}{4}}\left(1 + \sin^2\vartheta\cos^2\varphi\right) \\ + \frac{\sigma A}{2}\left[\cos 2\vartheta + \sin^2\vartheta\cos^2\varphi + \frac{4\xi}{\sigma}\sin\vartheta\cos\varphi\right]\right\}, \quad (318)$$

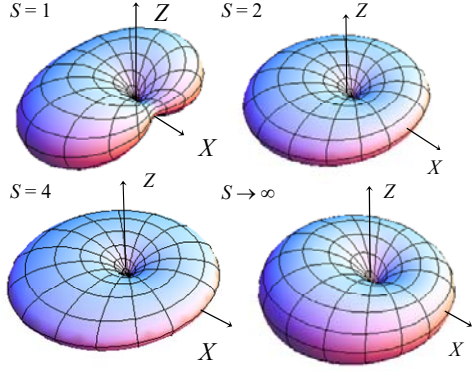

**Figure 8.** (Color on line) 3D plot of $\beta V(\vartheta,\varphi)$ for $\sigma = 10$, $h = 0.1$ and various values of $S = 1, 2, 4$, and $S \to \infty$ (classical limit).

etc. However, once again as the spin number $S$ increases, the analytical equations for the Wigner function $W_S(\vartheta,\varphi)$ rapidly become more and more complicated and thus rather impractical to use because $W_S(\vartheta,\varphi)$ for given spin may always be calculated much faster numerically from the general series expression Eq. (262).

Calculations of the "effective" free energy $\beta V(\vartheta,\varphi) = -\ln W_S(\vartheta,\varphi)$ are shown in Fig. 8 for various values of $S$ and $\sigma = 5$ and $h = \xi/(2\sigma) = 0.1$. In the classical limit, $S \to \infty$, the effective free energy function $V(\vartheta,\varphi)$ becomes the classical free energy $V_{cl}(\vartheta,\varphi)$ given by

$$\beta V_{cl}(\vartheta,\varphi) = -\sigma\left(\cos^2\vartheta + 2h\cos\varphi\sin\vartheta\right),$$

which is also shown in Fig. 8 for comparison. The "effective" potential $V(\vartheta,\varphi)$, just as the classical free energy $V_{cl}(\vartheta,\varphi)$, has two equivalent minima and one saddle point in the plane $\varphi = 0$ at $\vartheta = \pi/2$; the potential characteristics (such as the shape and barrier heights) strongly depend on $S$, e.g., the smallest barrier height increases with increasing $S$ from 0 (at $S = 1/2$) to its classical value $\sigma(1-h^2)$..

### 4. Biaxial anisotropy

We now calculate the Wigner function of a *biaxial* Hamiltonian $\hat{H}_S$ given by Eq. (269). Here the density matrix $\hat{\rho}_S = e^{-\beta \hat{H}_S}/Z_S$ can again be calculated in simple closed form for small $S$ using MATHEMATICA. For example, for $S = 1/2$, $S = 1$, etc., we have in terms of the spin operators [66]

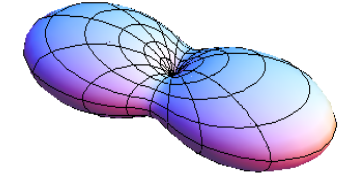

**Figure 9.** (Color on line) 3D plot of $\beta V(\vartheta,\varphi)$ for $S = 2$ and $\sigma = 5$ and $\delta = 2$.

$$\hat{\rho}_{1/2} = \frac{e^\sigma}{Z_{1/2}}\hat{I}, \qquad (319)$$

$$\hat{\rho}_1 = \frac{1}{Z_1}\left[\hat{I} + \left(e^\sigma \cosh\delta - 1\right)\hat{S}_Z^2 - e^\sigma \sinh\delta\left(\hat{S}_X^2 - \hat{S}_Y^2\right)\right], \qquad (320)$$

etc., where $Z_{1/2} = 2e^\sigma$ and $Z_1 = 1 + 2e^\sigma \cosh\delta$. The corresponding equations for the phase-space distribution $W_S(\vartheta,\varphi)$ are [66]

$$W_{1/2}(\vartheta,\varphi) = \frac{1}{2}, \qquad (321)$$

$$W_1(\vartheta,\varphi) = \frac{1}{2Z_1}\left[\sin^2\vartheta + e^\sigma \cosh\delta\left(1+\cos^2\vartheta\right) - e^\sigma \sinh\delta \sin^2\vartheta \cos 2\varphi\right], \qquad (322)$$

etc. As before with increasing $S$, $W_S(\vartheta,\varphi)$ may always be calculated numerically from the general expression Eq. (262).

The "effective" potential $\beta V(\vartheta,\varphi) = -\ln W_S(\vartheta,\varphi)$ is shown in Fig. 9 for $S = 2$, $\sigma = 5$, and $\delta = 5$. In the classical limit, $S \to \infty$, $V(\vartheta,\varphi)$ again tends to the classical free energy $V_{cl}(\vartheta,\varphi)$ given by

$$\beta V_{cl}(\vartheta,\varphi) = -\sigma\cos^2\vartheta + \delta\cos 2\varphi\sin^2\vartheta.$$

The "effective" potential $V(\vartheta,\varphi)$ [just as $V_{cl}(\vartheta,\varphi)$] has two equivalent minima and two saddle points in the plane $XZ$ at $\vartheta = \pi/2$; potential characteristics (such as the shape and barrier heights) again strongly depend on $S$. In particular, the barrier height increases with increasing $S$ from 0 (at $S = 1/2$) to its classical value $\sigma$.

### 5. Cubic anisotropy

Finally, we calculate the Wigner function of a cubic anisotropy free energy in the presence of a dc field with Hamiltonian $\hat{H}_S$ Eq. (270). For small $S$, the density matrix $\hat{\rho}_S = e^{-\beta \hat{H}_S}/Z_S$ can again be evaluated in closed form using MATHEMATICA. For example, for $S = 1/2$, $S = 1$, $3/2$, and 2, we have [66]

$$\hat{\rho}_{1/2} = \frac{e^{3\sigma_c/2}}{Z_{1/2}}\left(\hat{I}\cosh\xi + 2\hat{S}_z\sinh\xi\right), \quad (323)$$

$$\hat{\rho}_1 = \frac{e^{\sigma_c}}{Z_1}\left[\hat{I} + \sinh\xi\hat{S}_z + 2\sinh^2\frac{\xi}{2}\hat{S}_z^2\right], \quad (324)$$

$$\hat{\rho}_{3/2} = \frac{e^{41\sigma_c/54}}{Z_{3/2}}\left\{\frac{1}{8}\left(9\cosh\frac{\xi}{3} - \cosh\xi\right)\hat{I} + \frac{1}{12}\left(27\sinh\frac{\xi}{3} - \sinh\xi\right)\hat{S}_z \right.$$
$$\left. + 2\cosh\frac{\xi}{3}\sinh^2\frac{\xi}{3}\hat{S}_z^2 + \frac{4}{3}\sinh^3\frac{\xi}{3}\hat{S}_z^3\right\}, \quad (325)$$

$$\hat{\rho}_2 = \frac{1}{Z_2}\left\{\left(e^{3\sigma_c/4} - \frac{3\sigma_c}{4\xi}R\right)\hat{I} + \frac{1}{6}\left(8e^{9\sigma_c/16}\sinh\frac{\xi}{2} - R\right)\hat{S}_z \right.$$
$$\frac{1}{12}\left(16e^{9\sigma_c/16}\cosh\frac{\xi}{2} - 15e^{3\sigma_c/4} - P - \frac{93\sigma_c}{32\xi}R\right)\hat{S}_z^2 - \frac{1}{6}\left(2e^{9\sigma_c/16}\sinh\frac{\xi}{2} - R\right)\hat{S}_z^3 \quad (326)$$
$$\left. + \frac{1}{12}\left(3e^{3\sigma_c/4} - 4e^{9\sigma_c/16}\cosh\frac{\xi}{2} + P - \frac{9\sigma_c}{32\xi}R\right)\hat{S}_z^4 + \frac{\sigma_c}{32\xi}R\left(\hat{S}_X^4 + \hat{S}_Y^4\right)\right\},$$

where

$$Z_{1/2} = 2e^{3\sigma_c/2}\cosh\xi,$$

$$Z_1 = e^{\sigma_c}(1 + 2\cosh\xi),$$

$$Z_{3/2} = 4e^{41\sigma_c/54}\cosh\frac{\xi}{3}\cosh\frac{2\xi}{3},$$

$$Z_2 = e^{3\sigma_c/4} + 2e^{9\sigma_c/16}\cosh\frac{\xi}{2} + 2e^{21\sigma_c/32}\cosh\left[\sqrt{\xi^2 + (3\sigma_c/32)^2}\right],$$

$$P = e^{21\sigma_c/32}\cosh\left[\sqrt{\xi^2 + (3\sigma_c/32)^2}\right],$$

$$R = \frac{\xi e^{21\sigma_c/32}}{\sqrt{\xi^2 + (3\sigma_c/32)^2}}\sinh\left[\sqrt{\xi^2 + (3\sigma_c/32)^2}\right].$$

The corresponding equations for $W_S(\vartheta, \varphi)$ are [66]

$$W_{1/2}(\vartheta, \varphi) = \frac{e^{3\sigma_c/2}}{Z_{1/2}}\left(\cosh\frac{\xi}{2} + \sinh\frac{\xi}{2}\cos\vartheta\right), \quad (327)$$

$$W_1(\vartheta, \varphi) = \frac{e^{\sigma_c}}{Z_1}\left(\cosh\frac{\xi}{2} + \sinh\frac{\xi}{2}\cos\vartheta\right)^2, \quad (328)$$

$$W_{3/2}(\vartheta, \varphi) = \frac{e^{41\sigma_c/54}}{Z_{3/2}}\left(\cosh\frac{\xi}{2} + \sinh\frac{\xi}{2}\cos\vartheta\right)^3, \quad (329)$$

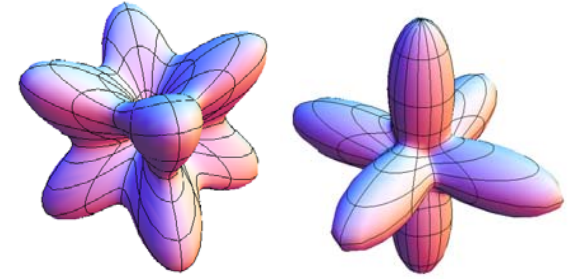

**Figure 10.** (Color on line) 3D plot of $\beta V(\vartheta, \varphi)$ for positive (left) and negative (right) cubic anisotropies for $S = 4$ and $\xi = 0$.

$$W_2(\vartheta, \varphi) = \frac{1}{8Z_2}\left\{2e^{9\sigma_c/16}\sin^2\vartheta\left[4\cos\vartheta\sin\frac{\xi}{2} + (3+\cos 2\vartheta)\cosh\frac{\xi}{2}\right] \right.$$
$$+ 3e^{3\sigma_c/4}\sin^4\vartheta + P(8\cos^2\vartheta + \sin^4\vartheta) \quad (330)$$
$$\left. + 4R\left(\cos\vartheta + \cos^3\vartheta + \frac{3\sigma}{128\xi}\cos 4\varphi\sin^4\vartheta\right)\right\}.$$

For $\xi = 0$, Eq. (330) yields

$$W_2(\vartheta, \varphi) = \frac{1}{2(3 + 2e^{3\sigma_c/16})}\left[1 + e^{3\sigma_c/16} + \frac{1}{4}\left(1 - e^{3\sigma_c/16}\right)\left(\sin^2 2\vartheta + \sin^4\vartheta\sin^2 2\varphi\right)\right]. \quad (331)$$

For large S, $W_S(\vartheta, \varphi)$ may always be calculated numerically from the general Eq. (262).

The normalized "effective" potential $\beta V(\vartheta, \varphi) = -\ln W_S(\vartheta, \varphi)$ is shown in Fig. 10 for $\sigma_c = \pm 8$, $\xi = 0$, and $S = 4$. The potential characteristics (such as the shape and barrier heights) again strongly depend on S. In the classical limit, $S \to \infty$, that potential once more tends to the classical free energy $\beta V_{cl}(\vartheta, \varphi)$ given by

$$\beta V_{cl}(\vartheta, \varphi) = \frac{\sigma_c}{4}\left(\sin^2 2\vartheta + \sin^4\vartheta\sin^2 2\varphi\right).$$

For positive anisotropy constant $\sigma_c > 0$, the cubic potential has 6 minima (wells), 8 maxima and 12 saddle points. For $\sigma_c < 0$, the maxima and minima are interchanged.

Clearly the Wigner-Stratonovich transformation yields in principle the equilibrium phase-space distribution via its finite series representation for any given anisotropy free energy. In particular, these results may be used to estimate the spin reversal time from transition state theory (TST) just as for particles in Sec. II.B.2.

### 6. TST reversal time

As we have seen in Sec II.B.2, TST because it is based on *equilibrium distributions* affords the simplest possible description of quantum corrections to the thermally activated escape rate. Now in applying TST to classical spins (i.e., classical magnetic moments $\mu$) to determine the escape rate due to thermal agitation from one metastable orientation say $A$ to another metastable orientation say $B$, we suppose that the free energy $V_{cl}(\vartheta,\varphi)$ has a multistable structure. Such a structure has minima at $\mathbf{n}_A$ and $\mathbf{n}_B$ separated by a potential barrier with a saddle point at $\mathbf{n}_C$. In the high barrier approximation, as far as TST is concerned, the classical escape rate $\Gamma_{cl}$ may be estimated via the flux over the barrier [20]

$$\Gamma_{cl} \sim \frac{I_C^{cl}}{Z_A^{cl}}, \qquad (332)$$

where the well partition function $Z_A^{cl}$ and the total current $I_C^{cl}$ of the (spin) representative points at the saddle point $C$ are, respectively,

$$Z_A^{cl} \sim \iint_{well} e^{-\beta V_{cl}(\vartheta,\varphi)} \sin\vartheta\, d\vartheta\, d\varphi \qquad (333)$$

and

$$I_C^{cl} \sim \iint_{saddle} J_C(\vartheta,\varphi) \sin\vartheta\, d\vartheta\, d\varphi \qquad (334)$$

[$J_C(\vartheta,\varphi)$ is the current density near the saddle point $C$].

Near the metastable minimum $\mathbf{n}_A$, the spin precesses about a uniform "effective" field which may be represented as the gradient of a potential, viz.,

$$\mathbf{H}_A = -\mu_0^{-1} \frac{\partial V_{cl}}{\partial \mathbf{\mu}}, \qquad (335)$$

so that the equation of motion of the magnetic moment $\mu$ is

$$\frac{d\mathbf{\mu}}{dt} = \gamma[\mathbf{\mu} \times \mathbf{H}_A]. \qquad (336)$$

Then to evaluate the classical partition function $Z_A^{cl}$ and the total current $I_C^{cl}$, we simply suppose [5,6,24] that the free energy $V_{cl}$ near the minimum $\mathbf{n}_A$ and the saddle point $\mathbf{n}_C$ can be approximated by the first two terms of its Taylor expansion, viz.,

$$V_{cl}(u_1^{(A)},u_2^{(A)}) = V_{cl}(\mathbf{n}_A) + \frac{1}{2}\left[c_1^{(A)}\left(u_1^{(A)}\right)^2 + c_2^{(A)}\left(u_2^{(A)}\right)^2\right], \qquad (337)$$

and

$$V_{cl}(u_1^{(C)},u_2^{(C)}) = V_{cl}(\mathbf{n}_C) + \frac{1}{2}\left[c_1^{(C)}\left(u_1^{(C)}\right)^2 + c_2^{(C)}\left(u_2^{(C)}\right)^2\right], \qquad (338)$$

where $(u_1^{(A)},u_2^{(A)},u_3^{(A)})$ and $(u_1^{(C)},u_2^{(C)},u_3^{(C)})$ denote the direction cosines of a magnetic moment $\mathbf{\mu}$ near $\mathbf{n}_A$ and $\mathbf{n}_C$, respectively, $c_1^{(A)} = \partial^2 V_{cl}/\partial u_1^{(A)2} > 0$ and $c_2^{(A)} = \partial^2 V_{cl}/\partial u_2^{(A)2} > 0$, i.e., the well has the form of an elliptic paraboloid, while $c_1^{(C)} = \partial^2 V_{cl}/\partial u_1^{(C)2} < 0$ and $c_2^{(C)} = \partial^2 V_{cl}/\partial u_2^{(C)2} > 0$, i.e., the saddle has the form of an hyperbolic paraboloid. Hence, in order to estimate $Z_A^{cl}$ from Eq. (333) in the high barrier limit, we have via Gaussian integrals using the Taylor expansion Eq. (337) the well partition function

$$Z_A^{cl} \sim \iint_{well} e^{-\beta V_{cl}(u_1^{(A)},u_2^{(A)})} du_1^{(A)} du_2^{(A)}$$

$$\approx \int_{-\infty}^{\infty}\int_{-\infty}^{\infty} e^{-\beta V_{cl}(u_1^{(A)},u_2^{(A)})} du_1^{(A)} du_2^{(A)} \qquad (339)$$

$$= \frac{2\pi e^{-\beta V_{cl}(\mathbf{n}_A)}}{\beta\sqrt{c_1^{(A)}c_2^{(A)}}} = \frac{2\pi\gamma e^{-\beta V_{cl}(\mathbf{n}_A)}}{\beta\mu\mu_0\omega_A},$$

where

$$\omega_A = \frac{\gamma}{\mu_0\mu}\sqrt{c_1^{(A)}c_2^{(A)}} \qquad (340)$$

is the well (precession) frequency playing the role of the attempt angular frequency in TST [20]. Here the limits of integration may be formally extended to $\pm$ infinity without significant error since the spins are almost all at $\mathbf{n}_A$. The total current $I_C^{cl}$ of representative points at the saddle point $\mathbf{n}_C$ may then be estimated as follows. We initially suppose that the saddle region has the shape of a hyperbolic paraboloid and the $u_1$-axis of the local coordinate system at the saddle point $\mathbf{n}_C$ *lies in the same direction as the current density* $\mathbf{J}_C$ *over the saddle*. Next recall that in TST, the Boltzmann distribution $\sim e^{-\beta V_{cl}}$ holds *everywhere* and that the current density $J_C$ is given by at the saddle point $C$ [24]

$$J_C(u_1^{(C)},u_2^{(C)}) = -\frac{\gamma}{\mu_0\mu}\delta\left(\partial u_1^{(C)}\right)\frac{\partial V_{cl}}{\partial u_2^{(C)}} e^{-\beta V_{cl}(u_1^{(C)},u_2^{(C)})}$$

$$= \frac{\gamma}{\beta\mu_0\mu}\delta\left(\partial u_1^{(C)}\right)\frac{\partial}{\partial u_2^{(C)}} e^{-\beta V_{cl}(u_1^{(C)},u_2^{(C)})}. \qquad (341)$$

Thus, we must have for the current at the saddle point

$$I_C^{cl} \sim \iint_{saddle} J_C(u_1^{(C)},u_2^{(C)}) du_1^{(C)} du_2^{(C)} \approx \frac{\gamma}{\mu\mu_0\beta} e^{-\beta V_{cl}(\mathbf{n}_C)}. \qquad (342)$$

Hence, using Eqs. (339) and (342), the flux over barrier Eq. (332) yields the *classical* TST formula for spins

$$\Gamma_{cl} \sim \frac{\omega_A}{2\pi} e^{-\beta\Delta V_{cl}}, \qquad (343)$$

where

$$\Delta V_{cl} = V_{cl}(\mathbf{n}_C) - V_{cl}(\mathbf{n}_A) \tag{344}$$

is the potential barrier height the determination of which always involves a detailed knowledge of the energy landscape.

In like manner, the *quantum* escape rate $\Gamma$ for a spin from a metastable orientation $A$ to another metastable orientation $B$ via the saddle point $C$ as determined by quantum TST may be given by an equation similar to the classical Eq. (332), viz.,

$$\Gamma \sim \frac{I_C}{Z_A}. \tag{345}$$

However, the well quantum partition function $Z_A$ and the total current over the saddle point $I_C$ must now be evaluated using the equilibrium phase-space distribution function $W_S(\vartheta,\varphi)$ of the spin system with the quantum spin Hamiltonian $\hat{H}_S$ instead of the classical Boltzmann distribution $\sim e^{-\beta V_{cl}(\vartheta,\varphi)}$. Nevertheless, the dynamics of a spin $\hat{\mathbf{S}}$ still comprise steady precession with the angular frequency $\boldsymbol{\omega}_A^S = \gamma \mathbf{H}_A^S$ in the *effective* magnetic field $\mathbf{H}_A^S$ in the well near the metastable minimum $\mathbf{n}_A$ so that the spin Hamiltonian $\hat{H}_S$ may be approximated by the simple equation $\hat{H}_S \approx -\hbar(\boldsymbol{\omega}_A^S \cdot \hat{\mathbf{S}})$. Thus, the quantum dynamics of the spin $\hat{\mathbf{S}}$ obey the Larmor equation [2,37] [cf. Eq. (336)]

$$\frac{d\hat{\mathbf{S}}}{dt} = \frac{i}{\hbar}\left[\hat{H}_S, \hat{\mathbf{S}}\right] = \gamma\left[\hat{\mathbf{S}} \times \mathbf{H}_A^S\right]. \tag{346}$$

However, near the metastable minimum $\mathbf{n}_A$, the distribution $W_S$ can be approximated simply by the Zeeman energy distribution for a spin in a uniform "effective" field $\mathbf{H}_A^S$ given by Eq. (287), viz., [cf. Eq. (306) for the uniaxial nanomagnet]

$$W_S(\vartheta,\varphi) \approx W_S(\mathbf{n}_A) e^{-S\xi_A} \left[\cosh\frac{\xi_A}{2} + F_A(\vartheta,\varphi)\sinh\frac{\xi_A}{2}\right]^{2S}, \tag{347}$$

where $\xi_A = \beta\hbar|\boldsymbol{\omega}_A^S|$ in accordance with Eq. (287),

$$F_A(\vartheta,\varphi) = \gamma_{X_A}\sin\vartheta\cos\varphi + \gamma_{Y_A}\sin\vartheta\sin\varphi + \gamma_{Z_A}\cos\vartheta,$$

and $\gamma_{X_A}, \gamma_{Y_A}, \gamma_{Z_A}$ are the direction cosines of the "effective" field $\mathbf{H}_A^S$ at the minimum $\mathbf{n}_A$. This effective field distribution approximation Eq. (347) is just the rotational analog of the harmonic oscillator distribution approximation for the well dynamics in the particle case (see Sec II.B.2). Just as with the classical case, the precession frequency $\boldsymbol{\omega}_A^S$ can be estimated from the well angular frequency Eq. (340), however the coefficients $c_1^{(A)}$ and $c_2^{(A)}$ are now determined from the truncated Taylor series expansion of the Weyl symbol $H_S$ of the Hamiltonian $\hat{H}_S$ of the spin, viz.,

$$\beta H_S = \beta H_S(\mathbf{n}_A) + \frac{1}{2}\left[c_1^{(A)}\left(u_1^{(A)}\right)^2 + c_2^{(A)}\left(u_2^{(A)}\right)^2\right]. \tag{348}$$

Now by using the partition function defined by the left-hand side of Eq. (289), we can approximate the well partition function in the context of quantum TST as

$$\begin{aligned}
Z_A &\sim W_S(\mathbf{n}_A)e^{-S\xi_A}\int_{well}\left[\cosh(\xi_A/2) + \sinh(\xi_A/2)F_A(\vartheta,\varphi)\right]^{2S}\sin\vartheta\, d\vartheta\, d\varphi \\
&\approx 4\pi W_S(\mathbf{n}_A)e^{-S\xi_A}\frac{\sinh\left[(S+1/2)\xi_A\right]}{(2S+1)\sinh(\xi_A/2)} \\
&= \frac{2\pi\left[1-e^{-(2S+1)\xi_A}\right]}{(S+1/2)(1-e^{-\xi_A})}W_S(\mathbf{n}_A).
\end{aligned} \tag{349}$$

Finally, the total spin current $I_C$ from the well may also be estimated just as with the classical Eq. (342) by defining the current density $J_C$ at the vicinity of the saddle point $\mathbf{n}_C$ of the "effective" potential $V_{ef}(u_1^{(C)}, u_2^{(C)}) = -\ln W_S(u_1^{(C)}, u_2^{(C)})$. Thus, we have [cf. Eq. (341)]

$$I_C \sim \frac{1}{\beta\hbar S}W_S(\mathbf{n}_C), \tag{350}$$

since the magnetic moment is now given by $\mu = \gamma\hbar S/\mu_0$. Hence, we obtain the TST escape rate as determined from Eqs. (345), (349), and (350), viz.,

$$\Gamma \sim \frac{(S+1/2)\left(1-e^{-\xi_A}\right)}{2\pi\beta\hbar S\left[1-e^{-(2S+1)\xi_A}\right]}\frac{W_S(\mathbf{n}_C)}{W_S(\mathbf{n}_A)}. \tag{351}$$

To compare this equation with the classical TST Eq. (343), we rewrite it in the form of the quantum TST Eq. (155) for particles, viz.,

$$\Gamma \sim \frac{\omega_A}{2\pi}\Xi_S e^{-\beta\Delta V_{cl}} = \Xi_S \Gamma_{cl}, \tag{352}$$

where

$$\Xi_S = \frac{\omega_A^S(S+1/2)\left(1-e^{-\xi_A}\right)}{\omega_A \xi_A S\left[1-e^{-(2S+1)\xi_A}\right]}\frac{W_S(\mathbf{n}_C)}{W_S(\mathbf{n}_A)}e^{\beta\Delta V_{cl}}, \tag{353}$$

represents the *quantum correction* factor strongly depending on the spin number $S$ and yielding $\Xi_S \to 1$ and $\Gamma \to \Gamma_{cl}$ in the classical limit, $S \to \infty$.

For example, for a uniaxial nanomagnet with the Hamiltonian $\beta\hat{H}_S = -\sigma\hat{S}_z^2/S^2$, the Weyl symbol $\beta H_S(\vartheta)$ and equilibrium phase-space distribution $W_S(\vartheta)$ are given by Eqs. (248) and (299), respectively. Furthermore, $W_S(\mathbf{n}_A) = W_S(\vartheta=0)$ and $W_S(\mathbf{n}_C) = W_S(\vartheta=\pi/2)$ are given by Eqs. (312) and (313). Thus, noting that in the classical limit

$$\omega_A = \frac{2\sigma\gamma}{\mu_0\mu\beta}, \tag{354}$$

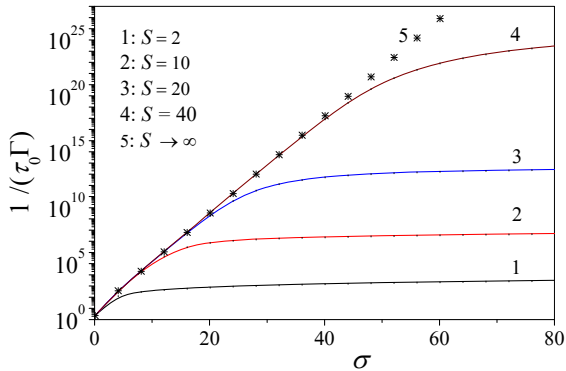

**Figure 11.** (Color on line) Normalized inverse escape rate $(\tau_0 \Gamma)^{-1}$ as a function of the inverse temperature parameter $\sigma \sim 1/T$ for various values of $S = 2, 10, 20, 40$, and $S \to \infty$ (classical limit).

$$\beta \Delta V_{cl} = \sigma \tag{355}$$

while

$$\omega_A^S = \omega_A \left(1 - \frac{1}{2S}\right), \tag{356}$$

the quantum correction factor $\Xi_S$ is given by

$$\Xi_S = \frac{(S+1/2)\left[1 - e^{-\sigma(2S-1)/S^2}\right]}{\sigma 2^{2S+1}\left[1 - e^{-\sigma(4S^2-1)/S^2}\right]} \sum_{m=-S}^{S} \frac{(2S)! e^{\sigma m^2/S^2}}{(S+m)!(S-m)!}. \tag{357}$$

In the limit $S \to \infty$, Eq. (352) reduces to the well-known classical TST (Néel) formula for a uniaxial nanomagnet, viz.,

$$\Gamma_{cl} \sim \frac{1}{2\pi\tau_0} e^{-\sigma}, \tag{358}$$

where $\tau_0 = 1/\omega_A$ is a normalizing time. The normalized inverse escape rate $(\tau_0 \Gamma)^{-1}$ as a function of the inverse temperature parameter $\sigma \sim 1/T$ is shown in Fig. 11 for various values of $S$. Clearly, the qualitative behavior of the quantum escape rates $\Gamma$ for finite $S$ strongly deviates from the Arrhenius behavior of the classical escape rates $\Gamma_{cl}$ at low temperatures. This difference is due to the tunneling effect.

Like the classical case, having evaluated the escape rate $\Gamma$ for a particular anisotropy, we have the reversal time at finite temperatures. In particular, by *equating the reversal time to the measuring time of a switching time experiment* one may estimate the switching field curves at *finite* temperatures just as with the classical theory [5]. Although TST always implies that the dissipation to the bath does not affect the escape rate, nevertheless, the results should still apply in a *wide range of dissipation*. The latter may be defined as wide enough to ensure that thermal noise is sufficiently strong to thermalize the escaping system yet not so wide as to disturb the thermal equilibrium in the well, i.e., an equilibrium distribution still prevails everywhere including the saddle point. In classical Kramers escape rate theory [6], this represents the so called *intermediate damping* case.

Now we shall now demonstrate how the phase-space representation for a given spin Hamiltonian may be used to calculate switching field curves and/or surfaces as a function of spin number $S$ at zero temperature.

### 7. Switching field curves

We recall that the first calculation of the magnetization reversal of single-domain ferromagnetic particles with uniaxial anisotropy subjected to an applied field was made by Stoner and Wohlfarth [17] with the hypothesis of coherent rotation of the magnetization and zero temperature so that *thermally induced switching between the potential minima is ignored*. In the simplest uniaxial anisotropy as considered by them, the magnetization reversal consequently occurs at that particular value of the applied field (called the switching field) which destroys the bistable nature of the potential. The parametric plot of the parallel vs. the perpendicular component of the switching field then yields the famous astroids. As mentioned in the Introduction, Thiaville [19] later developed a geometrical method for the calculation of the energy of a particle allowing one to determine the switching field for all values of the applied magnetic field yielding the critical switching field surface analogous to the Stoner-Wohlfarth astroids. This surface, as it generalizes the critical curves of the 2D problem of Stoner and Wohlfarth [17], is called the *limit of metastability surface*. In the general approach to the calculation of switching curves via the geometrical method of Thiaville [19], these curves or surfaces may be constructed for particles with *arbitrary* anisotropy at zero temperature. By fitting experimental switching field curves and surfaces, one can in particular determine the free energy of a nanomagnet and the corresponding anisotropy constants (see Fig. 12).

In order to generalize Thiaville's geometrical method [19] to include quantum effects in switching field curves and surfaces of a spin system with a model spin Hamiltonian $\hat{H}_S$, we must first determine the *Weyl symbol* $H_S(\vartheta, \varphi)$ corresponding to $\hat{H}_S$, which is defined by the map onto phase space given by Eq. (247). Then one may, in principle, again calculate the switching fields using Thiaville's method [19]. The starting point of this calculation is the normalized energy of the spin $\bar{V}(\mathbf{u})$ in the presence of a dc magnetic field $\mathbf{H}$ defined as

$$\bar{V}(\mathbf{u}) = G(\mathbf{u}) - 2(\mathbf{u} \cdot \mathbf{h}), \tag{359}$$

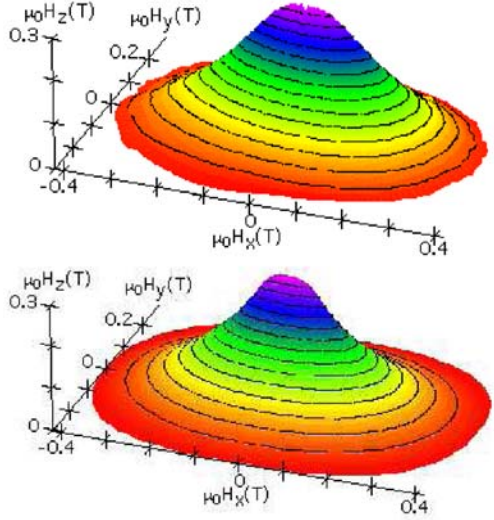

**Figure 12.** (Color on line) 3D theoretical and experimental (measured at $T = 35$ mK with the microSQUID; upper figure) switching field surfaces of a 3 nm cobalt cluster. These surfaces are symmetrical with respect to the $H_x - H_y$ plane and only the upper part ($\mu_0 H_z > 0$) is shown. Continuous lines on the surface are contour lines on which $\mu_0 H_z$ is constant. The theoretical switching field surface is calculated via Thiaville's method [19] the free energy $V = -K_1 \cos^2 \vartheta + K_2 \sin^2 \vartheta \sin^2 \varphi - (K_4/4)(\sin^2 2\vartheta + \sin^4 \vartheta \cos^2 2\varphi)$ with the anisotropy constants $K_1 = 2.2 \cdot 10^5$ J/m$^3$, $K_2 = 0.9 \cdot 10^5$ J/m$^3$, $K_4 = 0.1 \cdot 10^5$ J/m$^3$. Reprinted from M. Jamet, W. Wernsdorfer, C. Thirion, D. Mailly, V. Dupuis, P. Mélinon, and A. Pérez, Phys. Rev. Lett. 86, 4676 (2001), with the permission of the American Physical Society.

where $\mathbf{u} = (\sin\vartheta\cos\varphi, \sin\vartheta\sin\varphi, \cos\vartheta)$ is the unit vector specifying the representative point $(\vartheta, \varphi)$ in phase space (see Fig. 45 in Appendix D), $\mathbf{h}$ is the normalized external field $\mathbf{H}/H_A$ ($H_A$ is a normalizing constant which has the meaning of the effective anisotropy field), and $G(\mathbf{u}) = H_S(\vartheta, \varphi)/H_A$ is the normalized Hamiltonian in the absence of the external field $\mathbf{H}$. The switching field is characterized by the requirement that both the first and second derivatives of the normalized energy $\bar{V}$ with respect to $\vartheta$ and $\varphi$ must vanish, indicating that one metastable minimum and one saddle point in the potential $\bar{V}$ merge, giving rise to a point of inflexion. These conditions correspond to a *switching field surface* in 3D space. At any point of that surface, $\bar{V}$ must satisfy the stationary conditions

$$\frac{\partial \bar{V}}{\partial \vartheta} = \frac{\partial G}{\partial \vartheta} - 2(\mathbf{h} \cdot \mathbf{e}_\vartheta) = 0,$$

$$\frac{\partial \bar{V}}{\partial \varphi} = \frac{\partial G}{\partial \varphi} - 2(\mathbf{h} \cdot \mathbf{e}_\varphi)\sin\vartheta = 0,$$

so that the field vector $\mathbf{h}$ can be described by a parameter $\lambda$, viz.,

$$\mathbf{h} = \lambda \mathbf{e}_r + \frac{1}{2}\frac{\partial G}{\partial \vartheta}\mathbf{e}_\vartheta + \frac{1}{2\sin\vartheta}\frac{\partial G}{\partial \varphi}\mathbf{e}_\varphi, \tag{360}$$

where the unit vectors $\mathbf{e}_r$, $\mathbf{e}_\vartheta$, and $\mathbf{e}_\varphi$ forming the orthonormal direct basis are defined as

$$\mathbf{e}_r = \begin{pmatrix} \sin\vartheta\cos\varphi \\ \sin\vartheta\sin\varphi \\ \cos\vartheta \end{pmatrix}, \mathbf{e}_\vartheta = \begin{pmatrix} \cos\vartheta\cos\varphi \\ \cos\vartheta\sin\varphi \\ -\sin\vartheta \end{pmatrix}, \mathbf{e}_\varphi = \begin{pmatrix} -\sin\varphi \\ \cos\varphi \\ 0 \end{pmatrix}. \tag{361}$$

The switching conditions are now determined by the equation

$$\frac{\partial^2 \bar{V}}{\partial \vartheta^2}\frac{\partial^2 \bar{V}}{\partial \varphi^2} - \left[\frac{\partial^2 \bar{V}}{\partial \vartheta \partial \varphi}\right]^2 = 0. \tag{362}$$

Because the second derivatives of $\bar{V}$ are given by

$$\frac{\partial^2 \bar{V}}{\partial \vartheta^2} = \frac{\partial^2 G}{\partial \vartheta^2} + 2\lambda,$$

$$\frac{\partial^2 \bar{V}}{\partial \varphi^2} = \frac{\partial^2 G}{\partial \varphi^2} + \left(\cot\vartheta\frac{\partial G}{\partial \vartheta} + 2\lambda\right)\sin^2\vartheta,$$

$$\frac{\partial^2 \bar{V}}{\partial \vartheta \partial \varphi} = \frac{\partial^2 \bar{V}}{\partial \varphi \partial \vartheta} = \sin\vartheta\frac{\partial}{\partial \vartheta}\left(\frac{1}{\sin\vartheta}\frac{\partial G}{\partial \varphi}\right).$$

Eq. (362) reduces to a quadratic equation in $\lambda$, viz.,

$$4\lambda^2 + 2\lambda\left[\frac{1}{\sin^2\vartheta}\frac{\partial^2 G}{\partial \varphi^2} + \cot\vartheta\frac{\partial G}{\partial \vartheta} + \frac{\partial^2 G}{\partial \vartheta^2}\right]$$

$$+ \left[\frac{1}{\sin^2\vartheta}\frac{\partial^2 G}{\partial \varphi^2} + \cot\vartheta\frac{\partial G}{\partial \vartheta}\right]\frac{\partial^2 G}{\partial \vartheta^2} - \left[\frac{\partial}{\partial \vartheta}\left(\frac{1}{\sin\vartheta}\frac{\partial G}{\partial \varphi}\right)\right]^2 = 0,$$

which has two roots $\lambda^+(\vartheta, \varphi)$ and $\lambda^-(\vartheta, \varphi)$ given by

$$\lambda^\pm = -\frac{1}{4}\left(\frac{1}{\sin^2\vartheta}\frac{\partial^2 G}{\partial \varphi^2} + \cot\vartheta\frac{\partial G}{\partial \vartheta} + \frac{\partial^2 G}{\partial \vartheta^2}\right)$$

$$\pm \frac{1}{4}\sqrt{\left[\frac{1}{\sin^2\vartheta}\frac{\partial^2 G}{\partial \varphi^2} + \cot\vartheta\frac{\partial G}{\partial \vartheta} - \frac{\partial^2 G}{\partial \vartheta^2}\right]^2 + 4\left[\frac{\partial}{\partial \vartheta}\left(\frac{1}{\sin\vartheta}\frac{\partial G}{\partial \varphi}\right)\right]^2}. \tag{363}$$

Now the semi axis, Eq. (360), described by $\lambda > \lambda^+$ is the locus of the fields for which the magnetization is stable. Moreover when $\lambda = \lambda^+$ the metastable minimum in the potential $\bar{V}$ disappears so that the spin vector $\mathbf{S}$ can then escape from the potential well. Thus, the switching field surface may be obtained from the vector $\mathbf{h}_S$ defined as [19]

$$\mathbf{h}_S = \lambda^+ \mathbf{e}_r + \frac{1}{2}\frac{\partial G}{\partial \vartheta}\mathbf{e}_\vartheta + \frac{1}{2\sin\vartheta}\frac{\partial G}{\partial \varphi}\mathbf{e}_\varphi. \tag{364}$$

Numerous examples of the calculation of switching field surfaces in 3D space and the 2D critical curves for various *classical* free energy densities have been given, e.g., in Refs.

[5,6,18,19]. Here, following [66], we calculate 2D critical curves for the uniaxial, biaxial, cubic, and mixed Hamiltonians defined by Eqs. (243)-(246). The Weyl symbols $H_S^{un}(\vartheta,\varphi)$, $H_S^{bi}(\vartheta,\varphi)$, $H_S^{cub}(\vartheta,\varphi)$, and $H_S^{mix}(\vartheta,\varphi)$ of these four model Hamiltonians are given by Eqs. (248)-(251), which bear an obvious resemblance to the corresponding classical free energy densities (see Secs. II.D1-II.D.5). If we further suppose that a uniform external magnetic field $\mathbf{H}$ is applied in the $x$-$z$ plane, the Zeeman term operator $-(\xi/S)\left(\sin\psi \hat{S}_X - \cos\psi \hat{S}_Z\right)$ just transforms to the simple phase-space expression $-\xi\cos(\vartheta-\psi)$, where $\psi$ is the angle between the applied field $\mathbf{H}$ and the $Z$ axis. Thus, the switching fields $\mathbf{h}_{un}$, $\mathbf{h}_{bi}$, and $\mathbf{h}_{cub}$ in the $x-z$ plane (i.e., for $\varphi=0$) can be calculated from the Weyl symbols Eqs. (248) - (250) and Eq. (364) yielding

$$\mathbf{h}_{un} = Q_{un}\mathbf{h}_{un}^{cl}, \qquad (365)$$

$$\mathbf{h}_{bi} = Q_{bi}\mathbf{h}_{bi}^{cl}, \qquad (366)$$

$$\mathbf{h}_{cub} = Q_{cub}\mathbf{h}_{cub}^{cl}, \qquad (367)$$

where

$$Q_{un} = Q_{bi} = 1 - \frac{1}{2S}, \qquad (368)$$

$$Q_{cub} = \frac{(S-1/2)(S-1)(S-3/2)}{S^3} \qquad (369)$$

are the quantum correction factors to the corresponding classical switching fields $\mathbf{h}_{un}^{cl}$, $\mathbf{h}_{bi}^{cl}$, and $\mathbf{h}_{cub}^{cl}$ in the $x$-$z$ plane given by the known equations [5,6]

$$\mathbf{h}_{un}^{cl} = \left(\sin^3\vartheta, -\cos^3\vartheta\right),$$

$$\mathbf{h}_{cub}^{cl} = \left(\sin^3\vartheta(3\cos 2\vartheta+2), \cos^3\vartheta(3\cos 2\vartheta-2)\right),$$

$$\mathbf{h}_{bi}^{cl} = \left(\sin\vartheta[2\Delta+f(\vartheta)], \cos\vartheta[2-f(\vartheta)]\right),$$

where $\Delta=\delta/\sigma$ and

$$f(\vartheta) = (1+\Delta)\sin^2\vartheta + \left|2\Delta - (1+\Delta)\sin^2\vartheta\right|.$$

For *mixed* anisotropy, however, the corresponding equation for the switching field $\mathbf{h}_{mix}$ is much more complicated and therefore must be calculated numerically. The parametric plots of the parallel $h_Z$ vs. the perpendicular $h_X$ component of the switching field for the above spin systems are shown in Fig. 13. In general, the figure indicates that the switching field amplitudes *increase* markedly with *increasing S* all the while tending to their classical limiting values as $S\to\infty$ corresponding to diminishing tunneling effects as that mechanism is gradually shut off.

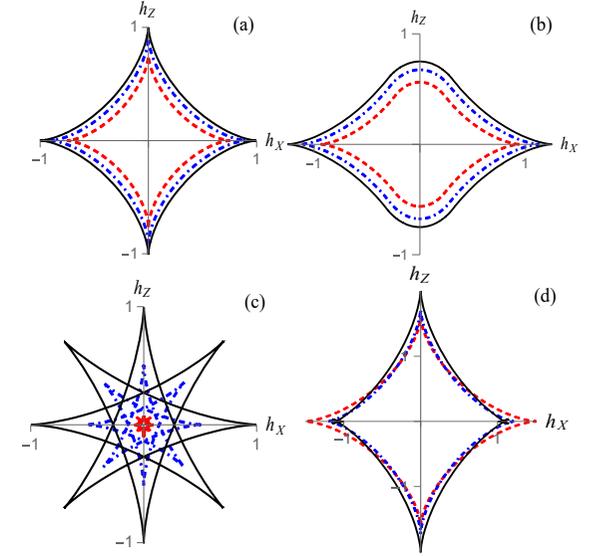

**Figure 13.** (Color on line) Spin dependence of switching field curves for uniaxial (a), biaxial at $\sigma/\delta = 0.25$ (b), cubic (c), and mixed at $\sigma_1/\sigma_2 = 0.5$ and $\chi = 0$ (d) anisotropies for $S = 2$ (red dashed lines), 5 (blue dash-dotted lines), and $S\to\infty$ (black solid lines; classical limit).

We emphasize that the above calculations because they are entirely based on the phase space representation of the Hamiltonian operator *ignore thermal effects* as in the original Stoner-Wohlfarth and Thiaville calculations. In order to account for these, it is necessary to estimate the temperature dependence of the spin reversal time, which may be accomplished, e.g., using the quantum TST, which we have described in Sec. II.D.6. This will again only involve the quantum equilibrium phase-space distributions, which we have calculated in the preceding sections.

*8. Discussion*

We have just shown how the phase space method may be used to construct equilibrium distribution functions in the configuration space of polar angles $(\vartheta,\varphi)$ for spin systems in the equilibrium state described by the equilibrium distribution $\hat{\rho}_S = e^{-\beta\hat{H}_S}/Z_S$. The Wigner function may be represented in all cases as a finite series of spherical harmonics like the corresponding classical orientational distribution and transparently reduces to the usual Fourier series Eq. (184) in the classical limit, $S\to\infty$. Moreover, relevant quantum mechanical averages (such as the magnetization) may be calculated in a manner analogous to the corresponding classical averages using the Weyl symbol of the appropriate quantum operator [see Eq. (238)]. The resulting Wigner functions can now be used to determine the spin dependence of the switching fields and hysteresis curves and may also be applied to other problems requiring only a knowledge of equilibrium distributions. This conclusion is significant particularly from an experimental point of view as the transition between magnetic molecular cluster and single domain ferromagnetic nanoparticle behavior is essentially demarcated via the hysteresis loops and the corresponding switching field curves [9]. Furthermore, such Wigner functions are important, in the interpretation of quantum tunneling phenomena in ferromagnetic nanoparticles and molecular magnets (see, e.g., [9]) and also in the investigation of the crossover region between reversal of the magnetization of these particles by thermal agitation and reversal by macroscopic quantum tunneling. For instance, by analogy with Néel's classical calculation [16] the simplest description of quantum effects in the magnetization reversal time of a nanoparticle is provided by the inverse escape rate from the wells of the magnetocrystalline and external field potential as rendered by quantum TST. Thus, the TST rate provides an important benchmark for both analytical calculations of the escape rate, which account for dissipation using quantum rate theory and for the numerical results obtained from the appropriate quantum master equation as well as allowing one to incorporate thermal effects in the switching fields. Now TST ignores the disturbance to the equilibrium distribution in the wells created by the loss of the magnetization due to escape over the barrier and so involves the equilibrium distribution only as that is assumed to prevail everywhere. Nevertheless via TST as corrected for quantum effects [e.g., Eq. (351)] which stems from the phase space representation it is possible to predict the temperature dependence of the switching fields and corresponding hysteresis loops within the limitations imposed by that theory. Therefore the results should be relevant to experiments seeking evidence for macroscopic quantum tunneling where the temperature dependence of the loops is crucial as they are used [9] to differentiate tunneling from thermal agitation behavior. The equilibrium quantum distribution is also essential in the inclusion of *nonequilibrium* effects in the quantum escape rate. For example, a master equation describing the time evolution of the quasiprobability density in the representation space is required in generalizing the classical escape rate calculations pioneered by Kramers [28] for point particles and by Brown [23,24] for single domain ferromagnetic particles using the Fokker-Planck equation.

In Section III, we shall show that a knowledge of the equilibrium phase-space distribution is also important in two other fundamental aspects of nonequilibrium phenomena involving master equations in the phase space representation. The first is in formulating the initial conditions for their solution as the appropriate quantum equilibrium distribution, which must now play the role of the Boltzmann distribution in the corresponding classical problem. Secondly, the equilibrium quantum distribution plays a vital role in the determination of the diffusion coefficients in a quantum master equation because this distribution must be the stationary solution of that equation. This fact, analogous to Einstein and Smoluchowski's imposition of the Maxwell-Boltzmann distribution as the stationary solution of the Fokker-Planck equation in order to determine drift and diffusion coefficients, will also allow one to calculate these coefficients in the quantum case. This is illustrated for the particular case of a spin in a uniform field in Sec. III.B.3 below (see also [62]) indicating clearly how all the solution methods developed for the classical Fokker-Planck equation apply to the quantum case just as the corresponding solutions for particles [48] (see Sec. II.B.3). We remark, however, that calculation of the drift and diffusion coefficients for axially symmetric potentials is much simpler than that for nonaxial symmetry since only the single polar angle $\vartheta$ is involved rather than the two angles $\vartheta$ and $\varphi$. The restriction to axial symmetry also gives rise to further mathematical simplifications, since the quantum master equation now has essentially the same form as the classical Fokker-Planck equation in the single coordinate $\vartheta$ implying that formulas for the mean first passage time, integral relaxation time, etc., may be directly carried over to the quantum case. This is not so for nonaxially symmetric potentials as the two variables involved give rise to a perturbation problem similar to that encountered in solving the Wigner problem for particles in classical phase space.

## III. MASTER EQUATION IN PHASE SPACE FOR AXIALLY SYMMETRIC SYSTEMS

### A. Master equation for a uniaxial nanomagnet subjected to a dc magnetic field

We shall now apply, as an illustrative example, the phase space method to a uniaxial nanomagnet of arbitrary spin number $S$ in an external *constant* magnetic field **H** applied along the $Z$-axis, i.e., the easy axis, where the Hamiltonian operator $\hat{H}_S$ has the axially symmetric form Eq. (267), namely,

$$\beta \hat{H}_S = -\frac{\sigma}{S^2}\hat{S}_Z^2 - \frac{\xi}{S}\hat{S}_Z. \qquad (370)$$

This Hamiltonian comprises a uniaxial anisotropy term $-\sigma \hat{S}_Z^2/S^2$ plus the Zeeman coupling to the external field $-\xi \hat{S}_Z/S$, constituting a generic model for relaxation phenomena in uniaxial spin systems such as molecular magnets, nanoclusters, etc. (see, e.g., Refs. 3 and 80 and references cited therein). In the standard basis of spin functions $|S,m\rangle$ (see Appendix A), which describe the states with definite spin $S$ and spin projection $m$ onto the $Z$-axis, i.e., $\hat{S}_Z|S,m\rangle = m|S,m\rangle$, the Hamiltonian $\hat{H}_S$, Eq. (370), has an energy spectrum with a double-well structure and two minima at $m = \pm S$ separated by a potential barrier. Notice that in strong bias fields, $\xi_0 > \sigma(2S-1)/S$, the barrier disappears. Now generally speaking, spin reversal can take place either by thermal activation or by tunneling or a combination of both. The tunneling may occur from one side of the barrier to the other between resonant, equal-energy states coupled by transverse fields or high-order anisotropy terms [80-82]. Now Garanin [80] and García-Palacios and Zueco [81,82] by using the spin density matrix in the second order of perturbation theory in the spin bath coupling have studied the longitudinal relaxation of quantum uniaxial nanomagnets with the Hamiltonian Eq. (370). In other words, they gave a concise treatment of the spin dynamics by directly proceeding from the quantum Hubbard operator representation of the evolution equation for the spin density matrix. This axially symmetric problem has also been treated via the phase space method in Ref. 64 and may be summarized as follows.

*1. Explicit form of the master equation*

Using the collision operator in the symmetrized Hubbard form (60) as written for the particular Hamiltonian given by Eq. (370), we have from the general reduced density matrix evolution Eq. (40) the evolution equation for the reduced density matrix of a uniaxial nanomagnet [64]

$$\frac{\partial \hat{\rho}_S}{\partial t} - \frac{i}{\hbar\beta}\left\{\frac{\sigma}{S^2}\left[\hat{S}_0^2, \hat{\rho}_S\right] + \frac{\xi}{S}\left[\hat{S}_0, \hat{\rho}_S\right]\right\} = \mathrm{St}(\hat{\rho}_S). \tag{371}$$

Thus written explicitly the collision kernel operator $\mathrm{St}(\hat{\rho}_S)$, characterizing the spin-bath interaction, is given by

$$\mathrm{St}(\hat{\rho}_S) = D_0\left(\left[\hat{S}_0, \hat{\rho}_S\hat{S}_0\right] + \left[\hat{S}_0\hat{\rho}_S, \hat{S}_0\right]\right)$$

$$-D_{-1}\left(\left[\hat{S}_{-1}, \hat{\rho}_S e^{-\frac{\sigma}{2S^2}\hat{S}_0^2 - \frac{\xi}{2S}\hat{S}_0}\hat{S}_1 e^{\frac{\sigma}{2S^2}\hat{S}_0^2 + \frac{\xi}{2S}\hat{S}_0}\right] + \left[e^{\frac{\sigma}{2S^2}\hat{S}_0^2 + \frac{\xi}{2S}\hat{S}_0}\hat{S}_1 e^{-\frac{\sigma}{2S^2}\hat{S}_0^2 - \frac{\xi}{2S}\hat{S}_0}\hat{\rho}_S, \hat{S}_{-1}\right]\right) \tag{372}$$

$$-D_1\left(\left[\hat{S}_1, \hat{\rho}_S e^{-\frac{\sigma}{2S^2}\hat{S}_0^2 - \frac{\xi}{2S}\hat{S}_0}\hat{S}_{-1} e^{\frac{\sigma}{2S^2}\hat{S}_0^2 + \frac{\xi}{2S}\hat{S}_0}\right] + \left[e^{\frac{\sigma}{2S^2}\hat{S}_0^2 + \frac{\xi}{2S}\hat{S}_0}\hat{S}_{-1} e^{-\frac{\sigma}{2S^2}\hat{S}_0^2 - \frac{\xi}{2S}\hat{S}_0}\hat{\rho}_S, \hat{S}_1\right]\right).$$

Because of the operator relations

$$e^{\frac{\sigma}{2S^2}\hat{S}_0^2 + \frac{\xi}{2S}\hat{S}_0}\hat{S}_{\pm 1}e^{-\frac{\sigma}{2S^2}\hat{S}_0^2 - \frac{\xi}{2S}\hat{S}_0} = e^{-\frac{\sigma}{2S^2} \pm \frac{\xi}{2S}}e^{\pm\frac{\sigma}{S^2}\hat{S}_0}\hat{S}_{\pm 1}, \tag{373}$$

$$\hat{S}_{\pm 1}e^{\mp\frac{\sigma}{S^2}\hat{S}_0} = e^{\frac{\sigma}{S^2}}e^{\mp\frac{\sigma}{S^2}\hat{S}_0}\hat{S}_{\pm 1}, \tag{374}$$

we have from Eq. (372) a simplified form of $\mathrm{St}(\hat{\rho}_S)$, viz.,

$$\mathrm{St}(\hat{\rho}_S) = D_{\|}\left(\left[\hat{S}_0, \hat{\rho}_S\hat{S}_0\right] + \left[\hat{S}_0\hat{\rho}_S, \hat{S}_0\right]\right)$$

$$-2D_{\perp}\left\{e^{\frac{\sigma}{2S^2}+\frac{\xi}{2S}}\left(\left[\hat{S}_1 e^{\frac{\sigma}{S^2}\hat{S}_0}\hat{\rho}_S, \hat{S}_{-1}\right] + \left[\hat{S}_1, \hat{\rho}_S e^{\frac{\sigma}{S^2}\hat{S}_0}\hat{S}_{-1}\right]\right)\right. \tag{375}$$

$$\left. + e^{\frac{\sigma}{2S^2}-\frac{\xi}{2S}}\left(\left[\hat{S}_{-1}, \hat{\rho}_S e^{-\frac{\sigma}{S^2}\hat{S}_0}\hat{S}_1\right] + \left[\hat{S}_{-1}e^{-\frac{\sigma}{S^2}\hat{S}_0}\hat{\rho}_S, \hat{S}_1\right]\right)\right\},$$

where we have introduced the notation $D_{\perp} = D_{\pm 1}/2$ and $D_{\|} = D_0$ for the diffusion coefficients. Now Eq. (371) describes the evolution of the spin system in contact with the thermal bath at temperature $T$. Thus one important property of the collision kernel operator, namely, Eq. (56) above, is satisfied by the $\mathrm{St}(\hat{\rho}_S)$ given by Eq. (375) namely that the equilibrium spin density matrix $\hat{\rho}_{eq} = e^{-\beta\hat{H}_S}/\mathrm{Tr}\{e^{-\beta\hat{H}_S}\}$ renders the collision kernel equal to zero, i.e., $\mathrm{St}(\hat{\rho}_{eq}) = 0$. Conditions for the validity of the reduced density matrix evolution Eq. (371) have been discussed in detail in Sec. II.A.2.

We now proceed to the phase space representation of Eq. (371), which is accomplished by writing that equation as the inverse map of a Weyl symbol (see also [76]). By substituting the density matrix $\hat{\rho}_S$ so rendered into the reduced density matrix evolution Eq. (371), we have the inverse map (dropping the parameter $s$)

$$\int\left(\hat{w}\frac{\partial}{\partial t}W_S - W_S\left\{\frac{i}{\hbar\beta}\left(\frac{\sigma}{S^2}\left[\hat{S}_0^2, \hat{w}\right] + \frac{\xi}{S}\left[\hat{S}_0, \hat{w}\right]\right) + \mathrm{St}(\hat{w})\right\}\right)d\Omega = 0, \tag{376}$$

where $d\Omega = \sin\vartheta d\vartheta d\varphi$. Now as it stands the formal inverse map Eq. (376) does not have the standard form, Eq. (235), of the usual inverse Wigner-Stratonovich map with kernel $\hat{w}(\vartheta,\varphi)$. Therefore in order to facilitate this objective we must first transform the various commutators occurring in the integrand of Eq. (376) into the phase-space representation These will then appear as configuration space differential operators acting on the Wigner-Stratonovich kernel $\hat{w}(\vartheta,\varphi)$ [cf. Eq. (C11) *et seq.* in Appendix C]. This procedure, which involves lengthy operator manipulations for each commutator occurring in Eq. (376) as fully described in Appendix C, will then allow one to express the integrand above in the standard phase-space form demanded by Eq. (235). In this way, we will ultimately have the master equation for the phase-space distribution $W_S(\vartheta,\varphi,t)$, viz.,

$$\frac{\partial}{\partial t}W_S - \frac{\sigma}{\hbar\beta S^2}\left(2S\cos\vartheta - \sin\vartheta\frac{\partial}{\partial\vartheta} + \frac{S\xi}{\sigma}\right)\frac{\partial}{\partial\varphi}W_S$$

$$= D_{\|}\frac{\partial^2}{\partial^2\varphi}W_S + D_{\perp}\frac{\cot\vartheta}{\sin\vartheta}\left[\cos\vartheta\frac{\partial^2}{\partial^2\varphi}R_+^{\prime(S)}W_S + \frac{\partial^2}{\partial^2\varphi}R_-^{\prime(S)}W_S\right]$$

$$+ \frac{D_{\perp}}{\sin\vartheta}\left[\frac{\partial}{\partial\vartheta}\sin\vartheta\left(\frac{\partial}{\partial\vartheta}R_+^{\prime(S)}W_S + \cos\vartheta\frac{\partial}{\partial\vartheta}R_-^{\prime(S)}W_S + 2S\sin\vartheta R_-^{\prime(S)}W_S\right)\right] \tag{377}$$

$$+ D_{\perp}\left[\left(\sin\vartheta\frac{\partial}{\partial\vartheta} - 2S\cos\vartheta\right)\frac{\partial}{\partial\varphi}R_+^{\prime\prime(S)}W_S - \frac{\partial}{\partial\varphi}R_-^{\prime\prime(S)}W_S\right],$$

where the phase-space differential operators $R_{\pm}^{(S)} = R_{\pm}^{\prime(S)} + iR_{\pm}^{\prime\prime(S)}$ are defined in Appendix C. The left-hand side of Eq. (377) is just the quantum analog of the classical Liouville equation for a uniaxial nanomagnet, while the collision operator given by the right-hand side of Eq. (377) is the quantum analog of the Fokker-Planck operator for classical spins Eq. (4). In summary, the master Eq. (377) follows from the equation of motion of the reduced density matrix Eq. (371) written as the standard form Eq. (235) of the inverse Wigner-Stratonovich map of a Weyl symbol (see Appendix C). Everywhere the interactions between the spin and the heat bath are taken small enough to use the weak coupling limit and the correlation time characterizing the bath is taken short enough to regard the stochastic process originating in the bath as Markovian.

In purely longitudinal relaxation, when the azimuthal angle dependence of $W_S$ may be ignored, the Liouville term vanishes in Eq. (377) and the corresponding phase space evolution equation for $W_S(z = \cos\vartheta, t)$ then simplifies to the axially symmetric form

$$\frac{\partial W_S}{\partial t} = \frac{\partial}{\partial z}\left(D^{(2)}W_S + D^{(1)}W_S\right), \quad (378)$$

where

$$D^{(1)} = 2D_\perp S(1-z^2)R_-^{\prime(S)} \quad (379)$$

and

$$D^{(2)} = D_\perp (1-z^2)\left(\frac{\partial}{\partial z}R_+^{\prime(S)} + z\frac{\partial}{\partial z}R_-^{\prime(S)}\right). \quad (380)$$

The phase space master Eq. (378) is then *formally* similar to the single spatial variable Fokker-Planck equation for the orientation distribution function $W(z,t)$

$$\frac{\partial W}{\partial t} = D_\perp \frac{\partial}{\partial z}\left[(1-z^2)\left(\frac{\partial W}{\partial z} + \beta W \frac{\partial V}{\partial z}\right)\right] \quad (381)$$

describing rotational diffusion of a classical spin in an axially symmetric potential [5]

$$\beta V = -\sigma \cos^2\vartheta - \xi \cos\vartheta \quad (382)$$

and in the classical limit, Eq. (378) reduces after lengthy manipulations to it as we shall demonstrate in Appendix C.

One of the major computational difficulties associated with the phase space master Eq. (377) now appears. By inspection of that equation even for axial symmetry high order spin number dependent differential operators occur apart from two notable exceptions. The first of these comprises noninteracting spins in a uniform field where all the higher order derivatives in the operators $R_\pm^{(S)}$ vanish. Thus they become closed transcendental functions (see for example Eq. (411) et. seq. below). In this case we will have differential recurrence relations (see Eq. (418) et. seq. below) which are essentially similar to those occurring in the corresponding classical problem (see for example Eq. (434) et. seq. below). The other exception is that in the absence of any potential whatsoever, i.e., when $\sigma = 0$ and $\xi = 0$, the differential operators merely reduce to $R_+^{(S)} = 1$ and $R_-^{(S)} = 0$, respectively (see Appendix C), so that Eq. (377) becomes (setting $D_\parallel = D_\perp$)

$$\frac{\partial W_S}{\partial t} = D_\perp \left[\frac{1}{\sin\vartheta}\frac{\partial}{\partial \vartheta}\left(\sin\vartheta \frac{\partial W_S}{\partial \vartheta}\right) + \frac{1}{\sin^2\vartheta}\frac{\partial^2 W_S}{\partial \varphi^2}\right]. \quad (383)$$

Equation (383) corresponds to the classical Fokker-Planck equation [5] in the single coordinate $\vartheta$ for the orientational distribution function of free magnetic dipole moments on the unit sphere. Hence, like the free quantum translational Brownian motion (see Sec. II.B.3), the phase-space distribution $W_S$ of free quantum spins obeys the classical Fokker-Planck equation Eq. (383) for the rotational diffusion of free classical spins [5].

Thus it would appear that the phase-space master Eq. (377) is in general of limited practical use. However, this does not preclude one from deriving differential-recurrence relations for observables via the density matrix evolution equation by using the one-to-one correspondence between the averages polarization operators in Hilbert space and the averages of the spherical harmonics (see Section II.C) as we now describe. This will automatically yield the recurrence relations Eq. (389) below for the averages of the polarization operators which may be mapped onto averages of the spherical harmonics via Eq. (392). This procedure is of course just a special case of the formal one outlined in Section II.C.3 [Eq. (252) et seq.]

*2. Differential-recurrence relations for the statistical moments*

Recalling the last paragraph of the previous section the formal solutions of the axially symmetric density matrix evolution Eq. (371) corresponding to the Hamiltonian (370) and the corresponding phase space Eq. (377) for *arbitrary S* may be written as

$$\hat{\rho}_S(t) = \hat{\rho}_{eq} + \sum_{L=0}^{2S} c_L(t)\hat{T}_{L0}^{(S)} \quad (384)$$

and

$$W_S(\vartheta,\varphi,t) = W_S^{eq}(\vartheta) + \sum_{L=0}^{2S} b_L(t)Y_{L0}(\vartheta,\varphi), \quad (385)$$

respectively. The equilibrium phase-space distribution $W_S^{eq}(\vartheta)$ is the stationary solution of both of the phase space Eqs. (377) and (378). We emphasize that $W_S^{eq}(\vartheta)$ corresponds to the equilibrium spin density matrix $\hat{\rho}_{eq}$ and is defined by Eq. (229), that is

$$W_S^{eq}(\vartheta) = \text{Tr}\{\hat{\rho}_{eq}\hat{w}(\vartheta)\}. \quad (386)$$

The distribution $W_S^{eq}(\vartheta)$ defined by the map Eq. (386) has already been calculated in Sec. II.D.2 and is given by the finite series of Legendre polynomials Eq. (299) while the collision kernel of Eq. (377) satisfies $\text{St}(W_S^{eq}) = 0$, i.e., the distribution $W_S^{eq}$ defined by Eq. (299) is indeed the stationary solution of the phase space master Eqs. (377) and (378). The coefficients $c_L(t)$ and $b_L(t)$ (corresponding to the statistical moments) are, in turn, the averages of the polarization operators $\hat{T}_{L0}^{(S)}$ and the spherical harmonics $Y_{L0}$, respectively, viz.,

$$c_L(t) = \left\langle \hat{T}_{L0}^{(S)} \right\rangle(t) - \left\langle \hat{T}_{L0}^{(S)} \right\rangle_{eq}, \tag{387}$$

$$b_L(t) = \left\langle Y_{L0} \right\rangle(t) - \left\langle Y_{L0} \right\rangle_{eq}. \tag{388}$$

By substituting the operator expansion Eq. (384) and the spherical harmonic expansion Eq. (385) into the density matrix evolution Eq. (371) and the phase space Eq. (377), respectively, we have in each case a *finite* hierarchy of differential-recurrence equations for the statistical moments (in contrast to the classical case, where the corresponding hierarchy is *infinite*).

Since either of the Eqs. (384) and (385) will yield similar hierarchies for the statistical moments (see Sec. II.C.3), we shall describe the derivation of their recurrence relations using the density matrix. This is accomplished by first substituting the operator expansion Eq. (384) into the explicit evolution Eq. (371). Next we use Eq. (C19) from Appendix C for the expansion of the matrix exponents $e^{\frac{\sigma}{2S^2} \pm \frac{\xi}{2S}} e^{\pm \frac{\sigma}{S^2} \hat{S}_0}$ in terms of the polarization operators $\hat{T}_{L0}^{(S)}$ and the product formula Eq. (A28) from Appendix A for the operators $\hat{T}_{LM}^{(S)}$ which allows products of these to be expressed as a sum. In this way, we ultimately have a hierarchy of multi-term differential-recurrence equations for the relaxation functions $c_L(t)$ in Eq. (384), namely,

$$\tau_N \frac{d}{dt} c_L(t) = \sum_{L'=0}^{2S} g_{L,L'}^{(S)} c_{L'}(t). \tag{389}$$

Here $\tau_N = (2D_\perp)^{-1}$ is the characteristic (free diffusion) time and we have for the expansion coefficients which are defined in the usual way by the average

$$g_{L,L'}^{(S)} = -\frac{1}{2} e^{\frac{\sigma}{2S^2}} L(L+1) \text{Tr} \left\{ \hat{T}_{L'0}^{(S)} \left[ \cosh\left( \frac{\sigma}{S^2} \hat{S}_0 + \frac{\xi}{2S} \right) \hat{T}_{L0}^{(S)} + \sinh\left( \frac{\sigma}{S^2} \hat{S}_0 + \frac{\xi}{2S} \right) \right. \right.$$
$$\left. \left. \times \left( \frac{\sqrt{(2S-L)(2S+L+2)}}{\sqrt{(2L+3)(2L+1)}} \hat{T}_{L+10}^{(S)} - \frac{\sqrt{(2S-L+1)(2S+L+1)}}{\sqrt{(2L-1)(2L+1)}} \hat{T}_{L-10}^{(S)} \right) \right] \right\}, \tag{390}$$

where $\cosh(\hat{A})$ and $\sinh(\hat{A})$ appearing in Eq. (390) are matrix functions. Likewise, in the phase space representation, we formally have the relevant system of differential-recurrence equations for the relaxation functions $b_L(t)$ from the general recurrence relation Eq. (253), the matrix elements Eq. (255) and the particular Eq. (389), viz.,

$$\tau_N \frac{d}{dt} b_L(t) = \sum_{L'} p_{L',L}^{(S)} b_{L'}(t), \tag{391}$$

where

$$p_{L',L}^{(S)} = \sqrt{\frac{(2S-L')!(2S+L'+1)!}{(2S-L)!(2S+L+1)!}} g_{L',L}^{(S)}. \tag{392}$$

Alternatively using the phase space method the recurrence relation Eqs. (391) could be derived by directly substituting the spherical harmonic expansion Eq. (385) into the phase space evolution Eq. (377) and then using the recurrence relations of the spherical harmonics, viz., Eqs. (B15)-(B17). However in general very detailed manipulations would be involved for the reasons we have outlined. It should be mentioned that the equilibrium averages $\left\langle \hat{T}_{L0}^{(S)} \right\rangle_{eq}$ and $\left\langle Y_{L0} \right\rangle_{eq}$ satisfy similar however time-independent recurrence relations, viz.,

$$\sum_{L'=0}^{2S} g_{L,L'}^{(S)} \left\langle \hat{T}_{L'0}^{(S)} \right\rangle_{eq} = 0 \tag{393}$$

and

$$\sum_{L'} p_{L',L}^{(S)} \left\langle Y_{L'0} \right\rangle_{eq} = 0. \tag{394}$$

The resulting system of Eq. (390) and/or (391), which we have just derived, can be solved by either direct matrix diagonalization which involves calculating the eigenvalues and eigenvectors of the system matrix (see Sec. II.A.4) or by the computationally efficient (matrix) continued fraction method [5,71]. As shown below, the solutions can be obtained both for the transient and ac stationary (linear and nonlinear) responses of spins in magnetic fields.

In the limiting case of zero anisotropy $\sigma = 0$, Eq. (390) can be further simplified by once again using the general formula for the product of polarization operators in terms of the Clebsch-Gordan series, i.e., Eq. (A28) of Appendix A, thereby yielding

$$g_{L,L'}^{(S)} = -\frac{1}{2} L(L+1) \left[ \delta_{L'L} \cosh\frac{\xi}{2S} + \sinh\frac{\xi}{2S} \right.$$
$$\left. \times \left( \delta_{L'L+1} \sqrt{\frac{(2S-L)(2S+L+2)}{(2L+3)(2L+1)}} - \delta_{L'L-1} \sqrt{\frac{(2S-L+1)(2S+L+1)}{(2L-1)(2L+1)}} \right) \right]$$

so that with the replacement

$$c_L(t) \to (-1)^L \frac{\sqrt{(2S-L)!(2S+L+1)!(2L+1)}}{4\pi(2S)!} f_L(t), \tag{395}$$

we have from Eq. (389) the simple three-term differential recurrence relation

$$\tau_N \frac{d}{dt} f_L(t) = q_L f_L(t) + q_L^- f_{L-1}(t) + q_L^+ f_{L+1}(t), \tag{396}$$

with

$$q_L = -\frac{L(L+1)}{2} \cosh\frac{\xi}{2S}, \tag{397}$$

$$q_L^\pm = \pm \frac{L(L+1)(2S \pm L + 3/2 \pm 1/2)}{2(2L+1)} \sinh\frac{\xi}{2S}. \tag{398}$$

The quantum relaxation function $f_L(t)$ defined by Eq. (395) for a given $S$ corresponds to

$$f_L(t) = \langle P_L \rangle(t) - \langle P_L \rangle_{eq},$$

where the $P_L(z)$ are the Legendre polynomials [105]. This limiting case exactly corresponds to the spin relaxation in a uniform field treated comprehensively in Section III.B below.

Returning to the general case, in the classical limit, $S \to \infty$, the Hamiltonian Eq. (370) corresponds to the classical free energy Eq. (382) while the quantum differential-recurrence relation Eq. (389) reduces to the usual five term differential-recurrence relation for a classical uniaxial nanomagnet subjected to a uniform longitudinal field, namely,

$$\tau_N \frac{d}{dt} f_L(t) = q_L f_L(t) + q_L^- f_{L-1}(t) + q_L^+ f_{L+1}(t) + q_L^{--} f_{L-2}(t) + q_L^{++} f_{L+2}(t), \quad (399)$$

where

$$f_L(t) = \langle P_L(\cos\vartheta) \rangle(t) - \langle P_L(\cos\vartheta) \rangle_{eq} \quad (400)$$

is now the *classical* relaxation function with

$$q_L = -\frac{L(L+1)}{2}\left(1 - \frac{2\sigma}{(2L-1)(2L+3)}\right),$$

$$q_L^\pm = \mp \xi \frac{L(L+1)}{2(2L+1)},$$

and

$$q_L^{--} = -q_{L-1}^{++} = \frac{\sigma L(L+1)(L-1)}{(2L-1)(2L+1)}.$$

This classical problem has been treated in detail in Refs. [5,6,23,150-155]. In particular, the exact solution of Eq. (399) is given in Ref. [5], Ch. 7. For zero anisotropy, i.e., $\sigma = 0$, we have from Eq. (399) the known result for relaxation of a classical spin in a uniform field [5,156-159]

$$\tau_N \frac{d}{dt} f_L(t) + \frac{L(L+1)}{2} f_L(t) = \frac{\xi L(L+1)}{2(2L+1)}\left[f_{L-1}(t) - f_{L+1}(t)\right], \quad (401)$$

which exact solution has been also given in Ref. [5], Ch. 7.

We have indicated (see Appendix C for details) how one may derive a master equation for the evolution of the phase-space quasiprobability distribution for a uniaxial nanomagnet in contact with a heat bath at temperature $T$. This is accomplished by first expressing the reduced density matrix master equation in Hilbert space in terms of an inverse Wigner-Stratonovich transformation according to Eq. (376). The various commutators in the integrand of Eq. (376) involving the spin operators may then be converted into phase space differential operators using the orthogonality and recurrence properties of the polarization operators and the corresponding spherical harmonics to ultimately yield via the standard form of the inverse transformation Eqs. (234) and (235) the desired master equation for the distribution function in the phase space of the polar angles. Despite the superficial resemblance of the quantum diffusion Eqs. (377) and (378) (governing the behavior of the phase space distribution) to the Fokker-Planck equations for classical spins, the problem is actually much more complicated. The difficulty lies in the collision kernel which involves powers of differential operators up to the spin number $S$ considered, only simplifying for large spin numbers ($S \to \infty$) when the high order derivatives occurring in the operators may be ignored.

We have illustrated the phase space representation of spin relaxation by treating a uniaxial nanomagnet in a uniform magnetic field of arbitrary strength directed along the easy axis, thereby realizing that only a master equation in configuration space akin to the Fokker-Planck equation for classical spins is involved. Hence for spins (just as particles), the existing classical solution methods [5,71] also apply in the quantum case indeed suggesting new closed form quantum results via classical ones. The magnetization, dynamic susceptibility, characteristic relaxation times, etc., for the uniaxial system may now be evaluated. Notice that the spin relaxation of this uniaxial system have already been treated using the quantum Hubbard operator representation of the evolution equation for the spin density matrix [80-82] and as shown in Ref. [62] for spins in an external field alone both the phase space and density matrix methods yield results in outwardly very different forms. Nevertheless, the numerical values from both methods for the same physical quantities (such as relaxation times and susceptibility) coincide thereby establishing a vital corollary between the phase space and the density matrix methods. Thus, the phase space representation, because it is closely allied to the classical representation, besides being complementary to the operator one, transparently illustrates how quantum distributions reduce to the classical ones. The analysis is carried out via the finite Fourier series representation embodied in the Wigner-Stratonovich map as we have illustrated for axially symmetric potentials. It may be extended in the appropriate limits to nonaxially symmetric systems such as biaxial, cubic, etc. However, the difficulties (e.g., the operator form of the diffusion coefficients in the master equation) encountered in our treatment of axially symmetric potentials are indicative of the even greater ones which would be faced when generalizing the phase space representation to such potentials, where the Lowville term never vanishes.

In the next Section, we first consider the linear and nonlinear longitudinal relaxation for the particular case of the model parameters $\sigma = 0$ and $\xi \neq 0$ corresponding to a spin in an external dc magnetic field **H** directed along the *Z*-axis. Then we shall consider in Sec. C the general case $\sigma \neq 0$ and $\xi \neq 0$, i.e., a uniaxial nanomagnet in an external magnetic field.

## B. Spin relaxation in a dc magnetic field

For noninteracting spins in an external dc magnetic field **H** directed along the *Z*-axis, the Hamiltonian $\hat{H}_S$ is simply Eq. (370) with $\sigma = 0$

$$\hat{H}_S = -\hbar\omega_0 \hat{S}_z, \qquad (402)$$

where $\omega_0 = \xi/(\beta\hbar S)$ is the precession (Larmor) frequency. We shall determine both the *exact* and approximate solutions yielding the averaged longitudinal component of a spin $\langle \hat{S}_z \rangle(t)$ for arbitrary *S*. Furthermore, we shall show how the solution of the corresponding classical problem [5,159] carries over into the quantum domain and how the *exact* solution for the integral relaxation time due to an *arbitrarily strong* sudden change in the uniform field may be obtained. We remark that the original treatment of this transient response problem via the master equation in phase space was first given by Shibata *et al.* [35-37] and was further developed by Kalmykov *et al.* [62]. In the *linear response approximation,* the solution reduces to that previously given by Garanin [80] using the spin density matrix in the second order of perturbation theory in the spin bath coupling. That result was later rederived by García-Palacios and Zueco [81] who (again using the density matrix solution) considered the linear response of the longitudinal relaxation of a spin for arbitrary *S*.

### *1. Basic equations*

Expanding on the introductory paragraph above following [62], we analyze the transient relaxation dynamics of a spin $\hat{\mathbf{S}}$ in an external dc magnetic field **H** directed along the *Z*-axis and a random field **h**(*t*) characterizing the collision damping (due to the heat bath) incurred by the precessional motion of the spin. For the Hamiltonian given by Eq. (402), the evolution equation for the reduced density matrix is simply

$$\frac{\partial \hat{\rho}_S}{\partial t} - i\omega_0 \left[ \hat{S}_0, \hat{\rho}_S \right] = \text{St}(\hat{\rho}_S). \qquad (403)$$

Equation (403) merely represents the *particular* case $\sigma = 0$ of the more general Eq. (371) for a uniaxial nanomagnet subjected to a uniform dc magnetic field treated in the previous Section. For the Hamiltonian $\hat{H}_S$ Eq. (402), the collision kernel operator $\text{St}(\hat{\rho}_S)$ [Eq. (375) with $\sigma = 0$] becomes

$$\text{St}(\hat{\rho}_S) = D_\parallel \left( \left[ \hat{S}_0, \hat{\rho}_S \hat{S}_0 \right] + \left[ \hat{S}_0 \hat{\rho}_S, \hat{S}_0 \right] \right)$$
$$-2D_\perp \left\{ e^{-\beta\hbar\omega_0/2} \left( \left[ \hat{S}_{-1}, \hat{\rho}_S \hat{S}_1 \right] + \left[ \hat{S}_{-1} \hat{\rho}_S, \hat{S}_1 \right] \right) + e^{\beta\hbar\omega_0/2} \left( \left[ \hat{S}_1 \hat{\rho}_S, \hat{S}_{-1} \right] + \left[ \hat{S}_1, \hat{\rho}_S \hat{S}_{-1} \right] \right) \right\}. \qquad (404)$$

Now, in the master Eq. (377) for the *particular* case $\sigma = 0$ *corresponding to a spin in a uniform field*, all the higher order derivatives will disappear so that the operators $R_{\pm}^{(S)}$ just become the closed transcendental expressions

$$R_+^{(S)} = \cosh \frac{\beta\hbar\omega_0}{2} \text{ and } R_-^{(S)} = \sinh \frac{\beta\hbar\omega_0}{2}.$$

Then the master Eq. (377) takes on a much simpler form, namely,

$$\frac{\partial W_S}{\partial t} = \omega_0 \frac{\partial W_S}{\partial \varphi} + \frac{D_\perp \sinh\left(\frac{1}{2}\beta\hbar\omega_0\right)}{\sin\vartheta} \left\{ \cot\vartheta \left[ \cos\vartheta \coth \frac{\beta\hbar\omega_0}{2} + 1 \right] \frac{\partial^2 W_S}{\partial \varphi^2} \right.$$
$$\left. + \frac{\partial}{\partial\vartheta} \left[ \sin\vartheta \left( \coth \frac{\beta\hbar\omega_0}{2} + \cos\vartheta \right) \frac{\partial W_S}{\partial \vartheta} + 2S \sin^2\vartheta W_S \right] \right\} + D_\parallel \frac{\partial^2 W_S}{\partial \varphi^2}. \qquad (405)$$

By introducing the renormalization of the diffusion coefficients $D_\perp e^{-\beta\hbar\omega_0/2} \to D_\perp$, Eqs. (404) and (405) yield the result previously obtained for a nonsymmetrized form of the collision kernel operator [35-37]. This master equation describing the time evolution of $W_S(\vartheta,\varphi,t)$ again has essentially the same form as the corresponding Fokker-Planck equation for the distribution function $W(\vartheta,\varphi,t)$ of classical spin orientations in the configuration space [35]

$$\frac{\partial W}{\partial t} = \omega_0 \frac{\partial W}{\partial \varphi} + D_\perp \left\{ \frac{\xi}{\sin\vartheta} \frac{\partial}{\partial \vartheta}(\sin^2\vartheta W) + \frac{1}{\sin\vartheta} \frac{\partial}{\partial \vartheta} \left( \sin\vartheta \frac{\partial W}{\partial \vartheta} \right) + \frac{1}{\sin^2\vartheta} \frac{\partial^2 W}{\partial \varphi^2} \right\} \qquad (406)$$

($\xi = \beta\hbar\omega_0 S$). Equation (405) serves as the simplest example of the phase space method for open spin systems. In fact, it is just the rotational analog of the quantum translational harmonic oscillator treated using the Wigner function by Agarwal [121]. In this instance, the evolution Eq. (168) for the Wigner distribution $W(q,p,t)$ in the phase space of positions and momenta has the same mathematical form as the Fokker-Planck equation for the classical oscillator (see Sec. II.B.3). For longitudinal relaxation, where the distribution function $W_S$ is independent of the azimuth, the Liouville term in the evolution equation vanishes and Eq. (405) reduces to an equation very similar to that governing a classical spin in a uniform magnetic field [5]

$$\frac{\partial W_S}{\partial t} = \frac{D_\perp \sinh(\beta\hbar\omega_0/2)}{\sin\vartheta} \frac{\partial}{\partial \vartheta} \left\{ \left[ \sin\vartheta \left( \cos\vartheta + \coth \frac{\beta\hbar\omega_0}{2} \right) \right] \frac{\partial W_S}{\partial \vartheta} + 2S \sin^2\vartheta W_S \right\}. \qquad (407)$$

This simplification arises naturally and is to be expected on intuitive grounds because precession of a spin in a uniform field is effectively the rotational analog of the translational harmonic oscillator.

Equation (405) applies in the *narrowing limit case* in which the correlation time $\tau_c$ of the random field **h**(*t*) acting on the spin satisfies the condition $\gamma H \tau_c \ll 1$, where *H* is the averaged amplitude of the random magnetic field. The left hand side of Eq. (405) is the quantum analog of the Liouville equation for a spin which now is the same as the classical case for particles with quadratic Hamiltonians, while the right-hand side (collision kernel) characterizes the interaction of the spin with the thermal bath at temperature $T$. The remaining conditions for the validity of Eq. (405) have already been discussed. We remark that for longitudinal relaxation, Eq. (405) may be plausibly derived by postulating (like in the phase space treatment of the quantum translational Brownian motion, see Sec. II.B.3) a master equation for the Wigner function $W_S$ with collision terms given by a Kramers-Moyal expansion truncated at the second term. The various drift and diffusion coefficients in the truncated expansion may then be calculated by requiring that the equilibrium Wigner distribution $W_{eq}$, corresponding to the equilibrium spin density matrix $\hat{\rho}_{eq} = e^{-\beta \hat{H}_S}/\text{Tr}\left(e^{-\beta \hat{H}_S}\right)$, renders the collision kernel equal to zero (see Sec III.B.3 below).

*2. Quantum analog of the magnetic Langevin equation*

The spin relaxation described by the master equation (405) may also be equivalently described using a quantum analog of the magnetic Langevin equation with multiplicative noise. To see this we use the Stratonovich definition [160] of such equations constituting the mathematical idealization of the spin relaxation process [161]. Thus, it is unnecessary to transform them to Itô stochastic differential equations (e.g., [161]). Moreover, one can then use conventional calculus [5,161]. The Langevin equations governing the two stochastic equations of motion for the variables $\vartheta$ and $\varphi$ corresponding to the phase space master Eq. (405) are in the Stratonovich interpretation [5,67]

$$\dot{\vartheta}(t) = D_\perp \cot \vartheta(t) \left( Q_1^S[\vartheta(t)] - \sqrt{Q_1^S[\vartheta(t)] Q_2^S[\vartheta(t)]} \right)$$
$$-2D_\perp(S+1/4)\sin\vartheta(t)\sinh\frac{\beta\hbar\omega_0}{2} + \sqrt{Q_1^S[\vartheta(t)]}\left[h_\vartheta(t) - \alpha^{-1}h_\varphi(t)\right], \quad (408)$$

$$\dot{\varphi}(t) = -\omega_0 + \sqrt{Q_2^S[\vartheta(t)]}\csc\vartheta(t)\left[h_\varphi(t) + \alpha^{-1}h_\vartheta(t)\right], \quad (409)$$

where

$$Q_1^S[\vartheta(t)] = \frac{D_\perp \beta}{\eta(1+\alpha^{-2})}\left(\cosh\frac{\beta\hbar\omega_0}{2} + \cos\vartheta(t)\sinh\frac{\beta\hbar\omega_0}{2}\right), \quad (410)$$

$$Q_2^S[\vartheta(t)] = \frac{D_\perp \beta}{\eta(1+\alpha^{-2})}\left\{\cos^2\vartheta(t)\cosh\frac{\beta\hbar\omega_0}{2} + \cos\vartheta(t)\sinh\frac{\beta\hbar\omega_0}{2} + \frac{D_\parallel}{D_\perp}\sin^2\vartheta(t)\right\}, \quad (411)$$

and the components $h_\vartheta(t)$, $h_\varphi(t)$ of the random field $\mathbf{h}(t)$ in the spherical coordinate system or basis are expressed in terms of the components $h_X(t), h_Y(t), h_Z(t)$, in the Cartesian basis as [5]

$$h_\vartheta(t) = h_X(t)\cos\vartheta(t)\cos\varphi(t) + h_Y(t)\cos\vartheta(t)\sin\varphi(t) - h_Z(t)\sin\vartheta,$$

$$h_\varphi(t) = -h_X(t)\sin\varphi(t) + h_Y(t)\cos\varphi(t),$$

with

$$\overline{h_i(t)} = 0, \quad \overline{h_i(t)h_j(t')} = \frac{2\eta}{\beta}\delta_{ij}\delta(t-t'). \quad (412)$$

Here the indices $i, j = 1, 2, 3$ in Kronecker's delta $\delta_{ij}$ correspond to the Cartesian axes $X,Y,Z$ of the laboratory coordinate system $OXYZ$, and $\alpha = \gamma\eta\mu$ is a dimensionless dissipation (damping) parameter, $\eta$ is a "friction" coefficient, and the overbar means the statistical average over the realizations of the random field. In the isotropic diffusion ($D_\perp = D_\parallel$) and classical ($S \to \infty$) limit, the Langevin Eqs. (408) and (409) reduce to those for isotropic diffusion of a classical spin in spherical coordinates [5], namely,

$$\dot{\vartheta}(t) = D_\perp \left\{-\xi\sin\vartheta(t) + \sqrt{\frac{D_\perp \beta}{\eta(1+\alpha^{-2})}}\left[h_\vartheta(t) - \alpha^{-1}h_\varphi(t)\right]\right\}, \quad (413)$$

$$\dot{\varphi}(t) = -\omega_0 + \sqrt{\frac{D_\perp \beta}{\eta(1+\alpha^{-2})}}\csc\vartheta(t)\left[h_\varphi(t) + \alpha^{-1}h_\vartheta(t)\right]. \quad (414)$$

To show that the Langevin equations (408) and (409) are equivalent to the master equation (405), we recall that by choosing Langevin equations for a set of two stochastic variables $\{\xi_1(t) = \vartheta(t), \xi_2(t) = \varphi(t)\}$ as

$$\dot{\xi}_i(t') = H_i[\xi_1(t'),\xi_2(t')] + \sum_{j=1}^{3} G_{ij}[\xi_1(t'),\xi_2(t')]h_j(t'),$$

($i = 1, 2, j = 1, 2, 3$) and subsequently interpreting them as Stratonovich stochastic differential equations, then the averaged equations for the drift, $D_i$, and diffusion, $D_{ij}$, coefficients time $t$ are [5,71]

$$D_i = \overline{\dot{\xi}_i} = \lim_{\tau \to 0}\frac{1}{\tau}\overline{[\xi_i(t+\tau) - \xi_i(t)]}$$
$$= H_i(x_1,x_2,t) + \frac{\eta}{\beta}\sum_{k=1}^{2}\sum_{j=1}^{3} G_{kj}(x_1,x_2)\frac{\partial}{\partial x_k}G_{ij}(x_1,x_2), \quad (415)$$

$$D_{ij} = \lim_{\tau \to 0}\frac{1}{2\tau}\overline{[\xi_i(t+\tau) - \xi_i(t)][\xi_j(t+\tau) - \xi_j(t)]}$$
$$= \frac{\eta}{\beta}\sum_{k=1}^{3} G_{ik}(x_1,x_2)G_{jk}(x_1,x_2). \quad (416)$$

Moreover due to Eqs. (415) and (416), we have from Eqs. (408) and (409) the drift and diffusion coefficients

$$D_1 = -D_\perp\left[(2S+1)\sin\vartheta\sinh\frac{\beta\hbar\omega_0}{2} - \cot\vartheta\left(\cosh\frac{\beta\hbar\omega_0}{2} + \cos\vartheta\sinh\frac{\beta\hbar\omega_0}{2}\right)\right],$$

$$D_2 = -\omega_0,$$

$$D_{11} = D_\perp \left( \cosh \frac{\beta \hbar \omega_0}{2} + \cos \vartheta \sinh \frac{\beta \hbar \omega_0}{2} \right),$$

$$D_{22} = D_\parallel + \frac{D_\perp}{\sin^2 \vartheta} \left( \cos^2 \vartheta \cosh \frac{\beta \hbar \omega_0}{2} + \cos \vartheta \sinh \frac{\beta \hbar \omega_0}{2} \right).$$

The Fokker-Planck equation for the probability density function $P(\vartheta, \varphi, t) = \sin \vartheta W_S(\vartheta, \varphi, t)$ corresponding to Eqs. (408) and (409) is

$$\frac{\partial P}{\partial t} = -\frac{\partial}{\partial \vartheta}(D_1 P) - \frac{\partial}{\partial \varphi}(D_2 P) + \frac{\partial^2}{\partial \vartheta^2}(D_{11} P) + \frac{\partial^2}{\partial \varphi^2}(D_{22} P). \qquad (417)$$

Equation (417) ultimately reduces to the master equation (405) for the phase-space distribution function $W_S(\vartheta, \varphi, t)$.

Moreover via the Langevin equations (408) and (409) for the stochastic variables $\vartheta(t)$ and $\varphi(t)$, we also have the Langevin equation for the evolution of the spherical harmonics $Y_{lm}(\vartheta, \varphi)$ rendered as [5]

$$\dot{Y}_{lm}[\vartheta(t), \varphi(t)] = \dot{\vartheta}(t) \frac{\partial}{\partial \vartheta} Y_{lm}[\vartheta(t), \varphi(t)] + \dot{\varphi}(t) \frac{\partial}{\partial \varphi} Y_{lm}[\vartheta(t), \varphi(t)],$$

where $\dot{\vartheta}(t)$ and $\dot{\varphi}(t)$ are given by Eqs. (408) and (409), respectively. Then by averaging the Langevin equation for $Y_{lm}(\vartheta, \varphi)$ over its realizations as described in Ref. 5, and using the recursion relations for the spherical harmonics (see Appendix B), we ultimately have a *closed* system of differential-recurrence equations for the statistical moments (averaged spherical harmonics), namely,

$$\frac{d}{dt}\langle Y_{lm}\rangle(t) = q^-_{l,m}\langle Y_{l-1\,m}\rangle(t) + q_{l,m}\langle Y_{lm}\rangle(t) + q^+_{l,m}\langle Y_{l+1\,m}\rangle(t), \qquad (418)$$

where $0 \leq l \leq 2S$,

$$q_{l,m} = -D_\parallel m^2 + im\omega_0 - D_\perp \left[ l(l+1) - m^2 \right] \cosh \frac{\beta \hbar \omega_0}{2},$$

$$q^-_{l,m} = D_\perp (l - 2S - 1) \sqrt{\frac{l^2 - m^2}{4l^2 - 1}} \sinh \frac{\beta \hbar \omega_0}{2},$$

$$q^+_{l,m} = D_\perp (l + 2S + 2) \sqrt{\frac{(l+1)^2 - m^2}{(2l+1)(2l+3)}} \sinh \frac{\beta \hbar \omega_0}{2}.$$

Here the number of recurrent equations is *finite* namely $2S + 1$ because $\langle Y_{Lm} \rangle(t) = 0$ for $L > 2S$ which is the main difference from the corresponding classical hierarchy of differential-recurrence equations for the moments, where the number of equations is infinite. Thus the Langevin and master equation treatments are now equivalent and yield the same results.

Here we have illustrated how a phase-space Langevin equation may be written by considering the simplest possible yet meaningful problem, viz., the relaxation of a spin of arbitrary number $S$ in a uniform magnetic field of arbitrary strength directed along the Z-axis. We emphasize that the Langevin equations are written down from *a priori* knowledge of the master equation *unlike the classical case* where they are written down independently of the Fokker-Planck equation. Consequently the results of each method in the classical case only coincide due to the Gaussian white noise properties of the random field, particularly Isserlis's (or Wick's) theorem [5] is satisfied. This theorem allows multiple time correlations of Gaussian random variables to be expressed as two time ones thereby leading directly to the correspondence between the Langevin and Fokker-Planck equations in the classical case.

*3. Exact solution of the master equation for longitudinal relaxation*

Having formulated the relevant evolution equations we shall now explicitly treat transient *nonlinear* spin relaxation by direct using the phase-space master equation because in this particular case it takes the Fokker-Planck form [see Eq. (419) below]. In order to accomplish this, we suppose that the magnitude of an externally uniform dc magnetic field is suddenly altered at time $t = 0$ from $\mathbf{H}_\mathrm{I}$ to a new value $\mathbf{H}_\mathrm{II}$ (the fields $\mathbf{H}_\mathrm{I}$ and $\mathbf{H}_\mathrm{II}$ are assumed to be applied parallel to the Z axis of the laboratory coordinate system). Thus we study as in the classical case [5], the transient longitudinal relaxation of a system of noninteracting spins starting from an equilibrium state I say with the initial distribution function $W^{H_\mathrm{I}}_{eq}$ ($t \leq 0$) to a new equilibrium state II say with the final distribution function $W^{H_\mathrm{II}}_{eq}$ ($t \to \infty$). Here the longitudinal component of the spin operator relaxes from the initial equilibrium value $\langle \hat{S}_Z \rangle_\mathrm{I}$ to the final equilibrium value $\langle \hat{S}_Z \rangle_\mathrm{II}$, the intervening transient being described by an appropriate relaxation function $\langle \hat{S}_Z \rangle(t) - \langle \hat{S}_Z \rangle_\mathrm{II}$ (see Fig. 14). The transient response so formulated is truly *nonlinear* because the change in amplitude $H_\mathrm{I} - H_\mathrm{II}$ of the external dc magnetic field is *arbitrary* (the linear response can be regarded as the particular case $|H_\mathrm{I} - H_\mathrm{II}| \to 0$). Here, the azimuthal angle dependence of the distribution function $W_S$ may be ignored. Thus, the master equation (407) becomes the single variable Fokker-Planck equation

$$\frac{\partial W_S}{\partial t} = \frac{\partial}{\partial z}\left( D_2(z) \frac{\partial W_S}{\partial z} + D_1(z) W_S \right), \qquad (419)$$

where $z = \cos \vartheta$,

$$D_1(z) = \frac{S(1 - z^2)}{\tau_\mathrm{N}} \sinh \frac{\xi}{2S}, \qquad (420)$$

$$D_2(z) = \frac{(1 - z^2)}{2\tau_\mathrm{N}} \left( \cosh \frac{\xi}{2S} + z \sinh \frac{\xi}{2S} \right), \qquad (421)$$

$\tau_N = 1/(2D_\perp)$ is the characteristic time of the free rotational "diffusion" of the spin, and the dimensionless field parameter $\xi$ is defined as

$$\xi = \beta\mu_0\mu H_{II}. \quad (422)$$

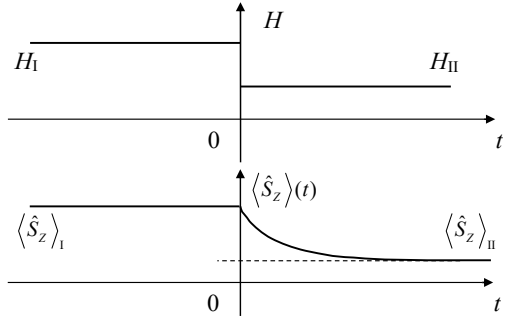

**Figure 14.** Schematic representation of the nonlinear transient response.

Hitherto the explicit expressions Eqs. (420) and (421) for $D_1(z)$ and $D_2(z)$ for spins subjected to a dc magnetic field $\mathbf{H}_0$ have been obtained (as we have just seen) by starting from the evolution equation for the density matrix $\hat{\rho}_S$ giving rise to lengthy calculations. However, these equations can also be obtained in far simpler fashion merely by knowing the functional form of the master Eq. (419) and the equilibrium phase-space distribution $W_{eq}^\xi(z)$ for spins. To illustrate this we shall again select the extension to the semiclassical case of the *Ansatz* of the imposition of a Boltzmann distribution originally used by Einstein, Smoluchowski, Langevin, and Kramers to determine drift and diffusion coefficients in the classical Brownian motion. We have already used this idea for the quantum translational Brownian motion in Section II and in Ref. [48].

Thus to determine $D_1(z)$ and $D_2(z)$ in Eq. (419) explicitly, we first recall that the equilibrium distribution $W_{eq}^\xi(z)$ given by Eq. (287) with $\gamma_X = \gamma_Y = 0$, viz.,

$$W_{eq}^\xi(z) = \frac{\sinh(\tfrac{1}{2}\xi/S)}{\sinh[(S+\tfrac{1}{2})\xi/S]}\left(\cosh\frac{\xi}{2S} + z\sinh\frac{\xi}{2S}\right)^{2S} \quad (423)$$

must also be the equilibrium solution of the generic master Eq. (419), i.e., it must satisfy

$$\frac{\partial}{\partial z}\left(D_2(z)\frac{\partial}{\partial z}W_{eq}^\xi(z) + D_1(z)W_{eq}^\xi(z)\right) = 0. \quad (424)$$

Now one is at liberty to seek $D_1(z)$ and $D_2(z)$ in series form as

$$D_1(z) = (1-z^2)(a_0^S + a_1^S z + a_2^S z^2 + \cdots), \quad (425)$$

$$D_2(z) = (1-z^2)(b_0^S + b_1^S z + b_2^S z^2 + \cdots). \quad (426)$$

By substituting Eqs. (425) and (426) into Eq. (424), then if $W_{eq}^\xi(z)$ from Eq. (423) is to satisfy Eq. (424), only the coefficients $a_0^S$, $b_0^S$, and $b_1^S$ can be nonzero so that $D_1(z)$ and $D_2(z)$ are, respectively,

$$D_1(z) = 2Sb_0^S(1-z^2)\tanh\frac{\xi}{2S}$$

and

$$D_2(z) = b_0^S(1-z^2)\left(1 + z\tanh\frac{\xi}{2S}\right).$$

In order to determine the normalizing coefficient $b_0^S$, we use the fluctuation-dissipation theorem [75] along with the additional requirements that in the classical limit $S \to \infty$, $D_1(z)$ and $D_2(z)$ must reduce to their classical counterparts for the rotational Brownian motion of a classical spin [5], viz.,

$$D_1(z) \to \frac{\xi}{2\tau_N}(1-z^2)$$

and

$$D_2(z) \to \frac{1}{2\tau_N}(1-z^2).$$

Thus, we obtain

$$b_0^S = \frac{1}{2\tau_N}\cosh\frac{\xi}{2S}$$

so that $D_1(z)$ and $D_2(z)$ are given by the closed form Eqs. (420) and (421). Now in the derivation of $D_1(z)$ and $D_2(z)$ we have imposed the stationary solution of the master equation as the equilibrium phase-space distribution Eq. (423) corresponding to the equilibrium density matrix $\hat{\rho}_{eq}$ given by Eq. (274) describing the system in *thermal equilibrium without coupling* to the thermal bath. However, from the theory of open quantum systems [120], the equilibrium state in general may deviate from the equilibrium density matrix $\hat{\rho}_{eq}$; the latter describes the thermal equilibrium of the system in the weak coupling and high temperature limits only. A detailed discussion is given, e.g., by Geva *et al*. [162]. Nevertheless, the imposition of the phase-space distribution Eq. (423) as the equilibrium solution of Eq. (424) so yielding $D_1(z)$ and $D_2(z)$, appears to be the exact analog of the *Ansatz* used by Gross and Lebowitz [133] in their formulation of quantum kinetic models of impulsive collisions. According to [133], for a system with a Hamiltonian $\hat{H}$, the equation governing the time behavior of the density matrix $\hat{\rho}_S$ is Eq. (40), where the collision kernel operator $\text{St}(\hat{\rho}_S)$ satisfies the condition $\text{St}(\hat{\rho}_{eq}) = 0$. Equation

(424) is entirely analogous to this condition. Moreover, as we have seen in Sec. II.A.2, the condition $\text{St}(\hat{\rho}_{eq}) = 0$ was also used by Redfield [14] to determine the matrix elements of the relaxation operator in his theory of quantum relaxation processes.

The time-dependent solution of the axially symmetric evolution Eq. (419) is obtained as usual by expanding the distribution function $W_S(z,t)$ in Legendre polynomials $P_n(z)$

$$W_S(z,t) = W_{eq}^{\xi}(z) + \sum_{n=0}^{2S}(n+1/2)P_n(z)f_n(t), \qquad (427)$$

where

$$f_n(t) = \langle P_n \rangle(t) - \langle P_n \rangle_{eq}^{\xi}$$

are the relaxation functions, the angular brackets $\langle \ \rangle(t)$ and $\langle \ \rangle_{eq}^{\xi}$ designate statistical averaging defined as

$$\langle P_n \rangle(t) = \left(S+\tfrac{1}{2}\right)\int_{-1}^{1} P_n(z)W_S(z,t)dz \qquad (428)$$

and

$$\langle P_n \rangle_{eq}^{\xi} = \left(S+\tfrac{1}{2}\right)\int_{-1}^{1} P_n(z)W_{eq}^{\xi}(z)dz. \qquad (429)$$

In particular, Eqs. (429) and (431) below yield the *equilibrium* average of $P_1(z)$, viz.,

$$\langle P_1 \rangle_{eq}^{\xi} = \left(S+\tfrac{1}{2}\right)\int_{-1}^{1} zW_{eq}^{\xi}(z)dz = \frac{S}{S+1}B_S(\xi), \qquad (430)$$

where $B_S(\xi)$ is the Brillouin function defined by Eq. (278). The equilibrium distribution $W_{eq}^{\xi}(z)$ can be equivalently defined as [cf. Eq. (307)]

$$W_{eq}^{\xi}(z) = \sum_{n=0}^{2S}(n+1/2)P_n(z)\langle P_n \rangle_{eq}^{\xi}. \qquad (431)$$

Substituting Eq. (427) into Eq. (421) and using the orthogonality and recurrence properties of Legendre polynomials $P_n(z)$, we have as in the classical case [5] a differential-recurrence relation for the relaxation functions $f_n(t)$, viz., [cf. the three-term recurrence relation Eq. (396)]

$$\tau_n \dot{f}_n(t) = q_n^- f_{n-1}(t) + q_n f_n(t) + q_n^+ f_{n+1}(t), \qquad (432)$$

where $1 \le n \le 2S$,

$$f_0(t) = f_{2S+1}(t) = 0,$$

$$\tau_n = \frac{2\tau_N}{n(n+1)},$$

$$q_n = -\cosh\frac{\xi}{2S},$$

$$q_n^{\pm} = \mp\frac{2S \pm n + (3\pm 1)/2}{(2n+1)}\sinh\frac{\xi}{2S}.$$

Since the initial value of the distribution function is $W_S(z,0) = W_{eq}^{\xi+\delta}(z)$, where the transient parameter $\delta = \beta\mu_0\mu(H_{\text{II}} - H_{\text{I}})$, the initial values of the relaxation function $f_n(t)$ are

$$f_n(0) = \langle P_n \rangle_{eq}^{\xi+\delta} - \langle P_n \rangle_{eq}^{\xi}. \qquad (433)$$

Equation (432) is a quantum counterpart of the recurrence relation Eq. (401) for a classical spin and has been solved in Refs. [38,39] for the *particular* spin numbers $S = 1/2, 1$, and $3/2$. Now we give the exact solution for the transient quantum nonlinear longitudinal relaxation governed by Eq. (432) for *arbitrary S*.

Using the one-sided Fourier transform, we have from Eq. (432)

$$(i\omega\tau_n - q_n)\tilde{f}_n(\omega) - q_n^-\tilde{f}_{n-1}(\omega) - q_n^+\tilde{f}_{n+1}(\omega) = \tau_n f_n(0), \qquad (434)$$

where

$$\tilde{f}_n(\omega) = \int_0^{\infty} e^{-i\omega t}f_n(t)dt.$$

The inhomogeneous algebraic three-term recurrence Eq. (434) can be solved exactly for the one sided Fourier transform of the relaxation function $\tilde{f}_1(\omega)$ using continued fractions like the corresponding classical problem (see for details the general solution of three-term recurrence relations given in Ref. [5], Chap. 2) yielding

$$\tilde{f}_1(\omega) = \tau_N \operatorname{csch}\frac{\xi}{2S}\sum_{n=1}^{2S}\frac{f_n(0)}{n(n+1)(S+1)}\prod_{k=1}^{n}\frac{q_{k-1}^+}{q_k^-}\Delta_k(\omega,\xi)$$

$$= \tau_N \operatorname{csch}\frac{\xi}{2S}\sum_{n=1}^{2S}(-1)^{n+1}\frac{(2n+1)(2S+n+1)!(2S-n)!}{n(n+1)(S+1)(2S+1)!(2S)!}f_n(0)\prod_{k=1}^{n}\Delta_k(\omega,\xi). \qquad (435)$$

Here the *finite* continued fraction $\Delta_n(\omega,\xi)$ is defined by the two term recurrence relation

$$\Delta_n(\omega,\xi) = \frac{q_n^-}{i\omega\tau_n - q_n - q_n^+\Delta_{n+1}(\omega,\xi)}$$

with $\Delta_{2S+1}(\omega,\xi) = 0$, moreover, we have the product

$$\prod_{k=1}^{n}\frac{q_{k-1}^+}{q_k^-} = (-1)^{n+1}\frac{(2n+1)(2S+n+1)!(2S-n)!}{(2S+1)!(2S)!}.$$

The equilibrium averages $\langle P_n \rangle_{eq}^{\xi}$, Eq. (429), can also be evaluated in terms of the continued fraction $\Delta_n(0,\xi)$ since $\langle P_n \rangle_{eq}^{\xi}$ satisfies the three-term recurrence relation

$$q_n^-\langle P_{n-1}\rangle_{eq}^{\xi} + q_n\langle P_n\rangle_{eq}^{\xi} + q_n^+\langle P_{n+1}\rangle_{eq}^{\xi} = 0, \qquad (436)$$

so that

$$\Delta_n(0,\xi) = \frac{\langle P_n \rangle_{eq}^{\xi}}{\langle P_{n-1} \rangle_{eq}^{\xi}}.$$

Consequently, we have

$$\langle P_n \rangle_{eq}^{\xi} = \prod_{k=1}^{n} \Delta_k(0,\xi). \tag{437}$$

Equation (435) is the exact solution for the one-sided Fourier transform of the nonlinear relaxation function $f_1(t)$ in terms of continued fractions. Having determined $f_1(t)$, various transient nonlinear responses of the longitudinal component of the magnetic moment may always be evaluated because in terms of averages of spin operators

$$\langle \hat{S}_Z \rangle(t) - \langle \hat{S}_Z \rangle_{II} = (S+1)f_1(t), \tag{438}$$

where in terms of the Brillouin function the final equilibrium value is

$$\langle \hat{S}_Z \rangle_{II} = (S+1)\langle P_1 \rangle_{eq}^{\xi} = SB_S(\xi).$$

In particular, we mention the rise, decay, and rapidly reversing field transient responses. The general relaxation equation (435) can often be simplified. For example, to treat the rise transient we suppose that a *strong* constant field $\mathbf{H}_{II}$ is suddenly switched on at time $t=0$ (so that $\mathbf{H}_I = 0$). Thus we require the nonlinear relaxation behavior of a system of spins starting from an equilibrium state I with the *isotropic* distribution function $W_{eq}^0 = 1/(2S+1)$ ($t \leq 0$) to another equilibrium state II with the *final* distribution function $W_{eq}^{H_{II}} = W_{eq}^{\xi}(z)$ ($t \to \infty$). Using Eq. (437), Eq. (435) simplifies to

$$\tilde{f}_1(\omega) = \frac{i}{\omega}\left[\Delta_1(0,\xi) - \Delta_1(\omega,\xi)\right]$$
$$= \frac{i}{\omega}\left[\langle P_1 \rangle_{eq}^{\xi} - \Delta_1(\omega,\xi)\right] \tag{439}$$

yielding the spectrum of the relaxation function $\tilde{f}_1(\omega)$ for the rise transient.

*4. Nonlinear longitudinal relaxation time*

The overall transient behavior of the relaxation function $f_1(t)$ is characterized by the integral relaxation time [5] (see Appendix E)

$$\tau_{int} = \frac{1}{f_1(0)}\int_0^{\infty} f_1(t)dt = \frac{\tilde{f}_1(0)}{f_1(0)} \tag{440}$$

[i.e., the area under the normalized relaxation function $f_1(t)/f_1(0)$] which can be evaluated in series form from the zero frequency limit [5] of Eq. (435) and (437) is

$$\tau_{int} = \tau_N \operatorname{csch}\frac{\xi}{2S}\sum_{n=1}^{2S} \frac{(-1)^{n+1}(2n+1)(2S+n+1)!(2S-n)!f_n(0)\langle P_n \rangle_{eq}^{\xi}}{n(n+1)(S+1)(2S+1)!(2S)!f_1(0)}. \tag{441}$$

Moreover, the latter result can also be written in an equivalent integral form because the master Eq. (419) is actually just a single variable Fokker-Planck equation, which may be integrated by quadratures. Now for any system with dynamics governed by a single variable Fokker-Planck equation, e.g., Eq. (419), the integral relaxation time $\tau_{int}$, characterizing the nonlinear relaxation behavior of $f_1(t)$, can be obtained in integral form in terms of the equilibrium distribution and the diffusion coefficient $D_2(z)$ only (see Appendix E for details) [5]. Hence, with Eqs. (419) and (E22) from Appendix E, we have like the classical case [5] an exact integral expression for $\tau_{int}$, viz.,

$$\tau_{int} = \frac{1}{(S+\tfrac{1}{2})f_1(0)}\int_{-1}^{1}\frac{\Phi(z)\Psi(z)}{D_2(z)W_{eq}^{\xi}(z)}dz, \tag{442}$$

where

$$\Psi(z) = (S+\tfrac{1}{2})\int_{-1}^{z}\left(x - \langle P_1 \rangle_{eq}^{\xi}\right)W_{eq}^{\xi}(x)dx,$$

$$\Phi(z) = (S+\tfrac{1}{2})\int_{-1}^{z}\left[W_{eq}^{\xi+\delta}(x) - W_{eq}^{\xi}(x)\right]dx,$$

$$f_1(0) = \langle P_1 \rangle_{eq}^{\xi+\delta} - \langle P_1 \rangle_{eq}^{\xi}$$
$$= \frac{S}{S+1}\left[B_S(\xi+\delta) - B_S(\xi)\right].$$

For the limiting case $S = 1/2$, $\tau_{int}$ is independent of the parameter $\delta$ and is given by

$$\tau_{int} = \frac{\tau_N}{\cosh\xi}, \tag{443}$$

while in the classical limit $S \to \infty$, one has

$$\tau_{int} = \frac{\tau_N \xi \operatorname{csch}\xi}{\langle P_1 \rangle^{\xi} - \langle P_1 \rangle^{\xi+\delta}}\int_{-1}^{1}\frac{\phi(z)\psi(z)e^{-\xi z}}{1-z^2}dz, \tag{444}$$

where

$$\langle P_1 \rangle^{\xi} = \coth\xi - \frac{1}{\xi},$$

$$\phi(z) = \int_{-1}^{z}\left[e^{\xi z'} - e^{(\xi+\delta)z'}\right]dz',$$

$$\psi(z) = \int_{-1}^{z}\left(\cos z' - \langle P_1 \rangle^{\xi}\right)e^{\xi z'}dz'$$

agreeing entirely with the established classical result ([5], Ch. 7).

Numerical calculations show that both the series expression Eq. (441) and the integral Eq. (442) yield identical results. Thus $\tau_{int}$ for various nonlinear transient responses (such as the rise, decay, and rapidly reversing field transients) may be easily evaluated from Eq. (442). The normalized relaxation time $\tau_{int}/\tau_N$ from Eq. (442) is plotted in Fig. 15 for various values of the transient strength $\delta$, the field strength parameter $\xi$ and spin number $S$. The figure indicates that the relaxation time decreases with increasing field strength $\xi$ with a strong dependence on *both S and the transient strength $\delta$*. The nonlinear effect *comprising accelerated relaxation in the external field* also exists for classical dipoles [5]. An explanation may be given as follows. In the absence of the field ($\xi = 0$), the relaxation time of the spin is just the *free diffusion relaxation time* $\tau_N$ so that $\tau_{int} = \tau_N$. However, in a *strong* field ($\xi \gg 1$) and $S \gg 1$, the relaxation time of the spin is determined by the *damped diffusion* of the spin in the field $\mathbf{H}_{\parallel}$ and the characteristic frequency is now the frequency of the spin oscillation about $\mathbf{H}_{\parallel}$ (in the vicinity of $z = 0$). Thus $\tau_{int}$ is of the order of $\sim 1/[2D_1(0)] = \tau_N/\xi$ so that

$$\tau_{int} \sim \tau_N/\xi. \quad (445)$$

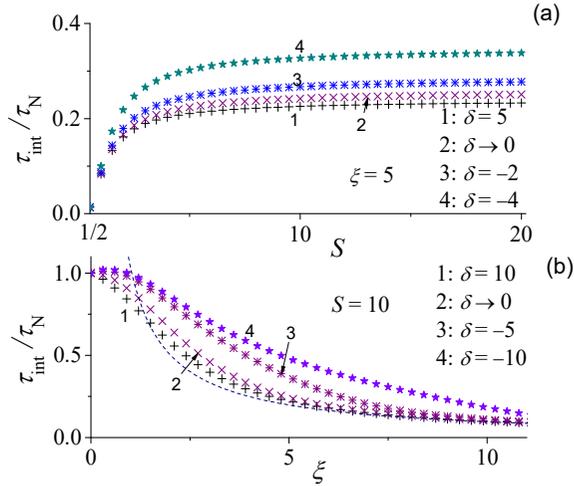

**Figure 15.** (Color on line) Normalized relaxation time $\tau_{int}/\tau_N$ from Eq. (442) as a function of $S$ (a) and $\xi$ (b) for various values of $\delta$ (symbols). Dashed line: Eq. (447).

This asymptotic formula may be used to estimate $\tau_{int}$ for $\xi \gg 1$ and $\delta > 0$ and $|\delta| \ll \xi$. However, for $\delta \sim -\xi$, a more accurate formula is

$$\tau_{int} \sim \frac{\tau_N}{\xi - 1 - \xi(\xi + \delta)}. \quad (446)$$

The influence of the transient parameter $\delta$ entering into the relaxation time owing to the initial distribution function $W_{eq}^{\xi+\delta}$ is more pronounced for field strengths $\xi \sim 2 \div 7$ (see Fig. 15). The enhanced dependence of $\tau_{int}$ on $\delta$ for negative values of $\delta$ can be understood because these situations correspond to the situation of rise and rapidly reversing transients, where the initial and final distributions differ considerably. As far as the *spin dependence* of $\tau_{int}$ for $\xi \gg 1$ and $\delta > 0$ and $|\delta| \ll \xi$ is concerned a simple asymptotic formula for $\tau_{int}$ is (see Fig. 14)

$$\tau_{int} \sim \frac{1}{2D_1(0)} = \frac{\tau_N}{2S}\operatorname{csch}\frac{\xi}{2S}. \quad (447)$$

### 5. Linear response

Now using the above results, we may also evaluate the *linear response* of a spin system to *infinitesimally small* changes in the magnitude of the dc field $\mathbf{H}_{\parallel}$. This is of particular interest as the corresponding integral relaxation time now becomes the correlation time, which has been previously evaluated [80-82] from the spin density matrix. Thus we again suppose that the uniform dc field $\mathbf{H}_{\parallel}$ is directed along the $Z$-axis of the laboratory coordinate system and that a *small* probing field $\mathbf{H}_1$ having been applied to the assembly of noninteracting spins in the distant past ($t = -\infty$) so that equilibrium conditions obtain at time $t = 0$, is suddenly switched off at $t = 0$. Here the normalized relaxation function $f_1(t)/f_1(0)$ reduces to the longitudinal equilibrium correlation function $C(t)$, that is [33,75]

$$C(t) = \lim_{\delta \to 0} \frac{f_1(t)}{f_1(0)} = \frac{1}{\beta\chi}\left\langle\int_0^\beta \left[\hat{S}_Z(-i\lambda\hbar) - \left\langle\hat{S}_Z\right\rangle_{\parallel}\right]\left[\hat{S}_Z(t) - \left\langle\hat{S}_Z\right\rangle_{\parallel}\right]d\lambda\right\rangle_{\parallel}, \quad (448)$$

where $\chi$ is the static susceptibility defined in terms of the Brillouin function Eq. (278) as

$$\chi = S^2\frac{\partial}{\partial\xi}B_S(\xi) = \frac{1}{4}\left[\operatorname{csch}^2\left(\frac{\xi}{2S}\right) - (2S+1)^2\operatorname{csch}^2\left(\frac{2S+1}{2S}\xi\right)\right]. \quad (449)$$

According to linear response theory (see, e.g., [75] and Appendix D), the dynamic susceptibility $\chi(\omega) = \chi'(\omega) - i\chi''(\omega)$ is defined as [75]

$$\frac{\chi(\omega)}{\chi} = 1 - i\omega\tilde{C}(\omega), \quad (450)$$

where $\tilde{C}(\omega)$ is the one-sided Fourier transform of $C(t)$, namely,

$$\tilde{C}(\omega) = \int_0^\infty C(t)e^{-i\omega t}dt. \quad (451)$$

In linear response, the integral relaxation time, that is, the correlation time $\tau_{int}|_{\delta \to 0} = \tau_{cor} = \tilde{C}(0)$ of $C(t)$, follows from the general nonlinear Eq. (442) in the limit of a very small transient strength parameter $\delta \to 0$ and is given by

$$\tau_{cor} = \frac{S(S+1)(S+\frac{1}{2})}{\chi} \int_{-1}^{1} \frac{\frac{\partial}{\partial \xi} W_{eq}^{\xi}(x) dx \int_{-1}^{z} \left( y - \frac{SB_S(\xi)}{S+1} \right) W_{eq}^{\xi}(y) dy}{D_2(z) W_{eq}^{\xi}(z)} dz, \qquad (452)$$

where

$$\frac{\partial}{\partial \xi} W_{eq}^{\xi}(z) = \frac{\operatorname{csch}(\xi/2S) - (2S+1)\operatorname{csch}(\xi + \xi/2S)(\cosh \xi - z\sinh \xi)}{2S[\cosh(\xi/2S) + z\sinh(\xi/2S)]} W_{eq}^{\xi}(z).$$

For the limiting case $S=1/2$, $\tau_{cor}$ is equal to $\tau_{int}$ as yielded by the closed expression Eq. (443), while in the classical limit $S \to \infty$, Eq. (452) becomes

$$\tau_{cor} = \frac{\tau_N \xi \operatorname{csch} \xi}{1 + \xi^{-2} - \coth^2 \xi} \int_{-1}^{1} \left[ z - \coth \xi + e^{-\xi(1+z)}(1 + \coth \xi) \right]^2 \frac{e^{\xi z} dz}{1-z^2} \qquad (453)$$

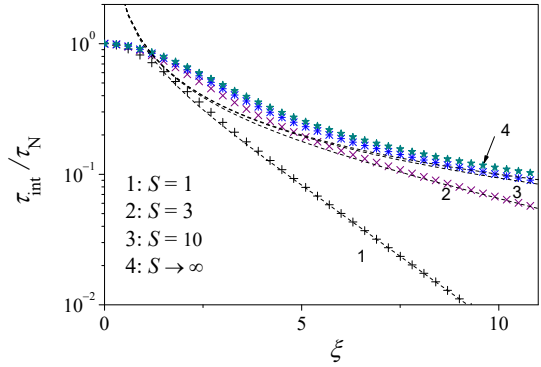

**Figure 16.** (Color on line) Normalized correlation time $\tau_{cor}/\tau_N$ of $C_{\parallel}(t)$ from Eq. (452) as a function of $\xi$ for various $S$ (symbols). Dashed lines: Eq. (447).

concurring with the result for classical spins ([5], Chap. 7). As far as the spin number dependence of $\tau_{cor}$ for $\xi \gg 1$ is concerned, an asymptotic formula for $\tau_{cor} = \tau_{int}|_{\delta \to 0}$ is given by Eq. (447). Generally, $\tau_{cor}$ varies smoothly from power law like behavior ($\tau_{cor} \sim \tau_N/\xi$) as $S \to \infty$ to exponential decrease $\tau_{cor} \sim \tau_N \operatorname{csch} \xi$ for $S=1/2$ and is plotted from the exact Eq. (452) in Fig. 16 as a function of $\xi$ for various $S$; the asymptotic Eq. (447) is also shown for comparison.

We remarked above that the linear response has been studied previously by Garcia-Palacios and Zueco [81] using the spin density matrix. They also gave an explicit expression for the linear response integral relaxation time, which was first derived by Garanin [80]. He derived his formula pertaining to a *uniaxial nanomagnet in a uniform field* with the Hamiltonian $\hat{H}_S = -\hbar \omega_0 \hat{S}_Z - D\hat{S}_Z^2$, which is also valid in the limit $D \to 0$, corresponding to our case. By applying his method [80] to the symmetrized form of the collision kernel given by Eq. (558) below, the corresponding equations are [see for details Sec. III.C.3 below, Eq. (521)]

$$\tau_{int} = \frac{2\tau_N}{\langle \hat{S}_Z \rangle_I - \langle \hat{S}_Z \rangle_{II}} \sum_{k=1-S}^{S} \frac{\sum_{m=k}^{S}(\rho_m^I - \rho_m^{II}) \sum_{m'=k}^{S} \left( m' - \langle \hat{S}_Z \rangle_{II} \right) \rho_{m'}^{II}}{[S(S+1)-k(k-1)]\sqrt{\rho_k^{II} \rho_{k-1}^{II}}}, \qquad (454)$$

$$\tau_{cor} = \frac{2\tau_N}{\chi} \sum_{k=1-S}^{S} \frac{\left( \sum_{m=k}^{S}(m - \langle \hat{S}_Z \rangle_0) \rho_m^{II} \right)^2}{[S(S+1)-k(k-1)]\sqrt{\rho_k^{II} \rho_{k-1}^{II}}}. \qquad (455)$$

where

$$\rho_n^i = \frac{e^{\xi n/S}}{\sum_{m=-S}^{S} e^{\xi m/S}}, \quad \langle \hat{S}_Z \rangle_i = \sum_{m=-S}^{S} m \rho_m^i, \quad \chi = \sum_{m=-S}^{S} m^2 \rho_m - \langle \hat{S}_Z \rangle_{II}^2.$$

Although the integral and series expressions Eqs. (442), (452) and (454), (455), respectively, have outwardly very different forms, nevertheless numerical calculation shows that both yield *identical* results establishing an essential corollary between the phase-space and density matrix methods.

### 6. Single mode approximation

Although the continued fraction solution given above is effective in numerical calculations, it has one significant drawback; namely, the qualitative behavior of the system is not at all obvious in a physical sense. Thus to gain a physical understanding of the relaxation process, we shall use the single mode approximation suggested by Shibata *et al*. [38,39] and Kalmykov *et al*. [62] for the relaxation of quantum and classical spins. We first recall that the spectrum $\tilde{f}_1(\omega)$ from Eq. (435) on Fourier inversion indicates that the time behavior of the relaxation function $f_1(t)$ in general comprises $2S$ exponentials

$$f_1(t) = f_1(0) \sum_{k=1}^{2S} c_k e^{-\lambda_k t}, \qquad (456)$$

where the $\lambda_k$ are the eigenvalues of the tridiagonal system matrix $\mathbf{X}$ with the matrix elements

$$(\mathbf{X})_{qp} = \delta_{pq+1} q_p^- + \delta_{pq} q_p + \delta_{pq-1} q_p^+.$$

In the frequency domain, the corresponding spectrum $\tilde{f}_1(\omega)$ is thus the series of $2S$ Lorentzians

$$\tilde{f}_1(\omega) = f_1(0) \sum_{k=1}^{2S} \frac{c_k}{\lambda_k + i\omega}. \qquad (457)$$

According to Eq. (457), the finite number of relaxation modes (corresponding to the eigenvalues $\lambda_k$) each contribute to the spectrum $\tilde{f}_1(\omega)$. However, as we shall see below, these *near degenerate* individual modes are indistinguishable in the spectrum $\tilde{f}_1(\omega)$ appearing merely as a single band suggesting that $\tilde{f}_1(\omega)$ may be approximated by the *single* Lorentzian

$$\frac{\tilde{f}_1(\omega)}{f_1(0)} \approx \frac{\tau_{\text{int}}}{1 + i\omega\tau_{\text{int}}}, \qquad (458)$$

where $\tau_{\text{int}}$ is given by Eq. (442). In the time domain, the single-mode approximation Eq. (458) amounts to the *Ansatz* that the relaxation function $f_1(t)$ as determined by Eq. (456) (comprising $2S$ exponentials) may be approximated by a *single* exponential, viz.,

$$f_1(t) \approx f_1(0) e^{-t/\tau_{\text{int}}}. \qquad (459)$$

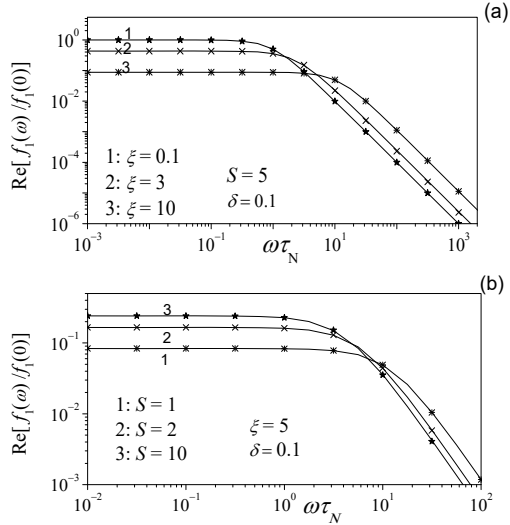

**Figure 17.** The real parts of the normalized spectra $\tilde{f}_1(\omega)/f_1(0)$ vs. the normalized frequency $\omega\tau_N$ evaluated from the exact continued fraction solution [Eq. (435): solid lines] for (a) $S = 5$, $\delta = 0.1$, and various $\xi$ and for (b) $\xi = 3$, $\delta = 0.1$ and various $S$ compared with those calculated from the single Lorentzian approximation Eq. (458) (symbols).

Now García-Palacios and Zueco [81,82] have also used the single mode approximation to evaluate the *linear* response of an isotropic spin system. In linear response, Eqs. (458) and (459) can be reformulated for the susceptibility $\chi(\omega)$ and correlation function $C(t)$ as

$$C(t) \approx e^{-t/\tau_{\text{cor}}} \qquad (460)$$

and

$$\chi(\omega) \approx \frac{\chi}{1 + i\omega\tau_{\text{cor}}}, \qquad (461)$$

where $\tau_{\text{cor}}$ is given by Eq. (455).

In order to test the single mode approximation, we plot in Fig. 17 the real parts of the normalized spectra $\tilde{f}_1(\omega)/f_1(0)$ as calculated both from the exact continued fraction solution [Eq. (435): solid lines] and from that approximation Eq. (458). Clearly, no practical difference exists between the exact continued fraction solution and the single mode one [the maximum deviation between the corresponding curves does not exceed a few percent]. Like the classical case ([5], Chap. 7), the single mode approximation is accurate because the finite number ($2S$) of relaxation modes are *near degenerate* again appearing merely as a *single* high-frequency band in the spectrum. Thus, they may be effectively approximated by a single mode, i.e., both the linear and nonlinear longitudinal relaxation for all $S$ is accurately described by the Bloch equation

$$\frac{d}{dt}\langle\hat{S}_Z\rangle(t) + \frac{1}{T_\parallel}\left(\langle\hat{S}_Z\rangle(t) - \langle\hat{S}_Z\rangle_\text{II}\right) = 0, \qquad (462)$$

where $T_\parallel = \tau_{\text{int}}$ is the longitudinal relaxation time, $\langle\hat{S}_Z\rangle_\text{II} = SB_S(\xi)$ is the equilibrium average of the operator $\hat{S}_Z$, and $B_S(x)$ is the Brillouin function Eq. (278),

We have treated nonlinear spin relaxation of noninteracting spins using phase space quasiprobability density evolution equations in configuration space via the extension of Wigner's phase space formulation of quantum mechanics to *open* systems. The calculations show that in particular limiting cases [e.g., the correlation time Eq. (452)] the results reduce to established ones obtained using the evolution equation for the density matrix in the second order of perturbation theory in the spin-bath coupling. Thus, we have an important check on the validity of our approach by demonstrating the equivalence of the two methods. Both exact (continued fraction) and approximate (single mode) solutions are given. The continued fraction solution yields the dependence of the longitudinal spin relaxation on the spin number $S$ in closed form. This solution is dominated by a single exponential having as time constant the integral relaxation time $\tau_{\text{int}}$, which strongly depends on both $S$ and the field strength for arbitrary $S$. Hence, *an accurate description in terms of a Bloch equation* (462) *holds even for the nonlinear response of a giant spin*.

Thus, we have explicitly demonstrated for noninteracting spins in an external magnetic field that the existing methods of solution of the classical Fokker-Planck equation (continued fractions, which can be evaluated by iterating a simple algorithm, integral representation of relaxation times, etc.) seamlessly carry over to the quantum case. Again, the methods suggest new closed form quantum results via the corresponding classical ones. An example being the quantum integral relaxation time, Eq. (442) above. We have illustrated the phase space method via the rudimentary problem of the longitudinal relaxation of a spin in a uniform magnetic field of arbitrary strength directed along the Z-axis [the relaxation of the transverse components of the magnetization can be treated in like manner using the master Eq. (405) and the associated quantum recurrence Eq. (418)]. This problem is the simplest example of the phase space method for spins and may be considered as the rotational analog of the Agarwal problem for the translational harmonic oscillator model [121] described by the master equation Eq. (168). Just as with translational oscillators, the phase space master Eq. (419) for spins has a Fokker-Planck equation form. This is not, however, true in general, e.g., for nonaxially symmetric magnetocrystalline anisotropy and external field potentials, where the corresponding master equation may have a very complicated form. Nevertheless, the simple noninteracting spins problem indicates how one may treat the influence of spin number $S$ on the relaxation behavior using the phase space method. A factor which is both essential in the formation of magnetic clusters and for nanomagnets in the quest for macroscopic quantum tunneling.

### C. Longitudinal relaxation of uniaxial nanomagnets

In contrast to the phase space approach used above, we shall now apply the density matrix method. We shall consider a uniaxial nanomagnet of arbitrary spin number $S$ in an external *constant* magnetic field **H** applied along the Z axis, i.e., the axis of symmetry, with the Hamiltonian $\hat{H}_S$ defined by Eq. (370) with $\sigma \neq 0$ and $\xi \neq 0$. Furthermore, we shall use the method in the form based on the relation between the averages of polarization operators and averages of spherical harmonics as described in Section III.A.2 (see also Section II.A.4). However as mentioned above, Garanin and García-Palacios *et al.* [80-82] have also treated a uniaxial nanomagnet in a uniform longitudinal field via the quantum Hubbard operator representation of the evolution equation for the spin density matrix. Now in the axially symmetric Hamiltonian Eq. (370), the diagonal terms of the density matrix *decouple* from the non-diagonal ones. Hence, only the former contribute to the time evolution of the longitudinal component of the spin operator so facilitating a treatment of the problem. Thus in order to describe the longitudinal relaxation of a uniaxial nanomagnet, in which case only the *diagonal* terms of the density matrix are involved, the evolution equation Eq. (371) with the collision kernel given by (375) simplifies to

$$\frac{\partial \hat{\rho}_S}{\partial t} = -\frac{1}{\tau_N} \left\{ e^{-\frac{\sigma}{2S^2} - \frac{\xi}{2S}} \left[ \hat{S}_{-1} e^{-\frac{\sigma}{S^2} \hat{S}_0} \hat{\rho}_S, \hat{S}_{+1} \right] + e^{\frac{\sigma}{2S^2} + \frac{\xi}{2S}} \left[ \hat{S}_{+1} e^{\frac{\sigma}{S^2} \hat{S}_0} \hat{\rho}_S, \hat{S}_{-1} \right] \right\}. \tag{463}$$

This simplified method will be described in detail in Section III.C2 below. Now the associated evolution equation for the phase-space distribution function $W_S(z = \cos\vartheta, t)$ corresponding to Eq. (371) is then given by Eq. (378), namely,

$$\frac{\partial W_S}{\partial t} = \frac{1}{2\tau_N} \frac{\partial}{\partial z} \left( (1-z^2) \left[ \frac{\partial}{\partial z} \left( R''^{(S)}_+ W_S \right) + z \frac{\partial}{\partial z} \left( R''^{(S)}_- W_S \right) + 2SR'^{(S)}_- W_S \right] \right) \tag{464}$$

(because the azimuthal angle $\varphi$ dependence of $W_S$ may be ignored in longitudinal relaxation). In the classical limit, Eq. (464) further reduces to the Fokker-Planck equation for a classical uniaxial nanomagnet in a dc magnetic field, viz. [5,6,23],

$$\frac{\partial W}{\partial t} = \frac{1}{2\tau_N} \frac{\partial}{\partial z} \left[ (1-z^2) \left( \frac{\partial W}{\partial z} + W \frac{\partial V}{\partial z} \right) \right], \tag{465}$$

where $V(z) = -\sigma z^2 - \xi z$ is the normalized classical free energy.

#### 1. Calculation of the observables

As before, we suppose that the magnitude of an external uniform dc magnetic field is suddenly altered at time $t = 0$ from $\mathbf{H}_I$ to $\mathbf{H}_{II}$ (the magnetic fields $\mathbf{H}_I$ and $\mathbf{H}_{II}$ are applied parallel to the Z axis of the laboratory coordinate system in order to preserve axial symmetry). Thus we study as in the classical case [5], the nonlinear transient longitudinal relaxation of a system of spins starting from an equilibrium state I with density matrix $\hat{\rho}^I_{eq}$ ($t \leq 0$) say to a new equilibrium state II with density matrix $\hat{\rho}^{II}_{eq}$ ($t \to \infty$) say, see Fig. 14. Simultaneously the longitudinal component of the spin $\langle \hat{S}_Z \rangle(t)$ relaxes from the equilibrium value $\langle \hat{S}_Z \rangle_I$ to the new value $\langle \hat{S}_Z \rangle_{II}$, the ensuing transient response being described by the relaxation function $\langle \hat{S}_Z \rangle(t) - \langle \hat{S}_Z \rangle_{II}$. The transient response so formulated is again truly *nonlinear* because the *change* in amplitude $H_I - H_{II}$ of the external dc magnetic field is now *arbitrary* (the linear response is the particular case $\beta\mu_0\mu|H_I - H_{II}| \to 0$). Now the equilibrium phase space distributions $W^I_{eq}$ and $W^{II}_{eq}$ corresponding to the equilibrium spin density matrixes $\hat{\rho}^I_{eq}$ and $\hat{\rho}^{II}_{eq}$ comprise the appropriate stationary (time independent) solutions of Eq. (464). These equilibrium distributions have been extensively studied in Sec. II D and are given by Eq. (299), viz.,

$$(S+1/2)W^i_{eq}(\vartheta) = \sum_{L=0}^{2S} (L+1/2)\langle P_L \rangle_i P_L(\cos\vartheta), \tag{466}$$

where $i$ = I, II and $\langle P_L \rangle_i$ are the equilibrium averages of the Legendre polynomials $P_L$ defined by Eq. (300).

As far as the transient response is concerned, according to the multi-term differential-recurrence relation Eq. (389) for the relaxation functions $c_L(t) = \langle \hat{T}_{L0}^{(S)} \rangle(t) - \langle \hat{T}_{L0}^{(S)} \rangle_{\text{II}}$ in terms of polarization operators, the behavior of any selected $c_L(t)$ is coupled to that of all the others so forming as usual a *finite* hierarchy of the averages of operators (because the index $L$ ranges only between 0 and $2S$). The solution of such a multi-term recurrence relation may always be obtained (as we saw) by rewriting it as a first-order linear matrix differential equation with constant coefficients. Thus we first construct the column vector $\mathbf{C}(t)$ such that

$$\mathbf{C}(t) = \begin{pmatrix} c_1(t) \\ c_2(t) \\ \vdots \\ c_{2S}(t) \end{pmatrix}. \qquad (467)$$

The column vector $\mathbf{C}(t)$ formed by Eq. (467) now contains just $2S$ rows (the index $L$ ranges between 1 and $2S$) since the evolution equation for the function $c_0(t)$ is simply $\partial_t c_0(t) = 0$ with the trivial solution $c_0(t) = \text{const}$. The initial conditions for the relaxation functions $c_L(t)$ are

$$c_L(0) = \langle \hat{T}_{L0}^{(S)} \rangle_{\text{I}} - \langle \hat{T}_{L0}^{(S)} \rangle_{\text{II}}. \qquad (468)$$

Hence the matrix representation of the recurrence equations for the functions $c_L(t)$ becomes the linear matrix differential equation

$$\dot{\mathbf{C}}(t) + \mathbf{X}\mathbf{C}(t) = 0, \qquad (469)$$

where $\mathbf{X}$ is the $2S \times 2S$ system matrix with matrix elements given by

$$(\mathbf{X})_{n,m} = -\tau_N^{-1} g_{n,m}^S \qquad (470)$$

with $g_{n,m}^S$ given by Eq. (390). For example, for $S = 1$, the system matrix $\mathbf{X}$ takes the simple two by two form

$$\mathbf{X} = \frac{e^{-\sigma/2}}{\tau_N} \begin{pmatrix} \cosh\frac{\xi_{\text{II}}}{2} & \frac{1}{\sqrt{3}}(2e^\sigma - 1)\sinh\frac{\xi_{\text{II}}}{2} \\ -\sqrt{3}\sinh\frac{\xi_{\text{II}}}{2} & (2e^\sigma + 1)\cosh\frac{\xi_{\text{II}}}{2} \end{pmatrix}. \qquad (471)$$

Now in general the solution of the homogeneous matrix Eq. (469) is [96]

$$\mathbf{C}(t) = e^{-\mathbf{X}t}\mathbf{C}(0), \qquad (472)$$

which may be written in a more useful form as

$$\mathbf{C}(t) = \mathbf{U} e^{-\mathbf{\Lambda} t} \mathbf{U}^{-1} \mathbf{C}(0), \qquad (473)$$

where $\mathbf{\Lambda}$ is a diagonal matrix composed of the eigenvalues $\lambda_1, \lambda_2, \ldots \lambda_{2S}$ of the system matrix $\mathbf{X}$ and $\mathbf{U}$ is a right eigenvector matrix composed of all the eigenvectors of $\mathbf{X}$, namely,

$$\mathbf{U}^{-1}\mathbf{X}\mathbf{U} = \mathbf{\Lambda}.$$

*All* the $\lambda_k$ are real and positive. The one sided Fourier transform of Eq. (472) also yields the spectrum $\tilde{\mathbf{C}}(\omega)$ of the column vector, viz.,

$$\tilde{\mathbf{C}}(\omega) = \int_0^\infty \mathbf{C}(t) e^{-i\omega t} dt = (\mathbf{X} + i\omega \mathbf{I})^{-1} \mathbf{C}(0). \qquad (474)$$

The function $\tilde{c}_1(0)$ is the first row of the column vector $\tilde{\mathbf{C}}(0)$ which itself in accordance with Eq. (474) is given by

$$\tilde{\mathbf{C}}(0) = (\mathbf{X})^{-1}\mathbf{C}(0). \qquad (475)$$

The formal matrix solutions (472) and (474) will then yield the longitudinal relaxation function $\langle \hat{S}_Z \rangle(t) - \langle \hat{S}_Z \rangle_{\text{II}}$ [cf. Eq. (86)], viz.

$$\langle \hat{S}_Z \rangle(t) - \langle \hat{S}_Z \rangle_{\text{II}} = \sqrt{\frac{S(S+1)(2S+1)}{3}} c_1(t), \qquad (476)$$

and its spectrum as well as the effective and integral relaxation times from their definitions (see Appendix E and Ref. [5])

$$\tau_{\text{ef}} = -\frac{c_1(0)}{\dot{c}_1(0)}, \qquad (477)$$

$$\tau_{\text{int}} = \frac{1}{\langle \hat{S}_Z \rangle_{\text{I}} - \langle \hat{S}_Z \rangle_{\text{II}}} \int_0^t \left[ \langle \hat{S}_Z \rangle(t) - \langle \hat{S}_Z \rangle_{\text{II}} \right] dt = \frac{\tilde{c}_1(0)}{c_1(0)}, \qquad (478)$$

where

$$\tilde{c}_1(\omega) = \int_0^\infty c_1(t) e^{-i\omega t} dt.$$

In accordance with the matrix Eqs. (473) and (475), the relaxation function $c_1(t)$ and the effective integral relaxation times are given by [5]

$$c_1(t) = \sum_{k=1}^{2S} u_{1k} r_k e^{-\lambda_k t}, \qquad (479)$$

$$\tau_{\text{ef}} = \frac{\sum_{k=1}^{2S} u_{1,k} r_k}{\sum_{k=1}^{2S} u_{1,k} r_k \lambda_k}, \qquad (480)$$

$$\tau_{\text{int}} = \frac{\sum_{k=1}^{2S} u_{1k} r_k \lambda_k^{-1}}{\sum_{k=1}^{2S} u_{1k} r_k}, \qquad (481)$$

where the $u_{lk}$ are the matrix elements of the eigenvector matrix $\mathbf{U}$ defined above and $r_k$ are those of the associated column vector $\mathbf{U}^{-1}\mathbf{C}(0)$. As usual, both the integral and effective relaxation times each contain contributions from *all* the eigenvalues $\lambda_k$ and so they characterize the overall relaxation behavior, while the inverse of the smallest nonvanishing eigenvalue $\lambda_1$ characterizes the spin reversal time. Furthermore, because the influence of the high-frequency relaxation modes on the low-frequency relaxation may often be ignored, $\lambda_1$ usually provides adequate information concerning the low-frequency dynamics of the system (see Sec.III.C.3).

Obviously, the matrix method also allows us to evaluate the *linear response* of a spin system due to infinitesimally small changes in the magnitude of the dc field, evaluated in [80-82] via the spin density matrix. Thus we again suppose that the uniform dc field $\mathbf{H}_{\text{II}}$ is directed along the Z axis of the laboratory coordinate system and that a small probing field $\mathbf{H}_1$ ($\mathbf{H}_1 \parallel \mathbf{H}_{\text{II}}$) having been applied to the assembly of spins in the distant past ($t = -\infty$) so that equilibrium conditions obtain at time $t = 0$, is switched off at $t = 0$. The only difference lies in the initial conditions. Instead of the *general* Eq. (468) pertaining to the transient response of *arbitrary strength*, in *linear response*, $\xi_{\text{I}} - \xi_{\text{II}} = \varepsilon \ll 1$, they become

$$c_L(0) = \left\langle \hat{T}_{L0}^{(S)} \right\rangle(0) - \left\langle \hat{T}_{L0}^{(S)} \right\rangle_{\text{II}}$$
$$\approx \frac{\varepsilon}{\beta} \text{Tr} \left\{ \hat{\rho}_{eq}^{\text{II}} \int_0^\beta \hat{S}_0(-i\lambda\hbar) d\lambda \left( \hat{T}_{L0}^{(S)} - \left\langle \hat{T}_{L0}^{(S)} \right\rangle_{\text{II}} \hat{I}^{(S)} \right) \right\}. \quad (482)$$

Here we have used the following identity concerning an exponential function of two operators [75]

$$e^{\beta(\hat{a}+\hat{b})} = e^{\beta\hat{a}} \left( 1 + \int_0^\beta e^{-\lambda\hat{a}} \hat{b} e^{\lambda(\hat{a}+\hat{b})} d\lambda \right). \quad (483)$$

Furthermore, $c_1(t)/c_1(0)$ reduces to the normalized equilibrium longitudinal correlation function $C(t)$ given by [33,75] [cf. Eq. (448)]

$$C(t) = \lim_{\varepsilon \to 0} \frac{c_1(t)}{c_1(0)}$$
$$= \frac{1}{\beta\chi} \left\langle \int_0^\beta \left[ \hat{S}_Z(-i\lambda\hbar) - \left\langle \hat{S}_Z \right\rangle_{\text{II}} \right] \left[ \hat{S}_Z(t) - \left\langle \hat{S}_Z \right\rangle_{\text{II}} \right] d\lambda \right\rangle_{\text{II}}, \quad (484)$$

where

$$\chi = \frac{1}{\beta} \left\langle \int_0^\beta \left[ \hat{S}_Z(-i\lambda\hbar) - \left\langle \hat{S}_Z \right\rangle_{\text{II}} \right] \left[ \hat{S}_Z(0) - \left\langle \hat{S}_Z \right\rangle_{\text{II}} \right] d\lambda \right\rangle_{\text{II}}$$
$$\approx \left\langle \hat{S}_Z^2 \right\rangle_{\text{II}} - \left\langle \hat{S}_Z \right\rangle_{\text{II}}^2 \quad (485)$$

is the normalized static susceptibility. Then we have the dynamic susceptibility $\chi(\omega) = \chi'(\omega) - i\chi''(\omega)$ [75], viz.,

$$\frac{\chi(\omega)}{\chi} = 1 - i\omega\tilde{C}(\omega), \quad (486)$$

where the one-sided Fourier transform $\tilde{C}(\omega)$ is defined by Eq. (451). We have also as in the classical case [5] the integral and effective relaxation times,

$$\tau_{\text{cor}} = \tilde{C}(0), \quad (487)$$

$$\tau_{\text{ef}} = -\frac{1}{\dot{C}(0)}, \quad (488)$$

which now represent the characteristic times governing the behavior of the autocorrelation function $C(t)$.

According to the formal definitions embodied in Eqs. (479) and (486), the dynamic susceptibility is once more a finite sum of Lorentzians, viz.,

$$\frac{\chi(\omega)}{\chi} = \sum_{p=1}^{2S} \frac{c_p}{1 + i\omega/\lambda_p}, \quad (489)$$

where

$$c_p = \frac{u_{1,p} r_p}{\sum_{m=1}^{2S} u_{1,m} r_m}$$

and

$$\sum_{p=1}^{2S} c_p = 1.$$

Moreover, in the low- ($\omega \to 0$) and high- ($\omega \to \infty$) frequency limits, the behavior of the dynamic susceptibility can be easily evaluated as in the classical case [5]. For example by means of Eqs. (481) and (480), we have from the general Eq. (489) for the limits $\omega \to 0$ and for $\omega \to \infty$ respectively

$$\chi(\omega) \approx \chi\left(1 - i\omega\tau_{\text{cor}} + ...\right), \quad \omega \to 0, \quad (490)$$

$$\chi(\omega) \sim \chi\left(i\omega\tau_{\text{ef}}\right)^{-1} + ..., \quad \omega \to \infty. \quad (491)$$

Furthermore, the equilibrium averages $\left\langle \hat{S}_Z \right\rangle_{\text{I}}$, $\left\langle \hat{S}_Z \right\rangle_{\text{II}}$, and $\left\langle \hat{S}_Z^2 \right\rangle_{\text{II}}$ can all be expressed in terms of either the density matrix or the phase-space distribution as

$$\left\langle \hat{S}_Z \right\rangle_i = \sum_{m=-S}^{S} m \rho_m^i, \quad (492)$$

$$\left\langle \hat{S}_Z^2 \right\rangle_i = \sum_{m=-S}^{S} m^2 \rho_m^i, \quad (493)$$

$$\left\langle \hat{S}_Z \right\rangle_i = (S + \tfrac{1}{2})(S+1) \int_{-1}^{1} z W_S^i(z) dz, \quad (494)$$

$$\left\langle \hat{S}_Z^2 \right\rangle_{\text{II}} = (S + \tfrac{1}{2})(S+1) \int_{-1}^{1} \left[ (S + \tfrac{3}{2}) z^2 - \tfrac{1}{2} \right] W_S^{\text{II}}(z) dz, \quad (495)$$

because the Weyl symbols of the operators $\hat{S}_Z$ and $\hat{S}_Z^2$ are, respectively,

$$S_Z = (S+1)\cos\vartheta$$

and

$$S_Z^2 = (S+1)\left[(S+\tfrac{3}{2})\cos^2\vartheta - \tfrac{1}{2}\right].$$

Our method allows one to calculate numerically the integral ($\tau_{ef}$ and $\tau_{cor}$), effective ($\tau_{ef}$), and longest ($\tau = 1/\lambda_1$) relaxation times as well as the dynamic susceptibility $\chi(\omega)$ for a uniaxial nanomagnet. Moreover, all these observables can be calculated analytically.

*2. Analytic equations for the characteristic relaxation times and dynamic susceptibility*

Although in general the method of determining observables based on the correspondence between averages of polarization operators discussed in Section II A4 and III A2 circumvent the phase space equation a much simpler method of treating axially symmetric problem exists. This is so because as already mentioned, for the axially symmetric Hamiltonian Eq. (370), the diagonal elements of the density matrix *decouple* from the non-diagonal ones. Hence, only the former contribute to the time evolution so forming the basis of our simple treatment which we now explain. To appreciate this, we first transform the reduced density-matrix evolution equation Eq. (463) into an evolution equation for its *individual* matrix elements. Thus, we have directly from Eq. (463) a three-term differential-recurrence equation for the *diagonal* matrix elements $\rho_m = \rho_{mm}$ of the density matrix, viz.,

$$\tau_N \frac{\partial \rho_m(t)}{\partial t} = q_m^- \rho_{m-1}(t) + q_m \rho_m(t) + q_m^+ \rho_{m+1}(t), \quad t > 0, \quad (496)$$

where $m = -S, -S+1, \ldots, S$,

$$q_m = -a_m^- e^{-(2m-1)\frac{\sigma}{2S^2} - \frac{\xi_{\parallel}}{2S}} - a_m^+ e^{(2m+1)\frac{\sigma}{2S^2} + \frac{\xi_{\parallel}}{2S}}, \quad (497)$$

$$q_m^\pm = a_m^\pm e^{\mp(2m\pm 1)\frac{\sigma}{2S^2} \mp \frac{\xi_{\parallel}}{2S}}, \quad (498)$$

$$a_m^\pm = -S_{m m\pm 1}^{\mp 1} S_{m\pm 1 m}^{\pm 1} = (S \mp m)(S \pm m + 1)/2. \quad (499)$$

Here

$$S_{m\pm 1 m}^{\pm 1} = \mp\sqrt{(S \mp m)(S \pm m + 1)/2}$$

[cf. Eq. (A7)] are the matrix elements of the spherical spin operators $\hat{S}_{\pm 1}$. Equation (496) is accompanied by the initial condition $\rho_m(0) = \rho_m^I$. Because of our usual *Ansatz* that the equilibrium spin density matrix $\hat{\rho}_{eq}$ must render the collision kernel zero, substitution of the final equilibrium matrix element $\rho_m^{II} = e^{\sigma m^2/S^2 + \xi_{\parallel} m/S}/Z_S^{II}$ with partition function $Z_S^{II} = \sum_{m=-S}^{S} e^{\sigma m^2/S^2 + \xi_{\parallel} m/S}$ into the right-hand side of Eq. (496) requires

$$q_m^- \rho_{m-1}^{II} + q_m \rho_m^{II} + q_m^+ \rho_{m+1}^{II} = 0. \quad (500)$$

Consequently $\rho_m^{II}$ is by inspection the stationary solution of Eq. (496).

To determine the integral relaxation time as defined by Eq. (478), we introduce the set of relaxation functions $f_m(t)$ defined by

$$f_m(t) = \rho_m(t) - \rho_m^{II}. \quad (501)$$

Then the $f_m(t)$ also satisfy the recurrence Eq. (496) with the initial conditions

$$f_m(0) = \rho_m^I - \rho_m^{II}. \quad (502)$$

Because

$$\langle \hat{S}_Z \rangle(t) - \langle \hat{S}_Z \rangle_{II} = \sum_{m=-S}^{S} m f_m(t)$$

and

$$\langle \hat{S}_Z \rangle(0) - \langle \hat{S}_Z \rangle_{II} = \langle \hat{S}_Z \rangle_I - \langle \hat{S}_Z \rangle_{II},$$

the Fourier-Laplace transform $\tilde{c}_1(\omega)/c_1(0)$ of the normalized relaxation function $c_1(t)/c_1(0)$ is

$$\frac{\tilde{c}_1(\omega)}{c_1(0)} = \frac{1}{\langle \hat{S}_Z \rangle_I - \langle \hat{S}_Z \rangle_{II}} \sum_{m=-S}^{S} m \tilde{f}_m(\omega), \quad (503)$$

so that the integral relaxation time is as usual by definition [cf. Eq.(478)]

$$\tau_{int} = \frac{1}{\langle \hat{S}_Z \rangle_I - \langle \hat{S}_Z \rangle_{II}} \sum_{m=-S}^{S} m \tilde{f}_m(0), \quad (504)$$

where

$$\langle \hat{S}_Z \rangle_i = \sum_{m=-S}^{S} m \rho_m^i. \quad (505)$$

The spectrum $\tilde{c}_1(\omega)/c_1(0)$ and the integral relaxation time $\tau_{int}$ can now be calculated analytically using continued fractions, which starting from Eq. (496) we describe as follows. For convenience, we first introduce a new index $n$ defined as $n = m + S$. Thus the differential recurrence Eq. (496) can then be rearranged as one for the relaxation functions $f_n(t)$ defined by Eq. (501), viz.,

$$\tau_N \frac{\partial f_n}{\partial t} = p_n^- f_{n-1} + p_n f_n + p_n^+ f_{n+1}, \quad (506)$$

where the new coefficients $p_n^\pm$ and $p_n$ are [cf. Eqs. (497)-(499)]

$$p_n = -\frac{n(2S-n+1)}{2}e^{-(2n-2S+1)\frac{\sigma}{2S^2}-\frac{\xi_\parallel}{2S}}$$
$$-\frac{(n+1)(2S-n)}{2}e^{(2n-2S+1)\frac{\sigma}{2S^2}+\frac{\xi_\parallel}{2S}}$$
(507)

$$p_n^+ = \frac{1}{2}(2S-n)(n+1)e^{-(2n-2S-1)\frac{\sigma}{2S^2}-\frac{\xi_\parallel}{2S}},$$
(508)

$$p_n^- = \frac{n}{2}(2S-n+1)e^{(2n-2S-1)\frac{\sigma}{2S^2}+\frac{\xi_\parallel}{2S}}.$$
(509)

Consequently, the new recurrence relation Eq. (506) can be written in the *homogeneous* matrix form

$$\tau_N \dot{\mathbf{F}}(t) = \mathbf{\Pi} \cdot \mathbf{F}(t),$$
(510)

where the column vector $\mathbf{F}(t)$ and the tridiagonal system matrix $\mathbf{\Pi}$ are

$$\mathbf{F}(t) = \begin{pmatrix} f_0(t) \\ f_1(t) \\ \vdots \\ f_{2S}(t) \end{pmatrix},$$
(511)

$$\mathbf{\Pi} = \begin{pmatrix} p_0 & p_0^+ & 0 & \cdots & 0 \\ p_1^- & p_1 & p_1^+ & \cdots & \vdots \\ \vdots & \vdots & \vdots & \ddots & p_{2S-1}^+ \\ 0 & \cdots & 0 & p_{2S}^- & p_{2S} \end{pmatrix}.$$
(512)

The tridiagonal system matrix $\mathbf{\Pi}$ has exactly the same eigenvalues as the actual system matrix $\mathbf{X}$ given by Eq. (470) save that it possesses an additional zero eigenvalue $\lambda_0 = 0$ corresponding to the thermal equilibrium state. Clearly, the matrix recurrence Eq. (510) could again be solved numerically by the matrix methods described in Sec. II.A.4. Rather, we prefer to obtain the *exact analytic solution* in terms of continued fractions. Applying the general method of solution of inhomogeneous three term recurrence relations to the Fourier-Laplace transform of the scalar Eq. (506) ([5], Section 2.7.3), we have the solution

$$\tilde{f}_n(\omega) = \Delta_n(\omega) p_n^- \tilde{f}_{n-1}(\omega) + \tau_N \left(p_{n-1}^+\right)^{-1} \sum_{l=n}^{2S} \prod_{k=n}^{l} \left(p_{k-1}^+ \Delta_k(\omega)\right) f_l(0),$$
(513)

where $\Delta_n(\omega)$ are the continued fractions defined by the two term recurrence equation

$$\Delta_n(\omega) = \frac{1}{i\omega\tau_N - p_n - p_n^+ p_{n+1}^- \Delta_{n+1}(\omega)}$$

with $0 \le n \le 2S$ and $\Delta_{2S+1}(\omega) = 0$. The spectrum $\tilde{c}_1(\omega)/c_1(0)$ from the definition Eq. (503) is then given by

$$\frac{\tilde{c}_1(\omega)}{c_1(0)} = \frac{1}{\langle \hat{S}_Z \rangle_I - \langle \hat{S}_Z \rangle_{II}} \sum_{n=1}^{2S} n\tilde{f}_n(\omega).$$
(514)

131

For $\omega = 0$, Eq. (513) simplifies yielding the two-term recurrence equation

$$\tilde{f}_n(0) = d_n \tilde{f}_{n-1}(0) + r_n,$$
(515)

where the coefficients are

$$d_n = e^{(2n-2S-1)\frac{\sigma}{S^2}+\frac{\xi_\parallel}{S}},$$

$$r_n = \frac{2\tau_N e^{(2n-2S-1)\frac{\sigma}{2S^2}+\frac{\xi_\parallel}{2S}}}{n(2S-n+1)} \sum_{l=n}^{2S} f_l(0)$$

because $\Delta_n(0) = \left(p_{n-1}^+\right)^{-1}$. However, the three term recurrence equations for the relaxation functions defined by Eq. (506) are no longer *linearly independent*, because the determinant of the matrix $\mathbf{\Pi}$ generated from Eq. (506) is zero ($\det \mathbf{\Pi} = 0$). Thus all the subsequent $\tilde{f}_n(0)$ needed to calculate the integral relaxation time can be determined only in terms of $\tilde{f}_0(0)$. However, in order to calculate $\tilde{f}_0(0)$, we can now utilize the normalization properties of the density matrix, namely,

$$\sum_{m=-S}^{S} f_m(t) = \sum_{m=-S}^{S} \left(\rho_m(t) - \rho_m^{II}\right) = 0,$$

so that

$$\sum_{m=-S}^{S} \tilde{f}_m(0) = 0.$$
(516)

Now because of Eqs. (515) and (516), we have the identity

$$\tilde{f}_0(0) + \sum_{n=1}^{2S} \left(d_n \tilde{f}_{n-1}(0) + r_n\right)$$
$$= \tilde{f}_0(0) + d_1 \tilde{f}_0(0) + r_1 + \sum_{n=2}^{2S} \left(d_n d_{n-1} \tilde{f}_{n-2}(0) + d_n r_{n-1} + r_n\right)$$
(517)
$$= \tilde{f}_0(0) + \tilde{f}_0(0) \sum_{n=1}^{2S} \prod_{l=1}^{n} d_l + \sum_{n=1}^{2S} \frac{r_n}{d_n} \sum_{k=n}^{2S} \prod_{l=n}^{k} d_l = 0,$$

where the products are given by

$$\frac{1}{d_n} \prod_{l=n}^{k} d_l = \frac{e^{(k-S)^2 \frac{\sigma}{S^2}+(k-S)\frac{\xi_\parallel}{S}}}{e^{(n-S)^2 \frac{\sigma}{S^2}+(n-S)\frac{\xi_\parallel}{S}}},$$

$$\prod_{l=1}^{n} d_l = e^{\xi_\parallel - \sigma} e^{(n-S)^2 \frac{\sigma}{S^2}+(n-S)\frac{\xi_\parallel}{S}}.$$

However

$$e^{(2k-1)\frac{\sigma}{2S^2}+\frac{\xi_\parallel}{2S}} = \sqrt{\frac{\rho_k^{II}}{\rho_{k-1}^{II}}},$$
(518)

so that Eq. (517) immediately yields a closed form expression for $\tilde{f}_0(0)$, viz.,

132

$$\tilde{f}_0(0) = -\frac{2\tau_N e^{\sigma-\xi_{II}}}{Z} \sum_{k=1-S}^{S} \frac{\sum_{m=k}^{S}(\rho_m^{I}-\rho_m^{II})\sum_{j=k}^{S}\rho_j^{II}}{[S(S+1)-k(k-1)]\sqrt{\rho_k^{II}\rho_{k-1}^{II}}}. \tag{519}$$

Thus we have

$$\sum_{n=1}^{2S} n\tilde{f}_n(0) = \tilde{f}_0(0)e^{\xi_{II}-\sigma}\left\langle \hat{S}_Z \right\rangle_{II} \tag{520}$$
$$+\sum_{n=1}^{2S} r_n e^{-(n-S)^2\frac{\sigma}{S^2}-(n-S)\frac{\xi_{II}}{S}} \sum_{k=n}^{2S} k e^{(k-S)^2\frac{\sigma}{S^2}+(k-S)\frac{\xi_{II}}{S}}.$$

Hence, by substituting Eqs. (519) and (520) into the definition Eq. (504), we finally have the integral relaxation time rendered in explicit series form as

$$\tau_{int} = \frac{\sum_{n=1}^{2S} n\tilde{f}_n(0)}{\left\langle \hat{S}_Z \right\rangle_{I}-\left\langle \hat{S}_Z \right\rangle_{II}} \tag{521}$$
$$= \frac{2\tau_N}{\left\langle \hat{S}_Z \right\rangle_{I}-\left\langle \hat{S}_Z \right\rangle_{II}} \sum_{k=1-S}^{S} \frac{\sum_{m=k}^{S}(\rho_m^{I}-\rho_m^{II})\sum_{j=k}^{S}(j-\left\langle \hat{S}_Z \right\rangle_{II})\rho_j^{II}}{[S(S+1)-k(k-1)]\sqrt{\rho_k^{II}\rho_{k-1}^{II}}}.$$

Both the eigensolution given by the formal Eq. (481) and the explicit Eq. (521) as determined from the definition of the integral relaxation time via the zero frequency limit of the normalized relaxation function $\tilde{c}_1(\omega)/c_1(0)$ yield exactly the same numerical result. Thus $\tau_{int}$ for various nonlinear transient responses (such as the rise, decay, and rapidly reversing field transients) may be easily evaluated from the explicit Eq. (521). Equation (521) is also valid for an *arbitrary* axially symmetric potential $\hat{H}_S(\hat{S}_Z)$, because the precise form of the potential is involved only in the *equilibrium* matrix elements of the density operator $\rho_m^{II}$ and in the averages $\left\langle \hat{S}_Z \right\rangle_{I}$ and $\left\langle \hat{S}_Z \right\rangle_{II}$. Furthermore, it is useful to recall that in the classical limit $S \to \infty$, the nonlinear integral relaxation time $\tau_{int}$ of the longitudinal relaxation function $c_1(t) = \left\langle \cos\vartheta \right\rangle(t) - \left\langle \cos\vartheta \right\rangle_{II}$ of a classical uniaxial nanomagnet with a free energy density

$$\beta V(\vartheta) = -\sigma \cos^2\vartheta - \xi_{II}\cos\vartheta \tag{522}$$

is given by [5] (see Appendix D for details)

$$\tau_{int} = \frac{2\tau_N}{\left\langle \cos\vartheta \right\rangle_{I}-\left\langle \cos\vartheta \right\rangle_{II}} \int_{-1}^{1} \frac{\Phi(z)\Psi(z)e^{-\sigma z^2-\xi_{II}z}}{1-z^2} dz, \tag{523}$$

where

$$\Phi(z) = \int_{-1}^{z} [W_{I}(x)-W_{II}(x)] dx$$
$$= \frac{\pi^{1/2}e^{-\sigma h_{II}^2}}{2\sigma^{1/2}Z_{II}} \left\{ \mathrm{erfi}[(z+h_{II})\sqrt{\sigma}]+\mathrm{erfi}[(1-h_{II})\sqrt{\sigma}] \right\}$$
$$-\frac{\pi^{1/2}e^{-\sigma h_{I}^2}}{2\sigma^{1/2}Z_{I}} \left\{ \mathrm{erfi}[(z+h_{I})\sqrt{\sigma}]+\mathrm{erfi}[(1-h_{I})\sqrt{\sigma}] \right\},$$

$$\Psi(z) = \int_{-1}^{z} (x-\left\langle \cos \right\rangle_{II})e^{\sigma(x^2+2h_{II}x)} dx$$
$$= \frac{1}{2\sigma}\left[ e^{\sigma(z^2+2h_{II}z)} - e^{\sigma(1-2h_{II})} \right] - e^{\sigma(1-h_{II}^2)}\frac{\sqrt{\pi}\sinh(2\sigma h_{II})}{2\sqrt{\sigma}^3 Z_{II}}$$
$$\times \left\{ \mathrm{erfi}[(z+h_{II})\sqrt{\sigma}]+\mathrm{erfi}[(1-h_{II})\sqrt{\sigma}] \right\},$$

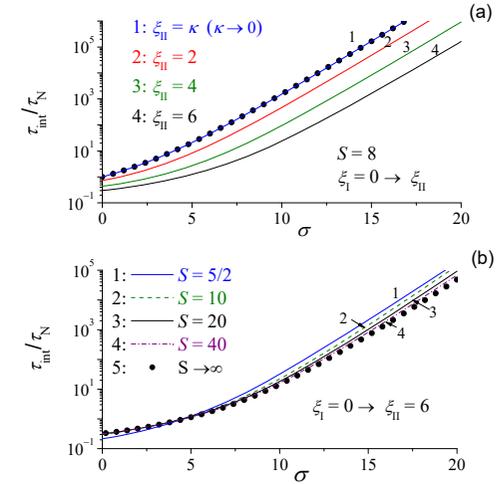

**Figure 18.** (Color on line) Nonlinear integral relaxation time $\tau_{int}/\tau_N$ for the rise transient response as a function of the barrier parameter $\sigma$ (a) for $S = 8$, $\xi_I = 0$, and various $\xi_{II} = \kappa \to 0, 2, 4, 6$ and (b) for $\xi_I = 0$ and $\xi_{II} = 6$ and various values of spin $S = 5/2, 10, 20, 40$, and $S \to \infty$. Solid lines: calculations from Eq. (521); circles: Eq. (531).

$$\left\langle \cos\vartheta \right\rangle_i = \frac{e^\sigma \sinh(2\sigma h_i)}{\sigma Z_i} - h_i, \tag{524}$$

$$Z_i = \frac{e^{-\sigma h_i^2}}{2}\sqrt{\frac{\pi}{\sigma}}\left\{ \mathrm{erfi}[(1+h_i)\sqrt{\sigma}]+\mathrm{erfi}[(1-h_i)\sqrt{\sigma}] \right\}, \tag{525}$$

where $h_i = \xi_i/(2\sigma)$ and the error function of imaginary argument erfi(x) is [105]

$$\mathrm{erfi}(x) = \frac{2}{\sqrt{\pi}} \int_0^x e^{t^2} dt. \tag{526}$$

The nonlinear relaxation time $\tau_{\mathrm{int}}$ for the rise transient response as a function of the anisotropy parameter $\sigma$ and the spin number $S$ is plotted in Fig. 18, indicating a pronounced dependence of this time on the field ($\xi_{\mathrm{II}}$), anisotropy ($\sigma$), and spin ($S$) parameters; in particular that time decreases with increasing field strength $\xi_{\mathrm{II}}$. It is apparent from Fig. 18 that for large $S$, the quantum solutions reduce to the corresponding classical ones. Typical values of $S$ for the quantum-classical crossover are ~20-40. *The smaller the anisotropy $\sigma$ the smaller the S value required for convergence of the quantum equations to the classical ones.*

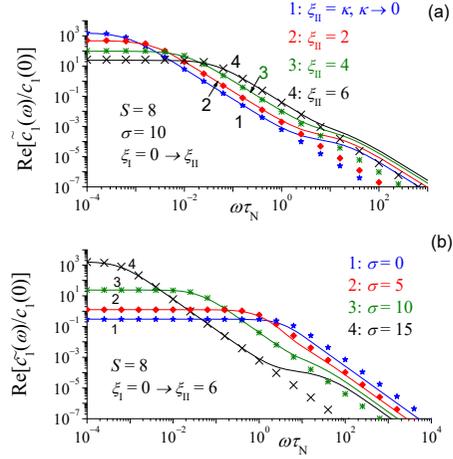

**Figure 19.** (Color on line) Real parts of the relaxation function spectrum $\tilde{c}_1(\omega)/c_1(0)$ vs. the normalized frequency $\omega\tau_{\mathrm{N}}$ for the rise transient response (a) for $S = 8$, $\sigma = 10$, $\xi_{\mathrm{I}} = 0$, and various $\xi_{\mathrm{II}} = \kappa \to 0, 2, 4, 6$ and (b) for $\xi_{\mathrm{I}} = 0$, $\xi_{\mathrm{II}} = 6$, $S = 8$, and various anisotropy parameters $\sigma$. Solid lines: calculations from Eq. (474); stars: Eq. (527).

In Figs. 19, we have plotted the real part of the normalized relaxation function spectrum $\tilde{c}_1(\omega)/c_1(0)$ vs. the normalized frequency $\omega\tau_N$ for the rise transient response, $\xi_{\mathrm{I}} = 0 \to \xi_{\mathrm{II}} \neq 0$. Like the classical case [5], knowledge of $\lambda_1$ alone is enough to accurately predict the low-frequency part of $\tilde{c}_1(\omega)/c_1(0)$ as well as the long time behavior of the relaxation function $c_1(t)/c_1(0)$. Here the single-mode approximation

$$\frac{\tilde{c}_1(\omega)}{c_1(0)} \approx \frac{\tau_{\mathrm{int}}}{1 + i\omega/\lambda_1}, \tag{527}$$

where $\tau_{\mathrm{int}}$ is given by Eqs. (521) and $\lambda_1$ is the smallest nonvanishing eigenvalue of the system matrix $\mathbf{X}$ defined by Eq. (470), is shown for comparison indicating that $\tau_{\mathrm{int}}$ and $\lambda_1$ comprehensively describe the low-frequency behavior of the spectrum $\tilde{c}_1(\omega)/c_1(0)$ as in the classical case. In the time domain, the single-mode approximation Eq. (527) amounts to assuming that the relaxation function $c_1(t)$ as determined by Eq. (479) (comprising 2$S$ exponentials) may be approximated for $t > 0$ by a *single* exponential. Consequently, the longtime relaxation behavior may once again be accurately approximated by a *single* exponential with relaxation time $T_1 = 1/\lambda_1$ and thus is again governed by the Bloch Eq. (462).

We may also evaluate the *linear response* of a uniaxial nanomagnet corresponding to infinitesimally small changes in the magnitude of the dc field, so that the integral relaxation time now becomes the correlation time. In linear response, i.e., considering transient relaxation between the states I and II with respective Hamiltonians

$$\hat{H}_S^{\mathrm{I}} = \frac{\sigma}{S^2}\hat{S}_Z^2 + \frac{\xi_{\mathrm{II}} + \varepsilon}{S}\hat{S}_Z \text{ and } \hat{H}_S^{\mathrm{II}} = \frac{\sigma}{S^2}\hat{S}_Z^2 + \frac{\xi_{\mathrm{II}}}{S}\hat{S}_Z,$$

where $\varepsilon$ is a small external field parameter, the initial conditions $f_m(0)$ and $\langle \hat{S}_Z \rangle_{\mathrm{I}} - \langle \hat{S}_Z \rangle_{\mathrm{II}}$ reduce to

$$f_m(0) = \frac{1}{Z_S^{\mathrm{I}}} e^{\frac{\sigma}{S^2}m^2 + \frac{\xi_{\mathrm{II}}+\varepsilon}{S}m} - \frac{1}{Z_S^{\mathrm{II}}} e^{\frac{\sigma}{S^2}m^2 + \frac{\xi_{\mathrm{II}}}{S}m} \approx \frac{\varepsilon}{S}\left(m - \langle \hat{S}_Z \rangle_{\mathrm{II}}\right)\rho_m^{\mathrm{II}}, \tag{528}$$

$$\langle \hat{S}_Z \rangle_{\mathrm{I}} - \langle \hat{S}_Z \rangle_{\mathrm{II}} \approx \frac{\varepsilon}{S}\chi, \tag{529}$$

where

$$\chi = \langle \hat{S}_Z^2 \rangle_{\mathrm{II}} - \langle \hat{S}_Z \rangle_{\mathrm{II}}^2 = \sum_{m=-S}^{S} m^2 \rho_m^{\mathrm{II}} - \left(\sum_{m=-S}^{S} m\rho_m^{\mathrm{II}}\right)^2. \tag{530}$$

Thus, in the limit $\varepsilon \to 0$, Eq. (521) yields the correlation time $\tau_{\mathrm{cor}}$ in the explicit series form

$$\tau_{\mathrm{cor}} = \frac{2\tau_{\mathrm{N}}}{\chi} \sum_{k=1-S}^{S} \frac{\left(\sum_{m=k}^{S}\left(m - \langle \hat{S}_Z \rangle_{\mathrm{II}}\right)\rho_m^{\mathrm{II}}\right)^2}{[S(S+1) - k(k-1)]\sqrt{\rho_k^{\mathrm{II}}\rho_{k-1}^{\mathrm{II}}}}. \tag{531}$$

Equation (531) in the limit $S \to \infty$ yields numerical results concurring with the classical ones [cf. Eq. (552) below].

Furthermore, for the model embodied in Eq. (496), we can also calculate the effective relaxation time $\tau_{\mathrm{ef}}$ given by Eq. (E2) from Appendix E. Thus, $\tau_{\mathrm{ef}}$ yielding precise information on the initial decay of $c_1(t)$ in the time domain and defined as usual by [5]

$$\tau_{\text{ef}} = -\frac{c_1(0)}{\dot{c}_1(0)} = -\frac{S \partial_\varepsilon \langle \hat{S}_Z \rangle_1 \big|_{\varepsilon=0}}{\sum_{m=-S}^{S} m\dot{f}_m(0)} = -\frac{\chi}{\sum_{m=-S}^{S} m\dot{f}_m(0)}, \tag{532}$$

is given by

$$\tau_{\text{ef}} = \frac{2\chi\tau_N}{\sum_{k=1-S}^{S}[S(S+1)-k(k-1)]\sqrt{\rho_k^{\text{II}}\rho_{k-1}^{\text{II}}}}, \tag{533}$$

where noting Eqs. (508), (509), and (528) we have used and that

$$\sum_{m=-S}^{S} m\dot{f}_m(0) = \sum_{n=1}^{2S} n\left[ p_n^- f_{n-1}(0) + p_n f_n(0) + p_n^+ f_{n+1}(0) \right]$$
$$= \sum_{n=1}^{2S}\left[ (n-1)p_{n-1}^+ \rho_n^{\text{II}} - np_n^- \rho_{n-1}^{\text{II}} \right]$$
$$= \frac{1}{2\tau_N}\sum_{k=1-S}^{S}[S(S+1)-k(k-1)]\sqrt{\rho_k^{\text{II}}\rho_{k-1}^{\text{II}}}.$$

Finally, the longest relaxation time $\tau$, which is associated with the spin reversal time, can be calculated via the smallest *nonvanishing* eigenvalue $\lambda_1$ of the matrix $\tau_N^{-1}\mathbf{\Pi}$ given by Eq. (512) as $\tau = \lambda_1^{-1}$ from the deterministic equation

$$\det\left(\tau_N^{-1}\mathbf{\Pi} - \lambda \mathbf{I}\right) = 0. \tag{534}$$

The left-hand side of Eq. (534) represents the polynomial of the order $2S+1$, viz.,

$$\left( k_{2S+1}\lambda^{2S} + k_{2S}\lambda^{2S-1} + \ldots + k_2\lambda + k_1 \right)\lambda = 0, \tag{535}$$

where

$$k_1 = -\sum_{i=0}^{2S} M_i^i, \tag{536}$$

$$k_2 = \sum_{i=0}^{2S-1}\sum_{j=i+1}^{2S} M_{ij}^{ij}, \tag{537}$$

and so on and we have noticed that $\det(\mathbf{\Pi}) = 0$. Here the $M_i^{i'}$ are the first minors of the matrix $\mathbf{\Pi}$, which are in tern the determinants of square matrixes as reduced from $\tau_N^{-1}\mathbf{\Pi}$ by removing the *i*th row and the *i'*th column of $\tau_N^{-1}\mathbf{\Pi}$ while the $M_{ij}^{i'j'}$ are the minors of the matrix $\tau_N^{-1}\mathbf{\Pi}$, which are the determinants of the square matrix reduced from $\tau_N^{-1}\mathbf{\Pi}$ by removing two (the *i*th and the *j*th) of its rows and two (the *i'*th and the *j'*th) columns. Now the smallest *nonvanishing* eigenvalue $\lambda_1$ can be readily evaluated numerically from Eq. (534), e.g., using MATHEMATICA. However, in the high barrier approximation when $\lambda_1 \ll 1$, it can be evaluated analytically by neglecting all higher powers of $\lambda^n$ with $n > 2$ in Eq. (535). Thus, we have

$$\lambda_1 \approx -\frac{k_1}{k_2}. \tag{538}$$

Equation (538) can be written equivalently in matrix form as

$$\lambda_1 \approx \frac{\text{Tr}(\mathbf{M}^{(1)})}{\text{Tr}(\mathbf{M}^{(2)})}, \tag{539}$$

where $\mathbf{M}^{(1)}$ is the matrix formed from the all first minors

$$\mathbf{M}^{(1)} = \begin{pmatrix} M_{2S}^{2S} & M_{2S}^{2S-1} & \cdots & M_{2S}^{0} \\ M_{2S-1}^{2S} & M_{2S-1}^{2S-1} & \cdots & M_{2S-1}^{0} \\ \vdots & \vdots & \ddots & \vdots \\ M_{0}^{2S} & M_{0}^{2S-1} & \cdots & M_{0}^{0} \end{pmatrix}, \tag{540}$$

and the matrix $\mathbf{M}^{(2)}$ contains all the $M_{ij}^{i'j'}$ minors

$$\mathbf{M}^{(2)} = \begin{pmatrix} M_{2S,2S-1}^{2S,2S-1} & M_{2S,2S-1}^{2S,2S-2} & \cdots & M_{2S,2S-1}^{0,0} \\ M_{2S,2S-2}^{2S,2S-1} & M_{2S,2S-2}^{2S,2S-2} & \cdots & M_{2S,2S-2}^{0,0} \\ \vdots & \vdots & \ddots & \vdots \\ M_{0,0}^{2S,2S-1} & M_{0,0}^{2S,2S-2} & \cdots & M_{0,0}^{0,0} \end{pmatrix}. \tag{541}$$

The matrixes $\mathbf{M}^{(1)}$ and $\mathbf{M}^{(2)}$ have, respectively, the dimensions $n \times n$ and $n(n-1)/2 \times n(n-1)/2$, where $n = 2S+1$. Furthermore, the ordering of the elements in the matrix $\mathbf{M}^{(2)}$ is such that reading across or down the final matrix, the successive lists of positions appear in lexicographic order. Now, $\text{Tr}(\mathbf{M}^{(1)})$ and $\text{Tr}(\mathbf{M}^{(2)})$ can be calculated analytically as

$$\text{Tr}(\mathbf{M}^{(1)}) = \frac{(-1)^{2S}}{\tau_N^{2S}}\sum_{i=0}^{2S}\prod_{s=1}^{i} p_s^- \prod_{r=i}^{2S-1} p_r^+ = \frac{(2S)!}{2^{2S}\tau_N^{2S}}\sum_{i=0}^{2S} e^{(i^2-2iS)\frac{\sigma}{S^2}+(i-S)\frac{\xi_\parallel}{S}}$$
$$= \frac{(2S)!}{2^{2S}\tau_N^{2S}}\sum_{k=-S}^{S} e^{(k^2-S^2)\frac{\sigma}{S^2}+k\frac{\xi_\parallel}{S}} = \frac{(2S)!e^{-\sigma}}{2^{2S}\tau_N^{2S}} Z_S \tag{542}$$

and

$$\text{Tr}(\mathbf{M}^{(2)}) = \frac{(-1)^{2S+1}}{\tau_N^{2S-1}}\sum_{i=0}^{2S-1}\sum_{j=i+1}^{2S}\left(\prod_{s=1}^{i} p_s^- \prod_{r=j}^{2S-1} p_r^+ \sum_{m=1}^{j-i} \prod_{u=j+2-m}^{j} p_u^- \prod_{v=i}^{j-m-1} p_v^+ \right)$$
$$= \frac{(2S)!e^{-\sigma}}{2^{2S-1}\tau_N^{2S-1}}\sum_{k=-S}^{S-1}\sum_{n=k+1}^{S}\sum_{m=1}^{n-k} \frac{e^{[2k^2-2n-1+2m(2n-m+1)]\frac{\sigma}{2S^2}+(2k+2m-1)\frac{\xi_\parallel}{2S}}}{(S+n-m+1)(S-n+m)}. \tag{543}$$

Here we have noticed that $\prod_{m=a}^{b} p_m^\pm = 1$ if $b < a$. Thus in the high barrier approximation, the longest relaxation time $\tau \approx \lambda_1^{-1}$ is given by the following approximate equation

$$\tau \approx \frac{2\tau_N}{Z_S}\sum_{k=-S}^{S-1}\sum_{n=k+1}^{S}\sum_{m=1}^{n-k}\frac{e^{[2k^2-2n-1+2m(2n-m+1)]\frac{\sigma}{2S^2}+(2k+2m-1)\frac{\xi_\parallel}{2S}}}{(S+n-m+1)(S-n+m)}. \tag{544}$$

We remarked above that the linear response has been previously studied by Garanin [80] and Garcia-Palacios and Zueco [81] using the spin density matrix whereby they also gave analytic expressions for $\tau_{cor}$, $\tau_{ef}$, and $\tau \approx \lambda_1^{-1}$ for more general models of linear and bilinear spin-bath interactions with superohmic damping. Using Garanin's method [80], the longest relaxation time $\tau$ can be found in a similar manner for the model embodied in Eq. (496) yielding

$$\tau = \frac{2\tau_N}{\chi_\Delta} \sum_{k=1-S}^{S} \frac{\left(\sum_{m=k}^{S}\left(m-\langle \hat{S}_Z \rangle_{II}\right)\rho_m^{II}\right)\left(\sum_{m=-S}^{k-1}[\Delta - \mathrm{sgn}(m-m_b)]\rho_m^{II}\right)}{[S(S+1)-k(k-1)]\sqrt{\rho_k^{II}\rho_{k-1}^{II}}}, \quad (545)$$

where $m_b$ is the quantum number corresponding to the top of the barrier, with

$$\Delta = \sum_{m=-S}^{S} \mathrm{sgn}(m-m_b)\rho_m^{II}$$

and

$$\chi_\Delta = \sum_{m=-S}^{S} m\,\mathrm{sgn}(m-m_b)\rho_m^{II} - \left(\sum_{m=-S}^{S} m\rho_m^{II}\right)\left(\sum_{m=-S}^{S} \mathrm{sgn}(m-m_b)\rho_m^{II}\right).$$

Furthermore, via the replacement $k+1 \to k$ and then via

$$\sum_{m=k}^{S}\left(m-\langle \hat{S}_Z \rangle_{II}\right)\rho_m^{II} = -\sum_{m=-S}^{k-1}\left(m-\langle \hat{S}_Z \rangle_{II}\right)\rho_m^{II},$$

one can rearrange Eq. (545) as

$$\tau = \frac{2\tau_N}{\chi_\Delta} \sum_{k=-S}^{S-1} \frac{\left(\sum_{m=-S}^{k}\left(m-\langle \hat{S}_Z \rangle_{II}\right)\rho_m^{II}\right)\left(\sum_{m=-S}^{k}[\mathrm{sgn}(m-m_b)-\Delta]\rho_m^{II}\right)}{[S(S+1)-k(k+1)]\sqrt{\rho_k^{II}\rho_{k+1}^{II}}}. \quad (546)$$

For $\xi_{II} < \sigma$, the relative deviation of $\tau$ given Eq. (546) from $\lambda_1^{-1}$ calculated numerically does not exceed 1%.

All the foregoing expressions have been derived via the density matrix method. They can also be obtained using the phase space formalism thereby exemplifying how they may reduce to the classical expressions. For example, $\tau_{ef}$ as rendered by Eq. (533) can be written as

$$\tau_{ef} = 2\tau_N e^{\xi/(2S)} \frac{\langle \hat{S}_Z^2 \rangle_{II} - \langle \hat{S}_Z \rangle_{II}^2}{\langle \hat{\mathbf{S}}^2 - \hat{S}_Z^2 + \hat{S}_Z \rangle_{II}}, \quad (547)$$

where $\langle \hat{S}_Z \rangle_{II}$ and $\langle \hat{S}_Z^2 \rangle_{II}$ can also be given via the phase space Eqs. (494)-(495) and the denominator is given by the phase space average

$$\langle \hat{\mathbf{S}}^2 - \hat{S}_Z^2 + \hat{S}_Z \rangle_{II} = (S+1)(S+\tfrac{1}{2})\int_{-1}^{1}\left[S(1-z^2)+\tfrac{1}{2}+z-\tfrac{3}{2}z^2\right]W_S^{II}(z)dz, \quad (548)$$

on recalling that the Weyl symbol of the operator $\hat{\mathbf{S}}^2 - \hat{S}_Z^2 + \hat{S}_Z$ is

$$\hat{\mathbf{S}}^2 - \hat{S}_Z^2 + \hat{S}_Z \to (S+1)\left[S(1-\cos^2\vartheta)+\cos\vartheta - \tfrac{3}{2}\cos^2\vartheta + \tfrac{1}{2}\right]. \quad (549)$$

Clearly, Eq. (547) is simply a quantum analog of the long established equation for the longitudinal effective relaxation time $\tau_{ef}$ of classical macrospins, viz. [5,6],

$$\tau_{ef} = 2\tau_N \frac{\langle \cos^2\vartheta \rangle_{II} - \langle \cos\vartheta \rangle_{II}^2}{1 - \langle \cos^2\vartheta \rangle_{II}}, \quad (550)$$

where

$$\langle \cos^2\vartheta \rangle_{II} = \frac{1}{Z_{II}}\int_{-1}^{1} x^2 e^{\sigma(x^2+2h_{II}x)}dx = \frac{e^\sigma[\cosh(2\sigma h_{II})-h\sinh(2\sigma h_{II})]}{\sigma Z_{II}} + h_{II}^2 - \frac{1}{2\sigma} \quad (551)$$

and $\langle \cos\vartheta \rangle_{II}$ and the partition function $Z_{II}$ are defined by Eqs. (524) and (525), respectively. Furthermore, the corresponding integral (correlation) time $\tau_{cor}$ is given by [5,6,151,153]

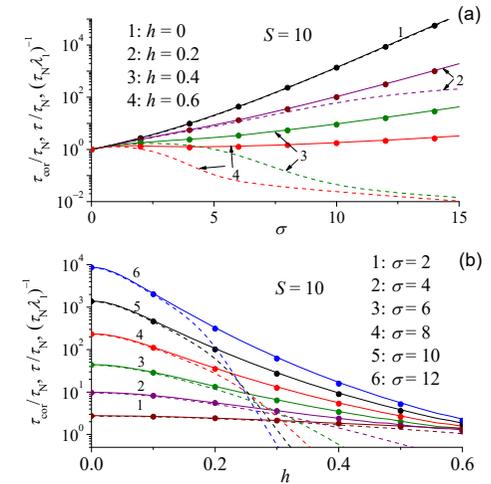

**Figure 20.** (Color on line) The correlation time $\tau_{cor}$ [Eq. (531); dashed lines], the inverse of the smallest eigenvalue $\lambda_1$ (filled circles) and its approximation $\tau$ [Eq. (545); solid line] (a) as a function of the barrier height parameter $\sigma$ for various field parameters $h = \xi_{II}/(2\sigma)$ and (b) as a function of the field parameter $h$ for various $\sigma$ ($S = 10$).

$$\tau_{cor} = \frac{2\tau_N}{Z_{II}\left(\langle \cos^2\vartheta \rangle_{II} - \langle \cos\vartheta \rangle_{II}^2\right)}\int_{-1}^{1}\left[\int_{-1}^{z}\left(x - \langle \cos\vartheta \rangle_{II}\right)e^{\sigma(x^2+2h_{II}x)}dx\right]^2 \frac{e^{\sigma(z^2+2h_{II}z)}}{1-z^2}dz. \quad (552)$$

For linear response, the correlation time $\tau_{cor}$ and overbarrier time $\lambda_1^{-1}$, are plotted in Fig. 20 as a function of the field parameter $h = \xi_{II}/(2\sigma)$ and the barrier height parameter $\sigma$ for $S =$

10. Like the classical case, the behavior of $\tau_{cor}$ and $\lambda_1^{-1}$ is similar only for *small* external fields. In a strong external field, $h \geq 0.2$, $1/\lambda_1$ can diverge exponentially from $\tau_{cor}$. This divergence effect for a classical uniaxial nanomagnet was discovered numerically by Coffey *et al.* [152] and later explained quantitatively by Garanin [153] (Ref. 5, Chap. 1). He showed analytically that the contribution of relaxation modes other than the overbarrier one to either the integral relaxation time becomes significant for high external fields due to population depletion of the shallower of the two potential wells of a bistable potential under the action of an external applied field. The field is far less than that needed to destroy the bistable nature of the potential [5]. Furthermore, $\tau_{cor}$ and $1/\lambda_1$, are also plotted in Figs. 21 as a function of $S$ for various values of $\xi_{\|}$ and $\sigma$. Clearly even for relatively small $S \sim 20$, the quantum formulas are very close to the classical ones.

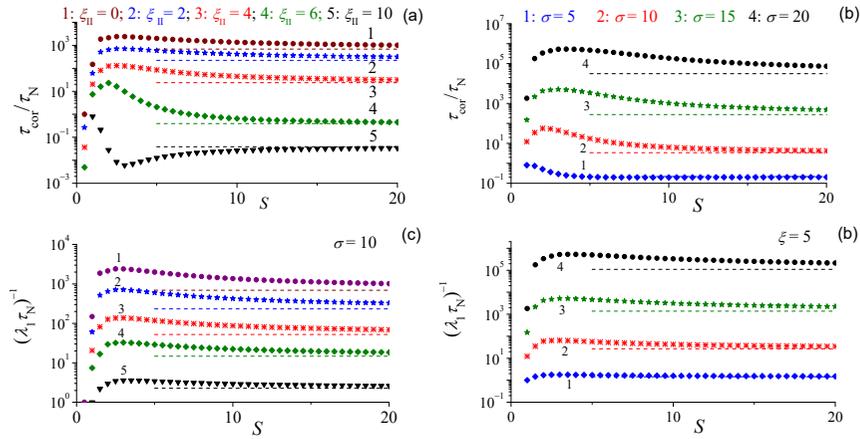

**Figure 21**. (Color on line) Correlation time $\tau_{cor}/\tau_N$ (a), (b) and $(\lambda_1 \tau_N)^{-1}$ (c), (d) vs. the spin number $S$ for various field parameters $\xi_{\|}$ for anisotropy barrier parameter $\sigma = 10$ and for different $\sigma$ and field parameters $\xi_{\|} = 5$. Dashed lines: classical limit.

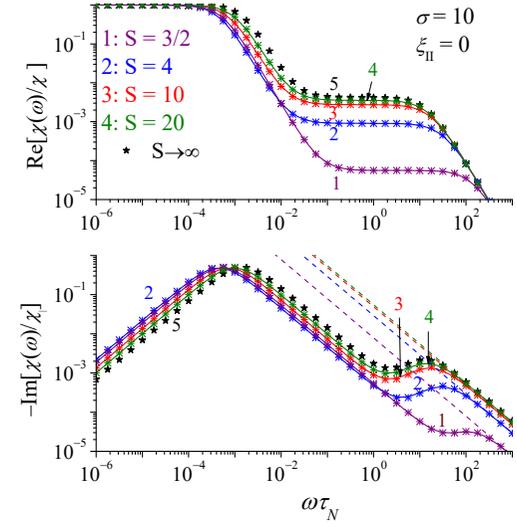
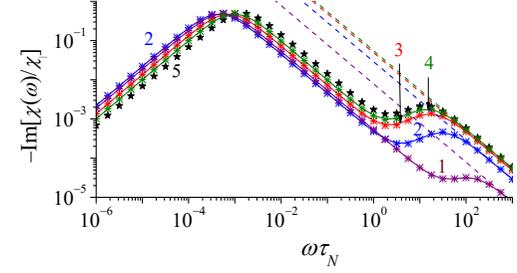

**Figure 22**. (Color on line) Normalized susceptibility $\chi(\omega)/\chi$, Eq. (489), vs. $\omega \tau_N$ for barrier parameter $\sigma = 10$, the uniform field parameter $\xi_{\|} = 0$ (*symmetrical wells*), and various spin numbers $S$. Asterisks: the two-mode approximation, Eq. (553). Straight dashed lines: the high-frequency asymptote, Eqs. (491) and (533). Stars: the classical limit, Eqs. (491) and (550).

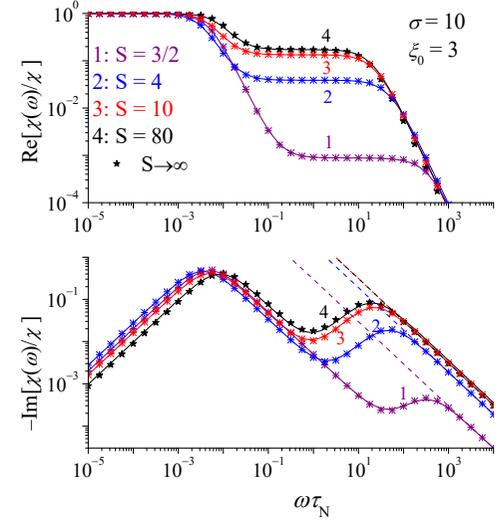
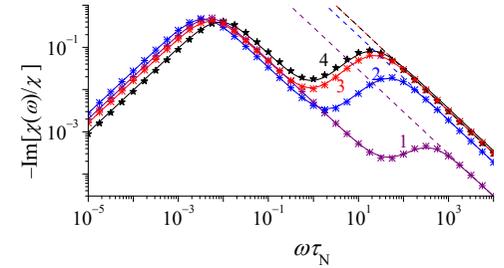

**Figure 23**. (Color on line) The normalized susceptibility $\chi(\omega)/\chi$, Eq. (489), vs normalized frequency $\omega\tau_N$ for barrier parameter $\sigma = 10$, the uniform field parameter $\xi_\parallel = 3$, and various spin numbers *S*. Asterisks: the two-mode approximation, Eq. (553). Straight dashed lines: the high-frequency asymptote as rendered by Eq. (491) and the effective relaxation time Eq. (533). Stars: the classical limit as rendered by Eq. (491) and the effective relaxation time Eq. (550).

In Figs. 22 and 23, we show the real and imaginary parts of the dynamic susceptibility $\chi(\omega)/\chi$. Clearly, like the classical case, two bands appear in the spectrum of the imaginary part $-\text{Im}[\chi(\omega)]$. The low-frequency band is due to the slowest "overbarrier" relaxation mode and can be described by a *single* Lorentzian. The characteristic frequency and the half-width of this band are determined by $\lambda_1$. The high-frequency band of $-\text{Im}[\chi(\omega)]$ is due to high-frequency well modes corresponding to the near degenerate eigenvalues $\lambda_k$ ($k \geq 2$). These individual "intrawell" modes are again indistinguishable in the spectrum of $-\text{Im}[\chi(\omega)]$ appearing merely as a single high-frequency Lorentzian band. Thus, like in the classical case [5], we may describe the behavior of $\chi(\omega)$ via the two-mode approximation [5], i.e., by supposing that it is given as a sum of two Lorentzians, viz.,

$$\frac{\chi(\omega)}{\chi} \approx \frac{1-\delta}{1+i\omega\tau} + \frac{\delta}{1+i\omega\tau_W}, \qquad (553)$$

where $\tau_W$ is a characteristic relaxation time of the near degenerate high-frequency well modes and $\delta$ in the present context is a parameter characterizing the contribution of these high-frequency modes to the susceptibility defined as [155]

$$\delta = \frac{\frac{\tau_{cor}}{\tau} + \frac{\tau}{\tau_{ef}} - \frac{\tau_{cor}}{\tau_{ef}} - 1}{\frac{\tau_{cor}}{\tau} + \frac{\tau}{\tau_{ef}} - 2}, \quad \tau_W = \frac{\tau_{cor} - \tau}{1 - \frac{\tau}{\tau_{ef}}} \qquad (554)$$

and $\tau_{cor}$, $\tau_{ef}$, and $\tau \approx \lambda_1^{-1}$ are given by the quantum expressions Eqs. (531), (533), and (546), respectively. The parameters $\delta$ and $\tau_W$ in Eqs. (553) and (554) have been determined by imposing the condition that the approximate two mode Eq. (553) must obey the *exact* asymptotic Eqs. (490) and (491). Now the longest relaxation time $\tau$ must be related to the frequency $\omega_{max}$ of the low frequency peak in the magnetic loss spectrum $-\text{Im}[\chi(\omega)]$, where it attains a maximum, and/or the half-width $\Delta\omega$ of the spectrum of the real part of the susceptibility $\text{Re}[\chi(\omega)]$ via

$$\tau \approx \omega_{max}^{-1} \approx \Delta\omega^{-1}. \qquad (555)$$

In the low frequency region ($\omega\tau \leq 1$), where the effect of the high frequency modes may be ignored, $\chi(\omega)$ may be approximated as

$$\frac{\chi(\omega)}{\chi} \approx 1 - \frac{i\omega\tau_{int}}{1+i\omega\tau}. \qquad (556)$$

We remark that Garcia-Palacios and Zueco [82] have shown that the two-mode approximation which was originally developed for classical systems [5,155] accurately describes the linear response of quantum uniaxial nanomagnets at all frequencies of interest.

In order to illustrate the accuracy of the two mode approximation for the quantum behavior, we plot in Figs. 22 and 23 the real and imaginary parts of $\chi(\omega)/\chi$ as calculated from the matrix solution, Eq. (489), representing a finite sum of Lorentzians and the approximate two mode Eq. (553) for zero dc field, $\xi_\parallel = 0$ (symmetrical wells) and for nonzero dc field, $\xi_\parallel = 3$ (asymmetrical wells). It is apparent from Figs. 22 and 23 that at low frequencies no *practical difference exists between the numerical solution and the two-mode approximation* (the maximum relative deviation between the corresponding curves does not exceed a few percent). In the classical limit, $S \to \infty$, the axially symmetric Hamiltonian Eqs. (370) or (557) below correspond to a free energy *V* given by Eq. (522). Here both $\tau_{cor}$ and $\tau_{ef}$ can be expressed in closed form, viz. Eqs. (552) and (550). The classical limit is also shown in Figs. 22 and 23 for comparison.

We have studied here the transient nonlinear longitudinal relaxation of a quantum uniaxial nanomagnet of arbitrary spin *S* in the high temperature and weak spin-bath coupling limit. The principal result is that one may once again determine the transition from quantum elementary spin relaxation to the classical superparamagnetic relaxation pertaining to a giant classical spin as a function of the spin number *S*. Furthermore, one may accurately estimate the value of *S* (typically in the range 20-40) wherein the crossover to classical superparamagnetic behavior takes place. Thus, one may assign a range of validity as a function of the spin number *S* to the classical Néel-Brown treatment of magnetic nanoparticles with the simplest uniaxial magnetocrystalline anisotropy and Zeeman energy given above. The exact continued fraction solution based on the diagonal elements of the density matrix yields in closed form the dependence of the longitudinal spin relaxation function on the spin number *S*, which is dominated by a single exponential with time constant the longest relaxation time $1/\lambda_1$. Thus, a *simple description of the long time behavior of the longitudinal relaxation function as a Bloch equation* (462) *again holds for the nonlinear response of a quantum nanomagnet for arbitrary spin S*. In linear response, the approach so developed reproduces (with some modifications due to the symmetrized collision kernel used) the results previously obtained by Garanin [80] and Garcia-Palacios and Zueco [81,82].

We have treated the longitudinal relaxation in two superficially distinct ways, viz., the phase space formalism embodied in the Wigner-Stratonovich bijective mapping and the density matrix. *The high temperature and weak spin-bath coupling limit being understood in each case.* Thus, we have provided a vital check on the validity of both methods by explicitly demonstrating their equivalence. Again we emphasize that a very useful feature of the phase space representation is that existing powerful computational techniques for the Fokker-Planck equation may be extended to the quantum domain which also suggest new closed form quantum results via corresponding classical ones. For example, the integral and effective relaxation times, Eqs. (521) and (547), are clearly quantum analogs of the corresponding classical expressions Eqs. (523) and (550).

Next, we shall apply our methods to the nonlinear ac stationary responses of quantum nanomagnets by generalizing the known solutions for classical spins driven by a strong ac field to treat quantum effects in their ac nonlinear response ([5], Ch. 9). The investigation is prompted by the fact that it has been shown experimentally (e.g., [163]) for the molecular magnet $Mn_{12}$ with $S = 10$ that the nonlinear susceptibility of quantum nanomagnets differs from that of classical spins.

*3. Nonlinear longitudinal relaxation in superimposed ac and dc magnetic fields*

Now the spin reversal process in quantum systems with finite spin number $S$ has a strong field dependence causing nonlinear effects in the dynamic susceptibility [68], stochastic resonance [69], etc. In general, the nonlinear response to an external field poses an extremely difficult problem because the response now always depends on the precise nature of the stimulus [5]. Thus, no *unique* response function valid for all stimuli exists unlike in linear response. These difficulties are compounded in quantum systems so that the literature available on them is relatively sparse. The nonlinear longitudinal relaxation of a quantum nanomagnet arising from a sudden change in the magnitude of a strong external dc field was treated in Sec. III.C.2 using an evolution equation for the reduced density matrix. The solution of the evolution equation was then written as a finite series of the polarization operators, where the coefficients of the series (statistical averages of the polarization operators) were found from differential recurrence relations. Moreover, it was shown that the matrix solution simplifies for axially symmetric Hamiltonians Eq. (557) because the *diagonal terms of the density matrix decouple from the non-diagonal ones so that only the former partake in the time evolution*. We now show how this technique is also applied to the the nonlinear dynamic magnetic susceptibility of a quantum nanomagnet with arbitrary $S$ in superimposed ac and dc magnetic uniform fields amounting to the calculation of the nonlinear ac stationary response of the nanomagnet to an arbitrary ac field in the presence of the thermal agitation. Now calculations of the *nonlinear* ac response of quantum uniaxial nanomagnets have hitherto been made via perturbation theory (e.g., Refs. [163]) by supposing that the potential energy of a spin in external magnetic fields is less than the thermal energy so that a small parameter exists. In the response to an ac field of *arbitrary strength*, however, such small parameters do not exist. The approach we shall use is, in some respects, analogous to that used in Ref. [164] for nonlinear dielectric relaxation behavior of polar molecules in a strong ac electric field and in Refs. [165,166] for the nonlinear magnetization relaxation of magnetic nanoparticles in superimposed ac and dc magnetic fields. The difference as usual is that for a finite spin number $S$ the solution of the evolution equation is rendered as a finite sum of spherical harmonics in contrast to the classical case, where the solution of the evolution equation involves an infinite sum of them. We shall, in particular, demonstrate that our quantum results in the classical limit, $S \to \infty$, correspond with those of Ref. [165]. Moreover, for small values of the ac applied fields (linear response) they agree with the results of Sec. III.C.2 calculated via the switch-off of a small longitudinal uniform field.

As an explicit example, we consider a uniaxial nanomagnet of arbitrary spin number $S$ subjected to both a *uniform* external magnetic field $\mathbf{H}_0$ and to an ac external field $\mathbf{H}(t)$ applied along the $Z$ axis, i.e., the axis of symmetry. The Hamiltonian $\hat{H}_S$ has the axially symmetric form [cf. Eq. (370)]

$$\beta \hat{H}_S(t) = -\frac{\sigma}{S^2} \hat{S}_Z^2 - \frac{\xi_0 + \xi \cos \omega t}{S} \hat{S}_Z, \quad (557)$$

where $\sigma$ is again the dimensionless anisotropy parameter, $\xi_0 = \beta S \hbar \gamma H_0$ and $\xi = \beta S \hbar \gamma H$ are the dc bias and ac field parameters, respectively. This Hamiltonian as before comprises a uniaxial anisotropy term plus the Zeeman term, representing as usual a generic model for quantum spin relaxation phenomena in molecular magnets, nanoclusters, etc.

The density matrix evolution equation describing the longitudinal relaxation of a uniaxial nanomagnet with the Hamiltonian defined by Eq. (557) is similar to Eq. (463) and is given by

$$\frac{\partial \hat{\rho}_S}{\partial t} = -\frac{1}{\tau_N} \left\{ e^{-\frac{\sigma}{2S^2} - \frac{\xi_0 + \xi \cos \omega t}{2S}} \left[ \hat{S}_{-1} e^{-\frac{\sigma}{S^2} \hat{S}_0} \hat{\rho}_S, \hat{S}_{+1} \right] + e^{\frac{\sigma}{2S^2} + \frac{\xi_0 + \xi \cos \omega t}{2S}} \left[ \hat{S}_{+1} e^{\frac{\sigma}{S^2} \hat{S}_0} \hat{\rho}_S, \hat{S}_{-1} \right] \right\}. \quad (558)$$

For our purposes use of the symmetrized form of collision kernel Eq. (558) is very significant as it allows a correct description of the harmonics of spectral moments in the nonlinear response when an ac stimulus is imposed, namely the *absence* of the even harmonics for symmetric double well potentials. Here the magnitude of the ac field $\xi$ is supposed to be so large that the energy of a spin is either comparable to or higher than the thermal energy $kT$, i.e., $\xi \geq 1$, so that one is always faced with an intrinsically nonlinear problem which is solved as follows.

We recall that for axially symmetric Hamiltonians such as that given by Eq. (557), the transformation of the evolution equation (558) for the density matrix $\hat{\rho}_S$ into differential-recurrence equations for its individual matrix elements can be radically simplified because the diagonal entries of $\hat{\rho}_S$ decouple from the non-diagonal ones. Hence, only the former contribute to the longitudinal spin relaxation. As before, we have from Eq. (558) the following three-term differential-recurrence equation for the *diagonal* entries $\rho_m = \rho_{mm}$ [cf. Eqs. (496)-(498)]

$$\tau_N \frac{d\rho_m(t)}{dt} = q_m^-(t)\rho_{m-1}(t) + q_m(t)\rho_m(t) + q_m^+(t)\rho_{m+1}(t), \quad (559)$$

where $m = -S, -S+1, ..., S$, $\tau_N = (2D_\perp)^{-1}$ is the characteristic (free diffusion) time and now the time dependent coefficients are

$$q_m(t) = -a_m^- e^{-(2m-1)\frac{\sigma}{2S^2} - \frac{\xi_0 + \xi \cos\omega t}{2S}} - a_m^+ e^{(2m+1)\frac{\sigma}{2S^2} + \frac{\xi_0 + \xi \cos\omega t}{2S}}, \quad (560)$$

$$q_m^\pm(t) = a_m^\pm e^{\mp(2m\pm1)\frac{\sigma}{2S^2} \mp \frac{\xi_0 + \xi \cos\omega t}{2S}}, \quad (561)$$

$$a_m^\pm = (S \mp m)(S \pm m + 1)/2. \quad (562)$$

Since we are solely concerned with the ac response corresponding to the stationary state, which is independent of the initial conditions, in calculating the longitudinal component of the magnetization defined as

$$\langle \hat{S}_Z \rangle (t) = \sum_{m=-S}^{S} m\rho_m(t), \quad (563)$$

we may seek the diagonal elements $\rho_m(t)$ as the Fourier series, viz.,

$$\rho_m(t) = \sum_{k=-\infty}^{\infty} \rho_m^k(\omega) e^{ik\omega t}. \quad (564)$$

According to Eqs. (563) and (564), $\langle \hat{S}_Z \rangle (t)$ is then automatically rendered as a Fourier series, viz.,

$$\langle \hat{S}_Z \rangle (t) = \sum_{k=-\infty}^{\infty} S_Z^k(\omega) e^{ik\omega t}, \quad (565)$$

where the amplitudes $S_Z^k(\omega)$ are themselves given by the finite series

$$S_Z^k(\omega) = \sum_{m=-S}^{S} m\rho_m^k(\omega). \quad (566)$$

Next, the time dependent coefficients $q_m(t)$ and $q_m^\pm(t)$ in Eqs. (560) and (561) can also be expanded into the Fourier series using the known Fourier-Bessel expansion [105]

$$e^{\pm\frac{\xi}{2S}\cos\omega t} = \sum_{k=-\infty}^{\infty} I_k\left(\pm\frac{\xi}{2S}\right) e^{ik\omega t}, \quad (567)$$

where $I_k(z)$ are the modified Bessel functions of the first kind [105]. Thus by direct substitution of Eq. (564) and the Fourier series for $q_m(t)$ and $q_m^\pm(t)$ into Eq. (559), we obtain a recurrence relation between the Fourier coefficients $\rho_m^k(\omega)$, viz.,

$$ik\omega\tau_N \rho_m^k(\omega) = \sum_{k'=-\infty}^{\infty} \left\{ a_m^- e^{\frac{\sigma(2m-1)}{2S^2} + \frac{\xi_0}{2S}} I_{k-k'}\left(\frac{\xi}{2S}\right) \rho_{m-1}^{k'}(\omega) + a_m^+ e^{-\frac{\sigma(2m+1)}{2S^2} - \frac{\xi_0}{2S}} I_{k-k'}\left(-\frac{\xi}{2S}\right) \rho_{m+1}^{k'}(\omega) \right. \\ \left. - \left[ a_m^- e^{-\frac{\sigma(2m-1)}{2S^2} - \frac{\xi_0}{2S}} I_{k-k'}\left(-\frac{\xi}{2S}\right) + a_m^+ e^{\frac{\sigma(2m+1)}{2S^2} + \frac{\xi_0}{2S}} I_{k-k'}\left(\frac{\xi}{2S}\right) \right] \rho_m^{k'}(\omega) \right\}. \quad (568)$$

The recurrence relation Eq. (568) can be solved exactly for $\rho_m^k(\omega)$ via matrix continued fractions as follows. By introducing the column vector

$$\boldsymbol{\rho}_n = \begin{pmatrix} \vdots \\ \rho_n^{-1}(\omega) \\ \rho_n^0(\omega) \\ \rho_n^1(\omega) \\ \vdots \end{pmatrix}, \quad (569)$$

$(n = m + S)$, we have the following matrix recurrence equation between the $\boldsymbol{\rho}_n$, namely,

$$\mathbf{Q}_n^- \boldsymbol{\rho}_{n-1} + \mathbf{Q}_n \boldsymbol{\rho}_n + \mathbf{Q}_n^+ \boldsymbol{\rho}_{n+1} = \mathbf{0}, \quad (570)$$

where the matrix elements of the infinite matrixes $\mathbf{Q}_n$ and $\mathbf{Q}_n^\pm$ are given by

$$[\mathbf{Q}_n]_{k,k'} = -i\omega\tau_N k \delta_{kk'} - a_n^+ e^{(2n-2S+1)\frac{\sigma}{2S^2} + \frac{\xi_0}{S}} I_{k-k'}\left(\frac{\xi}{2S}\right) - a_n^- e^{-(2n-2S-1)\frac{\sigma}{2S^2} - \frac{\xi_0}{2S}} I_{k-k'}\left(-\frac{\xi}{2S}\right),$$

$$[\mathbf{Q}_n^\pm]_{k,k'} = a_n^\pm e^{\mp(2n-2S\pm1)\frac{\sigma}{2S^2} \mp \frac{\xi_0}{2S}} I_{k-k'}\left(\mp\frac{\xi}{2S}\right).$$

Now according to the general method of solution of three-term recurrence relations [5,71], all higher order column vectors $\boldsymbol{\rho}_n$ defined by Eq. (569) can be expressed in terms of the *lowest order* column vector $\boldsymbol{\rho}_0$ as

$$\boldsymbol{\rho}_n = \mathbf{S}_n \mathbf{S}_{n-1} \ldots \mathbf{S}_1 \boldsymbol{\rho}_0, \quad (571)$$

where the $\mathbf{S}_m$ are finite matrix continued fractions defined by the matrix recurrence relation

$$\mathbf{S}_m = \left[-\mathbf{Q}_m + \mathbf{Q}_m^+ \mathbf{S}_{m+1}\right]^{-1} \mathbf{Q}_m^-. \quad (572)$$

Now the zero order column vector $\boldsymbol{\rho}_0$ itself can be found from the normalization condition for the density matrix elements, viz.,

$$\sum_{n=0}^{2S} \rho_n(t) = \sum_{k=-\infty}^{\infty} \sum_{n=0}^{2S} \rho_n^k(\omega) e^{i\omega k t} = 1 \quad (573)$$

thereby immediately yielding the inhomogeneous equation for $\boldsymbol{\rho}_0$, viz.,

$$\sum_{n=0}^{2S}\boldsymbol{\rho}_n = \left(\mathbf{I}+\mathbf{S}_1+\mathbf{S}_2\mathbf{S}_1+...+\mathbf{S}_{2S}...\mathbf{S}_2\mathbf{S}_1\right)\boldsymbol{\rho}_0 = \mathbf{v}, \qquad (574)$$

where $\mathbf{I}$ is the unit matrix and the infinite column vector $\mathbf{v}$ has only the single nonvanishing element $v_k = \delta_{k0}$, $-\infty < k < \infty$. Consequently, we have

$$\boldsymbol{\rho}_0 = \left(\mathbf{I}+\mathbf{S}_1+\mathbf{S}_2\mathbf{S}_1+...+\mathbf{S}_{2S}...\mathbf{S}_2\mathbf{S}_1\right)^{-1}\mathbf{v}. \qquad (575)$$

Having calculated the zero-order column vector $\boldsymbol{\rho}_0$, we can then determine via Eq. (571) all the other column vectors $\boldsymbol{\rho}_n$ and thus we can evaluate all the $S_Z^k(\omega)$ from Eq. (566) yielding the nonlinear stationary ac response of a uniaxial nanomagnet.

Initially, we treat the *frequency-dependent* fundamental component of the magnetization $S_Z^1(\omega)$ in Eq. (565). For a weak ac field, $\xi \to 0$, the normalized fundamental component $S_Z^1(\omega)/S_Z^1(0)$ yields the normalized linear dynamic susceptibility, viz.,

$$\frac{S_Z^1(\omega)}{S_Z^1(0)} \to \frac{\chi(\omega)}{\chi}, \qquad (576)$$

concurring with the linear response solution given in Sec. III.C.3. In strong ac fields, $\xi > 1$, like in linear response, two distinct absorption bands again appear in the spectrum of $-\mathrm{Im}[S_Z^1(\omega)/S_Z^1(0)]$ so that two accompanying dispersion regions occur in the spectrum of $\mathrm{Re}[S_Z^1(\omega)/S_Z^1(0)]$. However, due to the pronounced nonlinear effects, the low-frequency parts of $\mathrm{Re}[S_Z^1(\omega)/S_Z^1(0)]$ and $-\mathrm{Im}[S_Z^1(\omega)/S_Z^1(0)]$ may no longer be approximated by a *single* Lorentzian (see Fig. 24). Nevertheless, the frequency $\omega_{\max}$ of the maximum loss and/or the half-width $\Delta\omega$ of the spectrum of $\mathrm{Re}[S_Z^1(\omega)/S_Z^1(0)]$ may still be used to estimate an effective reversal time $\tau$ as defined in Eq. (555). The behavior of the low-frequency peak of $-\mathrm{Im}[S_Z^1(\omega)/S_Z^1(0)]$ as *a function of the ac field amplitude* crucially depends on whether or not a dc field is applied. For strong dc bias, $\xi_0 > 1$ (see Fig. 24), the low-frequency peak shifts to lower frequencies reaching a maximum at $\xi \sim \xi_0$ thereafter shifting to higher frequencies with increasing $\xi_0$. In other words, as the dc field increases, the reversal time of the spin *initially increases* and having attained its maximum at some critical value $\xi \sim \xi_0$ thereafter decreases. For small dc bias, $\xi_0 < 0.5$, the low-frequency peak shifts monotonically to higher frequencies with increasing $\xi$. This behavior is very similar to that that observed in the classical case [165,166].

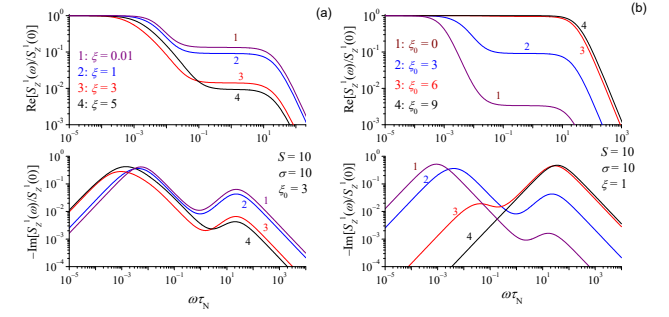

**Figure 24**. (Color on line) The real and imaginary parts of the normalized fundamental component $S_Z^1(\omega)/S_Z^1(0)$ vs. normalized frequency $\omega\tau_N$ (a) for various applied ac fields $\xi$ and the uniform field parameter $\xi_0 = 3$ and (b) for various dc field parameters $\xi_0$ and $\xi = 1$; the spin number $S = 10$ and anisotropy parameter $\sigma = 10$.

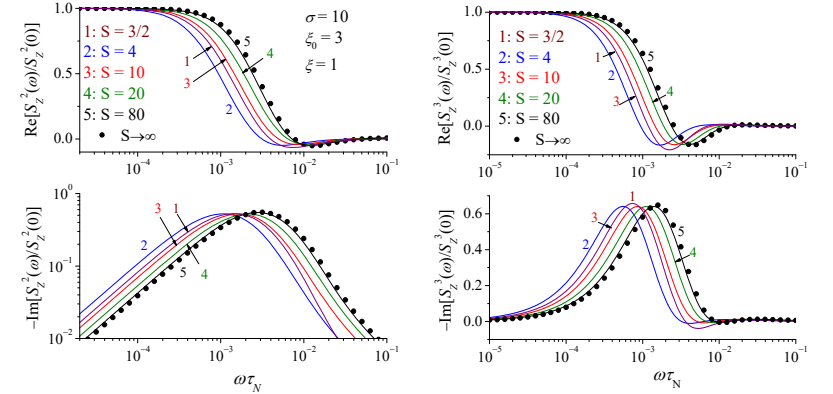

**Figure 25**. (Color on line) The real and imaginary parts of the normalized second and third harmonic components $S_Z^2(\omega)/S_Z^2(0)$ and $S_Z^3(\omega)/S_Z^3(0)$ of the nonlinear response vs. normalized frequency $\omega\tau_N$ for barrier parameter $\sigma = 10$, the uniform field parameter $\xi_0 = 3$ (*nonsymmetrical wells*), the ac field parameter $\xi = 1$ and various spin numbers $S$. Filled circles: the classical limit.

Now a striking feature of the nonlinear response is that the effective reversal time may also be evaluated from either the spectrum of the (now) frequency dependent dc component $S_Z^0(\omega)$ (only for nonzero dc bias, $\xi_0 \neq 0$) or those of the higher order harmonics $S_Z^k(\omega)$ with $k > 1$ because the low-frequency parts of these spectra are themselves, like the spectra of the fundamental, dominated by overbarrier relaxation processes. For illustration, the real and imaginary parts of the normalized second- and third-harmonic components $S_Z^2(\omega)/S_Z^2(0)$ and $S_Z^3(\omega)/S_Z^3(0)$ are shown in Fig. 25. Like the fundamental the behavior of both $-\mathrm{Im}\left[S_Z^2(\omega)/S_Z^2(0)\right]$ and $-\mathrm{Im}\left[S_Z^3(\omega)/S_Z^3(0)\right]$ depends on whether or not a dc field is applied. For *weak* dc bias $\xi_0 < 0.5$, the low-frequency peak shifts monotonically to higher frequencies. For *strong* dc bias, $\xi_0 > 1$, on the other hand the low-frequency peak shifts to lower frequencies reaching a maximum at $\xi \sim \xi_0$ thereafter decreasing with increasing $\xi$.

Thus in the nonlinear relaxation of a uniaxial quantum nanomagnet with arbitrary spin number *S* subjected to superimposed ac and dc magnetic fields in the high temperature and weak spin-bath coupling limit, we may determine once again the transition from elementary spin relaxation to that pertaining to a giant spin as a function of the spin number *S*. Here only uniaxial nanomagnets have been treated. Those with nonaxially symmetric anisotropies (cubic, biaxial, etc.) can be considered in like manner but with considerably more mathematical manipulation.

*4. Dynamic magnetic hysteresis*

We recall that nanoparticle magnetism has many novel applications, particularly in the (applied) area of information storage [167] and in medicine, e.g., in hyperthermia occasioned by induction heating of nanoparticles [168,169] with the dynamic magnetic hysteresis (DMH) induced in nanomagnets by an external ac field constituting a topic of special interest which we now study in the quantum case. Here the temperature directly influences the remagnetization conditions, strongly affecting the effective rates, so altering the loop shape, coercive force, and specific power loss in nanomagnets. The theory of DMH in single-domain magnetically *isotropic* nanoparticles subjected to thermal fluctuations having been proposed by Ignachenko and Gekht, [170] was later extended to uniaxial superparamagnetic particles with moderate to high internal barriers [171-175]. Our approach as applied to quantum spins is analogous to that of Refs. [174,175] for DMH of single-domain ferromagnetic particles, where perturbation theory cannot be used. For purposes of exposition, we take a uniaxial nanomagnet of arbitrary spin number *S* subjected to uniform external magnetic field **H**$_0$ and ac external field **H**(*t*) applied *along the Z axis*, i.e., the axis of symmetry, with the Hamiltonian $\hat{H}_S$ given by the axially symmetric Eq. (557) above. Again our results will coincide with the classical ones in the limit $S \to \infty$.

The stationary ac response of the longitudinal component $\left\langle \hat{S}_Z \right\rangle(t)$ was treated in the previous Sec. III.C.3 and is given by Eq. (565). Now, the DMH loop represents a parametric plot of the normalized magnetization as a function of the ac field, i.e.,

$m(t) = \left\langle \hat{S}_Z \right\rangle(t)/S$ vs. $h(t) = H(t)/H = \cos\omega t$.

Just as the classical case [172-174], the normalized area of the DMH loop $A_n$ (which is the energy loss per particle and per cycle of the AC field) is given by the exact equation

$$A_n = \frac{1}{4H}\oint m(t)dH(t) = -\frac{\pi}{2S}\mathrm{Im}\left[S_Z^1(\omega)\right]. \qquad (577)$$

In Figs. 26-32, we show the effects of ac and dc bias magnetic fields on the DMH loops in a uniaxial nanomagnet with arbitrary *S*. For a *weak* ac field, $\xi \to 0$, and low frequencies, $\omega\tau \leq 1$, the DMH loops are ellipses with normalized area $A_n$ given by Eq. (577); the behavior of $A_n \sim -\mathrm{Im}\left[S_Z^1(\omega)\right] \sim \chi''(\omega)$ being similar [cf. Eq. (577)] to that of $\chi''(\omega)$ (see Figs. 22 and 23). Indeed, the two-mode approximation for the susceptibility given by Eq. (553) implies that the overall relaxation process consists of two distinct entities, namely, the slow thermally activated overbarrier (or interwell) process and the fast intrawell relaxation in the wells. Now, at low frequencies and for large barriers between the wells, only the first term on the right side in Eq. (556) for $-\mathrm{Im}\left[S_Z^1(\omega)\right]$ need be considered. Furthermore, for *weak* dc bias fields, $\xi_0/(2\sigma) \ll 1$, the approximation $\delta \approx 1$ may always be used in Eq. (553) so that the *normalized* magnetization $m(t) = \left\langle \hat{S}_Z \right\rangle(t)/S$ is given by the simple linear response formula [174]

$$m(t) = \frac{1}{S}\left\langle \hat{S}_Z \right\rangle_0 + \frac{\chi\xi}{S}\frac{\cos\omega t + \omega\tau\sin\omega t}{1+\omega^2\tau^2}. \qquad (578)$$

If we introduce the variables *x* and *y* defined as

$$x = h(t) = \cos\omega t \text{ and } y = \frac{Sm(t) - \left\langle \hat{S}_Z \right\rangle_0}{\chi\xi},$$

we then can conclude from Eq. (578) that in the linear response approximation, a low frequency DMH loop in the $(x, y)$ plane is an ellipse, namely, [174]

$$x^2 + \frac{1}{\omega^2\tau^2}\left[\left(1+\omega^2\tau^2\right)y - x\right]^2 = 1. \qquad (579)$$

This ellipse is centered at $(0,0)$ and its major axis is inclined to the *x*-axis at an angle $\frac{1}{2}\arctan\left[2/(\omega\tau)^2\right]$.

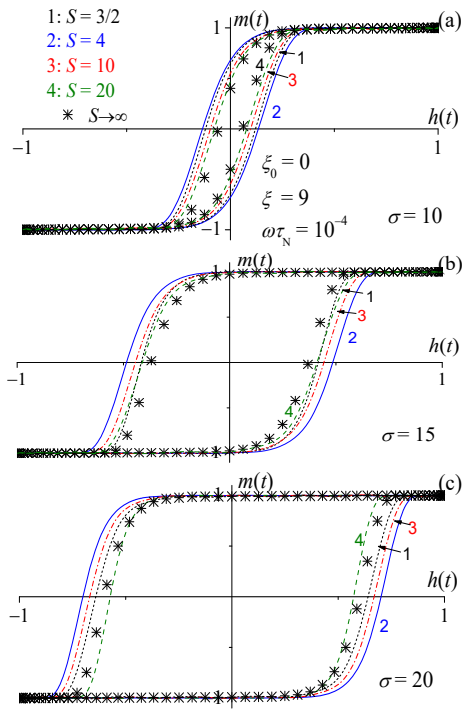

**Figure 26.** (Color on line) DMH loops for various anisotropy parameters $\sigma = 10$ (a), 15 (b), 20 (c) and various spin numbers $S = 3/2$ (1: short-dashed lines), 4 (2: solid lines), 10 (3: dashed-dotted lines), 20 (4: dashed lines), and $\infty$ (asterisks) at $\omega\tau_N = 10^{-4}$, $\xi_0 = 0$, and $\xi = 9$.

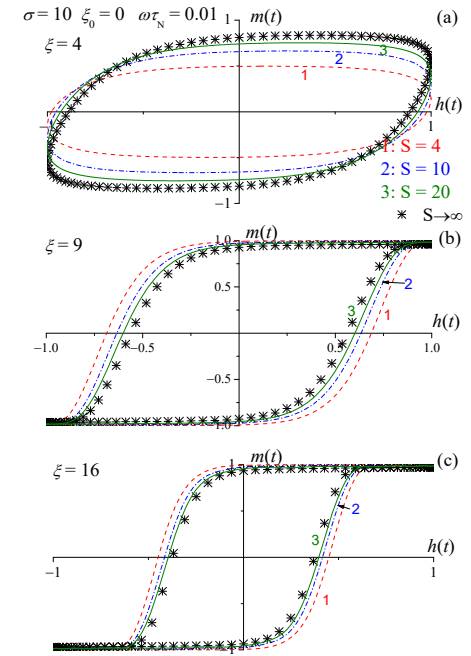

**Figure 27.** (Color on line) DMH loops for various ac external field parameters $\xi = 4$ (a), 9 (b), 16 (c) and various spin numbers $S = 4$ (1: dashed lines), 10 (2: dashed-dotted lines), 20 (3: solid lines), and $\infty$ (asterisks) at $\sigma = 10$, $\xi_0 = 0$, and $\omega\tau_N = 10^{-2}$.

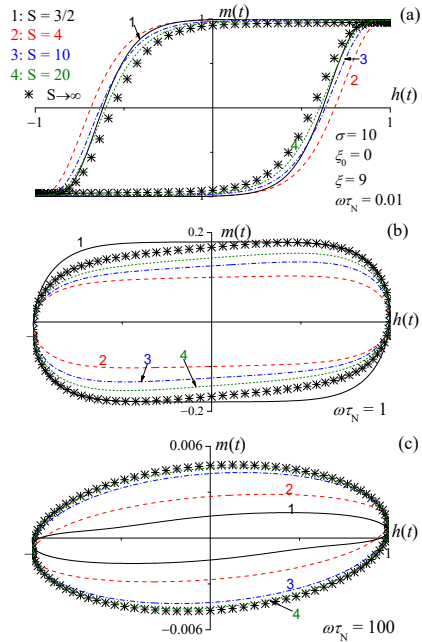

**Figure 28.** (Color on line) DMH loops for various frequencies $\omega\tau_N = 10^{-2}$ (a), 1 (b), $10^2$ (c) and various spin numbers $S = 3/2$ (1: solid lines), 4 (2: dashed lines), 10 (3: dashed-dotted lines), 20 (4: short-dashed lines), and $\infty$ (asterisks) at $\sigma = 10$, $\xi_0 = 0$, and $\xi = 9$.

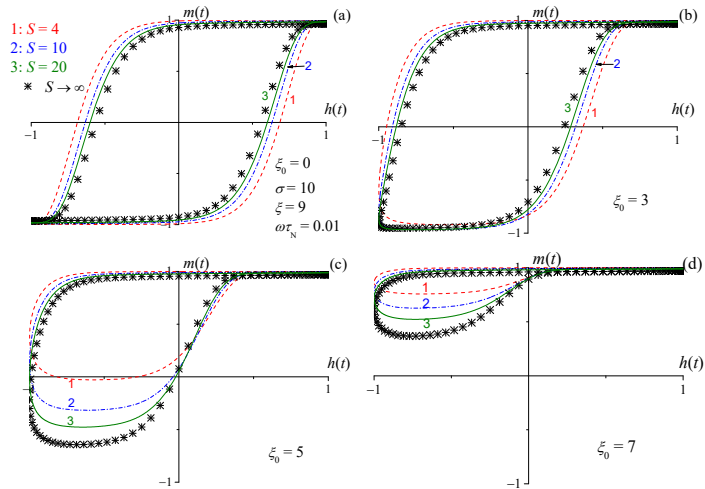

**Figure 29.** (Color on line) DMH loops for various dc bias field parameters $\xi_0 = 0$ (a), 3 (b), 5 (c), 7 (d) and various spin numbers $S = 4$ (1: dashed lines), 10 (2: dashed-dotted lines), 20 (3: solid lines), and $\infty$ (asterisks) at $\xi = 9$, $\sigma = 10$, and $\omega\tau_N = 10^{-2}$.

For *moderate* ac fields, $0.5 < \xi \leq 1$, the DMH loops still have approximately an ellipsoidal shape implying that only a few harmonics actually contribute to the weakly nonlinear response. In contrast, in *strong* ac fields, $\xi > 1$, the shape alters substantially and so the normalized area $A_n$ now exhibiting a pronounced dependence on the frequency $\omega$ and the ac and dc bias field amplitudes $\xi$ and $\xi_0$ as well as on the anisotropy parameter $\sigma$ and the spin number $S$ (see Figs. 26-29). In this regime, the external ac field is able to saturate the paramagnet magnetic moment as well as to induce its inversion (i.e., switching between the directions of the easy axis). In Figs. 26 and 25, we plot the loops for various $S$ and anisotropy ($\sigma$) and ac field ($\xi$) parameters exemplifying how their shapes (and consequently their areas) alter as these parameters vary. Clearly, the remagnetization time is highly sensitive to variations of these parameters. For example, with a strong ac driving field, the Arrhenius dependence of the reversal time on temperature $\log \tau \propto 1/T$, which accurately accounts for the linear response regime, is modified because the strong ac field intervenes so drastically reducing the effective response time of the nanomagnet. Thus, the nonlinear behavior facilitates re-magnetization regimes, which are never attainable with weak ac fields, the reason being that the dc bias component under the appropriate conditions *efficiently tunes* this effect by either *enhancing* or *blocking* the action of the strong ac field. The *pronounced* frequency dependence of the loops is highlighted in Fig. 28 for various $S$. At low frequencies, the field changes are *quasiadiabatic*, so that the magnetization reverses due to the *cooperative* shuttling action of thermal agitation combined with the ac field. The dc bias field effects on the DMH are illustrated in Fig. 29 showing the changes in the DMH caused by varying $\xi_0$ for various spin numbers $S$. In order to understand the effect of the dc bias field on the loop area, one must first recall that the magnetic relaxation time depends on the actual value of the applied field. Under the conditions of Fig. 29, the *positive limiting* (saturation) value of $m(t) \to 1$ corresponds to a total field $H_0 + H$, thus *favoring* the magnetization relaxation to the positive saturation value $m(t) \to 1$. However, for *negative* $h(t)$, the total field $H_0 - H$ is much weaker and so cannot induce relaxation to the negative saturation value $m(t) \to -1$. Therefore, the "center of area" of the loop moves upwards. In the classical limit, $S \to \infty$, our results concur with those for classical uniaxial nanomagnets [175].

The temperature dependence of the DMH is governed by the dimensionless anisotropy (inverse temperature) parameter $\sigma \propto 1/T$. The normalized DMH area $A_n$ as a function of $\sigma^{-1}$ is shown in Fig. 30 for various $S$ showing that the tuning action of the dc bias field described above is effective over a certain temperature interval. This conclusion once again indicates that the relaxation of the magnetization is mostly caused by thermal fluctuations, implying that the magnetic response time still retains a corresponding strong temperature dependence. The normalized area as a function of the frequency $\omega$ and ac field parameter $\xi/(2\sigma)$ is shown in Figs. 31 and 32, respectively. Clearly $A_n$ can invariably be represented as a nonmonotonic curve with a maximum the position of which is determined by $S$ as well as by the other model parameters. The peak in $A_n$ (Fig. 31) is caused by the field-induced modifications of the reversal time as strongly tuned by the dc bias field. As in Fig. 31, variation of the dc field strength shifts the frequency, where the maximum is attained, by several orders of magnitude. The normalized loop area presented in Fig. 32 illustrates the dependence of $A_n$ on the ac field amplitude, which is similar to that of classical nanomagnets.

The DMH in uniaxial nanomagnets has been treated above without any *a priori* assumptions regarding the potential barrier height, temperature, the magnetizing field strength and/or spin number $S$. In general, it appears that given appropriate conditions a small bias dc field (in comparison with the internal anisotropy field) can profoundly affect the shape of the DMH loops in nanomagnets accompanied by a strong dependence on the spin number $S$.

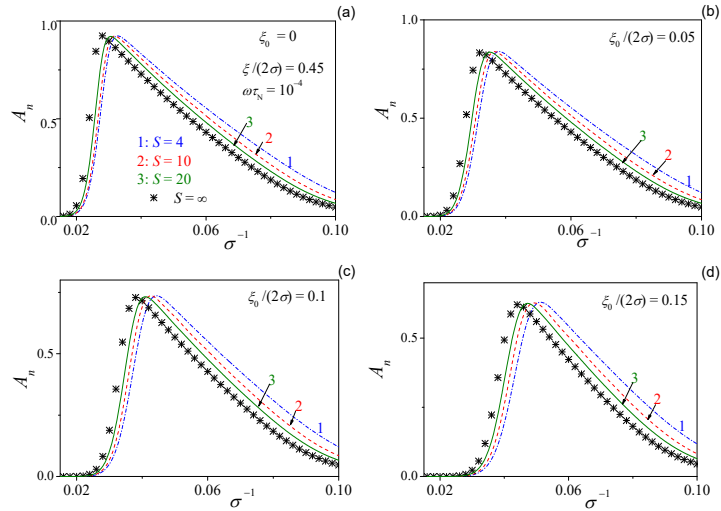

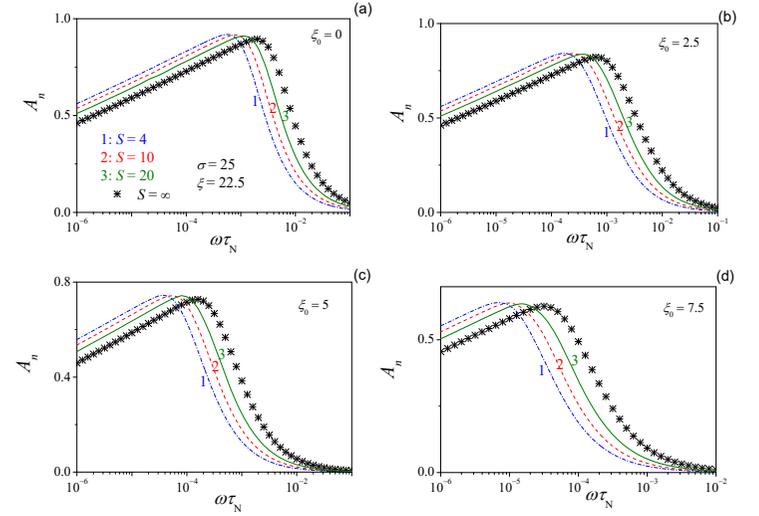

**Figure 31.** (Color on line) Normalized area of the DMH loop $A_n$ vs. the dimensionless frequency $\omega\tau_N$ under variation of the dc bias field $\xi_0 = 0$ (a), 2.5 (b), 5 (c), and 7.5 (d) for various spin numbers $S = 4$ (dashed-dotted lines), 10 (dashed lines), 20 (solid lines), and $\infty$ (asterisks); the anisotropy parameter $\sigma = 25$ and the ac field parameter $h = 0.45$.

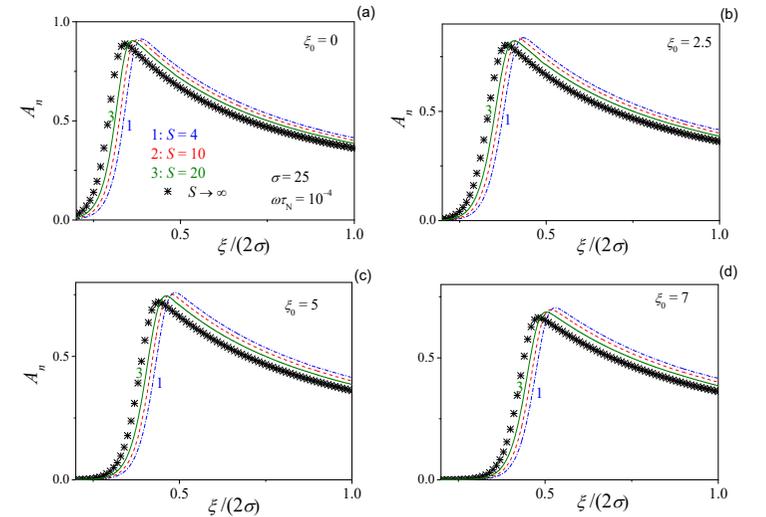

**Figure 30.** (Color on line) Normalized area of the DMH loop $A_n$ vs. the dimensionless temperature $\sigma^{-1}$ under variation of the dc bias field parameter $h_0 = \xi_0/(2\sigma) = 0$ (a), 0.05 (b), 0.1 (c), and 0.15 (d) for various spin numbers $S = 4$ (dashed-dotted lines), 10 (dashed lines), 20 (solid lines), and $\infty$ (asterisks) at the frequency $\omega\tau_N = 10^{-4}$ and the ac field amplitude $\xi/(2\sigma) = 0.45$.

**Figure 32.** (Color on line) Normalized area of the DMH loop $A_n$ as a function of the ac field amplitude $h = \xi/(2\sigma)$ under variation of the dc bias field parameter $\xi_0 = 0$ (a), 2.5 (b), 5 (c), and 7 (d) for various spin numbers $S = 4$ (dashed-dotted lines), 10 (dashed lines), 20 (solid lines), and $\infty$ (asterisks); the anisotropy parameter $\sigma = 25$ and the frequency $\omega\tau_N = 10^{-4}$.

## 5. Quantum effects in stochastic resonance

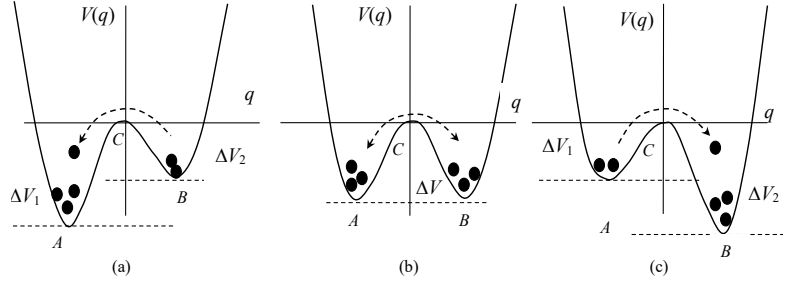

**Figure 33.** Double-well potential as used in stochastic resonance [5]. The minima are located at $A$ and $B$. In the absence of a periodic forcing function (b), the barrier heights $\Delta V_1$ and $\Delta V_2$ are equal to $\Delta V$, so that the potential is symmetric. The periodic forcing function causes the double-well potential to tilt back and forth, thereby raising and lowering the potential barriers of the right and left wells, respectively, in an antisymmetric cyclic fashion; (a) and (c).

Yet another aspect of the Brownian motion of particles and spins in a potential is the *stochastic* resonance [176] (SR) whereby a weak periodic forcing *synchronized* with the thermally activated hopping of particles or spins between the wells greatly enhances the rate of switching between them. The archetypal theoretical model of SR [5,176] is the motion of a heavily damped (so that inertial effects can be ignored) Brownian particle in a bistable potential $V(q)$ subjected to noise arising from a thermal bath (see Fig. 33). If we now apply a weak periodic forcing $f_0 \cos\Omega t$ of frequency $\Omega$, the double-well potential will be tilted up and down, thereby periodically raising and lowering the potential barriers $\Delta V$ [176]. Consequently if $\Omega$ is close to the rate of transitions (escape rates) between the wells despite the fact that the amplitude of the periodic forcing is insufficient to induce the transitions by *itself alone* noise-induced hopping between potential wells may become *synchronized* with it so facilitating the transition. This statistical synchronization takes place when the average waiting (escape) time between two noise-induced transitions is comparable with half the period $\tau_\Omega = 2\pi/\Omega$ of the periodic forcing. Consequently, switching may occur only by the *combined* effect of the regular ac force and the noise. The spectral density $\Phi(\omega)$ of the motion at the forcing frequency $\omega = \Omega$ is then evaluated, and the resulting SNR (or the spectral power amplification coefficient) is analyzed as a function of the noise intensity $D$. Now the curve of SNR versus $D$ has a bell-like shape, i.e., it passes through a maximum thus exhibiting *stochastic resonance*. The maximum in the SNR is interpreted as being due to the remarkable ability of noise to *enhance* the intensity of the interwell hopping's in the system. Stochastic resonance is, therefore, an important effect allowing one to control the behavior of periodic signals passing through noisy systems. As a manifestation of cross-coupling between stochastic and regular motions, the SR effect is universal in physics (e.g. optics, mechanics of solids, superconductivity, surface science), communications engineering (optimal detection and tracing of signals) as well as in various branches of chemistry and biology. Comprehensive reviews of diverse aspects of SR are available in Refs. [176-178].

Now the behavior of magnetic nanosystems (such as magnetic nanoparticles, nanoclusters, and molecular magnets) forced by a weak ac magnetic field is yet another important manifestation of SR. Here the magnetic anisotropy provides the multistable states for the magnetization **M** while the thermal fluctuations or random field due to the bath which is in perpetual thermal equilibrium at temperature $T$ are the source of the noise. These conditions give rise to *magnetic stochastic resonance* which again may be defined as the enhancement of the SNR due to noise [179]. The magnetic SR was first predicted theoretically [180-182] and shortly afterwards observed experimentally [183]. The SNR of the magnetic moment fluctuations is of some interest in information storage and in the crossover between classical and quantum behavior of the magnetization since we saw that single domain particles exhibit essentially classical behavior while smaller entities such as free nanoclusters made of many atoms, molecular clusters, and molecular magnets exhibit pronounced quantum behavior.

The main features of the magnetic SR in single-domain particles (classical spins) [184–188] may be completely understood in terms of the classical Brown (macrospin) model [23,24]. Here each particle behaves like a paramagnetic atom having a magnetic moment $\sim 10^4$ - $10^5$ $\mu_B$, i.e., $S \sim 10^4$ - $10^5$. In the presence of a dc bias field $\mathbf{H}_0$, the normalized magnetic free energy density $V$ of a *uniaxial* nanomagnet is given by the asymmetric bistable potential

$$\beta V(\vartheta) = -\sigma\left(\cos^2\vartheta + 2h\cos\vartheta\right), \qquad (580)$$

where $\sigma = vK/(kT)$ is the dimensionless barrier height parameter, $v$ is the volume of the particle, $K$ is the anisotropy constant, and $h = \mu_0 M_S H_0 /(2K)$ is the bias field parameter ($M_S$ is the saturation magnetization). Without the dc field, the magnetization of the uniaxial particle has two equivalent stable orientations at $\vartheta = 0$ and $\vartheta = \pi$, so that it is an ideal example of a bistable system subjected to noise. Here the reversal of the classical spin is due to thermal activation and the rate of transitions between the potential wells is controlled by the anisotropy or inverse temperature parameter $\sigma$. Thus one may regard $1/\sigma$ as the *dimensionless temperature*, i.e., the *noise intensity*. A dc bias field $\mathbf{H}_0$ when applied to the particle parallel to its anisotropy axis breaks the bidirectional symmetry of the potential. However, an asymmetric two-minima profile of the potential $V(\vartheta)$ survives as long as the bias field parameter $h \leq 1$. Now for $h = 0$, the basic concept of magnetic stochastic resonance has been well described by Raikher and Stepanov [184]. In the presence of noise, a weak alternating spatially uniform field of frequency $\Omega$ favoring the transitions between two equilibrium positions at $\vartheta = 0, \pi$ is applied. Under these conditions the SNR determined from the spectral density $\Phi_M(\omega)$ of the magnetic moment (i.e., the frequency response to the applied field) evaluated at the frequency $\Omega$ of the weak applied ac field, first *increases* with increasing noise strength $kT$, then on attaining a *pronounced maximum*, finally *decreases* again. This is the (magnetic) *stochastic resonance effect*, whereby the periodic response in both amplitude and phase may be manipulated by altering the noise strength.

In contrast to the classical case, in magnetic SR of nanomagnets with *smaller* spin numbers $S \sim 10\text{-}100$ both quantum effects and quantum-classical crossover appear. Here the spin reversal is either due to thermal activation or tunneling or a combination of both. The quantum effects are not the same as those in the SR for translational Brownian motion (see, e.g., [189,190] and references cited therein) because despite some analogies the quantum spin dynamics essentially differ from those of Brownian particles owing to the different symmetries of the rotational and translational groups. Here we shall treat quantum effects in the SR for magnetic spin systems modeled by a uniaxial nanomagnet of arbitrary $S$ in superimposed dc and ac external *uniform* magnetic fields $\mathbf{H}_0 + \mathbf{H}\cos\Omega t$ applied along the $Z$ axis, i.e., the axis of symmetry, so that the time dependent Hamiltonian $\hat{H}_S$ is [cf. Eq. (557)]

$$\beta \hat{H}_S(t) = -\frac{\sigma}{S^2}\hat{S}_Z^2 - \frac{\xi_0 + \xi\cos\Omega t}{S}\hat{S}_Z, \tag{581}$$

where $\sigma$ is the anisotropy constant, $\xi_0$ and $\xi$ are the external dc and ac magnetic field parameters. The longitudinal relaxation of uniaxial nanomagnets interacting with a thermal bath has been treated in Sec. III.C via the respective evolution equations for the reduced density matrix and phase-space distribution function using the methods already available for classical spins. In the large spin limit, we also saw in Sec. III.C that the quantum solutions reduce to those of the Fokker-Planck equation for a classical uniaxial nanomagnet while for linear response and finite $S$, the results agree with those predicted by the solutions of Garanin [80] and García-Palacios and Zueco [81,82]. Here we apply these findings to the SNR for uniaxial quantum nanomagnets.

Now we saw in Sec. III.C that the normalized longitudinal dynamic susceptibility of a quantum nanomagnet is defined in linear response as

$$\frac{\chi(\Omega)}{\chi} = 1 - i\Omega \int_0^\infty C(t)e^{-i\Omega t}dt, \tag{582}$$

where $C(t)$ is the normalized equilibrium correlation function defined by Eq. (484), $\chi = \beta\mu_0\mu^2 N\chi_0$ is the static susceptibility, $N$ is the number of nanomagnets per unit volume,

$$\chi_0 = \frac{1}{S^2}\left[\left\langle\hat{S}_Z^2\right\rangle_0 - \left\langle\hat{S}_Z\right\rangle_0^2\right]$$
$$= \frac{1}{S^2}\left[\sum_{m=-S}^S m^2\rho_m - \left(\sum_{m=-S}^S m\rho_m\right)^2\right] \tag{583}$$

is the normalized static susceptibility,

$$\rho_m = \frac{e^{\sigma(m^2/S^2 + 2hm/S)}}{\sum_{m=-S}^S e^{\sigma(m^2/S^2 + 2hm/S)}} \tag{584}$$

and $h = \xi_0/(2\sigma)$. For a uniaxial nanomagnet, both $C(t)$ and $\chi(\Omega)$ have been calculated in Sec. III.C.3. In particular, we recall that $C(t)$ may formally be written as the finite discrete set of relaxation modes, namely, [cf. Eq. (456)]

$$C(t) = \sum_{k=1}^{2S} c_k e^{-\lambda_k t}, \tag{585}$$

where $\lambda_k$ are the eigenvalues of the system matrix $\mathbf{X}$, Eq. (470), with the replacement $\xi_\| \to \xi_0$. Consequently, Eqs. (582) and (585) allow us to *formally* write $\chi(\Omega)$ as the finite discrete set of Lorentzians [cf. Eq. (489)]

$$\frac{\chi(\Omega)}{\chi} = \sum_{k=1}^{2S} \frac{c_k}{1 + i\Omega/\lambda_k}. \tag{586}$$

The asymptotic behavior of $\chi(\Omega)$ in the extreme cases of very low and very high frequencies is given as before by Eqs. (490) and (491), viz.,

$$\frac{\chi(\Omega)}{\chi} \sim \begin{cases} 1 - i\Omega\tau_{cor} + ..., & \Omega \to 0 \\ -i(\Omega\tau_{ef})^{-1} + ..., & \Omega \to \infty \end{cases}, \tag{587}$$

where $\tau_{cor}$ is the integral relaxation time and $\tau_{ef}$ is the effective relaxation time defined in terms of the eigenvalues $\lambda_k$ as

$$\tau_{cor} = \sum_{k=1}^{2S} c_k/\lambda_k \text{ and } \tau_{ef} = \left(\sum_{k=1}^{2S} c_k\lambda_k\right)^{-1}. \tag{588}$$

Furthermore, the smallest nonvanishing eigenvalue $\lambda_1$ may as usual be associated with the long time behavior of $C(t) \sim e^{t/\tau}$, $t \gg \tau = 1/\lambda_1$, which is characterized by the longest relaxation (or the reversal) time $\tau$. Now in Sec. III.C.3 above, it has been shown that all these times are given by simple analytic formulas, Eqs. (531), (533), and (546), respectively.

Again (see Sec. III.C.3), two distinct bands appear in the spectrum of the imaginary part $\chi''(\Omega)$ of the susceptibility for a uniaxial quantum nanomagnet. As usual the low-frequency band is due to the slowest "interwell" relaxation mode. The characteristic frequency and the half-width of this band are determined by the smallest nonvanishing eigenvalue $\lambda_1$ and as usual the latter is associated with the long time behavior of $C(t) \sim e^{-t/\tau}$, $t >> \tau$, which is dominated by the longest relaxation (or the reversal) time $\tau$. The high-frequency band in $\chi''(\Omega)$ is due to the individual near degenerate high-frequency modes corresponding to the eigenvalues $\lambda_k >> \lambda_1$ ($2S \geq k \geq 2$). Thus, if one is interested solely in the low frequency region ($\Omega\tau \leq 1$), where their effect may be ignored, $\chi(\Omega)$ may be approximated as the single Lorentzian [69] [cf. Eq. (556)]

$$\frac{\chi(\Omega)}{\chi} \approx 1 - \frac{i\Omega\tau_{\text{cor}}}{1+i\Omega\tau}, \tag{589}$$

where $\tau_{\text{cor}}$ and $\tau$ are defined by Eqs. (531) and (545), respectively.

Now magnetic SR may be generally described using linear response theory as follows [5]. The Fourier component $M_\omega$ of the longitudinal components of the magnetic moment is related to that of the applied ac field $H_\omega$ via the complex magnetic susceptibility $\chi(\omega)$ as

$$M_\omega = \chi(\omega)H_\omega. \tag{590}$$

The spectral density $\Phi_M^{(s)}(\Omega)$ of the forced magnetic oscillations in the ac field $H(t) = H\cos\Omega t$ at the excitation frequency $\Omega$ is [176]:

$$\Phi_M^{(s)}(\Omega) = \frac{1}{2}\lim_{\Delta\Omega\to 0}\int_{\Omega-\Delta\Omega}^{\Omega+\Delta\Omega}\left(H\chi_0|\chi(\Omega)|\right)^2\left[\delta(\omega+\Omega)+\delta(\omega-\Omega)\right]d\omega,$$

where the parity condition $\chi^*(\omega) = \chi(-\omega)$ is used. The noise-induced part $\Phi_M^{(n)}(\Omega)$ is obtained using the fluctuation-dissipation theorem as [184]

$$\Phi_M^{(n)}(\Omega) = \frac{\chi''(\Omega)}{\pi\beta\Omega}.$$

Thus on combining the above equations, we have the SNR of the magnetic moment fluctuations, viz.,

$$\text{SNR} = \frac{\Phi_M^{(s)}(\Omega)}{\Phi_M^{(n)}(\Omega)} = \frac{\beta\pi\Omega H^2|\chi(\Omega)|^2}{2\chi''(\Omega)}. \tag{591}$$

The linear response theory result Eq. (591) is very useful on account of its generality because it automatically reduces the calculation of the SNR to that of the dynamic susceptibility, which is a fundamental dynamical characteristic of any relaxing system. By analogy with the SNR for a classical nanomagnet [5,184], Eq. (591) can be written as

$$\text{SNR} = \frac{\pi\beta\chi_0 H^2}{2\tau_N\sigma}R_\Omega, \tag{592}$$

where the dimensionless SNR factor $R_\Omega$ is given by

$$R_\Omega = \frac{\sigma\tau_N\Omega|\chi(\Omega)|^2}{\chi''(\Omega)}. \tag{593}$$

Thus the relevant quantity is $R_\Omega$. In general, $R_\Omega$, besides the obvious dependence on the noise intensity (temperature), the constant (bias) field strength parameter $h$, and the frequency of the exciting field $\Omega$ depends on the spin number $S$. In the adiabatic limit, $\Omega \to 0$, with Eqs. (586), (588), and the correlation time Eq. (531), the SNR factor Eq. (593) simplifies yielding

$$R_0 = \frac{\tau_N\sigma\chi_0}{\tau_{\text{cor}}} = \frac{\frac{\sigma}{2S^2}\left[\sum_{m=-S}^{S}m^2\rho_m - \left(\sum_{m=-S}^{S}m\rho_m\right)^2\right]^2}{\sum_{k=1-S}^{S}\frac{\left(\sum_{m=k}^{S}\left(m-\sum_{n=-S}^{S}n\rho_n\right)\rho_m\right)^2}{[S(S+1)-k(k-1)]\sqrt{\rho_k\rho_{k-1}}}}. \tag{594}$$

On the other hand in the opposite very high-frequency limit, $\Omega \to \infty$, Eq. (593) reduces to

$$R_{\Omega\to\infty} = \frac{\tau_N\sigma\chi_0}{\tau_{\text{ef}}} = \frac{\sigma}{2S^2}\sum_{m=1-S}^{S}[S(S+1)-m(m-1)]\sqrt{\rho_m\rho_{m-1}} \tag{595}$$

with Eqs. (550), (586), and (588). In the classical limit, $S \to \infty$, the normalized $\chi(\Omega)$ is also given by the linear response Eq. (582), where $\chi_0 = \langle\cos^2\vartheta\rangle_0 - \langle\cos\vartheta\rangle_0^2$ and the correlation function $C(t)$ becomes

$$C(t) = \frac{\langle\cos\vartheta(t)\cos\vartheta(0)\rangle_0 - \langle\cos\vartheta\rangle_0^2}{\langle\cos^2\vartheta\rangle_0 - \langle\cos\vartheta\rangle_0^2}. \tag{596}$$

Here

$$\langle\cos\vartheta\rangle_0 = \frac{1}{Z}\int_{-1}^{1}xe^{\sigma(x^2+2hx)}dx$$

and

$$\langle\cos^2\vartheta\rangle_0 = \frac{1}{Z}\int_{-1}^{1}x^2e^{\sigma(x^2+2hx)}dx$$

are given by the analytical Eqs. (524) and (551), respectively. The classical analogs of the quantum Eqs. (594) and (595) are simply

$$R_0 = \frac{\tau_N\sigma}{\tau_{\text{cor}}}\left(\langle\cos^2\vartheta\rangle_0 - \langle\cos\vartheta\rangle_0^2\right), \tag{597}$$

$$R_\infty = \frac{\tau_N\sigma}{\tau_{\text{ef}}}\left(\langle\cos^2\vartheta\rangle_0 - \langle\cos\vartheta\rangle_0^2\right). \tag{598}$$

Here the classical relaxation times $\tau_{\text{ef}}$ and $\tau_{\text{cor}}$ are given by Eqs. (550) and (552).

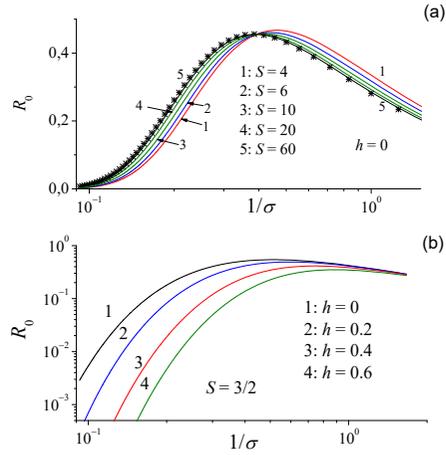

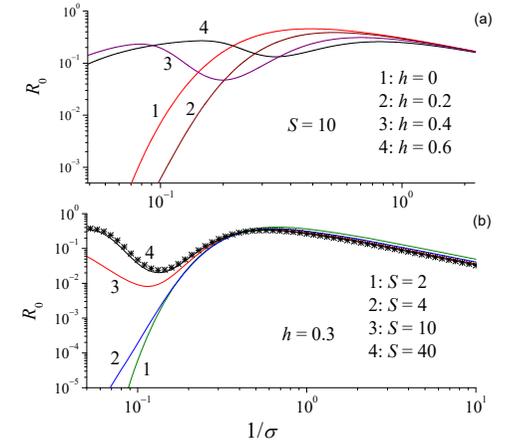

**Figure 34.** (Color on line) Signal-to-noise ratio $R_0$ vs. the dimensionless temperature parameter $\sigma^{-1}$ (a) for various spin numbers $S$ in the absence of the dc bias field ($h = 0$) and (b) for various field parameters $h$ and $S = 3/2$ in the adiabatic limit $\Omega = 0$. Asterisks: classical limit $S \to \infty$.

**Figure 35.** (Color on line) SNR $R_0$ as a function of dimensionless temperature $\sigma^{-1}$ for nonzero values of the applied constant field (a) for various values of $h$ and $S=10$ and (b) for various values of $S$ and $h=0.3$ in the adiabatic limit $\Omega = 0$. Solid lines: exact solution. Asterisks: classical limit $S \to \infty$.

The SNR factor $R_0$ in the adiabatic limit $\Omega = 0$ as a function of the dimensionless temperature parameter $\sigma^{-1}$ is shown in Figs. 34 and 35 for various spin numbers $S$ and field parameter $h$. Usually the maximum of $R_0$ is attained in the range $\sigma^{-1} \sim 0.3\text{-}0.5$ (corresponding to $T \sim 30$ K for the molecular magnet $Mn_{12}$ acetate with $S = 10$). Moreover, that maximum shifts to higher temperatures with increasing $h$ because the bias field radically alters the temperature dependence of the static susceptibility. In a nonzero bias field, the effect of saturation of the magnetization is crucial causing $R_0$ to tend to zero at zero temperature. Although in the low-temperature limit, $\sigma^{-1} \to 0$, $R_0 \to 0$ for *both small $S$ and $h$*; nevertheless, as long as $S$ increases at *finite $h$* or as $h$ increases at *finite $S$*, $R_0 \to$ constant (see Fig. 35). This is due to the temperature dependence of $\tau_{cor}$, which causes the latter to progressively lose its Arrhenius character with increasing $h$. In general, the quantum effects can lead to both amplification and attenuation of the SNR. The frequency dependent SNR $R_\Omega$ vs. the dimensionless temperature parameter $\sigma^{-1}$ is shown in Figs. 36 for various spin numbers $S$ and finite values of the forcing frequency $\Omega$. Clearly this figure exemplifies *the quantum effects via a pronounced deviation of the quantum SNR curves from the corresponding classical ones* (up to several orders of magnitude at low temperatures). Now, the SNR as a function of the dimensionless forcing frequency $\Omega\tau_N$ is also presented in Figs. 37. Here the SNR monotonically increases from its low-frequency limit Eq. (594) to its plateau value given by Eq. (595).

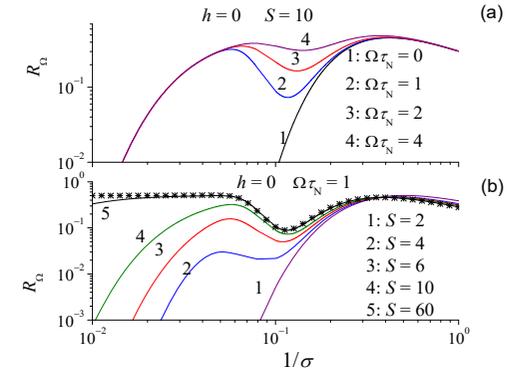

**Figure 36.** (Color on line) SNR $R_\Omega$ vs. the dimensionless temperature $\sigma^{-1}$ (a) for various normalized frequencies $\Omega\tau_N$ and $S = 10$ and (b) for various $S$ and $\Omega\tau_N = 1$ in the absence of the dc bias field ($h = 0$). Solid lines: exact solution Eqs. (582)-(588), and (593), Asterisks: classical limit, $S \to \infty$.

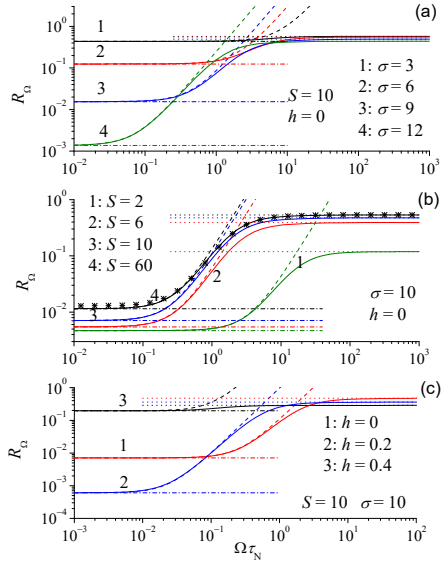

**Figure 37.** (Color on line) SNR $R_\Omega$ vs. the normalized frequency $\Omega\tau_N$ (a) for various $\sigma$, $S = 10$, and $h = 0$, (b) for various $S$, $\sigma=10$, and $h=0$, and (c) for various $h$, $S = 10$ and $\sigma = 10$. Solid lines: exact solution. Dashed lines: the low-frequency asymptote, Eqs. (593) and (589). Dashed-dotted lines: the adiabatic limit $\Omega = 0$, Eq. (594). Dotted lines: the high-frequency limit Eq. (595).

We have studied the magnetic SR of a quantum uniaxial nanomagnet of arbitrary spin $S$ in the high temperature and weak spin-bath coupling limit. The principal result is that one may determine the transition from the SR corresponding to quantum elementary spin relaxation to that pertaining to a giant spin as a function of the spin size $S$. Hence, one may accurately estimate the value of $S$ (typically in the range 20-40) at which the crossover from quantum to classical behavior takes place. Thus one may assign a range of validity to the classical Néel-Brown treatment of a nanomagnet with the simplest uniaxial anisotropy and Zeeman energy. The relatively elementary calculation outlined above is also fundamental towards gaining an understanding of the SR of spin systems characterized by nonaxially symmetric Hamiltonians. The extension to particular nonaxially symmetric spin systems such as biaxial, cubic, etc. would also allow one to include spin number effects in important technological applications of magnetic relaxation such as the magnetization reversal time, the switching field and hysteresis curves, etc.

## IV. MASTER EQUATION IN PHASE SPACE FOR NONAXIALLY SYMMETRIC SYSTEMS

### A. Uniaxial nanomagnet subjected to a dc bias field of arbitrary orientation

Clearly the analogy between the quantum and classical formulations provided by the phase space representation via the Wigner-Stratonovich map will also enable the powerful statistical moment method for classical Fokker-Planck equations [5,71] for an *arbitrary* Hamiltonian (which may be expanded in spherical harmonics) to be carried over into the quantum domain. Here our objective is to illustrate this method for the formal phase space master equation (252) pertaining to a Hamiltonian $\hat{H}_S$ as accomplished for the Fokker-Planck equation (3) for classical spins for arbitrary magnetocrystalline anisotropy-Zeeman energy potentials [5,145]. We shall illustrate how the magnetization and its relaxation times may be evaluated in the *linear response approximation* and how the solution of the corresponding classical problem [5,6] carries over into the quantum domain. In view of the formal difficulties associated with both the derivation and the direct solution of phase-space master equation we will again proceed indirectly using the method of Section II C. In this way the *explicit* solution is written for an *arbitrary* spin Hamiltonian $\hat{H}_S$ as a *finite* series of spherical harmonics analogous to the (infinite) Fourier series representation of the classical case governed by the Fokker-Planck equation (3). Therefore, the expansion coefficients, i.e., the statistical averages of the spherical harmonics, may be determined as before from a differential-recurrence relation yielding the stochastic spin dynamics for arbitrary spin number $S$. For large $S$ the differential-recurrence relations reduce to those generated by the Fokker-Planck equation. Thus the spin dynamics may once again be treated in a manner transparently linking to the classical representations, thereby providing quantum corrections to classical averages.

#### 1. Differential-recurrence equations for statistical moments

In accordance with the Wigner-Stratonovich map, the formal solution of the phase space master Eq. (252) for arbitrary $\hat{H}_S$ may be written as a *finite* linear combination of the spherical harmonics $Y_{L,M}(\vartheta,\varphi)$ embodied in the phase-space distribution Eq. (231). As shown in Section II C3, by substituting Eq. (231) into the master equation Eq. (252), we then formally have a *finite* set of differential-recurrence relations for the statistical moments $\langle Y_{LM}\rangle(t)$ which becomes for $s = -1$ (i.e., for the $Q$-function) [cf. Eq. (256)]

$$\frac{d}{dt}\langle Y_{LM}\rangle(t) = \sum_{L',M'} p'_{L'M';LM}\langle Y_{L'M'}\rangle(t), \qquad (599)$$

where

$$p'_{L'M';LM} = (-1)^{M-M'} \sqrt{\frac{(2S-L')!(2S+L'+1)!}{(2S-L)!(2S+L+1)!}} g_{L'-M';L-M} \quad (600)$$

with the expansion coefficients $g_{L'M';LM}$ given by Eq. (76). In the classical limit, $S \to \infty$, Eq. (599) reduces to the classical hierarchy, i.e., the recurrence relation given by Eq. (D18) from Appendix D [5,145]. Now the differential-recurrence relations Eq. (599) can be solved by direct matrix diagonalization, involving the calculation of the eigenvalues and eigenvectors of the system matrix. Thus, it is evident that it is relatively easy to generalize the phase space formalism to non-axially symmetric problems for an arbitrary spin Hamiltonian in a manner exactly analogous to that given for the classical Fokker-Planck equation (3) in Ref. 145.

Here we shall take as illustrative example a uniaxial nanomagnet of arbitrary $S$ in an external dc magnetic field $\mathbf{H}_0$ at an *arbitrary angle* to the $Z$-axis (i.e., the easy axis), which is a quantum analog of the most basic nonaxially symmetric model in superparamagnetism. Thus, the Hamiltonian $\hat{H}_S$ has the nonaxially symmetric form

$$\hat{H}_S = -vK\hat{S}_Z^2 - \gamma\hbar\left(\mathbf{H}_0 \cdot \hat{\mathbf{S}}\right). \quad (601)$$

Without loss of generality, we may suppose that $\mathbf{H}_0$ lies in the $XZ$ plane of the laboratory coordinate system so that the spin Hamiltonian (601) becomes

$$\beta\hat{H}_S = -\frac{\sigma}{S^2}\hat{S}_Z^2 - \frac{\xi}{S}\left(\cos\psi\,\hat{S}_Z + \sin\psi\,\hat{S}_X\right), \quad (602)$$

where

$$\xi = \beta\mu_0\mu H_0 \text{ and } \sigma = \beta vKS^2$$

are the dimensionless precession frequency and anisotropy constant, and $\psi$ is the angle between the constant field $\mathbf{H}_0$ and the $Z$ axis taken as the easy axis of the nanomagnet. Using MATHEMATICA for the matrix algebra with the polarization and spin operators as defined in Appendix B, the expansion coefficients $p'_{L'M';LM}$ in Eq. (599) can then be evaluated from the symmetrized Hubbard form of the collision operator Eq. (76) and consequently Eq. (600) as specialized to Eq.(602).

The calculation of the observables proceeds by solving the corresponding hierarchy of moment equations, Eq. (599). That hierarchy can conveniently be rewritten for the relaxation functions $c_{L,M}(t) = \langle Y_{LM}\rangle(t) - \langle Y_{LM}\rangle_0$ as

$$\frac{d}{dt}c_{L,M}(t) = \sum_{L',M'} p'_{L'M';LM}c_{L',M'}(t) \quad (603)$$

because the equilibrium averages $\langle Y_{LM}\rangle_0$ themselves satisfy the homogeneous recurrence relation

$$\sum_{L',M'} p'_{L'M';LM}\langle Y_{LM}\rangle_0 = 0. \quad (604)$$

Here the angular brackets $\langle\ \rangle_0$ designate the equilibrium average defined by

$$\langle Y_{LM}\rangle_0 = \frac{2S+1}{4\pi}\int_0^\pi\int_0^{2\pi} Y_{LM}(\vartheta,\varphi)W_S^{eq}(\vartheta,\varphi)\sin\vartheta\,d\vartheta\,d\varphi, \quad (605)$$

where $W_S^{eq}(\vartheta,\varphi)$ is the equilibrium quasiprobability distribution which can then be evaluated explicitly as described in Sec. II.D.2). Alternatively, $\langle Y_{LM}\rangle_0$ can be evaluated directly via the average polarization operators as

$$\langle Y_{LM}\rangle_0 = \sqrt{\frac{2S+1}{4\pi}}C^{SS}_{SSL0}\langle \hat{T}^{(S)}_{LM}\rangle_0 = \sqrt{\frac{2S+1}{4\pi}}C^{SS}_{SSL0}\mathrm{Tr}\left(\hat{\rho}_{eq}\hat{T}^{(S)}_{LM}\right), \quad (606)$$

where the equilibrium density matrix $\hat{\rho}_{eq}$ is given by

$$\hat{\rho}_{eq} = \frac{e^{\frac{\sigma}{S^2}\hat{S}_Z^2 + \frac{\xi}{S}\left(\cos\psi\hat{S}_Z + \sin\psi\hat{S}_X\right)}}{\mathrm{Tr}\left\{e^{\frac{\sigma}{S^2}\hat{S}_Z^2 + \frac{\xi}{S}\left(\cos\psi\hat{S}_Z + \sin\psi\hat{S}_X\right)}\right\}}. \quad (607)$$

We note that for the axially symmetric case $\psi = 0$ the hierarchy defined by Eq. (603) reduces to the particular recurrence relation Eq. (391) already derived in Sec III.A.2. For another particular case, viz., an individual spin in an external uniform magnetic field $\mathbf{H}_0$ directed along the $Z$-axis of the laboratory coordinate system, where $\beta\hat{H}_S = -\xi\hat{S}_Z/S$, Eq. (603) also yields the very simple *three-term* differential-recurrence relation Eq. (418) for the relaxation functions $c_{L,M}(t)$, which *decouple* for different $m$ and, as shown in Sec. III.B.3, can be solved *exactly* using continued fractions [62].

In the classical limit, $S \to \infty$ and $\hbar S \to$ constant, the hierarchy Eq. (603) associated with Eq. (602) reduces to the differential-recurrence equations for a classical uniaxial nanomagnet subjected to a dc bias field of arbitrary orientation (*v*. Refs. 166,175, and 191-194 for details). Here the number of recurring equations is *infinite* $(S \to \infty)$ again constituting the principal difference between the hierarchies for classical and quantum spins; in the latter case, the number of equations is *finite*. Actually, in the classical limit, $S \to \infty$, the Hamiltonian (602) corresponds to a free energy $V$ given by

$$\frac{V(\vartheta,\varphi)}{kT} = -\sigma\cos^2\vartheta - \xi(\cos\vartheta\cos\psi + \sin\psi\sin\vartheta\cos\varphi). \quad (608)$$

To describe the stochastic dynamics of a classical spin with magnetic moment $\boldsymbol{\mu}$, we may use Gilbert's equation [26] for the motion of the magnetic moment augmented by a random field, Eq. (2). In the weak coupling limit, $\alpha \ll 1$, the solution of the stochastic differential equation (2) with $V$ defined by Eq. (608) reduces to the solution of the *infinite* hierarchy of moment equations for the relaxation functions $c_{l,m}(t) = \langle Y_{lm}\rangle(t) - \langle Y_{lm}\rangle_0$, viz., [5,191,192]

$$\tau_N \frac{d}{dt}c_{n,m}(t) + \left[\frac{n(n+1)}{2} + \frac{i\xi m}{2\alpha}\cos\psi - \sigma\frac{n(n+1)-3m^2}{(2n-1)(2n+3)}\right]c_{n,m}(t)$$

$$= \left(\xi\frac{n+1}{2}\cos\psi - \frac{im\sigma}{\alpha}\right)\sqrt{\frac{n^2-m^2}{4n^2-1}}c_{n-1,m}(t) - \left(\xi\frac{n}{2}\cos\psi + \frac{im\sigma}{\alpha}\right)\sqrt{\frac{(n+1)^2-m^2}{(2n+3)(2n+1)}}c_{n+1,m}(t)$$

$$-\frac{\xi\sin\psi}{4}\left[n\sqrt{\frac{(n-m+1)(n-m+2)}{(2n+3)(2n+1)}}c_{n+1,m-1}(t)\right.$$

$$+i\frac{\sqrt{(n-m+1)(n+m)}}{\alpha}c_{n,m-1}(t) + (n+1)\sqrt{\frac{(n+m)(n+m-1)}{4n^2-1}}c_{n-1,m-1}(t)\right]$$

$$+\frac{\xi\sin\psi}{4}\left[n\sqrt{\frac{(n+m+1)(n+m+2)}{(2n+3)(2n+1)}}c_{n+1,m+1}(t)\right.$$

$$-i\frac{\sqrt{(n+m+1)(n-m)}}{\alpha}c_{n,m+1}(t) + (n+1)\sqrt{\frac{(n-m)(n-m-1)}{4n^2-1}}c_{n-1,m+1}(t)\right]$$

$$+\frac{\sigma(n+1)}{2n-1}\sqrt{\frac{(n^2-m^2)[(n-1)^2-m^2]}{(2n-3)(2n+1)}}c_{n-2,m}(t) - \frac{n\sigma}{2n+3}\sqrt{\frac{[(n+2)^2-m^2][(n+1)^2-m^2]}{(2n+5)(2n+1)}}c_{n+2,m}(t).$$

(609)

For $S\to\infty$, the hierarchy of quantum relaxation equations (603) reduces to the classical Eq. (609) so that all the quantum results agree with the classical ones. By solving Eq. (609) for the one-sided Fourier transforms of $c_{1,0}(t)$ and $c_{1,\pm 1}(t)$ as described in detail in Refs. 5 and 192, we can determine all relevant observables (see, e.g., Figs. 39-41 below). We shall compare below the predictions of the classical model with those of the quantum one with finite $S$.

*2. Characteristic relaxation times and dynamic susceptibility*

We now evaluate the *linear response* of the uniaxial nanomagnet with the Hamiltonian Eq. (601) due to infinitesimally small changes in the magnitude of the dc field. Thus we suppose as usual that a small probing field $\mathbf{H}$ ($\mathbf{H}\parallel\mathbf{H}_0$) having been applied to the nonaxially symmetric system in the distant past ($t=-\infty$) so that equilibrium conditions obtain at time $t=0$, is switched off at $t=0$. By solving the hierarchy Eq. (603) for $c_{1,0}(t)$ and $c_{1,\pm 1}(t)$, we then have all relevant quantities, namely the integral, effective and reversal times of the magnetization, the dynamic susceptibility, DMH loops, etc. This conclusion again follows from linear response theory. Here the decay of the magnetization $\langle M_H\rangle(t)$ defined as

$$\langle M_H\rangle(t) \sim \gamma\hbar\langle(\mathbf{H}\cdot\hat{\mathbf{S}})\rangle(t)$$

when a small uniform external field $\mathbf{H}$ parallel to $\mathbf{H}_0$, $\beta\mu_0\mu H \ll 1$, has been switched off at time $t=0$, is given by

$$\langle M_H\rangle(t) - \langle M_H\rangle_0 = \chi H C(t). \quad (610)$$

In Eq. (610), $C(t)$ is the normalized relaxation function defined in this particular instance as

$$C(t) = \frac{\sqrt{2}\cos\psi c_{1,0}(t) + \sin\psi[c_{1,-1}(t) - c_{1,1}(t)]}{\sqrt{2}\cos\psi c_{1,0}(0) + \sin\psi[c_{1,-1}(0) - c_{1,1}(0)]}, \quad (611)$$

and $\chi$ is the static magnetic susceptibility, given by

$$\chi = \sqrt{\frac{2\pi}{3}}\left\{\sqrt{2}\cos\psi c_{1,0}(0) + \sin\psi[c_{1,-1}(0) - c_{1,1}(0)]\right\}\frac{\gamma\hbar(S+1)}{H}. \quad (612)$$

Furthermore in writing Eqs. (610)-(612), we have again used the correspondence rules of spin operators and *c*-numbers [cf. the Weyl symbols Eqs. (239) and (241) for $s=1$], viz.,

$$\hat{S}_Z \to \sqrt{\frac{4\pi}{3}}(S+1)Y_{10}(\vartheta,\varphi)$$

and

$$\hat{S}_X \to \sqrt{\frac{2\pi}{3}}(S+1)\left(Y_{1-1}(\vartheta,\varphi) - Y_{11}(\vartheta,\varphi)\right).$$

The initial conditions $c_{L,M}(0)$ in Eqs. (611) and (612) are

$$c_{L,M}(0) = \lim_{\xi_1\to 0}\xi_1^{-1}\left(\langle Y_{LM}\rangle_{\xi_1} - \langle Y_{LM}\rangle_0\right)$$

$$= \sqrt{\frac{2S+1}{4\pi}}C_{SSL0}^{SS}\lim_{\xi_1\to 0}\xi_1^{-1}\left(\langle\hat{T}_{LM}^{(S)}\rangle_{\xi_1} - \langle\hat{T}_{LM}^{(S)}\rangle_0\right). \quad (613)$$

Moreover, in calculating the equilibrium averages $\langle\hat{T}_{L,M}^{(S)}\rangle_{\xi_1}$, we have also used the equilibrium Boltzmann distribution function $\rho_{\xi_1}$ for the Hamiltonian Eq. (602), namely

$$\rho_{\xi_1} = \frac{e^{\frac{\sigma}{S^2}\hat{S}_Z^2 + \frac{\xi+\xi_1}{S}(\cos\psi\hat{S}_Z + \sin\psi\hat{S}_X)}}{\text{Tr}\left\{e^{\frac{\sigma}{S^2}\hat{S}_Z^2 + \frac{\xi+\xi_1}{S}(\cos\psi\hat{S}_Z + \sin\psi\hat{S}_X)}\right\}}.$$

The corresponding dynamic susceptibility $\chi(\omega) = \chi'(\omega) - i\chi''(\omega)$ is as usual

$$\frac{\chi(\omega)}{\chi} = 1 - i\omega\tilde{C}(\omega), \quad (614)$$

where the spectrum

$$\tilde{C}(\omega) = \int_0^\infty C(t)e^{-i\omega t}dt. \quad (615)$$

We can also evaluate the integral relaxation time $\tau_{\text{cor}} = \tilde{C}(0)$ (in linear response the correlation time of $C(t)$) and the effective relaxation time $\tau_{\text{ef}} = -1/\dot{C}(0)$. In the frequency domain, these characteristic times as usual determine the low- and high-frequency behavior of the dynamic susceptibility $\chi(\omega)$ via Eqs. (490) and (491), namely,

$$\frac{\chi(\omega)}{\chi} \approx \begin{cases} 1 - i\omega\tau_{cor} + ..., & \omega \to 0 \\ (i\omega\tau_{ef})^{-1} + ..., & \omega \to \infty \end{cases}. \qquad (616)$$

As before yet another relevant quantity is the inverse of the smallest nonvanishing eigenvalue $\lambda_1$ of the transition matrix of the system [Eq. (618) below] which we recall is the time constant associated with the *long time behavior* of the correlation function $C(t)$ comprising the slowest (lowest frequency) relaxation mode. Thus, $\lambda_1^{-1}$ may again be associated with the spin reversal time. Furthermore, because the influence of the high-frequency relaxation modes on the low-frequency relaxation may often be ignored, $\lambda_1$ again provides more or less complete information concerning the low-frequency dynamics of the system and may be extracted from the eigenvalues of the transition matrix **X** given by Eq. (618) below. In the low temperature limit and a weak external dc field, the relation between the time constants defined above is

$$\lambda_1^{-1} > \tau_{cor} \gg \tau_N \gg \tau_{ef}. \qquad (617)$$

Thus, to determine the magnetization kinetics, we require [cf. the response function Eq. (611)] the one-sided Fourier transforms of $c_{1,0}(t)$ and $c_{1,\pm 1}(t)$. According to the differential-recurrence Eq. (603), these relaxation functions are as usual coupled to all the others so forming (unlike the classical case) a *finite* hierarchy of averages as before (because the index $L$ in Eq. (603) ranges only between 0 and $2S$). Once again the solution of such a multi-term recurrence relation may always be obtained by rewriting it as a first-order linear matrix differential equation like Eq. (80) for a column vector $\mathbf{C}(t)$ given by Eq. (79) with a transition supermatrix **X** of dimension $4S(S+1) \times 4S(S+1)$ with matrix elements

$$(\mathbf{X})_{L,L'} = \mathbf{G}_{L,L'}^S, \quad \left[\mathbf{G}_{L,L'}^S\right]_{M,M'} = -p_{L'M';LM}. \qquad (618)$$

Having solved the matrix Eq. (80) as described in detail in Sec. II.A.4, we have the relaxation functions $c_{1,0}(t)$ and $c_{1,\pm 1}(t)$, their spectra, and all desired observables such as characteristic times $\tau_{cor}$, $\tau_{ef}$, $\tau = 1/\lambda_1$, and the dynamic susceptibility $\chi(\omega)$. For simplicity, we shall suppose that the diffusion coefficients are given by $D_1 = D_{-1} = 2D_0 = 2D$ (i.e., isotropic diffusion). To compare with the semiclassical case, we simply write $\tau_N = 1/(2D)$ and $\alpha = \hbar S \beta D$.

The relaxation time $\tau_{cor}$ and inverse of the smallest nonvanishing eigenvalue $1/\lambda_1$ (the longest relaxation time) as a function of the oblique angle $\psi$, the anisotropy parameter $\sigma$, and the dimensionless damping $\alpha$ are plotted in Figs. 38-40, respectively, for various spin numbers $S$ and of the field parameter $h = \xi/(2\sigma)$. The classical solutions [5,192] corresponding to $S \to \infty$ are also shown for comparison. Both $\tau_{cor}$ and $\tau = 1/\lambda_1$ exhibit a pronounced dependence on the oblique angle $\psi$, the field $h$, anisotropy $\sigma$, damping $\alpha$, and spin number $S$ parameters. It is apparent from Figs. 38-40 that for large $S$, the quantum solutions reduce to the corresponding classical ones while in contrast they differ markedly from each other for small $S$. Typical values of $S$ for the quantum-classical crossover are ~20-40. The smaller the anisotropy $\sigma$ the smaller the $S$ value required for convergence of the quantum results to the classical ones. Now the *intrinsic* damping ($\alpha$) dependence of these characteristic relaxation times for the oblique field configuration shown in Fig. 39 represents coupling between the longitudinal and precessional modes of the magnetization. Hence, it should be possible to determine $\alpha$ by fitting the theory to the experimental dependence of the reversal time on the angle $\psi$ and dc bias field strength. Here the sole fitting parameter is $\alpha$, which can be determined at different temperatures $T$, exposing its temperature dependence. As before, the behavior of $\tau_{cor}$ and $\tau = 1/\lambda_1$ is similar only for small external fields, $h \ll 1$; in a strong external field, $h > 0.2$, $\tau$ can diverge exponentially from $\tau_{cor}$ as for classical spins (see Fig. 40). This effect as we have previously explained was discovered numerically for classical spins by Coffey *et al.* [152] and later interpreted quantitatively by Garanin [153] (see also Ref. 5, Chap. 1 for details).

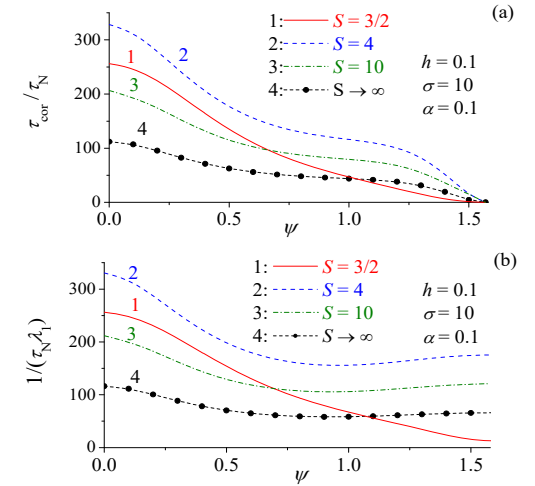

**Figure 38.** (Color on line) Correlation time $\tau_{cor}$ (a) and overbarrier time $1/\lambda_1$ (b) vs. the oblique angle $\psi$ for various spin numbers $S$ and anisotropy parameter $\sigma = 10$, dimensionless damping $\alpha = 0.1$, and field parameter $h = 0.1$.

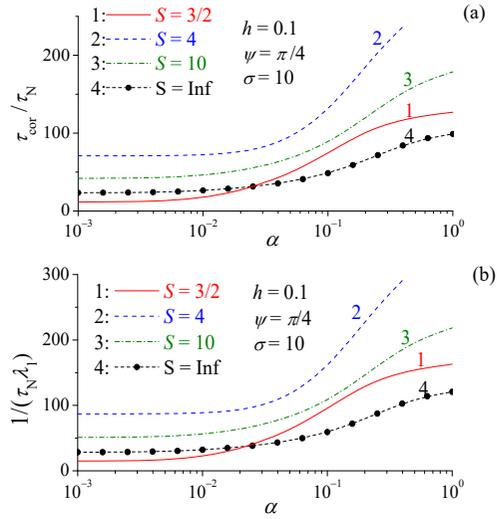

**Figure 39.** (Color on line) Correlation time $\tau_{cor}$ (a) and overbarrier time $1/\lambda_1$ (b) vs. dimensionless damping $\alpha$ for various spin numbers $S$ and oblique angle $\psi = \pi/4$, anisotropy parameter $\sigma = 10$, and field parameter $h = 0.1$.

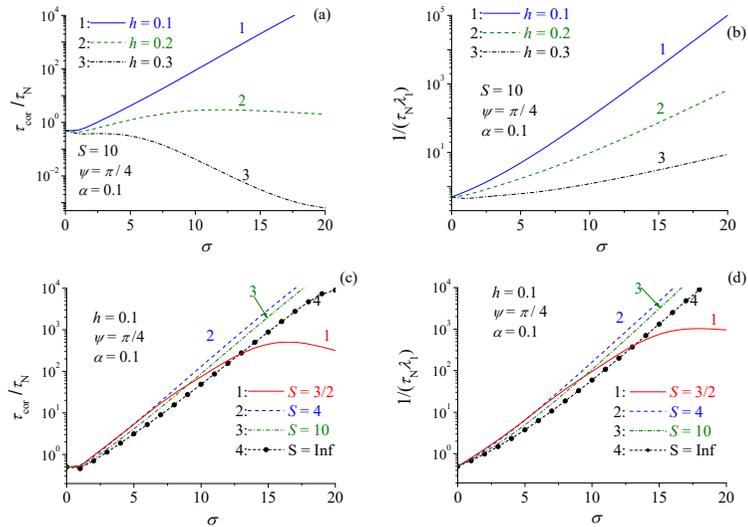

**Figure 40.** (Color on line) Correlation time $\tau_{cor}$ and overbarrier time $1/\lambda_1$ vs. the anisotropy parameter $\sigma$ for various field parameters $h$ and spin number $S = 4$, (a) and (b), for various $S$ with field parameter $h = 0.1$, (c) and (d); dimensionless damping $\alpha = 0.1$ and oblique angle $\psi = \pi/4$.

In Figs. 41-44, we have plotted the real and imaginary parts of $\chi(\omega)/\chi$ vs. normalized frequency $\omega/D$ for various model parameter values. Clearly three bands now appear in $\chi''(\omega)$ two of which are like those in axial symmetry while a third resonance band appears due to high-frequency precession of the spin in the effective field. The low-frequency relaxation band is as usual due to the slowest relaxation mode where the characteristic frequency and bandwidth are determined by $\lambda_1$. Like the classical case, $\lambda_1$ is sufficient to accurately predict the behavior of the low-frequency part of $\chi(\omega)$ as well as the long time behavior of $C(t)$. Thus, if one is interested solely in the low frequency region $(\omega/\lambda_1 \leq 1)$, where the effect of the high-frequency modes may be ignored, $\chi(\omega)$ may be again approximated by the single Lorentzian Eq. (556) which implies that $C(t)$ may be approximated for $t > 0$ by a *single* exponential with relaxation time $T_{\|} = 1/\lambda_1$. It is apparent from Figs. 41-44 since the influence of the high-frequency relaxation modes on the low-frequency relaxation may be ignored, that the simple Lorentzian formula Eq. (556) again accurately describes the entire low-frequency dynamics. The second far weaker high-frequency relaxation band in $\chi''(\omega)$ is once more due to high-frequency longitudinal "intrawell" modes. The individual "intrawell" modes are indistinguishable in the spectrum of $\chi''(\omega)$ appearing merely as a single high-frequency Lorentzian band. This "intrawell" relaxation band is more pronounced when the external field coincides with the easy axis, i.e., for $\psi = 0$. However, in general, it is masked by the third sharp resonance band due to excitation of transverse modes having frequencies close to the precession frequency of the spin which strongly manifests itself at high frequencies. For $\psi = 0$, the resonance peak disappears because the transverse modes are no longer excited. In contrast it is most pronounced when $\psi = \pi/2$.

In this Section, we have solved the differential-recurrence Eq. (599) for the evolution of the statistical moments (average spherical harmonics) for a *nonaxially symmetric* spin Hamiltonian, viz. Eq. (602). For purposes of illustration, the analysis was carried out for a uniaxial nanomagnet subjected to a dc external field applied at an *arbitrary* angle to the easy axis. In particular, we have evaluated the characteristic relaxation times along with the linear dynamic susceptibility via obvious generalizations of the methods previously used for classical spins [5,191,192]. Thus the phase space representation (because it is closely allied to the classical one) again transparently illustrates how quantum distributions reduce to the classical ones. When the direction of the external fields coincides with the easy axis, i.e., for $\psi = 0$, our method reproduces the results for nanomagnets subjected to a longitudinal field previously obtained in Sec. III.C.2. The method may also be extended to other non-axially symmetric multi-well systems such as biaxial, cubic, mixed, etc. Furthermore, the model can be generalized to time-dependent Hamiltonians as in Sec. III.C.3, so that we can also determine quantum effects in the nonlinear ac stationary response of quantum nanomagnets in the nonaxially symmetric problem we have just considered.

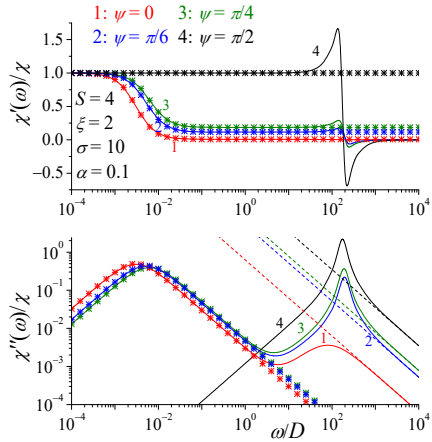

**Figure 41.** (Color on line) Real and imaginary parts of the dynamic susceptibility $\chi(\omega)$ vs. the normalized frequency $\omega/D$ for various oblique angles $\psi$ and field parameter $h = 0.1$, anisotropy parameter $\sigma = 10$, damping $\alpha = 0.1$, and spin $S = 4$. Asterisks: the single Lorentzian approximation, Eq. (556) while the straight dashed lines are the high-frequency asymptotes, Eq. (616).

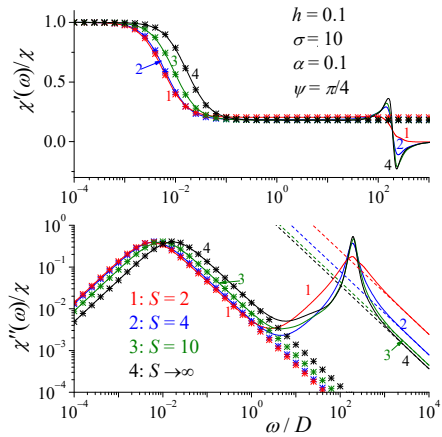

**Figure 42.** (Color on line) Real and imaginary parts of $\chi(\omega)$ vs. the normalized frequency parameter $\omega/D$ for various spin numbers $S$ and $h = 0.1$, $\sigma = 10$, $\alpha = 0.1$, and $\psi = \pi/4$. Asterisks are the single Lorentzian approximation, Eq. (556), while the straight dashed lines are the high frequency asymptotes, Eq. (616).

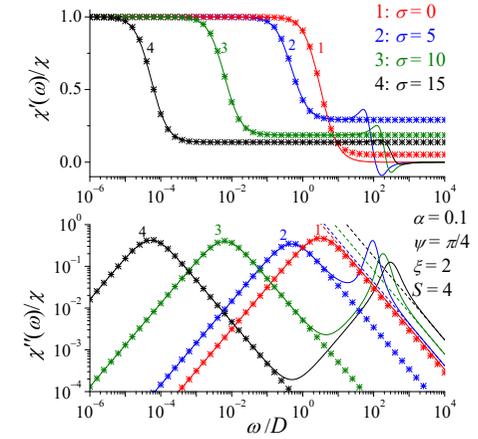

**Figure 43.** (Color on line) Real and imaginary parts of $\chi(\omega)$ vs. the normalized frequency $\omega/D$ for various anisotropy parameters $\sigma$ and spin number $S = 4$, field parameter $\xi = 2$, dimensionless damping $\alpha = 0.1$, and oblique angle $\psi = \pi/4$. Asterisks are the single Lorentzian approximation, Eq. (556), while the straight dashed lines are the high frequency asymptotes, Eq. (616).

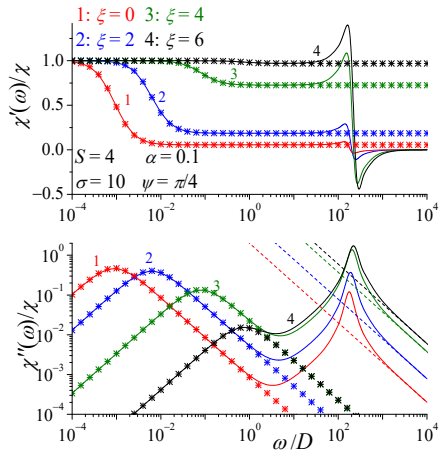

**Figure 44.** (Color on line) Real and imaginary parts of $\chi(\omega)$ vs. the normalized frequency $\omega/D$ for various field parameters $\xi$ and anisotropy parameter $\sigma = 10$, spin number $S = 4$, dimensionless damping $\alpha = 0.1$, and oblique angle $\psi = \pi/4$. Asterisks are the single Lorentzian approximation, Eq. (556), while the straight dashed lines are the high frequency asymptotes, Eq. (616).

## V. CONCLUSION

We have treated numerous illustrative examples of spin relaxation problems using Wigner's phase-space formulation of quantum mechanics of particles and spins. The one-to-one correspondence between the quantum state in the Hilbert space and a real representation space function first envisaged for the closed system in the spin context by Stratonovich [49], formally represents the quantum mechanics of spins as a statistical theory in the representation space of polar angles $(\vartheta, \varphi)$ which has a clear classical meaning. This procedure effectively generalizes the results of Wigner [41] who represented the quantum mechanics of a particle with Hamiltonian $\hat{H} = \hat{p}^2/2m + V(\hat{q})$ in Hilbert space as a statistical theory in a classically meaningful phase space with the canonical variables position and momentum $(q, p)$. Stratonovich [49] proceeded by introducing a quasiprobability density (Wigner) function on the sphere, defined as the linear invertible bijective map onto the representation space comprised of the trace of the product of the system density matrix and the irreducible tensor operators, the analysis being carried out via the finite series in spherical harmonics embodied in the bijective Wigner-Stratonovich map. Hence, the average value of a quantum spin operator may be calculated via its Weyl symbol just as the corresponding classical function in the representation space of polar angles $(\vartheta, \varphi)$. This may be accomplished essentially because the *polarization operators transform under rotation in the same way as the spherical harmonics*. Thus, the Stratonovich representation for spins [49] is well suited to the development of semiclassical methods of treatment of spin relaxation phenomena allowing one to obtain quantum corrections in a manner closely analogous to the classical case.

The merit of the phase space formalism as applied to spin relaxation problems is that only master equations for the phase-space distributions akin to Fokker-Planck equations for the evolution of classical phase-space distributions in configuration space are involved so that operators are unnecessary. The *explicit* solution of these equations can be expanded for an *arbitrary* spin Hamiltonian in a *finite* series of spherical harmonics Eq. (231) like in the classical case where an infinite number of spherical harmonics is involved. The expansion coefficients (statistical moments or averages of the spherical harmonics which are obviously by virtue of the Wigner-Stratonovich map the averages of the polarization operators) may be determined from a differential-recurrence relation Eq. (253) in a manner similar to the classical case. Although the form of the phase-space master equation is in general very complicated, we can circumvent the problem of determining differential recurrence relations by directly using the one-to-one correspondence between averages of polarization operators and those of spherical harmonics as outlined in Section II.C.3. Moreover, we have described this procedure via several illustrative examples. Thus, we still have a method of treating the spin relaxation for *arbitrary* spin number $S$ which is closely allied to the classical one even though a phase space master equation may not be explicitly involved. In the classical limit, the quantum differential-recurrence relation reduces to that yielded by the classical Fokker-Planck equation for *arbitrary* magnetocrystalline - Zeeman energy potentials. Furthermore, the phase space representation via the Weyl symbol of the relevant spin operator suggests how powerful computation techniques developed for Fokker-Planck equations (matrix continued fractions, mean first passage time, integral representation of relaxation times, etc. [5,71]) may be transparently extended to the quantum domain indeed suggesting new closed form quantum results via the corresponding classical ones. A specific example is the determination of the quantum integral relaxation time for a spin in a uniform magnetic field of arbitrary strength directed along the Z-axis, Eq. (442). Therefore, having solved the phase-space master equation, one can then, in principle, evaluate via obvious generalizations of the methods previously used for classical spins [5,6] all desired observables. These include the magnetization itself, the magnetization reversal time, the linear and nonlinear dynamic susceptibilities, the temperature dependence of the switching fields, the dynamic hysteresis loops etc.). In this way one can study the transition of the relaxation behavior from that of an elementary spin to molecular magnets ($S \sim 10$), to nanoclusters ($S \sim 100$), and to classical superparamagnets ($S \geq 1000$). Thus all *quantum effects in the spin relaxation phenomena can be treated in a manner linking directly to the classical representations*. Furthermore by treating a variety of spin relaxation problems, we have also amply demonstrated that although the density matrix and phase-space methods may yield results in *outwardly very different forms*, nevertheless, both approaches yield *identical* numerical values for the *same* physical quantities (such as relaxation times and susceptibility). Hence, we have established a vital corollary between the phase space and the density matrix methods thereby demonstrating that they are essentially equivalent while simultaneously providing an important check on the validity of the phase space method. Thus the phase space representation, because it is closely allied to the classical representation, besides being complementary to the operator one, transparently illustrates how quantum phase-space distributions reduce to the classical ones in the limit $S \to \infty$.

## ACKNOWLEDGEMENTS

We acknowledge the financial support of FP7-PEOPLE-Marie Curie Actions (project No. 295196 DMH). We would like to thank Paul Blaise, Declan Byrne, Pierre-Michel Déjardin, and Adina Ceausu-Velcescu who have carefully read the entire manuscript and proposed a number of corrections and improvements in the presentation. The review also forms part of the EU COST initiative MP1006 *Fundamental Principles in Quantum Physics* coordinated by Professor Angelo Bassi.

## APPENDIX A: SPIN AND POLARIZATION OPERATORS

The spin operator $\hat{\mathbf{S}}$ is Hermitian $\hat{\mathbf{S}}^\dagger = \hat{\mathbf{S}}$ and is usually represented by a set of three (since the spin vector $\hat{\mathbf{S}}$ has three components) square $(2S+1) \times (2S+1)$ matrixes with $S$ being the spin number [95]. For the Cartesian components of $\hat{\mathbf{S}}$, namely, for the operators $\hat{S}_X$, $\hat{S}_Y$, and $\hat{S}_Z$, the Hermitian property takes on the form $\hat{S}_i^\dagger = \hat{S}_i$, $(i = X, Y, Z)$, while for the spherical components $\hat{S}_{\pm 1}$ and $\hat{S}_0$ that property becomes $\hat{S}_\mu^\dagger = (-1)^\mu \hat{S}_{-\mu}$. The relations between the Cartesian and spherical components of $\hat{\mathbf{S}}$ are given by [95]

$$\hat{S}_{+1} = -\frac{1}{\sqrt{2}}(\hat{S}_X + i\hat{S}_Y), \quad \hat{S}_X = \frac{1}{\sqrt{2}}(\hat{S}_{-1} - \hat{S}_{+1}),$$

$$\hat{S}_{-1} = \frac{1}{\sqrt{2}}(\hat{S}_X - i\hat{S}_Y), \quad \hat{S}_Y = \frac{i}{\sqrt{2}}(\hat{S}_{-1} + \hat{S}_{+1}), \tag{A1}$$

$$\hat{S}_0 = \hat{S}_Z, \quad \hat{S}_Z = \hat{S}_0.$$

Polarization states of particles are described by spin functions $\chi_{Sm} = |S, m\rangle$, which depend on the spin variable $\sigma$ being the spin projection on the Z-axis. This variable takes $2S+1$ values, $\sigma = -S, -S+1, ..., S-1, S$. The dependence of the spin functions $\chi_{Sm}(\sigma)$ on the spin variable $\sigma$ is given by $\chi_{Sm}(\sigma) = \delta_{m\sigma}$ [95]. The spin functions $\chi_{Sm} = |S, m\rangle$ are eigenfunctions of the spin operators $\hat{\mathbf{S}}^2$ and $\hat{S}_Z$, viz., [95]

$$\hat{\mathbf{S}}^2 \chi_{Sm} = S(S+1) \chi_{Sm}, \tag{A2}$$

$$\hat{S}_Z \chi_{Sm} = m \chi_{Sm} \tag{A3}$$

with $m = -S, -S+1, ..., S-1, S$. The spin functions each have $2S+1$ components and describe polarization states of a particle with definite spin $S$ and spin projection $m$ onto the Z-axis. They can be rewritten as column matrixes [95]

$$\chi_{SS} = \begin{pmatrix} 1 \\ 0 \\ \vdots \\ 0 \end{pmatrix}, \quad \chi_{SS-1} = \begin{pmatrix} 0 \\ 1 \\ \vdots \\ 0 \end{pmatrix}, \quad ..., \quad \chi_{S-S} = \begin{pmatrix} 0 \\ 0 \\ \vdots \\ 1 \end{pmatrix}. \tag{A4}$$

Furthermore, the vector representation shows that the spin functions $\chi_{Sm}$ constitute a complete set of functions with the orthonormality and completeness conditions

$$\chi_{Sm'}^\dagger \chi_{Sm} = \delta_{mm'}, \tag{A5}$$

$$\sum_{m=-S}^{S} \chi_{Sm} \chi_{Sm}^\dagger = \hat{I}, \tag{A6}$$

where $\hat{I}$ is the unit $(2S+1) \times (2S+1)$ matrix.

Now the spherical components of the spin operator $\hat{\mathbf{S}}$ may be expressed using the Clebsch-Gordan coefficients in terms of the spin functions as [95]

$$\hat{S}_\mu = \sqrt{S(S+1)} \sum_{m,m'} C_{Sm'1\mu}^{Sm} \chi_{Sm} \chi_{Sm'}^\dagger, \quad (\mu = 0, \pm 1). \tag{A7}$$

The operators $\hat{S}_\mu$ satisfy the following commutation relation [95]

$$\left[ \hat{S}_\mu, \hat{S}_\nu \right] = -\sqrt{2} C_{1\mu 1\nu}^{1\lambda} \hat{S}_\lambda, \quad \left[ \hat{\mathbf{S}}^2, \hat{S}_\mu \right] = 0 \quad (\mu, \nu, \lambda = 0, \pm 1). \tag{A8}$$

The matrix elements of $\hat{S}_\mu$ are given by [95]

$$\left[ \hat{S}_\mu \right]_{m'm} = \chi_{Sm'}^\dagger \hat{S}_\mu \chi_{Sm} = \sqrt{S(S+1)} C_{Sm1\mu}^{Sm'} \tag{A9}$$

with the only nonvanishing elements being

$$\left[ \hat{S}_0 \right]_{mm} = m, \tag{A10}$$

$$\left[ \hat{S}_{\pm 1} \right]_{m \pm 1 m} = \mp \frac{1}{\sqrt{2}} \sqrt{(S \mp m)(S \pm m + 1)}. \tag{A11}$$

The Cartesian and spherical components of $\hat{\mathbf{S}}$ are related by Eq. (A1). For example, in the particular case $S = 1/2$, the operators $\hat{S}_i$ ($i = X, Y, Z$) and $\hat{S}_\mu$ ($\mu = 0, \pm 1$) are square 2×2 matrixes given by [95]

$$\hat{S}_X = \frac{1}{2}\begin{pmatrix} 0 & 1 \\ 1 & 0 \end{pmatrix}, \quad \hat{S}_Y = \frac{i}{2}\begin{pmatrix} 0 & -1 \\ 1 & 0 \end{pmatrix}, \quad \hat{S}_Z = \frac{1}{2}\begin{pmatrix} 1 & 0 \\ 0 & -1 \end{pmatrix}, \tag{A12}$$

$$\hat{S}_{+1} = \frac{1}{\sqrt{2}}\begin{pmatrix} 0 & -1 \\ 0 & 0 \end{pmatrix}, \quad \hat{S}_0 = \frac{1}{2}\begin{pmatrix} 1 & 0 \\ 0 & -1 \end{pmatrix}, \quad \hat{S}_{-1} = \frac{1}{\sqrt{2}}\begin{pmatrix} 0 & 0 \\ 1 & 0 \end{pmatrix}, \tag{A13}$$

while for $S = 1$, these operators are the square 3×3 matrixes given by [95]

$$\hat{S}_X = \frac{1}{\sqrt{2}}\begin{pmatrix} 0 & 1 & 0 \\ 1 & 0 & 1 \\ 0 & 1 & 0 \end{pmatrix}, \quad \hat{S}_Y = \frac{i}{\sqrt{2}}\begin{pmatrix} 0 & -1 & 0 \\ 1 & 0 & -1 \\ 0 & 1 & 0 \end{pmatrix}, \quad \hat{S}_Z = \begin{pmatrix} 1 & 0 & 0 \\ 0 & 0 & 0 \\ 0 & 0 & -1 \end{pmatrix}, \tag{A14}$$

$$\hat{S}_{+1} = \begin{pmatrix} 0 & -1 & 0 \\ 0 & 0 & -1 \\ 0 & 0 & 0 \end{pmatrix}, \quad \hat{S}_0 = \begin{pmatrix} 1 & 0 & 0 \\ 0 & 0 & 0 \\ 0 & 0 & -1 \end{pmatrix}, \quad \hat{S}_{-1} = \begin{pmatrix} 0 & 0 & 0 \\ 1 & 0 & 0 \\ 0 & 1 & 0 \end{pmatrix}. \tag{A15}$$

Now in order to describe a polarization (spin) state of a particle, the polarization operators are often used. These operators denoted by $\hat{T}_{LM}^{(S)}$, where $L = 0, 1,\ldots, 2S$ and $M = -L, -L+1,\ldots,L$, are matrixes, which act on spin functions. However, the explicit form of $\hat{T}_{LM}^{(S)}$ depends on the representation chosen for the spin functions. In particular, the matrix elements $T_{m'm} = \left[\hat{T}_{LM}^{(S)}\right]_{m'm}$ of the polarization operator $\hat{T}_{LM}^{(S)}$ which has the explicit form

$$\hat{T}_{LM}^{(S)} = \begin{pmatrix} T_{SS} & \cdots & T_{S-S} \\ \vdots & \ddots & \vdots \\ T_{-SS} & \cdots & T_{-S-S} \end{pmatrix} \tag{A16}$$

are related to those of the spin functions in the spherical basis representation via [95]

$$T_{m'm} = \chi_{Sm'}^\dagger \hat{T}_{LM}^{(S)} \chi_{Sm} = \sqrt{\frac{2L+1}{2S+1}} C_{SmLM}^{Sm'}. \tag{A17}$$

For example, the operator $\hat{T}_{00}^{(S)}$ is proportional to the unit $(2S+1)\times(2S+1)$ matrix

$$\hat{T}_{00}^{(S)} = \frac{1}{\sqrt{2S+1}}\hat{I}, \tag{A18}$$

while for $L = 1$ the operators $\hat{T}_{1\mu}^{(S)}$ are proportional to the spherical components of the spin operator

$$\hat{T}_{1\mu}^{(S)} = \frac{1}{\sqrt{2}a}\hat{S}_\mu, \tag{A19}$$

where

$$a = \sqrt{\frac{S(S+1)(2S+1)}{6}}. \tag{A20}$$

Likewise, the Cartesian components of $\hat{\mathbf{S}}$ given by Eq. (A1) may also be expressed via the polarization operators $\hat{T}_{10}^{(S)}$ and $\hat{T}_{1\pm 1}^{(S)}$ as

$$\hat{S}_X = a\left(\hat{T}_{1-1}^{(S)} - \hat{T}_{11}^{(S)}\right),$$

$$\hat{S}_Y = ia\left(\hat{T}_{1-1}^{(S)} + \hat{T}_{11}^{(S)}\right), \tag{A21}$$

$$\hat{S}_Z = \sqrt{2}a\hat{T}_{10}^{(S)}.$$

The polarization operators $\hat{T}_{LM}^{(S)}$ are normalized and satisfy the relations [95]

$$\hat{T}_{LM}^{\dagger(S)} = (-1)^M \hat{T}_{L-M}^{(S)} \tag{A22}$$

and

$$\text{Tr}\left(\hat{T}_{L,M}^{(S)}\right) = \sqrt{(2S+1)}\delta_{L0}\delta_{M0}, \tag{A23}$$

i.e., all $\hat{T}_{LM}^{(S)}$ have zero trace except of $\hat{T}_{00}^{(S)}$. Furthermore, the operators $\hat{T}_{LM}^{(S)}$ also constitute an orthonormal basis in the space of $(2S+1)\times(2S+1)$ matrices with $S$ integer or half-integer. Hence, it follows that an *arbitrary* square $(2S+1)\times(2S+1)$ matrix operator $\hat{A}$ may be expanded as a series of the polarization operators [95]

$$\hat{A} = \sum_{L=0}^{2S} \sum_{M=-L}^{L} A_{LM}\hat{T}_{LM}^{(S)}, \tag{A24}$$

where the expansion coefficients $A_{LM}$ are given by

$$A_{LM} = \text{Tr}\left(\hat{T}_{LM}^{\dagger(S)}\hat{A}\right). \tag{A25}$$

If the matrix $\hat{A}$ is Hermitian ($\hat{A}^\dagger = \hat{A}$) then

$$A_{LM}^* = (-1)^M A_{L-M}. \tag{A26}$$

Moreover, the matrix products of the spin functions $\chi_{Sm}$ and $\chi_{Sm'}^\dagger$ [as arranged in the form of Eq. (A4)] are themselves square $(2S+1)\times(2S+1)$ matrixes and, therefore, may also be expanded as a finite series of polarization operators, viz., [95]

$$\chi_{Sm}\chi_{Sm'}^\dagger = \sum_L \sqrt{\frac{2L+1}{2S+1}} C_{Sm'LM}^{Sm} \hat{T}_{LM}^{(S)}. \tag{A27}$$

Furthermore, products of two polarization operators may be written in the form of the Clebsch-Gordan series [95]

$$\hat{T}_{L_1M_1}^{(S)}\hat{T}_{L_2M_2}^{(S)} = \sqrt{(2L_1+1)(2L_2+1)}\sum_L (-1)^{2S+L}\begin{Bmatrix} L_1 L_2 L \\ S S S \end{Bmatrix} C_{L_1M_1L_2M_2}^{LM} \hat{T}_{LM}^{(S)}, \tag{A28}$$

where $\begin{Bmatrix} L_1 L_2 L \\ S S S \end{Bmatrix}$ is Wigner's 6j-symbol [95]. The polarization operators also satisfy the commutation relations [95]

$$\left[\hat{T}^{(S)}_{L_1M_1},\hat{T}^{(S)}_{L_2M_2}\right] = \sqrt{(2L_1+1)(2L_2+1)}\sum_L (-1)^{2S+L}\left[1-(-1)^{L_1+L_2+L}\right]\begin{Bmatrix} L_1 L_2 L \\ S S S \end{Bmatrix} C^{LM}_{L_1M_1L_2M_2}\hat{T}^{(S)}_{LM}. \quad (A29)$$

Equation (A29) then automatically yields the commutation relation for the spherical components of the spin operator $\hat{S}_\mu$ and polarization operator $\hat{T}^{(S)}_{LM}$, namely, [95]

$$[\hat{S}_\mu, \hat{T}^{(S)}_{LM}] = \sqrt{L(L+1)} C^{LM+\mu}_{LM1\mu} \hat{T}^{(S)}_{LM+\mu} \quad (A30)$$

or

$$[\hat{S}_0, \hat{T}^{(S)}_{LM}] = \sqrt{L(L+1)} C^{LM}_{LM10} \hat{T}^{(S)}_{LM} = M\hat{T}^{(S)}_{LM}, \quad (A31)$$

$$[\hat{S}_{\pm 1}, \hat{T}^{(S)}_{LM}] = \sqrt{L(L+1)} C^{LM\pm 1}_{LM1\pm 1} \hat{T}^{(S)}_{LM\pm 1}$$
$$= \mp\sqrt{\frac{L(L+1)-M(M\pm 1)}{2}} \hat{T}^{(S)}_{LM\pm 1}. \quad (A32)$$

Finally, traces of products of the polarization operators are given by [95]

$$\text{Tr}\left(\hat{T}^{(S)}_{L_1M_1}\hat{T}^{(S)}_{L_2M_2}\right) = (-1)^{M_1}\delta_{L_1L_2}\delta_{M_1-M_2}, \quad (A33)$$

$$\text{Tr}\left(\hat{T}^{(S)\dagger}_{LM}\hat{T}^{(S)}_{L'M'}\right) = \delta_{LL'}\delta_{MM'}, \quad (A34)$$

$$\text{Tr}\left(\hat{T}^{(S)}_{L_1M_1}\hat{T}^{(S)}_{L_2M_2}\hat{T}^{(S)}_{L_3M_3}\right) = (-1)^{2S+L_3+M_3}\sqrt{(2L_1+1)(2L_2+1)} C^{L_3-M_3}_{L_1M_1L_2M_2}\begin{Bmatrix} L_1 L_2 L_3 \\ S S S \end{Bmatrix}, \quad (A35)$$

etc.

Finally, under rotation of coordinate system defined by the Euler angles $\alpha,\beta,\gamma$, the basis spin functions $\chi_{Sm'}$ are transformed by the rotation operator $\hat{D}^S(\alpha,\beta,\gamma)$ given by [95]

$$\hat{D}^S(\alpha,\beta,\gamma) = e^{-i\alpha\hat{S}_X} e^{-i\beta\hat{S}_Y} e^{-i\gamma\hat{S}_Z}$$
$$= \sum_{L,M,m,m'}\frac{2L+1}{2S+1} C^{Sm}_{Sm'LM} D^S_{mm'}(\alpha,\beta,\gamma)\hat{T}^{(S)}_{LM}, \quad (A36)$$

yielding

$$\chi'_{Sm'} = \hat{D}^S(\alpha,\beta,\gamma)\chi_{Sm'} = \sum_m D^S_{mm'}(\alpha,\beta,\gamma)\chi_{Sm}, \quad (A37)$$

where $\chi'_{Sm'}$ describe quantum states with definite spin $S$ and spin projection $m'$ on the new $Z'$-axis and $D^S_{mm'}(\alpha,\beta,\gamma)$ are the Wigner D functions. Similarly, the polarization operators $\hat{T}^{(S)}_{LM}$, which are irreducible tensors of rank $L$, are transformed by the operator $\hat{D}^L(\alpha,\beta,\gamma)$ given by Eq. (A36) with $S = L$.

# APPENDIX B: SPHERICAL HARMONICS

A spherical harmonic $Y_{lm}(\vartheta,\varphi)$ is a complex function of two arguments, namely, the colatitude $0 \le \vartheta \le \pi$ and the azimuth $0 \le \varphi \le 2\pi$ and may be defined as [95]

$$Y_{lm}(\vartheta,\varphi) = \sqrt{\frac{(2l+1)(l-m)!}{4\pi(l+m)!}} e^{im\varphi} P_l^m(\cos\vartheta), \quad (B1)$$

$$Y_{l-m} = (-1)^m Y^*_{lm}, \quad (B2)$$

where $P_l^m(x)$ are the associated Legendre functions defined as [95]

$$P_l^m(\cos\vartheta) = \frac{(-1)^m}{2^l l!}(\sin\vartheta)^m \frac{d^{l+m}}{(d\cos\vartheta)^{l+m}}(\cos^2\vartheta-1)^l \quad (B3)$$

with $m = l, l-1, \ldots l$, and the asterisk denotes the complex conjugate. For the particular case $m = 0$, $Y_{l0}(\vartheta,\varphi)$ is given by

$$Y_{l0}(\vartheta,\varphi) = \sqrt{\frac{2l+1}{4\pi}} P_l(\cos\vartheta), \quad (B4)$$

where $P_l(\cos\vartheta)$ is the Legendre polynomial of order $l$ [95,105]. In particular, Eq. (B1) yields

$$Y_{10}(\vartheta,\varphi) = \sqrt{\frac{3}{4\pi}}\cos\vartheta, \; Y_{1\pm 1}(\vartheta,\varphi) = \mp\sqrt{\frac{3}{8\pi}} e^{\pm i\varphi}\sin\vartheta, \text{ etc.} \quad (B5)$$

In quantum mechanics, the spherical harmonics $Y_{lm}(\vartheta,\varphi)$ play an important role describing the distribution of particles which move in spherically symmetric field with the orbital angular momentum $l$ and projection on the quantization axis $m$ [95]. The spherical harmonics $Y_{lm}(\vartheta,\varphi)$ are the eigenfunctions of the square of the angular momentum operator $\hat{\mathbf{L}}$ and its projection $\hat{L}_Z$ onto the $Z$-axis, namely,

$$\hat{\mathbf{L}}^2 Y_{lm}(\vartheta,\varphi) = l(l+1)Y_{lm}(\vartheta,\varphi), \quad (B6)$$

$$\hat{L}_Z Y_{lm}(\vartheta,\varphi) = m Y_{lm}(\vartheta,\varphi), \quad (B7)$$

where

$$\hat{\mathbf{L}}^2 = -\frac{1}{\sin\vartheta}\frac{\partial}{\partial\vartheta}\left(\sin\vartheta\frac{\partial}{\partial\vartheta}\right) - \frac{1}{\sin^2\vartheta}\frac{\partial^2}{\partial\varphi^2} \quad (B8)$$

and

$$\hat{L}_Z = -i\frac{\partial}{\partial\varphi}. \quad (B9)$$

The completeness relation for the spherical harmonics is as follows [95]

$$\sum_{l=0}^{\infty}\sum_{m=-l}^{l} Y^*_{lm}(\vartheta,\varphi) Y_{lm}(\vartheta',\varphi') = \delta(\varphi-\varphi')\delta(\cos\vartheta-\cos\vartheta'), \quad (B10)$$

while the normalization and orthogonality relation of the spherical harmonics is given by [95]

$$\int_0^{2\pi}\int_0^{\pi} Y_{lm}(\vartheta,\varphi)Y_{l'm'}^*(\vartheta,\varphi)\sin\vartheta\, d\vartheta\, d\varphi = \delta_{ll'}\delta_{mm'}. \tag{B11}$$

Thus an arbitrary function $f(\vartheta,\varphi)$ defined on the interval $0 \le \vartheta \le \pi$ and $0 \le \varphi \le 2\pi$ (the unit sphere) which satisfies the square integrability condition

$$\int_0^{2\pi}\int_0^{\pi} |f(\vartheta,\varphi)|^2 \sin\vartheta\, d\vartheta\, d\varphi < \infty$$

can be expanded in a series of the spherical harmonics as [95]

$$f(\vartheta,\varphi) = \sum_{l=0}^{\infty}\sum_{m=-l}^{l} a_{lm}Y_{lm}(\vartheta,\varphi), \tag{B12}$$

where the expansion coefficients $a_{lm}$ are defined by

$$a_{lm} = \int_0^{2\pi}\int_0^{\pi} f(\vartheta,\varphi)Y_{lm}^*(\vartheta,\varphi)\sin\vartheta\, d\vartheta\, d\varphi. \tag{B13}$$

Moreover, a product of two spherical harmonics may be expanded in the Clebsch-Gordan series as [95]

$$Y_{l_1m_1}Y_{l_2m_2} = \sum_{L,M}\sqrt{\frac{(2l_1+1)(2l_2+1)}{4\pi(2L+1)}}C_{l_10l_20}^{L0}C_{l_1m_1l_2m_2}^{LM}Y_{LM}. \tag{B14}$$

Some useful recurrence relations for the spherical harmonics are [95]

$$\cos\vartheta\, Y_{lm} = \sqrt{\frac{(l+1)^2-m^2}{(2l+3)(2l+1)}}Y_{l+1m} - \sqrt{\frac{l^2-m^2}{(2l+1)(2l-1)}}Y_{l-1m}, \tag{B15}$$

$$\sin\vartheta\frac{\partial Y_{lm}}{\partial\vartheta} = l\sqrt{\frac{(l+1)^2-m^2}{(2l+3)(2l+1)}}Y_{l+1m} - (l+1)\sqrt{\frac{l^2-m^2}{(2l+1)(2l-1)}}Y_{l-1m}. \tag{B16}$$

The second derivative of $Y_{l,m}(\vartheta,\varphi)$ is given by [95]

$$\frac{\partial^2 Y_{lm}}{\partial \vartheta^2} = \left(\frac{m^2}{\sin^2\vartheta} - l(l+1)\right)Y_{lm} - \cot\vartheta\frac{\partial Y_{lm}}{\partial \vartheta}. \tag{B17}$$

One of the known trigonometric identities for the spherical harmonics, which has been used in the main text [cf. Eq. (211)] is [95]

$$Y_{LM}^*(\vartheta,\varphi) = \frac{\cos^{4S}(\vartheta/2)}{C_{SSL0}^{SS}}\sqrt{\frac{2L+1}{4\pi}}\sum_{m=-S}^{S}\frac{(2S)!C_{Sm-MLM}^{Sm}(\tan(\vartheta/2))^{2S-2m+M}e^{-iM\varphi}}{\sqrt{(S+m-M)!(S-m+M)!(S+m)!(S-m)!}}. \tag{B18}$$

Because the spherical harmonics $Y_{lm}(\vartheta,\varphi)$ are components of some irreducible tensor of rank $l$, under arbitrary rotation of the coordinate system described by the Euler angles $\alpha,\beta,\gamma$, the spherical harmonics $Y_{lm}(\vartheta,\varphi)$ are transformed according to the rule [95] [cf. Eq. (A37)]

$$Y_{lm'}(\vartheta',\varphi') = \sum_m Y_{lm}(\vartheta,\varphi)D_{mm'}^l(\alpha,\beta,\gamma). \tag{B19}$$

Here $\vartheta,\varphi$ and $\vartheta',\varphi'$ are polar angles of the position vector in the original and final coordinate systems and $D_{mm'}^l(\alpha,\beta,\gamma)$ are the Wigner D functions.

# APPENDIX C: DERIVATION OF THE MASTER EQUATION FOR A UNIAXIAL PARAMAGNET SUBJECTED TO A DC MAGNETIC FIELD

In order to find the phase space representation of the density matrix evolution equation Eq. (371), we must transform the following integrands of Eq. (376) into the phase space representation, viz.,

$$\frac{\sigma}{S^2}\left[\hat{S}_0^2,\hat{w}\right] + \frac{\xi}{S}\left[\hat{S}_0,\hat{w}\right], \tag{C1}$$

$$\left[\hat{S}_0,\hat{w}\hat{S}_0\right] + \left[\hat{S}_0\hat{w},\hat{S}_0\right], \tag{C2}$$

$$\left[\hat{S}_1 e^{\frac{\sigma}{S^2}\hat{S}_0}\hat{w},\hat{S}_{-1}\right] + \left[\hat{S}_1,\hat{w}e^{\frac{\sigma}{S^2}\hat{S}_0}\hat{S}_{-1}\right], \tag{C3}$$

$$\left[\hat{S}_{-1},\hat{w}e^{-\frac{\sigma}{S^2}\hat{S}_0}\hat{S}_1\right] + \left[\hat{S}_{-1}e^{-\frac{\sigma}{S^2}\hat{S}_0}\hat{w},\hat{S}_1\right]. \tag{C4}$$

We start with the commutation relation $\left[\hat{S}_0^2,\hat{w}\right]$ in Eq. (C1) and its analogous differential operator in configuration space. In order to accomplish this we observe that we have from the polarization operator expansion Eq. (230) of the Wigner-Stratonovich kernel $\hat{w}$, the following commutation relation indicated by the Liouville term of the integrand of Eq. (376), viz.,

$$\left[\hat{S}_0^2,\hat{w}\right] = \sqrt{\frac{4\pi}{2S+1}}\sum_{L=0}^{2S}\sum_{M=-L}^{L}\left(C_{SSL0}^{SS}\right)^{-1}Y_{LM}^*\left[\hat{S}_0^2,\hat{T}_{LM}^{(S)}\right]. \tag{C5}$$

However the spin operator $\hat{S}_0^2$ must first be written in terms of the polarization operators $\hat{T}_{LM}^{(S)}$ by using Eqs. (A20) and (A28), namely,

$$\hat{S}_0^2 = \frac{\sqrt{S(S+1)(2S+1)}}{3}\left(\frac{\sqrt{(2S-1)(2S+3)}}{\sqrt{5}}T_{20}^{(S)} + \sqrt{S(S+1)}T_{00}^{(S)}\right). \tag{C6}$$

Now the polarization operator $T_{0,0}^{(S)}$ is given by Eq. (A18) and is proportional to the unit matrix. However the commutator of any polarization operator with the unit matrix is zero. Thus the last term in Eq. (C6) can be discarded. Next, we can use Eq. (A29) regarding commutators of polarization operators from Appendix A to get

$$\left[\hat{T}_{20}^{(S)},\hat{T}_{LM}^{(S)}\right] = 2\sqrt{5(2L+1)}(-1)^{2S+L+1}\left(\begin{Bmatrix}2 & L & L+1\\ S & S & S\end{Bmatrix}C_{2,0,L,M}^{L+1,M}\hat{T}_{L+1,M}^{(S)} + \begin{Bmatrix}2 & L & L-1\\ S & S & S\end{Bmatrix}C_{20LM}^{L-1M}\hat{T}_{L-1M}^{(S)}\right)$$

from which we conclude that

$$\left[\hat{S}_0^2, \hat{T}_{LM}^{(S)}\right] = M\sqrt{\frac{\left[(L+1)^2 - M^2\right]\left[(2S+1)^2 - (L+1)^2\right]}{(2L+3)(2L+1)}}\hat{T}_{L+1M}^{(S)}$$
$$+ M\sqrt{\frac{(L^2 - M^2)\left[(2S+1)^2 - L^2\right]}{(2L-1)(2L+1)}}\hat{T}_{L-1M}^{(S)}. \tag{C7}$$

Therefore we have from Eqs. (C5)-(C7)

$$\left[\hat{S}_0^2, \hat{w}\right] = \sqrt{\frac{4\pi}{2S+1}} \sum_{L=0}^{2S}\sum_{M=-L}^{L}\left(C_{SSL0}^{SS}\right)^{-1} M Y_{LM}^*$$
$$\times \left\{\sqrt{\frac{\left[(L+1)^2 - M^2\right]\left[(2S+1)^2 - (L+1)^2\right]}{(2L+3)(2L+1)}}\hat{T}_{L+1M}^{(S)} + \sqrt{\frac{(L^2 - M^2)\left[(2S+1)^2 - L^2\right]}{(2L-1)(2L+1)}}\hat{T}_{L-1M}^{(S)}\right\}. \tag{C8}$$

Next by means of the replacement $L \pm 1 \to L$ in Eq. (C8) and subsequently using the explicit expression for the Clebsch-Gordan coefficients $C_{SSL0}^{SS}$ from Eq. (212), we then have

$$\int W_S \left[\hat{S}_0^2, \hat{w}\right] d\Omega = \sqrt{\frac{4\pi}{2S+1}} \int W_S \sum_{L=0}^{2S}\left(C_{SSL0}^{SS}\right)^{-1}\sum_{M=-L}^{L}\hat{T}_{LM}^{(S)} M$$
$$\times \left[L\sqrt{\frac{(L+1)^2 - M^2}{(2L+3)(2L+1)}}Y_{L+1M}^* - (L+1)\sqrt{\frac{L^2 - M^2}{4L^2 - 1}}Y_{L-1M}^*\right. \tag{C9}$$
$$\left. + 2(S+1)\left(\sqrt{\frac{(L+1)^2 - M^2}{(2L+3)(2L+1)}}Y_{L+1M}^* + \sqrt{\frac{L^2 - M^2}{4L^2 - 1}}Y_{L-1M}^*\right)\right] d\Omega.$$

In Eq. (C9), the terms containing the spherical harmonics $Y_{2S+1,M}$ are omitted because they vanish on averaging (due to the orthogonality relations and because the quasi-distribution function $W_S$ contains only the $Y_{LM}$ up to order $L = 2S$). By using the recursion relations of the $Y_{LM}$, Eqs. (B15) and (B16) from Appendix B, we then have the simplified expression

$$\int W_S \left[\hat{S}_0^2, \hat{w}\right] d\Omega = \sqrt{\frac{4\pi}{2S+1}}\int W_S \sum_{L=0}^{2S}\left(C_{SSL0}^{SS}\right)^{-1}\sum_{M=-L}^{L}\hat{T}_{LM}^{(S)} M$$
$$\times \left[\sin\vartheta\frac{\partial}{\partial\vartheta}Y_{LM}^* + 2(S+1)\cos\vartheta Y_{LM}^*\right] d\Omega. \tag{C10}$$

Thus, via Eq. (187), we obtain by inspection the closed form

$$\int W_S \left[\hat{S}_0^2, \hat{w}\right] d\Omega = i\int W_S \left[\sin\vartheta\frac{\partial}{\partial\vartheta} + 2(S+1)\cos\vartheta\right]\frac{\partial}{\partial\varphi}\hat{w} d\Omega, \tag{C11}$$

i.e., we have found in the *configuration representation* that the analog of the commutator $\left[\hat{S}_0^2, \hat{w}\right]$ is the differential operator

$$i\left[\sin\vartheta\partial_\vartheta + 2(S+1)\cos\vartheta\right]\partial_\varphi \hat{w}.$$

Finally, using integration by parts in Eq. (C11) in order to render it in the standard form of an inverse Wigner-Stratonovich transformation Eq. (235), we have the desired transformation via the inverse map of a Weyl symbol

$$\int W_S \left[\hat{S}_0^2, \hat{w}\right] d\Omega = i\int \hat{w}\left[\frac{1}{\sin\vartheta}\frac{\partial}{\partial\vartheta}\sin^2\vartheta - 2(S+1)\cos\vartheta\right]\frac{\partial}{\partial\varphi}W_S d\Omega. \tag{C12}$$

The above derivation has been given in detail merely as an illustration of how the inverse Wigner-Stratonovich map Eq. (235) ultimately leads to the phase space representation of the density matrix evolution equation via the integrand of Eq. (C12) and associated equations, which follow.

Next we have the following commutation relation indicated by the second Liouville term $\left[\hat{S}_0, \hat{w}_S\right]$ in Eq. (C1), viz.,

$$\left[\hat{S}_0, \hat{w}\right] = \sqrt{\frac{4\pi}{2S+1}}\sum_{L=0}^{2S}\sum_{M=-L}^{L}\left(C_{SSL0}^{SS}\right)^{-1}Y_{LM}^*\left[\hat{S}_0, \hat{T}_{LM}^{(S)}\right]$$
$$= \sqrt{\frac{4\pi}{2S+1}}\sum_{L=0}^{2S}\sum_{M=-L}^{L} M\left(C_{SSL0}^{SS}\right)^{-1}Y_{LM}^*\hat{T}_{LM}^{(S)} \tag{C13}$$
$$= i\sqrt{\frac{4\pi}{2S+1}}\sum_{L=0}^{2S}\sum_{M=-L}^{L}\left(C_{SSL0}^{SS}\right)^{-1}\frac{\partial Y_{LM}^*}{\partial\varphi}\hat{T}_{LM}^{(S)}.$$

Here we used Eqs. (230), (A1), (A21), (A29), (B7) and (187). Thus we have from Eq. (C13)

$$\left[\hat{S}_0, \hat{w}\right] = i\frac{\partial \hat{w}}{\partial\varphi}. \tag{C14}$$

Equations (C13) and (C14) (again via integration by parts) then yield the Liouville (deterministic) part of the master equation for $W_S$ via the inverse map of a Weyl symbol

$$\int W_S \left\{\frac{\sigma}{S^2}\left[\hat{S}_0^2, \hat{w}\right] + \frac{\xi}{S}\left[\hat{S}_0, \hat{w}\right]\right\} d\Omega$$
$$= i\frac{\sigma}{S^2}\int \hat{w}\left(2S\cos\vartheta - \sin\vartheta\frac{\partial}{\partial\vartheta} - \frac{S\xi}{\sigma}\right)\frac{\partial W_S}{\partial\varphi} d\Omega. \tag{C15}$$

Next, we consider the commutator $\left[\hat{S}_0\hat{w}, \hat{S}_0\right] + \left[\hat{S}_0, \hat{w}\hat{S}_0\right]$ and determine its phase space representation. We have

$$\left[\hat{S}_0\hat{w}, \hat{S}_0\right] + \left[\hat{S}_0, \hat{w}\hat{S}_0\right]$$
$$= -\sqrt{\frac{4\pi}{2S+1}}\sum_{L=0}^{2S}\sum_{M=-L}^{L}M^2\left(C_{SSL0}^{SS}\right)^{-1}Y_{LM}^*\hat{T}_{LM}^{(S)} \tag{C16}$$
$$= \sqrt{\frac{4\pi}{2S+1}}\sum_{L=0}^{2S}\sum_{M=-L}^{L}\left(C_{SSL0}^{SS}\right)^{-1}\frac{\partial^2 Y_{LM}^*}{\partial\varphi^2}\hat{T}_{LM}^{(S)} = \frac{\partial^2 \hat{w}}{\partial\varphi^2}$$

so that integrating by parts, we again obtain the standard inverse map of a Weyl symbol, viz.,

$$\int W_S \left[\hat{S}_0\hat{w}, \hat{S}_0\right] + \left[\hat{S}_0, \hat{w}\hat{S}_0\right] d\Omega = \int \hat{w}\frac{\partial^2 W_S}{\partial\varphi^2} d\Omega. \tag{C17}$$

Now we consider the remaining commutators (which are more difficult to treat) in the collision operator $\mathrm{St}(\hat{w})$, namely,

$$e^{\frac{\sigma}{2S^2}\pm\frac{\xi}{2S}}\left(\left[\hat{S}_{\pm 1}e^{\pm\frac{\sigma}{S^2}\hat{S}_0}\hat{w},\hat{S}_{\mp 1}\right]+\left[\hat{S}_{\pm 1},\hat{w}e^{\pm\frac{\sigma}{S^2}\hat{S}_0}\hat{S}_{\mp 1}\right]\right) \quad \text{(C18)}$$
$$=\left[\hat{S}_{\pm 1}\hat{P}_{\pm}^{(S)}\hat{w},\hat{S}_{\mp 1}\right]+\left[\hat{S}_{\pm 1},\hat{w}\hat{P}_{\pm}^{(S)}\hat{S}_{\mp 1}\right],$$

where for convenience we have introduced the matrix exponential operators $\hat{P}_{\pm}^{(S)} = e^{\frac{\sigma}{2S^2}\pm\frac{\xi}{2S}}e^{\pm\frac{\sigma}{S^2}\hat{S}_0}$ in Eq. (C18). However, they too can also be expanded as a series of the polarization operators (see Eq. (A34) et seq. in Appendix A), viz.,

$$\hat{P}_{\pm}^{(S)} = e^{\frac{\sigma}{2S^2}\pm\frac{\xi}{2S}}\sum_{l=0}^{2S}a_l^{\pm}\hat{T}_{l0}^{(S)}, \quad \text{(C19)}$$

where the scalar expansion coefficients $a_l^{\pm}$ can be found using the orthogonality property Eq. (A34) of the polarization operators and the explicit form of their matrix elements in terms of the Clebsch-Gordan coefficients Eq. (A17) (see Appendix A). The expansion coefficients are then as usual given by the trace

$$a_l^{\pm} = \mathrm{Tr}\left(e^{\pm\frac{\sigma}{S^2}\hat{S}_0}\hat{T}_{l0}^{(S)}\right) = \sqrt{\frac{2l+1}{2S+1}}\sum_{m=-S}^{S}C_{Sm l0}^{Sm}e^{\pm\frac{\sigma}{S^2}m}. \quad \text{(C20)}$$

However, Eq. (C20) may be further simplified for $l = 0$ using $C_{Sm00}^{Sm} = 1$ to yield the closed form

$$a_0^{\pm} = \sqrt{\frac{2l+1}{2S+1}}\sum_{m=-S}^{S}e^{\pm\frac{\sigma}{S^2}m} = \frac{\sinh\frac{\sigma(S+1/2)}{S^2}}{\sqrt{2S+1}\sinh\frac{\sigma}{2S^2}}. \quad \text{(C21)}$$

The higher order expansion coefficients may now be found because differential recurrence relations which allow one to determine $a_l^{\pm}$ may be derived as follows. We first use Eq. (A28) concerning products of polarization operators to write

$$\hat{T}_{10}^{(S)}\hat{T}_{l-10}^{(S)} = \frac{l\sqrt{3(2S-l+1)(2S+l+1)}}{2\sqrt{(2l+1)(2l-1)S(S+1)(2S+1)}}\hat{T}_{l0}^{(S)} \quad \text{(C22)}$$
$$+ \frac{(l-1)\sqrt{3(2S-l+2)(2S+l)}}{2\sqrt{(2l-3)(2l-1)S(S+1)(2S+1)}}\hat{T}_{l-20}^{(S)}.$$

Next by substituting $\hat{T}_{l0}^{(S)}$ as extracted from Eq. (C22) into the left hand side of Eq. (C20), we have

$$a_l^{\pm} = \mathrm{Tr}\left(e^{\pm\frac{\sigma}{S^2}\hat{S}_0}\hat{T}_{l0}^{(S)}\right)$$
$$= \frac{2\sqrt{(2l+1)(2l-1)S(S+1)(2S+1)}}{l\sqrt{3(2S-l+1)(2S+l+1)}}\mathrm{Tr}\left(e^{\pm\frac{\sigma}{S^2}\hat{S}_0}\hat{T}_{10}^{(S)}\hat{T}_{l-10}^{(S)}\right) \quad \text{(C23)}$$
$$- \frac{(l-1)\sqrt{(2l+1)(2S-l+2)(2S+l)}}{l\sqrt{(2l-3)(2S-l+1)(2S+l+1)}}\mathrm{Tr}\left(e^{\pm\frac{\sigma}{S^2}\hat{S}_0}\hat{T}_{l-20}^{(S)}\right).$$

Now

$$e^{\pm\frac{\sigma}{S^2}\hat{S}_0}\hat{T}_{10}^{(S)}\hat{T}_{l-10}^{(S)} = \pm S\sqrt{\frac{3S}{(S+1)(2S+1)}}\frac{\partial}{\partial\sigma}\left(e^{\pm\frac{\sigma}{S^2}\hat{S}_0}\hat{T}_{l-10}^{(S)}\right), \quad \text{(C24)}$$

therefore we have the desired differential recurrence relation for the expansion coefficients

$$a_l^{+} = \frac{2S^2}{l}\sqrt{\frac{4l^2-1}{(2S+1)^2-l^2}}\frac{\partial}{\partial\sigma}a_{l-1}^{+} + \frac{l-1}{l}\sqrt{\frac{(2l+1)[(2S+1)^2-(l-1)^2]}{(2l-3)[(2S+1)^2-l^2]}}a_{l-2}^{+} \quad \text{(C25)}$$

with

$$a_l^{+} = (-1)^l a_l^{-}. \quad \text{(C26)}$$

Next mindful of the matrix exponential operator $e^{\pm\frac{\sigma}{S^2}\hat{S}_0}$ embodied in Eq. (C19) and prompted by that equation we may regard the polarization operator $\hat{T}_{l0}^{(S)}$ as a (matrix) operator acting on the transformation kernel $\hat{w}$ and consequently may denote the corresponding differential operator by $P_l^{(S)}$ and its associated form $\overline{P}_l^{(S)}$. Both of these differential operators are defined in obvious fashion via

$$\int W_S \hat{T}_{l,0}^{(S)} \hat{w} d\Omega = \int W_S P_l^{(S)} \hat{w} d\Omega = \int \hat{w}\overline{P}_l^{(S)} W_S d\Omega. \quad \text{(C27)}$$

The last term in Eq. (C27), which serves to define the differential operator $\overline{P}_l^{(S)}$, has now the desired form, viz. an inverse Wigner-Stratonovich map. In Eq. (C27), the differential operator $\overline{P}_l^{(S)}$ is obtained from the differential operator $P_l^{(S)}$ using integration by parts in the middle term. The operator expansion defined by Eq. (C19) involving $\hat{P}_{\pm}^{(S)}$ now allows one to express both of the phase space differential operators $P_{\pm}^{(S)}$ and $\overline{P}_{\pm}^{(S)}$ as suggested by that equation in terms of the phase space differential operators $P_l^{(S)}$ and $\overline{P}_l^{(S)}$ [as defined in Eq. (C27)] corresponding to the polarization operators $\hat{T}_{l,0}^{(S)}$ as

$$\begin{pmatrix}P_{\pm}^{(S)}\\ \overline{P}_{\pm}^{(S)}\end{pmatrix} = e^{\frac{\sigma}{2S^2}\pm\frac{\xi}{2S}}\sum_{l=0}^{2S}a_l^{\pm}\begin{pmatrix}P_l^{(S)}\\ \overline{P}_l^{(S)}\end{pmatrix} \quad \text{(C28)}$$

with $a_l^{\pm}$ defined by the recurrence relations Eqs. (C25) and (C26). The differential operators $P_l^{(S)}$ and $\overline{P}_l^{(S)}$ on the right-hand side of Eq. (C28) may now be determined by upward iteration. In order to find them explicitly we first recall the expression for the product of the polarization operators embodied in Eq. (C22), which may be rearranged as the upward operator recurrence equation

$$\hat{T}_{l0}^{(S)} = A_{l-1}\hat{T}_{10}^{(S)}\hat{T}_{l-10}^{(S)} - A_{l-2}\hat{T}_{l-20}^{(S)},$$

with coefficients $A_{l-1}$ and $A_{l-2}$ given by

$$A_{l-1} = \frac{2}{l}\sqrt{\frac{S(S+1)(2S+1)(4l^2-1)}{3[(2S+1)^2 - l^2]}}, \quad (C29)$$

$$A_{l-2} = \frac{l-1}{l}\sqrt{\frac{(2l+1)[(2S+1)^2 - (l-1)^2]}{(2l-3)[(2S+1)^2 - l^2]}}. \quad (C30)$$

However, by virtue of the correspondence expressed in Eq. (C27) the phase space differential operators $P_l^{(S)}$ and $\overline{P}_l^{(S)}$ must also satisfy a similar recurrence equation, viz.,

$$\begin{pmatrix} P_l^{(S)} \\ \overline{P}_l^{(S)} \end{pmatrix} = A_{l-1} \begin{pmatrix} P_{l-1}^{(S)} P_1^{(S)} \\ \overline{P}_1^{(S)} \overline{P}_{l-1}^{(S)} \end{pmatrix} - A_{l-2} \begin{pmatrix} P_{l-2}^{(S)} \\ \overline{P}_{l-2}^{(S)} \end{pmatrix} \quad (C31)$$

so that we have explicitly for the first two members of the hierarchy of operator recurrence relations

$$P_0^{(S)} = \overline{P}_0^{(S)}(2S+1)^{-1/2} \quad (C32)$$

and

$$\begin{pmatrix} P_1^{(S)} \\ \overline{P}_1^{(S)} \end{pmatrix} = \frac{1}{2}\sqrt{\frac{3}{S(S+1)(2S+1)}}\left[(2S+1\pm 1)\cos\vartheta \pm \sin\vartheta\frac{\partial}{\partial\vartheta} \pm i\frac{\partial}{\partial\varphi}\right]. \quad (C33)$$

Thus it is now obvious that in general differential operators of arbitrary order will be involved. To establish the second member Eq. (C33), we made the following steps

$$\hat{T}_{10}^{(S)}\hat{w} = \sqrt{\frac{4\pi}{2S+1}}\sum_{L=0}^{2S}\sum_{M=-L}^{L}\left(C_{SSL0}^{SS}\right)^{-1}Y_{LM}^*\frac{1}{2}\sqrt{\frac{3}{S(S+1)(2S+1)}}\left[M\hat{T}_{LM}^{(S)}\right.$$
$$+\frac{\sqrt{(L-M+1)(L+M+1)(2S-L)(2S+L+2)}}{\sqrt{(2L+3)(2L+1)}}\hat{T}_{L+1M}^{(S)} \quad (C34)$$
$$\left.+\frac{\sqrt{(L-M)(L+M)(2S-L+1)(2S+L+1)}}{\sqrt{(2L-1)(2L+1)}}\hat{T}_{L-1M}^{(S)}\right].$$

Next by the replacement $L\pm 1 \to L$ in Eq. (C34), and then using the explicit expression for the Clebsch-Gordan coefficients $C_{SSL0}^{SS}$ from Eq. (212) we have

$$\hat{T}_{10}^{(S)}\hat{w} = \sqrt{\frac{4\pi}{2S+1}}\sum_{L=0}^{2S}\sum_{M=-L}^{L}\left(C_{SSL0}^{SS}\right)^{-1}\hat{T}_{LM}^{(S)}\sqrt{\frac{3}{S(S+1)(2S+1)}}$$
$$\times\left[\frac{M}{2}Y_{LM}^* + \frac{2S-L+1}{2}\sqrt{\frac{(L-M)(L+M)}{(2L+1)(2L-1)}}Y_{L-1M}^*\right.$$
$$\left.+\frac{2S+L+2}{2}\sqrt{\frac{(L-M+1)(L+M+1)}{(2L+1)(2L+3)}}Y_{L+1M}^*\right] \quad (C35)$$
$$= \sqrt{\frac{3}{S(S+1)(2S+1)}}\left[(S+1)\cos\vartheta + \frac{1}{2}\sin\vartheta\frac{\partial}{\partial\vartheta} + \frac{i}{2}\frac{\partial}{\partial\varphi}\right]\hat{w}$$
$$= P_1^{(S)}\hat{w}.$$

Consequently by substituting Eq. (C35) into the defining Eq. (C27) and integrating by parts we have Eq. (C33) above.

Returning to the phase-space representation of the commutators in Eq. (C18), the matrix exponential operators $\hat{P}_\pm^{(S)}$ act on the *polarization operators* in the expansion of transformation kernel $\hat{w}$ while its phase-space correspondents $P_\pm^{(S)}$ from Eq. (C28), in contrast, represent sets of differential operators acting on the *spherical harmonics* in $\hat{w}$ leading in the end to the same result, viz.,.

$$\int W_S[\hat{S}_{\pm 1}\hat{P}_\pm^{(S)}\hat{w},\hat{S}_{\mp 1}]d\Omega = \int \overline{P}_\pm^{(S)}W_S[\hat{S}_{\pm 1}\hat{w},\hat{S}_{\mp 1}]d\Omega. \quad (C36)$$

Because the commutator $[\hat{S}_{\pm 1}\hat{P}_\pm^{(S)}\hat{w},\hat{S}_{\mp 1}]$ gives the differential operator, which is complex conjugate to that corresponding to $[\hat{S}_{\pm 1},\hat{P}_\pm^{(S)}\hat{w}\hat{S}_{\mp 1}]$ in the right-hand side of Eq. (C18), we need to consider below only the commutator $[\hat{S}_{\pm 1}\hat{P}_\pm^{(S)}\hat{w},\hat{S}_{\mp 1}]$. We have

$$[\hat{S}_{\pm 1}\hat{w},\hat{S}_{\mp 1}] = \frac{\sqrt{4\pi}}{\sqrt{2S+1}}\sum_{L=0}^{2S}\sum_{M=-L}^{L}\left(C_{SSL0}^{SS}\right)^{-1}Y_{LM}^*(\vartheta,\varphi)\left[\hat{S}_{\pm 1}\hat{T}_{LM}^{(S)},\hat{S}_{\mp 1}\right], \quad (C37)$$

where

$$\left[\hat{S}_{\pm 1}\hat{T}_{LM}^{(S)},\hat{S}_{\mp 1}\right] = -\frac{1}{4}\left[(L\mp M)(L\pm M+1)\hat{T}_{LM}^{(S)}\right.$$
$$\mp \frac{(L\pm M+2)\sqrt{(L+M+1)(L-M+1)(2S-L)(2S+L+2)}}{\sqrt{(2L+3)(2L+1)}}\hat{T}_{L+1M}^{(S)} \quad (C38)$$
$$\left.\pm \frac{(L\mp M-1)\sqrt{(L+M)(L-M)(2S-L+1)(2S+L+1)}}{\sqrt{(2L-1)(2L+1)}}\hat{T}_{L-1M}^{(S)}\right].$$

Next via the replacement $L\pm 1 \to L$ in Eq. (C8) and then using the explicit expression for the Clebsch-Gordan coefficients $C_{SSL0}^{SS}$ from Eq. (212) we have

$$[\hat{S}_{\pm 1}\hat{w},\hat{S}_{\mp 1}] = -\frac{1}{4}\frac{\sqrt{4\pi}}{\sqrt{2S+1}}\sum_{L=0}^{2S}\sum_{M=-L}^{L}\left(C_{SSL0}^{SS}\right)^{-1}\hat{T}_{LM}^{(S)}\left[(L\mp M)(L\pm M+1)Y_{LM}^{*}\right.$$

$$\mp\frac{(2S+L+2)(L\pm M+1)\sqrt{(L+M+1)(L-M+1)}}{\sqrt{(2L-1)(2L+1)}}Y_{L-1M}^{*}$$  (C39)

$$\left.\pm\frac{(2S-L+1)(L\mp M)\sqrt{(L+M)(L-M)}}{\sqrt{(2L+3)(2L+1)}}Y_{L+1M}^{*}\right].$$

Now, we can ultimately write Eq. (C39) in differential form using properties of the angular momentum operators $\hat{L}_\mu$ and $\hat{L}^2$ (see Appendix B), viz.,

$$\hat{L}^2 Y_{LM}^* = -\left[\frac{1}{\sin\theta}\frac{\partial}{\partial\vartheta}\left(\sin\vartheta\frac{\partial}{\partial\vartheta}\right)+\frac{1}{\sin^2\vartheta}\frac{\partial^2}{\partial^2\varphi}\right]Y_{LM}^*$$  (C40)
$$= L(L+1)Y_{LM}^*,$$

$$\hat{L}_0 Y_{LM}^* = -i\frac{\partial}{\partial\varphi}Y_{LM}^* = -MY_{LM}^*,$$  (C41)

$$\hat{L}_{\pm 1}Y_{LM}^* = -\frac{1}{\sqrt{2}}e^{\pm i\varphi}\left(\frac{\partial}{\partial\vartheta}\pm i\cot\vartheta\frac{\partial}{\partial\varphi}\right)Y_{LM\mp 1}^*$$  (C42)
$$= \pm\sqrt{\frac{L(L+1)-M(M\mp 1)}{2}}Y_{LM\mp 1}^*,$$

and

$$\hat{L}_{\pm}\hat{L}_{\mp}Y_{LM}^* = \frac{1}{2}\left(\frac{\partial^2}{\partial\vartheta^2}+\cot\vartheta\frac{\partial}{\partial\vartheta}\pm i\frac{\partial}{\partial\varphi}+\cot^2\vartheta\frac{\partial^2}{\partial^2\varphi}\right)Y_{LM}^*$$  (C43)
$$= -\frac{1}{2}(L\mp M)(L\pm M+1)Y_{LM}^*.$$

Thus we can rearrange the commutator given by Eq. (C37) as the differential form

$$[\hat{S}_{\pm 1}\hat{w},\hat{S}_{\mp 1}] = \sqrt{\frac{\pi}{2S+1}}\sum_{L=0}^{2S}\sum_{M=-L}^{L}\left(C_{SSL0}^{SS}\right)^{-1}\hat{T}_{LM}^{(S)}$$

$$\times\left[\frac{1}{2}\left(\frac{\partial^2}{\partial\vartheta^2}+\cot\vartheta\frac{\partial}{\partial\vartheta}\pm i\frac{\partial}{\partial\varphi}+\cot^2\vartheta\frac{\partial^2}{\partial^2\varphi}\right)\right.$$

$$+i(S+1)\cos\vartheta\frac{\partial}{\partial\varphi}-\left(S+\frac{1}{2}\right)\sin\vartheta\frac{\partial}{\partial\vartheta}+\frac{i}{2}\sin\vartheta\frac{\partial}{\partial\vartheta}\frac{\partial}{\partial\varphi}$$  (C44)

$$\left.-\frac{1}{2}\cos\vartheta\left[\frac{1}{\sin\theta}\frac{\partial}{\partial\vartheta}\left(\sin\vartheta\frac{\partial}{\partial\vartheta}\right)+\frac{1}{\sin^2\vartheta}\frac{\partial^2}{\partial^2\varphi}\right]\right]Y_{LM}^*.$$

By substituting Eq. (C44) into Eq. (C36) and integrating the latter by parts, we then have that equation rendered as *the inverse map of a Weyl symbol*, viz.,

$$\int W_S[\hat{S}_{\pm 1}\hat{P}_{\pm}^{(S)}\hat{w},\hat{S}_{\mp 1}]d\Omega = -\frac{1}{4}\int\hat{w}\left\{i\left(\sin\vartheta\frac{\partial}{\partial\vartheta}-2S\cos\vartheta\mp 1\right)\frac{\partial}{\partial\varphi}\right.$$

$$\left.+\frac{1}{\sin\vartheta}\left[\frac{\partial}{\partial\vartheta}\left(\sin\vartheta(1\pm\cos\vartheta)\frac{\partial}{\partial\vartheta}\pm 2S\sin^2\vartheta\right)+\cot\vartheta\left(\cos\vartheta\pm 1\right)\frac{\partial^2}{\partial\varphi^2}\right]\right\}\bar{P}_{\pm}^{(S)}W_S d\Omega.$$  (C45)

In the derivation of Eq. (C45), the recurrence properties of the spherical harmonics given in Appendix B, namely, Eqs. (B14)-(B17), have been used. Finally, because the commutator $[\hat{S}_{\pm 1}\hat{P}_{\pm}^{(S)}\hat{w},\hat{S}_{\mp 1}]$ gives the differential operator which is complex conjugate to that corresponding to $[\hat{S}_{\pm 1},\hat{P}_{\pm}^{(S)}\hat{w}\hat{S}_{\mp 1}]$, we have once again the standard inverse Wigner-Stratonovich map of a Weyl symbol via

$$\int W_S\left([\hat{S}_{\pm 1}\hat{P}_{\pm}^{(S)}\hat{w},\hat{S}_{\mp 1}]+[\hat{S}_{\pm 1},\hat{w}\hat{P}_{\pm}^{(S)}\hat{S}_{\mp 1}]\right)d\Omega$$

$$= e^{\frac{\sigma}{2S^2}\pm\frac{\xi}{2S}}\int W_S\left([\hat{S}_{\pm 1}e^{\pm\frac{\sigma}{S^2}\hat{S}_0}\hat{w},\hat{S}_{\mp 1}]+[\hat{S}_{\pm 1},\hat{w}e^{\pm\frac{\sigma}{S^2}\hat{S}_0}\hat{S}_{\mp 1}]\right)d\Omega$$

$$= -\frac{1}{2}\int\hat{w}\operatorname{Re}\left\{\left[\frac{1}{\sin\vartheta}\frac{\partial}{\partial\vartheta}\left(\sin\vartheta(1\pm\cos\vartheta)\frac{\partial}{\partial\vartheta}\pm 2S\sin^2\vartheta\right)\right.\right.$$  (C46)

$$\left.\left.+\frac{1}{\sin\vartheta}\cot\vartheta\left(\cos\vartheta\pm 1\right)\frac{\partial^2}{\partial\varphi^2}+i\left(\sin\vartheta\frac{\partial}{\partial\vartheta}-2S\cos\vartheta\mp 1\right)\frac{\partial}{\partial\varphi}\right]\bar{P}_{\pm}^{(S)}\right\}W_S d\Omega.$$

Then due to Eq. (376), we have from the Weyl symbols of Eqs. (C15), (C17), and (C46) the master equation for the phase-space distribution $W_S(\vartheta,\varphi,t)$, viz.,

$$\frac{\partial W_S}{\partial t}-\frac{\sigma}{\hbar\beta S^2}\left(2S\cos\vartheta-\sin\vartheta\frac{\partial}{\partial\vartheta}-\frac{S\xi}{\sigma}\right)\frac{\partial W_S}{\partial\varphi}=\operatorname{St}\{W_S\},$$  (C47)

where the collision kernel $\operatorname{St}\{W_S\}$ is

$$\operatorname{St}\{W_S\} = D_{\parallel}\frac{\partial^2}{\partial^2\varphi}W_S + D_{\perp}\frac{\cot\vartheta}{\sin\vartheta}\left(\cos\vartheta\frac{\partial^2}{\partial^2\varphi}R_+^{\prime\prime(S)}+\frac{\partial^2}{\partial^2\varphi}R_-^{\prime\prime(S)}\right)W_S$$

$$+\frac{D_{\perp}}{\sin\vartheta}\left[\frac{\partial}{\partial\vartheta}\sin\vartheta\left(\frac{\partial}{\partial\vartheta}R_+^{\prime\prime(S)}+\cos\vartheta\frac{\partial}{\partial\vartheta}R_-^{\prime\prime(S)}+2S\sin\vartheta R_-^{\prime\prime(S)}\right)\right]W_S$$  (C48)

$$+D_{\perp}\left[\left(\sin\vartheta\frac{\partial}{\partial\vartheta}-2S\cos\vartheta\right)\frac{\partial}{\partial\varphi}R_+^{\prime\prime(S)}-\frac{\partial}{\partial\varphi}R_-^{\prime\prime(S)}\right]W_S$$

with the phase space differential operators $R_{\pm}^{(S)}$ generally involving differential operators of arbitrary order [cf. Eq. (C28)] defined as

$$R_{\pm}^{(S)} = R_{\pm}^{\prime(S)}+iR_{\pm}^{\prime\prime(S)} = \bar{P}_+^{(S)}\pm\bar{P}_-^{(S)}.$$  (C49)

In the classical limit, we see that since all the terms involving derivatives in the differential operator expansion [Eq. (C28) *et seq.*] now vanish, the operator $\bar{P}_l^{(S)}$ simply reduces to the Legendre polynomial of order $l$

$$\bar{P}_l^{(S\gg 1)} \to Y_{l0}(\vartheta,\varphi)\sqrt{\frac{4\pi}{2S+1}}.$$  (C50)

Thus in this limit by substituting the Legendre polynomials into the operator series Eq. (C28), we must then have

$$\bar{P}_{\pm}^{(S\to\infty)}=e^{\frac{\sigma}{2S^2}\pm\frac{\xi}{2S}}\sqrt{\frac{4\pi}{2S+1}}\sum_{l=0}^{2S}a_l^{\pm}Y_{l0}(\vartheta,\varphi)=e^{\frac{\sigma}{2S^2}\pm\frac{\xi}{2S}}F_{\pm}^{(0)}(\vartheta,\varphi),$$

in which the series

$$F_{\pm}^{(s)}(\vartheta,\varphi)=\sqrt{\frac{4\pi}{2S+1}}\sum_{l=0}^{2S}\left(C_{SSL0}^{SS}\right)^{-s}a_l^{\pm}Y_{l0}(\vartheta,\varphi)$$

represents the Weyl symbol of the Hilbert space operator $e^{\pm\sigma\hat{S}_0^2/S^2}$. For large $S$,

$$F_{\pm}^{(s)}(\vartheta,\varphi)\to e^{\pm(\sigma/S)\cos\vartheta}$$

so that proceeding to the limit $S\to\infty$, we have

$$\lim_{S\to\infty}S\left(e^{[\sigma\cos\vartheta+(\sigma/S+\xi)/2]/S}-e^{[-\sigma\cos\vartheta+(\sigma/S-\xi)/2]/S}\right)=\xi+2\sigma\cos\vartheta$$

and

$$\lim_{S\to\infty}e^{(\xi+\sigma/S+2\sigma\cos\vartheta)/S}=1.$$

Hence it follows that in the classical limit, $S\to\infty$, Eq. (C47) reduces to the corresponding classical Fokker-Planck equations for isotropic rotational diffusion of a magnetic dipole in the uniaxial potential [5] (if $D_\perp=D_\parallel$)

$$\frac{\partial W}{\partial t}=\frac{\gamma}{\mu\sin\vartheta}\frac{\partial V}{\partial\vartheta}\frac{\partial W}{\partial\varphi}$$
$$+D_\perp\left[\frac{1}{\sin\vartheta}\frac{\partial}{\partial\vartheta}\left(\sin\vartheta\frac{\partial W}{\partial\vartheta}\right)+\frac{1}{\sin^2\vartheta}\frac{\partial^2 W}{\partial\varphi^2}+\frac{\beta}{\sin\vartheta}\frac{\partial}{\partial\vartheta}\left(\sin\vartheta W\frac{\partial V}{\partial\vartheta}\right)\right],\quad\text{(C51)}$$

where $\mu=\gamma\hbar S/\mu_0$ is the magnetic dipole moment and

$$\beta V(\vartheta)=-\sigma\cos^2\vartheta-\xi\cos\vartheta$$

is the normalized free energy density.

The master equation (C47) for the evolution of the phase space quasi-probability distribution $W_S(\vartheta,\varphi,t)$ for a uniaxial spin system in contact with a heat bath at temperature $T$ was derived in the weak coupling limit, i.e., it was supposed that the correlation time characterizing the bath is so short that the stochastic process originating from it is Markovian so that one may assume frequency independent damping. This has been accomplished by expressing the reduced density matrix master equation (371) in terms of the inverse Wigner-Stratonovich transformation. In order to achieve this objective various commutators involving the spin operators occurring in the integrand of Eq. (376) have then been evaluated by means of the orthogonality and recurrence properties of the polarization operators and the corresponding spherical harmonics to yield their analogs in phase space. Thus, we have expressed the master equation as a partial differential equation for the distribution function in the phase space of the polar angles. Despite the superficial resemblance of the quantum diffusion Eq. (C47) to the corresponding classical Fokker-Planck equation for a classical spin in a uniaxial potential Eq.

(C51), it is in reality much more complicated. The complications arise because Eq. (C47) involves the complicated differential operators $R_\pm^{(S)}$ only simplifying for large spin numbers ($S\to\infty$) when the higher order derivatives, as indicated by Eq. (C49), occurring in the operators $R_\pm^{(S)}$ may be ignored. Consequently it is often much easier to use the density matrix formulation where for axially symmetric problems only the diagonal terms partake in the time evolution.

## APPENDIX D: BROWN'S THEORY OF THE BROWNIAN MOTION OF A CLASSICAL SPIN

The rigorous treatment of the magnetization dynamics of fine magnetic particles in the presence of thermal agitation was set in the context of the general theory of stochastic processes by W.F. Brown [23,24] via the classical theory of the Brownian motion using by analogy ideas originating in the Debye theory of dielectric relaxation of polar dielectrics [7,8]. The starting point of Brown's treatment [23,24] of the dynamical behavior of the magnetization **M** for a single-domain particle was Gilbert's equation [26], viz., [cf. Eq. (2)]

$$\dot{\mathbf{u}}=\gamma\left[\mathbf{u}\times\left(\mathbf{H}_{\text{ef}}-\alpha\dot{\mathbf{u}}/\gamma\right)\right] \quad\text{(D1)}$$

(here $\mathbf{u}=M_S^{-1}\mathbf{M}$ is a unit vector in the direction of **M**). In general,

$$\mathbf{H}_{\text{ef}}=-\partial V/\partial\mathbf{M}\text{ and }-\alpha\dot{\mathbf{u}}/\gamma \quad\text{(D2)}$$

represent the *conservative* and *dissipative* parts of an "effective field", respectively. Brown now supposes in order to treat thermal agitation that the dissipative "effective field" $-\alpha\dot{\mathbf{u}}/\gamma$ describes only the *statistical average* of the rapidly fluctuating random field due to thermal agitation, and that this term for an individual particle must become

$$-\alpha\dot{\mathbf{u}}/\gamma\to-\alpha\dot{\mathbf{u}}/\gamma+\mathbf{h}(t),$$

where the random field $\mathbf{h}(t)$ has the white noise properties

$$\overline{h_i(t)}=0,\ \overline{h_i(t_1)h_j(t_2)}=\frac{2kT\alpha}{v\gamma\mu_0 M_S}\delta_{ij}\delta(t_1-t_2). \quad\text{(D3)}$$

Here the indices $i,j=1,2,3$ in Kronecker's delta $\delta_{ij}$ and $h_i$ correspond to the Cartesian axes $X,Y,Z$ of the laboratory coordinate system $OXYZ$, $\delta(t)$ is the Dirac-delta function, and the overbar means the statistical average over an ensemble of particles which all have at time $t$ the *same* magnetization **M**. The random field accounts for the thermal fluctuations of the magnetization of an individual particle without which the random orientational motion would not be sustained.

Brown was then able to derive, after a long and tedious calculation using the methods of Wang and Uhlenbeck [195], the Fokker–Planck equation Eq. (3) for the distribution function $W(\vartheta,\varphi,t)$ of the orientations of the magnetic moment vector $\boldsymbol{\mu}=v\mathbf{M}$ ($v$ is the volume of the particle) on the surface of the unit sphere. This lengthy procedure may be circumvented, however, by using an alternative approach also given by him [23] which appears to be based on an argument originally due to Einstein [196] in order to heuristically derive the Smoluchowski equation for point particles. Einstein accomplished this by adding a diffusion current representing the effect of the heat bath on the deterministic drift current due to an external force. In order to illustrate this method, we first write (cross-multiplying vectorially by $\mathbf{u}$ and using the triple vector product formula) Gilbert's equation (D1) in the absence of thermal agitation (noiseless equation) as an explicit equation for $\dot{\mathbf{u}}$. Transposing the $\alpha$ term, we have

$$\dot{\mathbf{u}} + \alpha[\mathbf{u}\times\dot{\mathbf{u}}] = \gamma[\mathbf{u}\times\mathbf{H}_{\mathrm{ef}}]. \tag{D4}$$

Cross-multiplying vectorially by $\mathbf{u}$ in Eq. (D4), using the triple vector product formula

$$[[\mathbf{u}\times\dot{\mathbf{u}}]\times\mathbf{u}] = \dot{\mathbf{u}} - \mathbf{u}(\mathbf{u}\cdot\dot{\mathbf{u}}), \tag{D5}$$

we obtain

$$[\dot{\mathbf{u}}\times\mathbf{u}] = -\alpha\dot{\mathbf{u}} + \gamma[[\mathbf{u}\times\mathbf{H}_{\mathrm{ef}}]\times\mathbf{u}] \tag{D6}$$

because $(\mathbf{u}\cdot\dot{\mathbf{u}}) = 0$. Substituting Eq. (D6) into Eq. (D4) yields the explicit solution for $\dot{\mathbf{u}}$ in the Landau-Lifshitz form [25,94]

$$\dot{\mathbf{u}} = \alpha^{-1}h'M_S[\mathbf{u}\times\mathbf{H}_{\mathrm{ef}}] + h'M_S[[\mathbf{u}\times\mathbf{H}_{\mathrm{ef}}]\times\mathbf{u}], \tag{D7}$$

where $h'$ is Brown's parameter defined as $h' = \gamma/[(\alpha+\alpha^{-1})M_S]$. With Eq. (D2), Eq. (D7) becomes

$$\dot{\mathbf{u}} = -\frac{h'}{\alpha}\left[\mathbf{u}\times\frac{\partial V}{\partial \mathbf{u}}\right] + h'\left[\mathbf{u}\times\left[\mathbf{u}\times\frac{\partial V}{\partial \mathbf{u}}\right]\right]. \tag{D8}$$

Now the instantaneous orientation $(\vartheta,\varphi)$ of the magnetization $\mathbf{M}$ of a single-domain particle may be represented by a point on the unit sphere $(1,\vartheta,\varphi)$. As the magnetization changes its direction the representative point moves over the surface of the sphere. Following [5,23], consider now a statistical ensemble of identical particles and let $W(\vartheta,\varphi,t)d\Omega$ be the probability that $\mathbf{u}$ has orientation $(\vartheta,\varphi)$ within the solid angle $d\Omega = \sin\vartheta d\vartheta d\varphi$ (see Fig. 45). The time derivative of $W(\vartheta,\varphi,t)$ is then related to the probability current $\mathbf{J}$ of such representative points swarming over the surface $S$ of the sphere by the continuity equation

$$\dot{W} + \mathrm{div}\mathbf{J} = 0. \tag{D9}$$

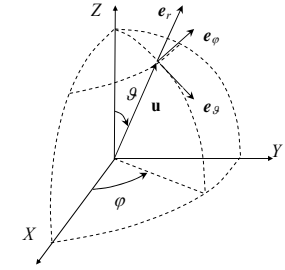

**Figure 45.** Spherical polar coordinate system.

Equation (D9) states that the swarming representative points are neither created nor destroyed, merely moving to new positions on the surface of the sphere. Now in the absence of thermal agitation, we have the deterministic drift current $\mathbf{J} = W\dot{\mathbf{u}}$, where $\dot{\mathbf{u}}$ is given by Eq. (D8). Next add to this deterministic $\mathbf{J}$ a diffusion term $-k'\partial_{\mathbf{u}}W$ ($k'$ is a proportionality constant to be determined later), which represents the effect of thermal agitation; its tendency is to smooth out the distribution, i.e., to make it more uniform. Recall the alternative and equivalent Langevin picture of a systematic retarding torque tending to slow down the spin superimposed on a rapidly fluctuating white noise random torque maintaining the rotational motion. This intuitive procedure essentially due to Einstein gives for the components of $\mathbf{J}$ (on evaluating $[\mathbf{u}\times\partial_{\mathbf{u}}V]$, etc. in spherical polar coordinates)

$$J_\vartheta = -h'\left[\left(\frac{\partial V}{\partial \vartheta} - \frac{1}{\alpha\sin\vartheta}\frac{\partial V}{\partial \varphi}\right)W + \frac{k'}{h'}\frac{\partial W}{\partial \vartheta}\right], \tag{D10}$$

$$J_\varphi = -h'\left[\left(\frac{1}{\alpha}\frac{\partial V}{\partial \vartheta} + \frac{1}{\sin\vartheta}\frac{\partial V}{\partial \varphi}\right)W + \frac{k'}{h'\sin\vartheta}\frac{\partial W}{\partial \varphi}\right]. \tag{D11}$$

Equations (D10) and (D11), when substituted into the continuity Eq. (D9), now yield Brown's Fokker–Planck equation for the surface density of magnetic moment orientations on the unit sphere, which may be written as

$$\frac{\partial W}{\partial t} = \frac{1}{2\tau_N \sin\vartheta}\left\{\left[\frac{\partial}{\partial \vartheta}\left(\sin\vartheta\frac{\partial}{\partial \vartheta}W\right) + \frac{1}{\sin\vartheta}\frac{\partial^2 W}{\partial \varphi^2}\right]\right.$$
$$\left. + \frac{v}{kT}\left[\frac{\partial}{\partial \vartheta}\left(\sin\vartheta\frac{\partial V}{\partial \vartheta} - \alpha^{-1}\frac{\partial V}{\partial \varphi}\right)W + \frac{\partial}{\partial \varphi}\left(\alpha^{-1}\frac{\partial V}{\partial \vartheta} + \frac{1}{\sin\vartheta}\frac{\partial V}{\partial \varphi}\right)W\right]\right\}, \tag{D12}$$

or, equivalently, in the compact vector form of Eq. (3). Here,

$$\tau_N = \frac{v\mu_0 M_S(\alpha+\alpha^{-1})}{2\gamma kT}$$

is the free diffusion time and the constant $k' = kTh'/v = (2\tau_N)^{-1}$ was evaluated by requiring that the Boltzmann distribution $W_{eq}(\vartheta,\varphi) = Ae^{-vV(\vartheta,\varphi)/(kT)}$ of orientations ($A$ is a normalizing constant) should be the stationary (equilibrium) solution of the Fokker–Planck equation Eq. (D12). Here we have given Brown's intuitive derivation of his magnetic Fokker–Planck equation, Eq. (2), for the isotropic Brownian motion of the classical spin. A rigorous derivation of that equation from the Gilbert-Langevin equation (2) is given elsewhere [5,23,194].

Now Brown's Fokker–Planck equation (2) for the probability density function $W(\vartheta,\varphi,t)$ of orientations of the unit vector **u** in configuration space $(\vartheta,\varphi)$, can be solved by separation of the variables. This gives rise to a Sturm–Liouville problem so that $W(\vartheta,\varphi,t)$ can be written as

$$W(\vartheta,\varphi,t) = W_0(\vartheta,\varphi) + \sum_{k=1}^{\infty} \Phi_k(\vartheta,\varphi) e^{-\lambda_k t}, \tag{D13}$$

where $\Phi_k(\vartheta,\varphi)$ and $\lambda_k$ are the eigenfunctions and eigenvalues of the Fokker–Planck operator $L_{FP}$ and $W_0(\vartheta,\varphi)$ is the stationary solution of that equation, i.e., $L_{FP}W_0 = 0$, corresponding to Boltzmann equilibrium. Then, the reversal time of the magnetization $\tau$ can be estimated [5,6] via the inverse of the smallest nonvanishing eigenvalue $\lambda_1$ of the operator $L_{FP}$ in Eq. (2), viz.,

$$\tau = 1/\lambda_1. \tag{D14}$$

An alternative method involving the observables directly is to expand $W(\vartheta,\varphi,t)$ as a Fourier series of appropriate orthogonal functions forming an orthonormal basis related to them; here these are the spherical harmonics $Y_{lm}(\vartheta,\varphi)$ (see Appendix B), viz.,

$$W(\vartheta,\varphi,t) = \sum_{l=0}^{\infty} \sum_{m=-l}^{l} Y_{lm}^*(\vartheta,\varphi) \langle Y_{lm} \rangle(t), \tag{D15}$$

where by orthogonality the expectation values of the spherical harmonics are given by

$$\langle Y_{lm} \rangle(t) = \int_0^{2\pi} \int_0^{\pi} W(\vartheta,\varphi,t) Y_{lm}(\vartheta,\varphi) \sin\vartheta \, d\vartheta \, d\varphi. \tag{D16}$$

Moreover, for *arbitrary* magnetocrystalline anisotropy, which can be expressed in terms of spherical harmonics as

$$\frac{vV(\vartheta,\varphi)}{kT} = \sum_{R=1}^{\infty} \sum_{S=-R}^{R} A_{R,S} Y_{RS}(\vartheta,\varphi), \tag{D17}$$

we have by assuming a solution in the form of the Fourier expansion Eq. (D15) for the Fokker-Planck equation (2), an infinite hierarchy of differential-recurrence equations for the statistical moments $\langle Y_{lm} \rangle(t)$, viz., (details are in Refs. 5 and 145)

$$\tau_N \frac{d}{dt}\langle Y_{lm} \rangle(t) = \sum_{s,r} e_{l,m,l+r,m+s} \langle Y_{l+r m+s} \rangle(t). \tag{D18}$$

In Eq. (D18), the $e_{l,m,l',m\pm s}$ are the matrix elements of the Fokker-Planck operator expressed as

$$e_{l,m,l',m\pm s} = -\frac{l(l+1)}{2}\delta_{ll'}\delta_{s0} + (-1)^m \frac{1}{4}\sqrt{\frac{(2l+1)(2l'+1)}{\pi}}$$
$$\times \sum_{r=s}^{\infty} A_{r,\pm s} \left\{ \frac{[l'(l'+1) - r(r+1) - l(l+1)]}{2\sqrt{2r+1}} C_{l0l'0}^{r0} C_{lml'-m\mp s}^{r\mp s} + \frac{i}{\alpha}\sqrt{\frac{(2r+1)(r-s)!}{(r+s)!}} \right.$$
$$\left. \times \sum_{\substack{L=s-\varepsilon_{r,s} \\ \Delta L=2}}^{r-1} \sqrt{\frac{(L+s)!}{(L-s)!}} C_{l0l'0}^{L0} \left[ mC_{lml'-m\mp s}^{L\mp s} \pm s\sqrt{\frac{(l\mp m)(l\pm m+1)}{(L+s)(L-s+1)}} C_{lm\pm 1l'-m\mp s}^{L\mp s\pm 1} \right] \right\}, \tag{D19}$$

where $s \geq 0$ and $C_{lml'm'}^{rs}$ are the Clebsch–Gordan coefficients. We remark that Eq. (D19) determines the coefficients of the linear combination $e_{l,m,l',m'}$ for *arbitrary* magnetocrystalline anisotropy and Zeeman energy densities. The Gilbert-Langevin equation, Eq. (2), can also be reduced to the moment system for $\langle Y_{lm} \rangle(t)$, Eq. (D18), by an appropriate transformation of variables and by direct averaging (without recourse to the Fokker–Planck equation) of the stochastic equation thereby obtained [5,145]. Examples of explicit calculations of the $e_{l,m,l',m'}$ for particular magnetocrystalline anisotropies, are available in Refs. 5 and 6 and further references therein.

The recurrence Eq. (D18) may always be written in matrix form as

$$\dot{\mathbf{X}}(t) = \mathbf{A}\mathbf{X}(t), \tag{D20}$$

where **A** is the system matrix and $\mathbf{X}(t)$ is an infinite column vector formed from $\langle Y_{lm} \rangle(t)$. The general solution of Eq. (E7) is determined by successively increasing the size of **A** until convergence is attained. Alternatively, we can always transform the moment systems, Eqs. (D18) into the tri-diagonal vector differential-recurrence equation

$$\tau_N \dot{\mathbf{C}}_n(t) = \mathbf{Q}_n^- \mathbf{C}_{n-1}(t) + \mathbf{Q}_n \mathbf{C}_n(t) + \mathbf{Q}_n^+ \mathbf{C}_{n+1}(t), \tag{D21}$$

where $\mathbf{C}_n(t)$ are column vectors arranged in an appropriate way from $\langle Y_{lm} \rangle(t)$ and $\mathbf{Q}_n^\pm, \mathbf{Q}_n$ are matrixes with elements $e_{l',m',l,m}$. As shown in Ref. 197 (see also Ref. 5, Chapter 2), the *exact matrix continued fraction solution* of Eq. (D21) for the Laplace transform of $\mathbf{C}_1(t)$ is given by

$$\tilde{\mathbf{C}}_1(s) = \tau_N \mathbf{\Delta}_1(s) \left\{ \mathbf{C}_1(0) + \sum_{n=2}^{\infty} \left[ \prod_{k=2}^{n} \mathbf{Q}_{k-1}^+ \mathbf{\Delta}_k(s) \right] \mathbf{C}_n(0) \right\}, \tag{D22}$$

where

$$\tilde{\mathbf{C}}_1(s) = \int_0^{\infty} \mathbf{C}_1(t) e^{-st} dt,$$

$\mathbf{\Delta}_n(s)$ is the matrix continued fraction defined by the recurrence equation

$$\boldsymbol{\Delta}_n(s) = \left[\tau_N s \mathbf{I} - \mathbf{Q}_n - \mathbf{Q}_n^+ \boldsymbol{\Delta}_{n+1}(s) \mathbf{Q}_{n+1}^-\right]^{-1}, \tag{D23}$$

and $\mathbf{I}$ is the unit matrix. Having determined $\tilde{\mathbf{C}}_1(s)$, one may evaluate all the relevant observables.

Hitherto we have used Gilbert's form of the Langevin equation, namely, Eq. (2) and its accompanying Fokker–Planck equation, Eq. (D12). Equations (2) and (D12) often occur in stochastic magnetization dynamics. Brown [23,24] justified his use of the Gilbert equation because all the terms in it can be derived from a Lagrangian function and a Rayleigh dissipation function. Moreover, Gilbert's equation fits naturally into escape rate theory in all damping ranges if the damping torque is regarded as the time average of a fluctuating torque, whose instantaneous value contains also a random term with statistical properties. However, in the literature, alternative forms of the Langevin equations governing the magnetization $\mathbf{M}(t)$ have also been proposed. Two other frequently used Langevin equations for stochastic spin dynamics are the Landau–Lifshitz (e.g., [25]) and Kubo [27,90,91] forms, respectively,

$$\dot{\mathbf{u}}(t) = \gamma \mathbf{u}(t) \times \left[\mathbf{H}_{ef}(t) + \mathbf{h}(t)\right] - \gamma \alpha \mathbf{u}(t) \times \left[\mathbf{u}(t) \times [\mathbf{H}_{ef}(t) + \mathbf{h}(t)]\right] \tag{D24}$$

and

$$\dot{\mathbf{u}}(t) = \gamma \mathbf{u}(t) \times \left[\mathbf{H}_{ef}(t) + \mathbf{h}(t)\right] - \gamma \alpha \mathbf{u}(t) \times \left[\mathbf{u}(t) \times \mathbf{H}_{ef}(t)\right]. \tag{D25}$$

The difference between these two models is that in the Kubo Eq. (D25) the random field $\mathbf{h}(t)$ appears only in the gyromagnetic term. In general, the explicit form of the infinite hierarchy of differential-recurrence equations for the statistical moments depends on the Langevin equation. Furthermore, the corresponding Fokker–Planck equation is also determined by that equation. Nevertheless, all the Langevin equations, Eqs. (2), (D24), and (D25), yield very similar hierarchies and Fokker–Planck equations, the only difference being in the definition of the free diffusion time $\tau_N$ (see for details Ref. [5,6]). Moreover, the Kubo and Landau–Lifshitz models, despite the different forms of the Langevin equations Eqs. (D24) and (D25), yield *identical* mathematical forms for the corresponding Fokker–Planck equations. Thus the Gilbert, Kubo, and Landau–Lifshitz models for Brownian motion of classical spins irrespective of the Langevin equations, yield the *same* form of the corresponding Fokker–Planck equations, as well as the same infinite hierarchy of differential-recurrence equations for the statistical moments, the only difference being in the free-diffusion time constant, a difference that is negligible at low damping [5,6] (the most interesting damping range from an experimental point of view). However, only the Gilbert model, where the systematic and random terms in the stochastic equation, Eq. (2), viewed as the kinematic relation $\dot{\mathbf{u}} = \boldsymbol{\omega} \times \mathbf{u}$, are in the original Langevin form, i.e., with the rate of change of the angular momentum systematically slowed down superimposed on which is a rapidly fluctuating white noise random torque, can be used in all damping ranges. In contrast, neither the Kubo nor the Landau–Lifshitz models can be used for high damping, because under this condition they may predict unphysical behavior of the observables (relaxation times, escape rates, etc.).

Finally we remark that in the more general treatment of the isotropic Brownian motion of the classical spin, the memoryless assumption, i.e.,

$$\phi(t_1 - t_2) = \alpha / (v \gamma \mu_0 M_S) \delta(t_1 - t_2)$$

is discarded. Thus the random field $\mathbf{h}(t)$ has no longer white noise properties, namely,

$$\overline{h_i(t)} = 0, \quad \overline{h_i(t_1) h_j(t_2)} = 2kT \delta_{ij} \phi(t_1 - t_2), \tag{D26}$$

and the generalized stochastic magnetic Langevin equation becomes [198,199]

$$\dot{\mathbf{u}}(t) = \gamma \left[\mathbf{u}(t) \times \left(\mathbf{H}_{ef}(t) + \mathbf{h}(t)\right)\right] - \left[\mathbf{u}(t) \times \int_0^t \phi(t - t') \dot{\mathbf{u}}(t') dt'\right]. \tag{D27}$$

Here Eq. (D27) takes into account memory effects and the random field correlation function $\phi(t - t')$ has the meaning of a memory function.

## APPENDIX E: CHARACTERISTIC TIMES OF RELAXATION AND CORRELATION FUNCTIONS

We have seen by solving the differential-recurrence equations for the statistical moments, how we can evaluate the characteristic times of the relaxation and/or correlation functions $C_i(t)$ $(i = X, Y, Z)$ of the longitudinal and transverse components of spin operators (or vectors in the classical case). Now, to characterize the overall time behavior of $C_i(t)$, we may formally introduce (see Ref. 5) the integral relaxation time $\tau_{int}^i$, viz.,

$$\tau_{int}^i = \frac{1}{C_i(0)} \int_0^\infty C_i(t) dt, \tag{E1}$$

which is the area under the decay curve of $C_i(t)$. Yet another time constant characterizing the time behavior of $C_i(t)$ is the effective relaxation time $\tau_{ef}^i$ defined by

$$\tau_{ef}^i = -\frac{C_i(0)}{\dot{C}_i(0)} \tag{E2}$$

(yielding precise information on the initial decay of $C_i(t)$ in the time domain). For spin systems with dynamics governed by Fokker-Planck equations, the times $\tau_{int}^i$ and $\tau_{ef}^i$ may equivalently be defined using the eigenvalues ($\lambda_k^i$) of the Fokker–Planck operator from the evolution equation Eq. (2) because (Ref. 5, Chapter 2) the normalized relaxation function $C_i(t) / C_i(0)$ may formally be written as

$$\frac{C_i(t)}{C_i(0)} = \sum_k c_k^i e^{-\lambda_k^i t}, \tag{E3}$$

so that, from Eqs. (E1), (E2) and (E3), we have

$$\tau_{\text{int}}^i = \sum_k c_k^i / \lambda_k^i \tag{E4}$$

and

$$\tau_{\text{ef}}^i = \sum_k \lambda_k^i c_k^i. \tag{E5}$$

Now the relaxation times $\tau_{\text{int}}^i$ and $\tau_{\text{ef}}^i$ each contain contributions from *all* the eigenvalues $\lambda_k^i$. Therefore in general, in order to evaluate both $C_i(t)$, $\tau_{\text{int}}^i$, and $\tau_{\text{ef}}^i$ numerically, a knowledge of each individual $\lambda_k$ and $c_k^i$ is required. However, in the low temperature (high barrier) limit, for the *longitudinal* relaxation of the magnetization, $\lambda_1^Z \ll |\lambda_k^Z|$ and $c_1^Z \approx 1 \gg c_k^Z$ ($k \neq 1$) provided the wells of the potential remain equivalent or nearly equivalent, the approximation $\tau_{\text{int}}^Z \approx 1/\lambda_1^Z$ is valid. In other words, the inverse of the smallest nonvanishing eigenvalue $\lambda_1^Z$ closely approximates the longitudinal relaxation time $\tau_{\text{int}}^Z$ in the low temperature limit for zero or very weak external fields. Furthermore, in the *longitudinal* relaxation of the magnetization, the smallest nonvanishing eigenvalue(s) $\lambda_1^Z$ of the Fokker–Planck operator characterizes the long-time behavior of

$$\langle \hat{S}_Z \rangle(t) - \langle \hat{S}_Z \rangle_{eq} \sim C_\parallel(t) \sim e^{-\lambda_1^Z t} = e^{-t/\tau}, \ t \gg \tau. \tag{E6}$$

Thus it may be associated with the longest relaxation (reversal) time of the magnetization. In order to evaluate the reversal time $\tau$ numerically, we note that the recurrence equations for the statistical moments may always be written in matrix form as

$$\dot{\mathbf{X}}(t) = \mathbf{A}\mathbf{X}(t), \tag{E7}$$

where $\mathbf{A}$ is the system matrix and $\mathbf{X}(t)$ is an infinite column vector formed from the statistical moments. The $\tau$ may then be determined from the smallest nonvanishing root of the characteristic equation

$$\det(\lambda \mathbf{I} - \mathbf{A}) = 0 \tag{E8}$$

by selecting a sufficiently large number of equations. The general solution of Eq. (E7) is determined by successively increasing the size of $\mathbf{A}$ until convergence is attained.

The integral relaxation times $\tau_{\text{int}}^i$ can also be calculated via the one-sided Fourier transform of the appropriate correlation function $\tilde{C}_i(-i\omega) = \int_0^\infty C_i(t) e^{i\omega t} dt$ as

$$\tau_{\text{int}}^i = \tilde{C}_i(0)/C_i(0).$$

Here we may evaluate the reversal time $\tau$ via the one-sided Fourier transform of the longitudinal correlation function $\tilde{C}_Z(-i\omega) = \int_0^\infty C_Z(t) e^{i\omega t} dt$ as follows. We consider the long-time behavior of $C_Z(t)$ which is dominated by an exponential, viz.,

$$C_Z(t) \approx C_0 e^{-t/\tau}. \tag{E9}$$

Then the longest relaxation time $\tau$ can then be extracted from $\tilde{C}_Z(-i\omega)$ (by eliminating $C_0$) as [5]

$$\tau = \lim_{\omega \to 0} \frac{C_Z(0) - \tilde{C}_Z(-i\omega)}{i\omega \tilde{C}_Z(-i\omega)}. \tag{E10}$$

In practical applications, such as to magnetization reversal, matrix continued fractions due to their rapid convergence are much better suited to numerical calculations than standard direct matrix inversion based on the matrix representation, Eq. (E7), of the infinite system of linear differential-recurrence relations for the averaged spherical harmonics.

In the general case, the integral relaxation time can only be evaluated numerically. However, for systems with dynamics governed by single-variable Fokker–Planck equations it can be calculated analytically for both linear and nonlinear transient responses.

Here, we first derive following Ref. 5, an exact analytic equation for the *nonlinear transient response* relaxation time of a system governed by a one-dimensional Fokker–Planck equation for the probability distribution function $W(z,t)$ of a single variable $z$, viz.,

$$\frac{\partial W}{\partial t} = \mathrm{L}_{\mathrm{FP}} W, \tag{E11}$$

where the Fokker–Planck operator $\mathrm{L}_{\mathrm{FP}}$ may be represented as [71]

$$\mathrm{L}_{\mathrm{FP}} W = \frac{\partial}{\partial z}\left[ D_2(z)\left(\frac{\partial W}{\partial z} + D_1(z) W\right)\right]. \tag{E12}$$

Here $D_1(z)$ and $D_2(z)$ are the coordinate dependent coefficients and $z$ is defined in the range $-1 \leq z \leq 1$). Moreover we assume that the relaxation dynamics of spins obey the single-variable Fokker–Planck equation, Eq. (E11). Suppose that at time $t = 0$, the external field $H$ is suddenly altered from $H_I$ to $H_{II}$ (see Fig. 14). We are interested in the relaxation of the system starting from an equilibrium (stationary) state I with the distribution function $W_I(z)$, which evolves under the action of the stimulus of arbitrary strength to another equilibrium (stationary) state II with the distribution function $W_{II}(z)$. This problem is intrinsically nonlinear, because *changes in the magnitude of the potential are arbitrary*. Thus, the concept of relaxation functions and relaxation times must now be used, rather than correlation functions and correlation times.

Following [200] we may define the relaxation function $f_A(t)$ of a dynamical variable $A$ by

$$f_A(t) = \begin{cases} \langle A \rangle(t) - \langle A \rangle_{\mathrm{II}}, & (t > 0), \\ \langle A \rangle_{\mathrm{I}} - \langle A \rangle_{\mathrm{II}}, & (t \leq 0), \end{cases} \quad (E13)$$

where $\langle A \rangle(t)$ is the time-dependent average and $\langle A \rangle_{\mathrm{I}}$ and $\langle A \rangle_{\mathrm{II}}$ are equilibrium (stationary) averages defined as

$$\langle A \rangle(t) = \int_{-1}^{1} A(z) W(z,t) dz, \quad (E14)$$

$$\langle A \rangle_i = \int_{-1}^{1} A(z) W_i(z) dz \quad (i = \mathrm{I}, \mathrm{II}). \quad (E15)$$

Our goal is to evaluate the integral relaxation time $\tau_{\mathrm{int}}$ of the relaxation function $f_A(t)$, which is defined as

$$\tau_{\mathrm{int}} = \frac{1}{f_A(0)} \int_0^\infty f_A(t) dt$$
$$= \frac{1}{f_A(0)} \lim_{s \to 0} \int_0^\infty e^{-st} f_A(t) dt = \frac{\tilde{f}_A(0)}{f_A(0)}, \quad (E16)$$

where $\tilde{f}_A(s)$ is the Laplace transform of $f_A(t)$. The relaxation time, Eq. (E16), may be written as

$$\tau_{\mathrm{int}} = \frac{1}{\langle A \rangle_{\mathrm{I}} - \langle A \rangle_{\mathrm{II}}} \int_{-1}^{1} [A(z) - \langle A \rangle_{\mathrm{II}}] \tilde{W}(z,0) dz, \quad (E17)$$

where

$$\tilde{W}(z,0) = \lim_{s \to 0} \tilde{W}(z,s)$$

and

$$\tilde{W}(z,s) = \int_0^\infty W(z,t) e^{-st} dt.$$

Now $\tilde{W}(z,0)$ in Eq. (E11) can be calculated analytically via the final-value theorem of Laplace transformation [105], namely,

$$\lim_{s \to 0} s \tilde{W}(z,s) = \lim_{t \to \infty} W(z,t) = W_{\mathrm{II}}(z).$$

Thus, we obtain, from the Fokker–Planck equation, Eq. (E11), for $t > 0$, the ordinary differential equation

$$W_{\mathrm{II}}(z) - W_{\mathrm{I}}(z) = \frac{d}{dz}\left[ D_2(z) \left( \frac{d}{dz} \tilde{W}(z,0) + D_1(z) \tilde{W}(z,0) \right) \right]. \quad (E18)$$

The particular solution of Eq. (E18) is

$$\tilde{W}(z,0) = W_{\mathrm{II}}(z) \int_{-1}^{1} \frac{\Phi(y) dy}{D_2(y) W_{\mathrm{II}}(y)}, \quad (E19)$$

where

$$\Phi(y) = \int_{-1}^{y} [W_{\mathrm{II}}(z) - W_{\mathrm{I}}(z)] dz \quad (E20)$$

and $W_{\mathrm{II}}(z)$ is the stationary solution of the equation

$$\frac{d}{dz}\left[ D_2(z) \left( \frac{d}{dz} W_{\mathrm{II}}(z) + D_1(z) W_{\mathrm{II}}(z) \right) \right] = 0. \quad (E21)$$

Hence, using the definitions Eq. (E17) and Eq. (E19), we have

$$\tau_{\mathrm{int}} = \frac{1}{\langle A \rangle_{\mathrm{I}} - \langle A \rangle_{\mathrm{II}}} \int_{-1}^{1} [A(x) - \langle A \rangle_{\mathrm{II}}] W_{\mathrm{II}}(x) \int_{-1}^{x} \frac{\Phi(y) dy}{D_2(y) W_{\mathrm{II}}(y)} dx$$

or, by integration by parts, [200]

$$\tau_{\mathrm{int}} = \frac{1}{\langle A \rangle_{\mathrm{II}} - \langle A \rangle_{\mathrm{I}}} \int_{-1}^{1} \frac{\Phi(x) \Psi(x)}{D_2(x) W_{\mathrm{II}}(x)} dx, \quad (E22)$$

where for convenience we have written

$$\Psi(x) = \int_{x_1}^{x} [A(y) - \langle A \rangle_{\mathrm{II}}] W_{\mathrm{II}}(y) dy. \quad (E23)$$

Equation (E22) is an *exact* equation for the *nonlinear transient response relaxation time*. Examples of applications of Eq. (E22) to nonlinear response problems have been given in the present review and in Ref. 5.

If we now suppose that the change in the magnitude of the external field $H$ from $H_{\mathrm{I}}$ to $H_{\mathrm{II}}$ is very small, i.e., $|H_{\mathrm{II}} - H_{\mathrm{I}}| \to 0$, the problem becomes intrinsically linear, because *changes in the magnitude of the potential are insignificant*. Thus, linear response theory and the concept of correlation functions and correlation times rather than relaxation functions and relaxation times can now be used. The *equilibrium* (stationary) correlation function $C_{AB}(t)$ is defined by

$$C_{AB}(t) = \langle A[z(0)] B[z(t)] \rangle_0 - \langle A \rangle_0 \langle B \rangle_0$$
$$= \int_{-1}^{1} [A(z) - \langle A \rangle_0] e^{L_{\mathrm{FP}} t} [B(z) - \langle B \rangle_0] W_0(z) dz. \quad (E24)$$

Here $W_0$ is the stationary (equilibrium) distribution function satisfying $L_{\mathrm{FP}} W_0 = 0$, the symbols $\langle \rangle$ and $\langle \rangle_0$ designate the statistical averages over $W$ and $W_0$, respectively. Now, the salient feature of one-dimensional systems is that an *exact integral formula* for the correlation time $\tau_{\mathrm{cor}}$

$$\tau_{\mathrm{cor}} = \frac{1}{C_{AB}(0)} \int_0^\infty C_{AB}(t) dt \quad (E25)$$

[defined as the area under the curve of the normalized correlation function $C_{AB}(t)$ exists because the relevant Fokker-Planck equation (E11) may be integrated by quadratures as with the nonlinear responses [5,71]. The details of the derivation are given in Refs. 5 and 71 so we merely quote the final analytic expression for $\tau_{\text{cor}}$, viz.,

$$\tau_{\text{cor}} = \frac{1}{C_{AB}(0)} \int_{-1}^{1} \frac{\psi_A(x)\psi_B(x)dx}{D_2(x)W_0(x)}, \quad \text{(E26)}$$

where

$$\psi_A(x) = \int_{-1}^{x} [A(y) - \langle A \rangle_0] W_0(y) dy, \quad \text{(E27)}$$

$$\psi_B(x) = \int_{-1}^{x} [B(y) - \langle B \rangle_0] W_0(y) dy. \quad \text{(E28)}$$

For $A = B$, Eq. (E26) reduces to

$$\tau_{\text{cor}} = \frac{1}{\langle A^2 \rangle_0 - \langle A \rangle_0^2} \int_{-1}^{1} \frac{\psi_A^2(x)dx}{D_2(x)W_0(x)}. \quad \text{(E29)}$$

The relaxation time in integral form, Eq. (E29), was first given by Szabo [201], and later reproduced by other authors under different guises [5,151,153,200,202-203].